\documentclass[aps,prb,showpacs,twocolumn,superscriptaddress]{revtex4-2}
\usepackage{bm,color,amsmath,amssymb,mathrsfs,latexsym,graphicx,psfrag,mathtools}
\usepackage{color}
\usepackage[dvipsnames]{xcolor}
\usepackage[hidelinks,colorlinks,linkcolor=blue,
citecolor=blue,urlcolor=blue]{hyperref}
\usepackage{dsfont}
\usepackage{booktabs}
\usepackage{relsize}
\usepackage{multirow}
\usepackage{empheq}

\newcommand{\bsl}[1]{\boldsymbol{#1}}

\def\normOrd#1{\mathop{:}\nolimits\!#1\!\mathop{:}\nolimits}

\newcommand{\bra}[1]{\langle #1|}
\newcommand{\ket}[1]{|#1 \rangle}

\DeclareRobustCommand{\Eq}[1]{Eq.~(\ref{#1})}
\newcommand{\ii}{\mathrm{i}}

\newcommand{\Tr}{\mathop{\mathrm{Tr}}}

\newcommand{\eqnref}[1]{Eq.\,\eqref{#1}}
\newcommand{\figref}[1]{Fig.\,\ref{#1}}
\newcommand{\tabref}[1]{Tab.\,\ref{#1}}
\newcommand{\secref}[1]{Sec.\,\ref{#1}}
\newcommand{\appref}[1]{Appendix.\,\ref{#1}}
\newcommand{\refcite}[1]{Ref.\,\cite{#1}}

\newcommand{\mat}[1]{\left(\begin{matrix}#1\end{matrix}\right)}

\newcommand{\eq}[1]{\begin{equation} #1 \end{equation}}

\newcommand{\eqa}[1]{\begin{align}\begin{split} #1 \end{split}\end{align}}

\let\oldAA\AA
\renewcommand{\AA}{\text{\normalfont\oldAA}}

\newcommand{\TR}{\mathcal{T}}

\newcommand{\V}{\mathcal{V}}

\newcommand{\K}{\text{K}}

\newcommand{\bea}{\begin{equation} \begin{aligned}}
\newcommand{\eea}{\end{aligned} \end{equation} }

\newcommand{\al}{\alpha}
\newcommand{\be}{\beta}
\newcommand{\bpm}{\begin{pmatrix}}
\newcommand{\epm}{\end{pmatrix}}

\renewcommand{\th}{\theta}

\newcommand{\lp}{\left(}
\newcommand{\rp}{\right)}

\newcommand{\mbf}[1]{\mathbf{#1}}

\usepackage{braket}
\usepackage{amsmath}
\usepackage{amssymb}
\usepackage{wasysym}
\usepackage{mathrsfs}

\usepackage{comment}

\usepackage{commath}    
\usepackage{amsmath}

\begin{document}

\title{Moir\'e Fractional Chern Insulators III: Hartree-Fock Phase Diagram, Magic Angle Regime for  Chern Insulator States, the Role of the Moir\'e Potential and Goldstone Gaps in Rhombohedral Graphene Superlattices}

\author{Yves H. Kwan}
\affiliation{Princeton Center for Theoretical Science, Princeton University, Princeton, NJ 08544}
\author{Jiabin Yu}
\affiliation{Department of Physics, Princeton University, Princeton, New Jersey 08544, USA}
\author{Jonah Herzog-Arbeitman}
\affiliation{Department of Physics, Princeton University, Princeton, New Jersey 08544, USA}
\author{Dmitri~K.~Efetov}
\affiliation{Faculty of Physics, Ludwig-Maximilians-University Munich, Munich 80799, Germany}
\affiliation{Munich Center for Quantum Science and Technology (MCQST), Ludwig-Maximilians-University Munich, Munich 80799, Germany}
\author{Nicolas Regnault}
\affiliation{Department of Physics, Princeton University, Princeton, New Jersey 08544, USA}
\affiliation{Laboratoire de Physique de l’Ecole normale sup\'erieure,
ENS, Universit\'e PSL, CNRS, Sorbonne Universit\'e,
Universit\'e Paris-Diderot, Sorbonne Paris Cit\'e, 75005 Paris, France}
\author{B. Andrei Bernevig}
\email{bernevig@princeton.edu}
\affiliation{Department of Physics, Princeton University, Princeton, New Jersey 08544, USA}
\affiliation{Donostia International Physics Center, P. Manuel de Lardizabal 4, 20018 Donostia-San Sebastian, Spain}
\affiliation{IKERBASQUE, Basque Foundation for Science, Bilbao, Spain}

\begin{abstract}

We investigate in detail the $\nu=+1$ displacement-field-tuned interacting phase diagram of $L=3,4,5,6,7$ layer rhombohedral graphene aligned to hBN (R$L$G/hBN). Our calculations account for the 3D nature of the Coulomb interaction, the inequivalent stacking orientations $\xi=0,1$, the effects of the filled valence bands, and the choice of `interaction scheme' for specifying the many-body Hamiltonian. We show that the latter has a dramatic impact on the Hartree-Fock phase boundaries and the properties of the phases, including for pentalayers (R5G/hBN) with large displacement field $D$ where recent experiments observed a Chern insulator at $\nu=+1$ and fractional Chern insulators for $\nu<1$. In this large $D$ regime, the low-energy conduction bands are polarized away from the aligned hBN layer, and 
are hence well-described by the folded bands of moir\'eless rhombohedral graphene at the non-interacting level.
Despite this, the filled valence bands develop moir\'e-periodic charge density variations which can generate an effective moir\'e potential, thereby explicitly breaking the approximate continuous translation symmetry in the conduction bands, and leading to contrasting electronic topology in the ground state for the two stacking arrangements. Within time-dependent Hartree-Fock theory, we further characterize the strength of the moir\'e pinning potential in the Chern insulator phase by computing the low-energy $\mathbf{q}=0$ collective mode spectrum, where we identify competing gapped pseudophonon and valley magnon excitations. Our results emphasize the importance of careful examination of both the single-particle and interaction model for a proper understanding of the correlated phases in R$L$G/hBN.

\end{abstract}
\maketitle

\section{Introduction}

After more than a decade since their initial theoretical proposals~\cite{neupert,sheng,regnault}, fractional Chern insulators (FCIs) have been recently observed in twisted bilayer MoTe$_2$ ($t$MoTe$_2$)~\cite{park2023observation,zeng2023integer,Xu2023FCItMoTe2,cai2023signatures} and rhombohedral pentalayer graphene twisted on hexagonal boron nitride~\cite{Ju2023PentalayerGraphenehBN}, following related observations of fractional quantum Hall like states at nonzero external magnetic fields~\cite{2018Sci...360...62S,2021Natur.600..439X}.
The experimental realizations of FCIs have attracted a large amount of theoretical attention~\cite{reddy2023fractional,wang2023fractional,Dong2023CFLtMoTe2,Goldman2023CFLtMoTe2,Reddy2023GlobalPDFCI,Xu2023MLWOFCItTMD,Zaletel2023tMoTe2FCI,Yu2023FCI,Fu2023BandMixingFCItMoTe2,Fengcheng2023tMoTe2HFnum1,firstMFCI,Zhang2023MoTe2,Xiao2023tMoTe2abinitio,dong2023theory,zhou2023fractional,dong2023anomalous,guo2023theory}. In both $t$MoTe$_2$ and pentalayers, the $|\nu| =1/3, 4/3$ FCI state is  so far absent, whereas the straightforward  diagonalization of projected Hamiltonians into a single Chern band is expected to produce (and indeed found to give)  a robust 1/3 state. In $t$MoTe$_2$ this was explained \cite{Yu2023FCI, reddy2023fractional} due to the weak/nonexistent spin gaps at those fillings. In pentalayers however, the presence of many single-particle dispersive bands almost degenerate with the flat band poses more immediate initial questions, related to the appearance of the $\nu=1$ Chern insulator. The phase diagram of $L$-layer rhombohedral graphene twisted on hBN (R$L$G/hBN), the regions of nontrivial Chern number and the gapless states,  the role of the moir\'e interaction and of the 3D Coulomb interaction, the stability of the Chern phase, and the collective modes of the system are all outstanding questions that are yet to be fully addressed.

In Ref.~\cite{MFCI-II}, we have performed ab initio calculations on the relaxation and the single-particle band structure of R$L$G/hBN, and provided a model that includes the effects of relaxation, trigonal distortion, moir\'e potential, and internal and external electric polarizations/fields. We have then developed a continuum model~\cite{MFCI-II} where the parameters were directly fitted to {\it ab initio} calculations which include relaxation. This is a generalized version of the models introduced in Refs.~\cite{MacDonald2014R5GhBN,Kushino2014R5GhBN,Senthil2019R3G,Park2023RMGhBNChernFlatBands}. 

In this work, we perform Hartree-Fock (HF) and time-dependent Hartree-Fock (TDHF) calculations at integer filling $\nu=+1$ of R$L$G/hBN based on the model in \refcite{MFCI-II} for $L=3-7$ layers. Crucially, we consider two different `interaction schemes' --- the charge neutrality (CN) scheme and the average scheme. These choices differ in the way that the filled valence bands influence the low-energy conduction bands by generating interaction-induced one-body potentials. In particular in the average scheme, the conduction electrons are sensitive to the moir\'e charge density of the valence bands, resulting in contrasting phase diagrams for the two inequivalent stacking orientations $\xi=0,1$. 
On the other hand, the CN scheme does not distinguish between the two stackings for large interlayer potential $V>0$. This is because the conduction bands, which are polarized away from the aligned hBN layer, couple neither to the hBN potential, nor to the electrostatic moir\'e charge background set up by the filled valence bands. The conflicting predictions in these two interaction schemes, combined with experimental characterizations of the stacking arrangement, can be used to narrow down the interaction scheme that most faithfully captures the physics of R$L$G/hBN. Because we include valence bands in our projected calculations and account for the layer-dependence of the electron interactions, which allows for internal screening of the external field, we are able to recover the experimental phenomenology of the $\nu=1$ state in R5G/hBN with $\xi=1$ stacking for a wide range of displacement fields $D$ using the average interaction scheme. In contrast, we do not find a sizable gapless region near $V=0$ in the CN interaction scheme. We also show how the position and topology of the gapped phases in the phase diagram change as the twist angle $\theta$ and number of layers $L$ is varied. Our results suggest the existence of a `magic-angle' regime for realizing correlated topological phases --- there is a tendency for topologically trivial insulators for small $\theta$, while the large displacement fields required to obtain the Chern insulator at large $\theta$ seem challenging to access experimentally.

Going beyond HF, we consider the low-lying collective modes of the various correlated insulating states, with the objective of understanding quantitatively how the moir\'e potential affects the Chern insulator at $V>0$. Despite the fact that the conduction band electrons are polarized away from the hBN and barely affected by the hBN coupling directly, we find that in the average interaction scheme, the moir\'e potential generated by the occupied valence bands is able to induce sizable gaps in the putative pseudophonons corresponding to the continuous translation symmetry present in the moir\'eless limit. Furthermore, in the Chern insulator phase, these pseudophonons are undercut in energy by a low-lying intervalley mode with energy $\simeq 2-3\,\text{meV}$ that is substantially smaller than the HF charge gap, and provides a more realistic scale for the stability of the $|C|=1$ state. On the other hand, the pseudophonon gap in the CN scheme remains relatively small, but still at meV level. Our findings clarify the phase diagram of R$L$G/hBN, demonstrate the interaction-induced injection of extrinsic moir\'e effects onto the conduction bands, and emphasize the importance of careful microscopic modelling in understanding the correlated fractional and integer Chern insulator states in rhombohedral graphene superlattices.

\section{Continuum models}

\begin{figure}[t]
    \centering
\includegraphics[width=1.0\linewidth]{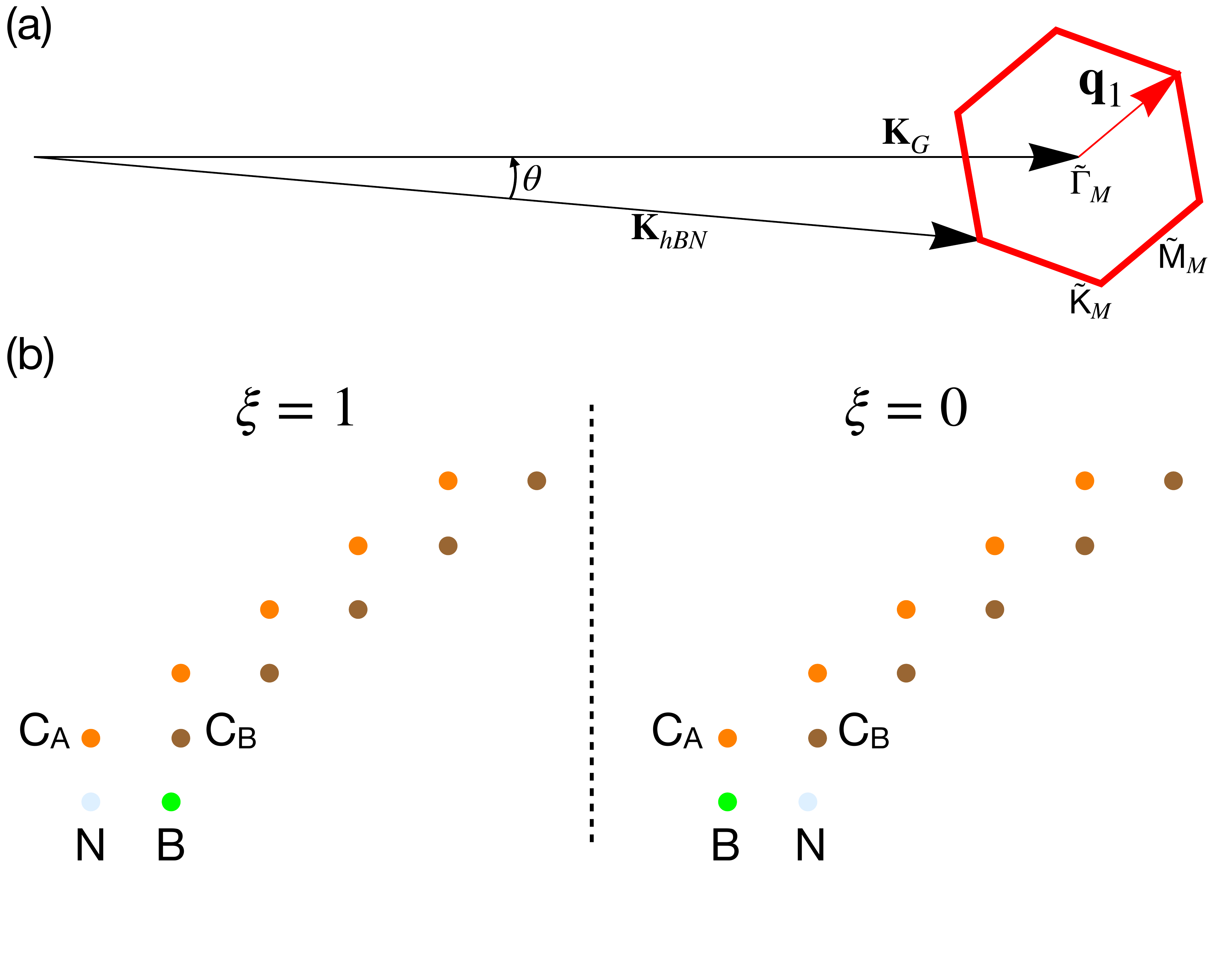}
    \caption{(a) The moir\'e Brillouin zone (mBZ). $\K_G$ and $\K_{hBN}$ are the K point of rhombohedral graphene and hBN, respectively. The red hexagon is the mBZ, the red arrow is $\mbf{q}_1 = \K_G - \K_{hBN}$, and $\theta$ is the twist angle.
    (b) Schematics of the two stacking configurations $\xi$ related by rotating only hBN by 180$^\circ$.
    C$_\text{A}$, C$_\text{B}$, B, N refer to the carbon atom at A sublattice, the carbon atom at B sublattice, boron atom in hBN, and nitrogen atom in hBN, respectively. 
    }
\label{fig:moireconventions}
\end{figure}

\subsection{Single-particle model}\label{sec:sp}

\begin{figure*}[t]
    \centering
    \includegraphics[width=\linewidth]{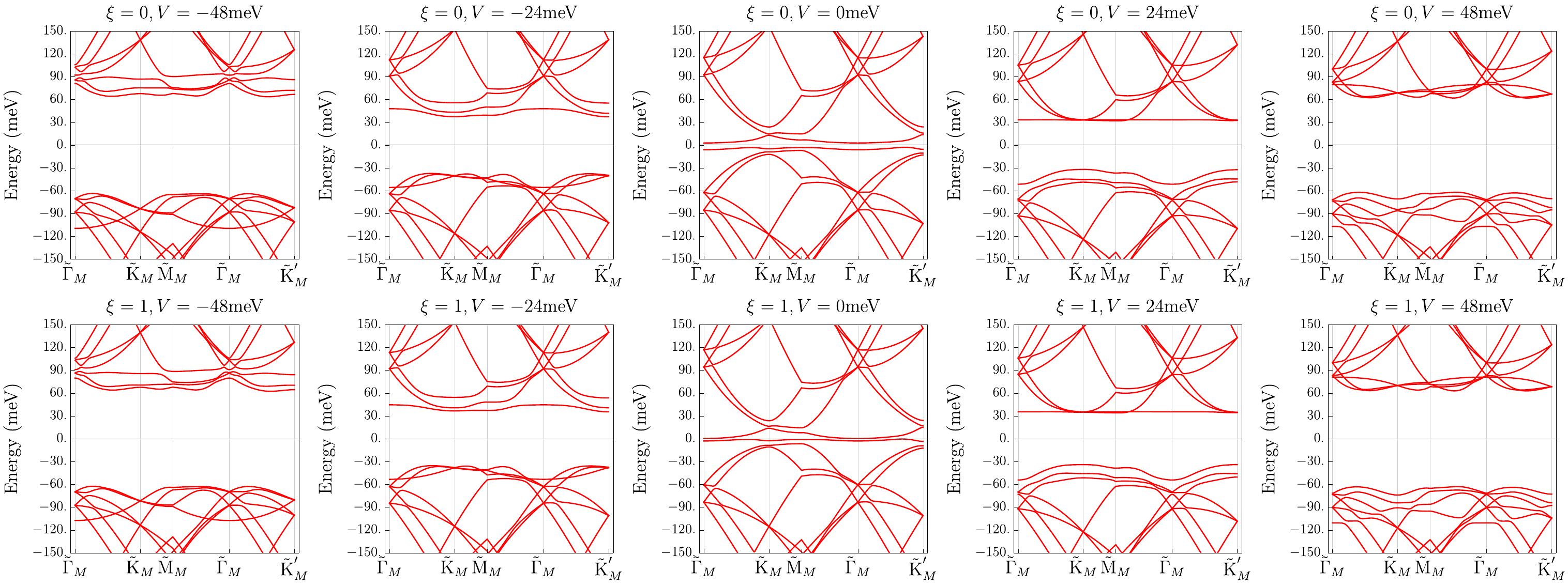}
    \caption{
    The band structures of the R5G/hBN single-particle model in $\K$ valley (Eq.~\ref{eq_main:H_K_nohBN}) and $\theta=0.77^\circ$. Top (bottom) row shows $\xi=0$ ($\xi=1$) stacking. See Tab.~\ref{tab:parameters_full} in App.~\ref{app:SP_model} for a list of parameters in the continuum model.
    }
    \label{fig:SP_bands}
\end{figure*}

We use the single-particle model discussed in \refcite{MFCI-II}. 
For self-consistency, we will first briefly review this model here (more details can be found in App.~\ref{app:SP_model} and \refcite{MFCI-II}). We then discuss the charge fluctuations of the filled valence bands.
The single-particle model consists of three parts: the pristine rhombohedral $L$-layer graphene (R$L$G), the moir\'e potential induced by hBN, and the displacement field.
As discussed in \refcite{MFCI-II}, we only consider one aligned hBN layer to simulate the single alignment in the experiment~\cite{Ju2023PentalayerGraphenehBN}.
The continuum model for pristine R$L$G has basis $ c^\dagger_{\bsl{r}, l\sigma \eta s} $, where $\bsl{r}=(x,y)$ is the continuum 2D position, $l=0,1,2,...,L-1$ is the layer index for $L$ total layers, $\sigma=A,B$ represents the sublattice, $\eta=\pm \K$ labels the valley, and $s=\uparrow,\downarrow$ is the spin index. 
Due to spin $SU(2)_S$ symmetry, in the discussion of the single-particle model, we neglect the spin index unless specified otherwise.

In the $\K$ valley, the matrix Hamiltonian for R$L$G reads
\eqa{ 
\label{eq_main:H_K}
H_{\K}(\bsl{p}) &= \bpm
v_F\mbf{p} \cdot \pmb{\sigma}  & t^\dag(\mbf{p}) & t'^\dagger &   &\\
t(\mbf{p}) & \ddots & \ddots & t'^\dagger \\
t' & \ddots & v_F\mbf{p} \cdot \pmb{\sigma} & t^\dagger(\mbf{p})\\
& t' & t(\mbf{p})  & v_F\mbf{p} \cdot \pmb{\sigma}
\epm + H_{ISP}, 
}
where $\bsl{p}=-\ii \nabla$, $\bsl{\sigma}=(\sigma_x,\sigma_y)$ are Pauli matrices in sublattice subspace, $t(\mathbf{p})$ and $t'$ are $2\times 2$ matrices that carry the sublattice index:
\eq{
t(\mbf{p}) = -\bpm v_4 p_+ & -t_1 \\ v_3 p_- &  v_4 p_+ \epm, \qquad  \qquad t' = \bpm 0 & 0 \\ t_2 & 0 \epm\ ,
}
$p_\pm = p_x \pm \ii p_y$,  $v_F$ is the Fermi velocity, $t_1,t_2,v_3,v_4$ are inter-layer hopping parameters (we set $v_3=v_4$ throughout this work), and the $2\times 2$ blocks in $H_{\K}(\bsl{p})$ are arranged according to the layer index.
$H_{ISP}$ is the inversion symmetric polarization, and characterizes the local chemical potential environment of each graphene layer:
\eq{
[H_{ISP}]_{l l'} = V_{ISP} \delta_{ll'} \left| l - \frac{L-1}{2} \right| \sigma_0\ ,
}
where $V_{ISP} = 16.65$meV is determined by fitting to the DFT calculated bands in \refcite{MFCI-II}, and $\sigma_0$ is the identity $2\times 2$ matrix for the sublattice index.

The hBN-induced moir\'e potential has the form
\eqa{
\label{eq_main:Vxifinal}
V_\xi(\mbf{r}) &= V_0 + \left[V_1 e^{i\psi_\xi}\sum_{j=1}^3 e^{i \mbf{g}_j\cdot\mbf{r}}\bpm 1& \omega^{-j} \\ \omega^{j+1} &\omega \epm + h.c.\right]\ ,
}
which only acts on the bottom layer of R$L$G.
Here $\mbf{g}_j = R(\frac{2\pi}{3}(j-1)) (\mbf{q}_2-\mbf{q}_3)$, with $R(\phi)$ the counterclockwise rotation matrix by $\phi$.
The $\mbf{q}$ vectors are defined as
\eq{
\label{eq_main:qvecmain}
\mbf{q}_1 = \mbf{K}_G - \mbf{K}_{hBN} = \frac{4\pi}{3 a_G}\left(1 - \frac{R(-\th)}{1+0.01673} \right)\hat{x}
}
with its $C_3$ partners $\mbf{q}_{j+1} = R(\frac{2\pi}{3}) \mbf{q}_j$, where $\th$ is the twist angle, $\mbf{K}_G$ and $\mbf{K}_{hBN}$ are the K vector of graphene and hBN, $a_G = 2.46\AA$ is the graphene lattice constant, and $(1+0.01673)a_G$ is the hBN lattice constant.
The moir\'e Brillouin zone (mBZ) is shown in \figref{fig:moireconventions}(a).
$\xi=0,1$ labels the two stacking configurations related by a 180$^\circ$ rotation of hBN, as shown in Fig.~\ref{fig:moireconventions}(b).
We only keep the first harmonics in the effective moir\'e potential $V_\xi(\mbf{r})$, as discussed in App.~\ref{app:SP_model}. 
Combined with the external-applied interlayer potential $V$, the single-particle Hamiltonian in the $\K$ valley reads
\eq{
\label{eq_main:H_K_nohBN}
H_{\K, \xi}(\mbf{r}) = H_{\K}(- i \pmb{\nabla}) + H_{\text{moir\'e},\xi}(\bsl{r})+ H_{D} \ ,
}
where
\eq{
\label{eq_main:H_V}
\null [H_{\text{moir\'e},\xi}(\bsl{r})]_{l \sigma,l' \sigma'} = \left[ V_\xi(\mbf{r}) \right]_{\sigma\sigma} \delta_{l0}\delta_{ll'}\ ,
}
\eq{
\label{eq_main:H_D}
\null [H_{D}]_{l \sigma,l' \sigma'}=V_l\delta_{ll'}\delta_{\sigma\sigma'} = V \left(l - \frac{n-1}{2} \right) \delta_{ll'} \delta_{\sigma \sigma'}.
}
The relation between $V$ and the displacement field $D$ reads $V= e D  d/\epsilon_\perp$, where $e$ is the charge of electron, $d\approx 3.33\AA$ is the interlayer distance, and $\epsilon_\perp$ is the perpendicular dielectric constant.
The parameter values of the model determined by fitting to the ab initio calculations in \refcite{MFCI-II} are listed in \tabref{tab:parameters_full} of \appref{app:SP_model}.

For later convenience, we introduce an artificial tuning knob $\kappa_\text{hBN}$ which modifies the moir\'e potential as $V_\xi(\mbf{r})\rightarrow \kappa_\text{hBN}V_\xi(\mbf{r})$. $\kappa_{\text{hBN}}=1$ is the physical limit (which is assumed in the following unless otherwise stated), while in the $\kappa_\text{hBN}=0$ limit the model possesses continuous translation symmetry.

\begin{figure}[t]
    \centering
\includegraphics[width=1.0\linewidth]{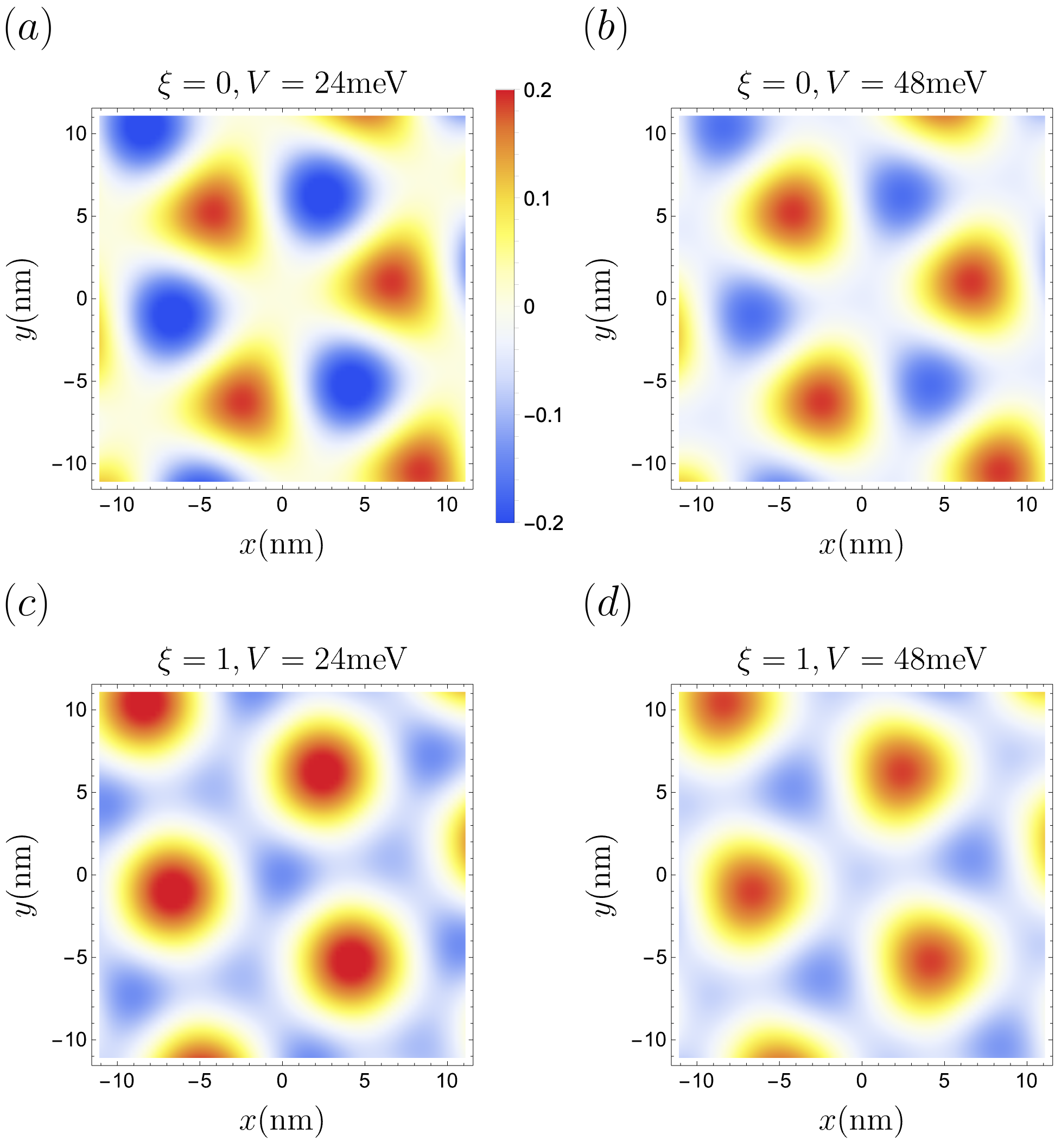}
    \caption{
    The dimensionless density fluctuation $\Delta\rho(\bsl{r})$ (\eqnref{eq_main:density_profile}, color bar) of all valence bands (up to the momentum cutoff) of the continuum model (\eqnref{eq_main:H_K_nohBN}) in the $\K$ valley for (a) $\xi=0, V=24$meV, (b) $\xi=0,V=48$meV, (c) $\xi=1,V=24$meV and (d) $\xi=1,V=48$meV.
    We choose $L=5$ layers, $\theta=0.77^\circ$, and 4 shells of reciprocal lattice vectors (in total 19 reciprocal lattice vectors, which give $95$ valence bands per valley per spin).
    }
\label{fig_main:chargedensity}
\end{figure}

\subsection{Charge density background}

Filling the valence bands can generate a charge density background, as shown by the plot of the dimensionless density fluctuation in \figref{fig_main:chargedensity}.
Here the dimensionless density fluctuation is defined as 
\eq{
\label{eq_main:density_profile}
\Delta\rho(\bsl{r}) = \rho(\bsl{r}) - \langle \rho(\bsl{r}) \rangle \ ,
}
where
\eq{
\rho(\bsl{r}) = \frac{1}{N}\sum_{n,\bsl{k},l,\sigma} \left| \sum_{\bsl{G}} e^{\ii (\bsl{k}+\bsl{G}) \cdot\bsl{r}} U^{\K}_{n,\bsl{G}l\sigma}(\bsl{k})\right|^2 
}
is the product of the real-space particle number density (per spin at $\K$ valley) and the area of the moir\'e unit cell $\Omega$, $U^{\K}_{n}(\bsl{k})$ is the eigenvector for the $n$th band of the single-particle Hamiltonian in the $\K$ valley, $\bsl{G}$ labels the moir\'e lattice vector, the summation of $n$ is over all the valence bands in the model, and
\eq{
\langle \rho(\bsl{r}) \rangle = \frac{1}{\Omega} \int d^2 r \rho(\bsl{r}),
}
where the real-space integral is taken over the moir\'e unit cell.
The charge density carries information of the moir\'e potential, and will change the effective one-body dispersion via the Hartree and Fock potentials as discussed in~\secref{subsec:projection}.
While a phenomenological treatment of the charge density was included in \refcite{LongJu2023FCIPentalayerGraphenehBN}, we will instead compute the charge density microscopically in this work.
As elaborated in App.~\ref{app:charge_density_band_cutoff}, the charge fluctuation cannot be accurately captured by only accounting for a subset of the valence bands obtained by diagonalizing the continuum model, due to the moir\'e coupling within the moir\'eless valence-band subspace. 
Instead, we must include the contribution of the charge density from all valence bands within the plane wave cutoff used to diagonalize the continuum model.
The charge fluctuations do not depend on the number of reciprocal lattice vectors (RLVs) within the plane wave cutoff (when it is large enough, e.g.~no smaller than 19 RLVs), as shown in App.~\ref{app:charge_density_band_cutoff}.

\subsection{Normal-ordered interaction}\label{subsec:interactions}

We incorporate interactions into the model by adding a long-range density-density interaction term to the single-particle continuum Hamiltonian $\hat{H}_{\text{s.p.}}(V)=\sum_{\eta}\hat{H}_{\eta,\xi}(V)$, which was described in Sec.~\ref{sec:sp}. We define the interaction Hamiltonian $\hat{H}_\text{int}$ to consist of any part of the many-body Hamiltonian $\hat{H}$ that depends on the interaction potential, including any effective one-body terms. As discussed later in Sec.~\ref{subsec:schemes} and App.~\ref{secapp:int_HF}, there is no unique prescription for $\hat{H}_\text{int}$, and there are several possible interaction `schemes' for specifying $\hat{H}_\text{int}$. In this subsection, we consider the following choice
\begin{gather}\label{eq:Hstart}
    \hat{H}=\hat{H}_{\text{s.p.}}(V)+\hat{H}_\text{int}\\
    \hat{H}_\text{int}=\frac{1}{2 N} \sum_{\mbf{q}\mbf{G}ll'} \frac{V_{ll'}(\mbf{q}+\mbf{G})}{\Omega} \delta \hat{\rho}_{\mbf{q}+\mbf{G},l} \delta \hat{\rho}_{-\mbf{q}-\mbf{G},l'},\label{eq:Hint_rho1/2}
\end{gather}
which will correspond to the `average' interaction scheme as explained in Sec.~\ref{subsec:schemes}. $N$ is the number of moir\'e unit cells of area $\Omega$, $\delta\hat{\rho}_{\mbf{q},l}$ is the relative density operator to be defined shortly, and we have allowed for layer dependence in the gate-screened interaction potential~\cite{kolar2023electrostatic} (see App.~\ref{secapp:layer_int} for a derivation)
\begin{widetext}
\begin{align}\label{eq:Vint_layer}
    V_{ll'}(\mbf{q})=\frac{e^2}{2\epsilon_0\epsilon_r q}\bigg[\frac{e^{-q(z_l+z_{l'})}\left(-e^{2q(d_\text{sc}+z_l+z_{l'})}-e^{2qd_\text{sc}}+e^{2qz_l}+e^{2qz_{l'}}\right)}{e^{4qd_\text{sc}}-1}
    +e^{-q|z_l-z_{l'}|}\bigg]
\end{align}
\end{widetext}
where $z_l=\left(l-\frac{L-1}{2}\right)d$ is the vertical coordinate of layer $l=0,\ldots,L-1$ with $d=3.33\,$\r{A} the interlayer distance between graphene layers, and the two metallic gates are positioned at $z=\pm d_{\text{sc}}$. If all layers are at $z_l=0$, this reduces to the 2d interaction $V(\mbf{q})=\frac{e^2}{2\epsilon_0 \epsilon_r}\tanh qd_\text{sc}$. Unless otherwise stated, we choose $d_\text{sc}=10\,\text{nm}$. We will consider different values of the relative permittivity $\epsilon_r$.  Since the low-energy valence and conduction bands tend to be layer-polarized on opposite sides of the R$n$G system, accounting for the layer dependence of $V_{ll'}(\mbf{q})$ is crucial for a quantitative treatment of their interactions. For instance, we have the interlayer suppression $V_{0,4}(\mbf{g}_1)/V_{2,2}(\mbf{g}_1)=43\%$ at finite wavevector $\mbf{q}=\mbf{g}_1$ for R5G/hBN at $\theta=0.77^\circ$. At $\mbf{q}=0$, the interaction between the total charges on each layer leads to 
\begin{equation}
    V_{ll'}(\mbf{q}=0)=-\frac{e^2|z_l-z_{l'}|}{2\epsilon_0\epsilon_r}.
\end{equation} 
For simplicity, we assume an isotropic $\epsilon_r$, though in principle the perpendicular and in-plane components of the dielectric tensor could differ. The layer dependence in $V_{ll'}(\mbf{q}=0)$ leads to internal electrostatic screening of the external interlayer potential $V$, which will be explained in Sec.~\ref{subsec:internal_screening} and App.~\ref{secapp:screening}.

To respect the approximate particle-hole symmetry of R$L$G, we measure the density relative to a uniform background at neutrality
\begin{equation}
    \delta \hat{\rho}_{\mbf{q}+\mbf{G},l} = \sum_{\eta s\mbf{G}'\sigma} \sum_{\mbf{k} } (c^\dag_{\mbf{k}+\mbf{q},\mbf{G}+\mbf{G}', l\sigma\eta s} c_{\mbf{k},\mbf{G}', l\sigma\eta s} - \frac{1}{2} \delta_{\mbf{q},\mbf{0}} \delta_{\mbf{G},\mbf{0}} ),
\end{equation}
where $\sigma=A,B$ indexes the graphene sublattice. 

We now perform a unitary transformation 
\begin{equation}\label{eq:basis_choice}
    c^\dag_{\mbf{k},n,\eta,s} = \sum_{\mbf{G}l\sigma} c^\dag_{\mbf{k},\mbf{G},l\sigma\eta s} U^\eta_{\mbf{G}l\sigma,n}(\mbf{k})
\end{equation}
to a moir\'e band basis with band label $n$. This new basis, whose single-particle states are indexed by $(\bm{k},n,\eta,s)$, is taken to be general, and not necessarily the eigenbasis of $\hat{H}_{\text{s.p.}}(V)$. [For instance in Sec.~\ref{subsec:internal_screening}, we will use the eigenbasis of $\hat{H}_{\text{s.p.}}(U)$ evaluated using a screened interlayer potential $U$.] We work in periodic gauge $U_{\mbf{G}-\mbf{G}',l\sigma,n}(\mbf{k}+\mbf{G}')=U_{\mbf{G},l\sigma,n}(\mbf{k})$. We define the form factors describing the overlap of Bloch functions~\cite{TBG3,2021PhRvB.103t5414L,PhysRevB.103.205415}
\begin{equation}
    M_{mn}^{l\eta}(\mbf{k},\mbf{q}+\mbf{G}) = \sum_{\mbf{G}'\sigma}U^{\eta *}_{\mbf{G}+\mbf{G}',l\sigma,m}(\mbf{k}+\mbf{q}) U^\eta_{\mbf{G}',l\sigma,n}(\mbf{k})
\end{equation}
which satisfies $M_{mn}^{l\eta}(\mbf{k},\mbf{q}+\mbf{G})=M_{nm}^{l\eta*}(\mbf{k}+\mbf{q},-\mbf{q}-\mbf{G})$. Using $\sum_{ln} M^{l\eta}_{nn}(\mbf{k},\mbf{G}) \delta_{mn} = \sum_{\mbf{G}'l\sigma} \delta_{\mbf{G},0}$, the density operator can be rewritten as
\begin{align}
\begin{split}
    \delta \hat{\rho}_{\mbf{q}+\mbf{G},l} =  &\sum_{\mbf{k}mn, \eta s} M^{l\eta}_{mn}(\mbf{k},\mbf{q}+\mbf{G})\\
    &\times(c^\dag_{\mbf{k}+\mbf{q},m, \eta s} c_{\mbf{k},n,\eta s} - \frac{1}{2} \delta_{\mbf{q},0} \delta_{mn}).
\end{split}
\end{align}

The next step in defining the interaction is to shuffle the operators in $\hat{H}_\text{int}$ so that the four-fermion term appears in normal-ordered form $\normOrd{\hat{H}_\text{int}}$
\begin{equation}\label{eq:Hint_decomp}
    \hat{H}_\text{int}=\normOrd{\hat{H}_\text{int}}-\hat{H}_{\text{HF,int}}[P^\text{ref,avg.}]
\end{equation}
\begin{align}\begin{split}\label{eq:Hint_nord}
    \normOrd{\hat{H}_\text{int}}=&\frac{1}{2N}\sum_{\mbf{q},\mbf{G}} \sum_{\mbf{k} mnl \eta s,\mbf{k}'l'm'n' \eta' s'} \frac{V_{ll'}(\mbf{q}+\mbf{G})}{\Omega} \\
    &\times M^{l\eta}_{mn}(\mbf{k},\mbf{q}+\mbf{G}) M^{l'\eta'*}_{n'm'}(\mbf{k}'-\mbf{q},\mbf{q}+\mbf{G})\\
    &\times
c^\dag_{\mbf{k}+\mbf{q},m, \eta s}  c^\dag_{\mbf{k}'-\mbf{q},m', \eta' s'} c_{\mbf{k}',n',\eta' s'} c_{\mbf{k},n,\eta s}
\end{split}
\end{align}
where 
\begin{equation}\label{eq:HFint}
\hat{H}_{\text{HF,int}}[P]=\hat{H}_\text{H,int}[P]+\hat{H}_\text{F,int}[P] 
\end{equation}
is the HF functional describing the mean-field decoupling of $\normOrd{\hat{H}_\text{int}}$ using the one-body density matrix 
\begin{equation}
    P_{m\eta,n\eta'}(\mbf{k},s)=\langle c^\dagger_{\mbf{k},m,\eta s}c_{\mbf{k},n,\eta' s} \rangle,
\end{equation}
and
\begin{align}\begin{split}\label{eq:HF_H}
\hat{H}_\text{H,int}[P]=&\frac{1}{N}\sum_{\mbf{k}l l' m n \eta s}  \sum_{\mbf{G}} \frac{V_{ll'}(\mbf{G})}{\Omega}M^{l\eta}_{mn}(\mbf{k},\mbf{G}) \\
&\times  \lp\sum_{\mbf{k}'m'n'\eta's'}M^{l'\eta' *}_{n'm'}(\mbf{k}',\mbf{G})P_{m'\eta',n'\eta'}(\mbf{k}',s')\rp  \\
&\times c^\dag_{\mbf{k},m,\eta s} c_{\mbf{k},n,\eta s}
\end{split}
\end{align}
\begin{align}
\begin{split}\label{eq:HF_F}
\hat{H}_\text{F,int}[P]=&-\frac{1}{N} \sum_{\mbf{k} mn \eta\eta' s} \sum_{\mbf{q},\mbf{G},m',n'} \frac{V_{ll'}(\mbf{q}+\mbf{G})}{\Omega} \\
&\times M^{l'\eta' *}_{n'm}(\mbf{k},\mbf{q}+\mbf{G}) M^{l\eta}_{m'n}(\mbf{k},\mbf{q}+\mbf{G})\\
&\times P_{m'\eta,n'\eta'}(\mbf{k}+\mbf{q},s)c^\dag_{\mbf{k},m, \eta' s}c_{\mbf{k},n,\eta s}.
\end{split}
\end{align}
In Eq.~\ref{eq:Hint_decomp}, we have introduced a density matrix
\begin{equation}\label{eq:Pref_average}
P^\text{ref,avg.}_{m\eta,n\eta'}(\mbf{k},s)=\frac{1}{2}\delta_{mn}\delta_{\eta\eta'}
\end{equation}
whose definition holds for any choice of the unitary transformation in Eq.~\ref{eq:basis_choice}. 
The properties of $P^\text{ref,avg.}$ are independent of Eq.~\ref{eq:basis_choice} since $P^\text{ref,avg.}$ is just the identity operator.

\subsection{Interaction schemes}\label{subsec:schemes}
The parameterization of $\hat{H}_\text{int}$ in Eq.~\ref{eq:Hint_decomp} implies that $P^\text{ref,avg.}$ is a `reference density' from which interactions are measured from. To see this, consider the expression for the mean-field HF Hamiltonian $\hat{H}_{\text{HF}}[P]$ for a physical Slater determinant state with density matrix $P$
\begin{align}
\begin{split}
\hat{H}_{\text{HF}}[P]&\equiv\hat{H}_\text{s.p.}-\hat{H}_{\text{HF,int}}[P^\text{ref,avg.}]+\hat{H}_\text{HF,int}[P]\\
&=\hat{H}_\text{s.p.}+\hat{H}_{\text{HF,int}}[P-P^\text{ref,avg.}],\label{eq:HFint_deltaP_average}
\end{split}
\end{align} 
where the mean-field decoupling is in the particle-hole channel. In the second line, we have used the fact that $\hat{H}_{\text{HF,int}}[P]$ is a linear functional of $P$, so that the HF Hamiltonian depends only on $P-P^\text{ref,avg.}$. If we choose $P=P^\text{ref,avg.}$, then we obtain $\hat{H}_{\text{HF}}[P^\text{ref,avg.}]=\hat{H}_\text{s.p.}$. 

We can generalize Eqs.~\ref{eq:Hstart}, \ref{eq:Hint_decomp} and \ref{eq:HFint_deltaP_average} using a general reference density matrix $P^\text{ref}$
\begin{gather}\label{eq:HFint_deltaP}
\hat{H}=\hat{H}_{\text{s.p.}}(V)+\hat{H}_\text{int}\\
    \hat{H}_\text{int}=\normOrd{\hat{H}_\text{int}}-\hat{H}_{\text{HF,int}}[P^\text{ref}]\\
    \hat{H}_{\text{HF}}[P]=\hat{H}_\text{s.p.}+\hat{H}_{\text{HF,int}}[\delta P]
\end{gather}
where $\delta P=P-P^\text{ref}$. This suggests the possibility of different interaction schemes (or `subtraction' schemes) parameterized by the choice of $P^\text{ref}$. As is the case for theoretical studies of moir\'e graphene~\cite{XieSub,TBG3,2021PhRvB.103t5414L,Bultinck_2020,hejazi2021,Liu2021nematic,Parker2021fieldtunedFCITBG,kwan2022skyrmions,faulstich2023interacting},  there is no unique prescription. Certainly $P^\text{ref}$ must satisfy the symmetries of $\hat{H}_{\text{s.p.}}$. In addition, it is desirable that the reference density respects the approximate particle-hole symmetry of rhombohedral graphene, and that the physical properties of $P^\text{ref}$ do not depend on parameters that could be tuned {\it in situ} experimentally, such as the displacement field $D$. The choice Eq.~\ref{eq:Pref_average} satisfies these conditions, and corresponds to the so-called average (or infinite-temperature) scheme, a variant of which is widely adopted in twisted multilayer graphene~\cite{TBG3,2021PhRvB.103t5414L,Parker2021fieldtunedFCITBG,Calugaru2021TSTG,Christos2022TSTG,wang2023kekule,kwan2023strongcoupling}.

Another scheme that has been utilized in the regime of large $D$ for R$L$G/hBN~\cite{dong2023anomalous,dong2023theory,zhou2023fractional} is the so-called charge neutrality (CN) scheme. In this scheme, the density matrix $P^\text{ref}$ is constructed by filling all valence bands of the non-interacting model $\hat{H}_\text{s.p.}(V)$ evaluated at the external interlayer potential $V$. Written explicitly in the non-interacting band basis of $\hat{H}_\text{s.p.}(V)$, the reference density is
\begin{equation}
    P^\text{ref,CN}_{m\eta,n\eta'}(\mbf{k},s)=\begin{cases}
        \delta_{mn}\delta_{\eta\eta'}, &\text{if $m$ is a valence band}\\
        0, &\text{otherwise.}
    \end{cases}
\end{equation}
Note that the non-interacting band structure and wavefunctions of $\hat{H}_\text{s.p.}(V)$ change with $V$. For example, the valence band subspace of $\hat{H}_\text{s.p.}(V)$ becomes increasingly polarized towards the hBN (which is adjacent to the lowest graphene layer $l=0$) for larger $V>0$. Therefore, the physical properties of $P^\text{ref,CN}$, such as its layer polarization, vary with $V$. 

As will be demonstrated in Sec.~\ref{sec:HF_results}, one consequence of the CN interaction scheme for large $|V|$ is that the conduction bands are only weakly affected by the properties of the valence bands for $\nu>0$. For example in a HF calculation of $\hat{H}$, because of the large single-particle gap induced by $V$, the valence bands are expected to be nearly fully occupied in the ground state, whose density matrix is $P$. As result, $\delta P=P-P^{\text{ref,CN}}$ is nearly vanishing in the valence band subspace. According to Eq.~\ref{eq:HFint_deltaP}, the interacting part of the HF Hamiltonian $\hat{H}_{\text{HF,int}}[P]$ (Eqs.~\ref{eq:HFint}, \ref{eq:HF_H} and \ref{eq:HF_F}) directly uses $\delta P$, which is nearly 0 in the valence band subspace, and hence any terms in $\hat{H}_{\text{HF,int}}[P]$ where the band indices $m',n'$ are valence band indices are suppressed. Note that this means any terms in $\hat{H}_{\text{HF,int}}[P]$ with form factors involving valence band indices are suppressed. Since the form factors in Eqs.~\ref{eq:HF_H} and \ref{eq:HF_F} are the only quantities in the HF Hamiltonian that encode the properties of the band wavefunctions, this implies that the HF calculation only depends weakly on the properties of the valence band wavefunctions. This effect is exacerbated in any calculation that projects only onto the conduction bands of $\hat{H}_\text{s.p.}(V)$ (see Sec.~\ref{subsec:projection} for an explanation of how calculations are projected), since the valence band occupations in $P$ are all forced to be 1 (see Eq.~\ref{eq:genPsiproj}, where $\mathscr{H}_{\text{rem.~val.}}$ is taken to be the single-particle Hilbert space of all valence bands). In this case, $\delta P$ is exactly 0 for the valence band subspace, and the conduction bands are therefore completely unaffected by the valence bands and their associated moir\'e charge density.

In this work, we perform calculations using both the average interaction scheme and CN interaction scheme. As is known from theoretical studies in twisted multilayer graphene~\cite{hejazi2021,kwan2022skyrmions,faulstich2023interacting}, the choice of interaction scheme can potentially have a qualitative impact on the phase diagram, and the properties of the ground states and their excitations. In the context of R$L$G/hBN, $\hat{H}_\text{int}$ in the average interaction scheme does not depend on $V$, and the interaction between valence and conduction band subspaces is not heavily suppressed (unlike the CN interaction scheme). We will therefore point out the differences and similarities between these two schemes.

\section{Hartree-Fock, projection, and collective modes}

\subsection{Projection onto active subspace}\label{subsec:projection}
In Secs.~\ref{subsec:interactions} and \ref{subsec:schemes}, we defined the many-body Hamiltonian
\begin{equation}\label{eq:H_beforeprojection}
    \hat{H}=\hat{H}_{\text{s.p.}}(V)+\normOrd{\hat{H}_\text{int}}-\hat{H}_{\text{HF,int}}[P^\text{ref}]
\end{equation}
and discussed how $P^\text{ref}$ depends on the choice of interaction scheme. We still need to specify the `total' single-particle Hilbert space $\mathscr{H}$, from which the many-body Hilbert space is constructed.  In moir\'e continuum models, $\mathscr{H}$ consists of plane wave momentum states that lie within a plane wave cutoff in each valley (for instance for valley $K$, plane waves that lie within a cutoff circle centered on the Dirac wavevector $\mbf{K}_G$). For a given mBZ momentum $\mbf{k}$, valley $\eta$ and spin $s$, if there are $N_\mbf{G}$ RLVs within the plane wave cutoff, then $\mathscr{H}$ has dimension $2LN_\mbf{G}$, where the factor $2L$ arises from the 2 sublattices and $L$ layers. 
For $N_\mbf{G}=19$ (i.e.~4 shells), diagonalizing $\hat{H}_{\text{s.p.}}(V)$ for R5G/hBN yields 190 bands per spin and valley. 

For practical calculations of the interacting many-body Hamiltonian $\hat{H}$, it is useful to effectively reduce the dimension of the single-particle Hilbert space in order to decrease the computational difficulty. In the non-interacting band structure, the majority of valence and conduction bands have kinetic energies that are far from the Fermi level. For the low-energy many-body states $\ket{\Psi}$ of interest, these valence (conduction) states are hence expected to be fully filled (empty). We are therefore justified in freezing their occupations in the many-body state $\ket{\Psi}$, and considering an effective interacting calculation involving fewer degrees of freedom. This procedure is formalized in a technique known as projection, which consists of two steps. 

The first step for projection is to identify an `active' subspace of single-particle states $\mathscr{H}_\text{act.}\subset\mathscr{H}$, which includes at least those states that we expect to participate non-trivially in the low-energy physics, and which might not have frozen occupation numbers. The rest of $\mathscr{H}$ is divided into a remote valence subspace $\mathscr{H}_{\text{rem. val.}}$ and a remote conduction subspace $\mathscr{H}_{\text{rem. con.}}$. The occupations of the single-particle states in these remote subspaces are \emph{frozen}. This means that we restrict the many-body Hilbert space to states of the form
\begin{equation}\label{eq:genPsiproj}
    \ket{\Psi}=\hat{\mathcal{O}}\prod_{\alpha\in\mathscr{H}_{\text{rem.~val.}}}c^\dagger_{\alpha}\ket{\text{vac}}
\end{equation}
where $\ket{\text{vac}}$ is the fermion vacuum, and $\hat{\mathcal{O}}$ is an operator consisting of an arbitrary combination of creation operators $c^\dagger_{\alpha}$ for single-particle states $\alpha\in\mathscr{H}_\text{act.}$ belonging to the active subspace. Hence in a projected calculation, the remote degrees of freedom are not allowed to fluctuate. 

$\mathscr{H}_\text{act.}$ is typically specified by diagonalizing some single-particle moir\'e continuum Hamiltonian $\hat{H}_0$ defined on $\mathscr{H}$, and selecting some contiguous subset of low-energy bands near the Fermi energy. The band basis of $\hat{H}_0$ will referred to as the \emph{projection band basis}. If $n_c$ conduction and $n_v$ valence bands of $\hat{H}_0$ are designated as active, then we refer to  this as $(n_v+n_c)$ projection in the basis of $\hat{H}_0$. The $n_v,n_c$ are referred to as active band cutoffs. All other valence (conduction) band states of $\hat{H}_0$ are assigned to $\mathscr{H}_{\text{rem. val.}}$ ($\mathscr{H}_{\text{rem. con.}}$). Usually, $\hat{H}_0$ is taken to be the non-interacting term of $\hat{H}$, i.e.~$\hat{H}_0=\hat{H}_\text{s.p.}(V)$. In Sec.~\ref{subsec:internal_screening} and App.~\ref{secapp:screening}, we will explain why this is not always a satisfactory choice, and describe how we obtain $\hat{H}_0$.

In the second step of projection, our goal is to obtain a many-body Hamiltonian $\hat{H}_{\text{act.}}$ that explicitly acts only on the many-body Hilbert space constructed from $\mathscr{H}_\text{act.}$. Crucially, the remote degrees of freedom are not completely ignored, since they can affect the physics in the active subspace by renormalizing the effective one-body potential felt by the active degrees of freedom. To see this, consider the interaction $\normOrd{\hat{H}_\text{int}}$ (see e.g.~Eq.~\ref{eq:Hint_nord}, which we consider to be written in the projection band basis). In terms of $\normOrd{\hat{H}_\text{int}}$ where one creation and one annihilation operator belong to $\mathscr{H}_{\text{rem. val.}}$, they can be replaced (after anticommuting to bring them together) by a delta function of their quantum numbers. If the other two operators belong to $\mathscr{H}_\text{act.}$, then this generates a one-body contribution acting on active states in $\mathscr{H}_{\text{act.}}$. These contributions are collected in the one-body term 
\eq{
\hat{H}^\text{eff}_{\text{rem. val.}} = \left.\hat{H}_\text{HF,int}[P_\text{rem. val.}]\right|_{\text{act.}}\ ,
}
where
\eq{
\left[P_\text{rem. val.}(\bsl{k},s)\right]_{m\eta,n \eta'} = \begin{cases}
        \delta_{mn}\delta_{\eta\eta'}, & \text{if $m$ in } \mathscr{H}_\text{rem.~val.}\\
        0, &\text{otherwise,}
    \end{cases}
}
and for any second-quantized operator $\hat{O}$ acting on the many-body Hilbert space constructed from $\mathscr{H}$, $\left.\hat{O}\right|_{\text{act.}}$ denotes the truncation to only terms that solely involve creation/annihilation operators belonging to $\mathscr{H}_\text{act.}$. $\hat{H}^\text{eff}_{\text{rem. val.}}$ captures the renormalization from the filled remote valence states. We then obtain
\begin{align}
\begin{split}\label{eq:Hact}
    \hat{H}_{\text{act.}}=&\left.\hat{H}_{\text{s.p.}}(V)\right|_{\text{act.}}+\left.\normOrd{\hat{H}_\text{int}}\right|_{\text{act.}}\\
    &-\left.\hat{H}_{\text{HF,int}}[P^\text{ref}]\right|_{\text{act.}}+\hat{H}^\text{eff}_{\text{rem. val.}}.
\end{split}
\end{align}
We can then perform computations using the projected interaction Hamiltonian $\hat{H}_{\text{act.}}$. Note that the energy expectation value of $\ket{\Psi_\text{act.}}=\hat{\mathcal{O}}\ket{\text{vac}}$ in $\hat{H}_\text{act.}$ is equal to that of $\ket{\Psi}=\hat{\mathcal{O}}\prod_{\alpha\in\mathscr{H}_{\text{rem.~val.}}}c^\dagger_{\alpha}\ket{\text{vac}}$ (Eq.~\ref{eq:genPsiproj}) in $\hat{H}$, up to constants that do not depend on $\hat{\mathcal{O}}$.

\subsection{Internal screening and screened basis}\label{subsec:internal_screening}

To correctly capture the low-energy physics of the unprojected Hamiltonian $\hat{H}$ (Eq.~\ref{eq:H_beforeprojection}), a projected calculation using $\hat{H}_\text{act.}$ (Eq.~\ref{eq:Hact}) should yield results that do not change upon increasing the active band cutoffs $n_v,n_c$. While the accuracy of projection can always be improved by increasing $n_v,n_c$, recall that the active subspace $\mathscr{H}_{\text{act.}}$ was defined in Sec.~\ref{subsec:projection} by diagonalizing $\hat{H}_0$ to obtain the projection band basis, and selecting the lowest $n_v$ valence and $n_c$ conduction bands to be active. A judicious choice of $\hat{H}_0$ will reduce the number of active bands required for satisfactory convergence. A natural choice is $\hat{H}_0=\hat{H}_{\text{s.p}}(V)$ since states whose non-interacting kinetic energies have a large magnitude are expected to be fully filled or empty, and hence they can be safely designated as remote (i.e.~not active) bands. 

However, the interacting terms in $\hat{H}$ could renormalize the effective one-body energies, such that they are not quantitatively (or even qualitatively) captured by the eigenenergies of $\hat{H}_{\text{s.p}}(V)$. If this is the case, then $\hat{H}_0=\hat{H}_{\text{s.p}}(V)$ does not provide a suitable projection band basis for constructing $\mathscr{H}_\text{act.}$. We now explain why this can occur for R$L$G/hBN in the average interaction scheme (see Sec.~\ref{subsec:interactions} and \ref{subsec:schemes}) and a layer-dependent interaction potential $V_{ll'}(\mbf{q})$. The physical reason is that the actual interlayer potential $U$ (which can be defined by the layer-diagonal $\mbf{q}=0$ piece in the mean-field Hamiltonian) in the many-body ground state is renormalized from the externally applied value $V$ due to $\mbf{q}=0$ interlayer electrostatic (Hartree) corrections. To see this, consider the $\mbf{q}=0$ Hartree Hamiltonian corresponding to the unprojected physical density matrix $P$ defined over all bands of $\mathscr{H}$. Using the Hartree decoupling (Eq.~\ref{eq:HF_H}) of $\normOrd{\hat{H}_\text{int}}$ in the many-body Hamiltonian $\hat{H}$ (Eq.~\ref{eq:H_beforeprojection}), and isolating just the $\mbf{q}=0$ Hartree contribution in all interacting terms, we obtain (see App.~\ref{secapp:screening} for details)
\begin{gather}\label{eq:H_Hq0_ldep}
    \hat{H}_{\text{H},\mbf{q}=0}[P]=\hat{H}_{\text{s.p}}\big(V_l+V_{\text{int},l}[\delta P]\big)\\
    V_{\text{int},l}[\delta P]=\frac{1}{N}\sum_{l'}\frac{V_{ll'}(\mbf{q}=0)}{\Omega}N_{l'}[\delta P],\label{eq:Vintl}
\end{gather}
where $\delta P=P-P^\text{ref,avg.}$, and $N_{l}[\delta P]$ is the expectation value of the total number of electrons in layer $l$ in the density matrix $\delta P$
\begin{equation}
    N_l[\delta P]
    =\sum_{\mbf{k}mn\eta s}M^{l\eta }_{mn}(\mbf{k},\mbf{0})\delta P_{m\eta,n\eta}(\mbf{k},s).
\end{equation}
We have explicitly indicated the layer dependence of the layer potential argument of $\hat{H}_\text{s.p.}$ in Eq.~\ref{eq:H_Hq0_ldep} (note that the external $V_l$ is linear in layer index according to Eq.~\ref{eq_main:H_D}). In Eq.~\ref{eq:Vintl}, we have defined the internal interlayer Hartree functional $V_{\text{int},l}[\delta P]$, which describes the interaction-induced potential generated by the total charges on each layer. 

Consider R$L$G/hBN in an external interlayer potential $V$ at neutrality $\nu=0$, and define $P_\text{s.p.}(V)$ as the density matrix obtained by filling all valence bands of $\hat{H}_{\text{s.p}}(V)$. $P_\text{s.p.}(V)$ serves as our initial `guess' of the ground state of the $\mbf{q}=0$ Hartree Hamitonian $\hat{H}_{\text{H},\mbf{q}=0}[P]$ at $\nu=0$; we will find shortly that $P_\text{s.p.}(V)$ needs to be adjusted to obtain the actual self-consistent ground state. The layer charge density $N_{l}[\delta P]$ is mostly localized (with opposite signs) on the outer layers (see Fig.~\ref{Nl_L5} in App.~\ref{secapp:screening}). This is because at zero interlayer potential, the low-energy states in R$L$G are superpositions of the outermost layers, and will quickly resolve into layer-definite states upon applying $V$. Therefore, we approximate $V_{\text{int},l}[\delta P_\text{s.p.}(V)]$ as being linear in layer and define (see Sec.~\ref{secapp:screening})
\begin{equation}
    V_{\text{int}}[\delta P]=(V_{\text{int},L-1}[\delta P]-V_{\text{int},0}[\delta P])/(L-1)
\end{equation}
so that we can drop the layer index in the argument of $\hat{H}_{\text{s.p}}$ in the Hartree Hamiltonian
\begin{equation}\label{eq:HHq0_linear}
    \hat{H}_{\text{H},\mbf{q}=0}[P]=\hat{H}_{\text{s.p}}\big(V+V_{\text{int}}[\delta P]\big).
\end{equation}

If $V_{\text{int}}[\delta P_\text{s.p.}(V)]$ vanished (which would be the case for a purely 2d interaction where there is no layer dependence, or the CN interaction scheme where $\delta P_\text{s.p.}(V)=0$), then self-consistency is achieved since $P_\text{s.p.}(V)$ would indeed correspond to the ground state of $\hat{H}_{\text{H},\mbf{q}=0}[P_\text{s.p.}(V)]$. However, this is generally not the case. For example, for R5G/hBN and $\epsilon_r=5$, an external $V=48\,\text{meV}$ corresponds to $V_{\text{int}}[\delta P_\text{s.p.}(V)]=-32\,\text{meV}$, so that the system effectively experiences a net interlayer potential $U=V+V_{\text{int}}=16\,\text{meV}$. The strong internal screening of the external potential means that $P_\text{s.p.}(V)$, the density matrix of the filled valence bands of $\hat{H}_{\text{s.p}}(V)$, does not closely resemble the ground state of the Hartree Hamiltonian. As a result, the single-particle energies and band wavefunctions of $\hat{H}_{\text{s.p}}(V)$ substantially differ from those of $\hat{H}_{\text{H},\mbf{q}=0}[P_\text{s.p.}(V)]$, and constructing $\mathscr{H}_{\text{act.}}$ using the band basis of $\hat{H}_{\text{s.p}}(V)$ is likely to yield poor convergence with respect to the projection band cutoffs $n_v,n_c$ (see App.~\ref{secapp:basiscomparison} for a demonstration of this in a HF calculation).

To remedy the above issue, we solve the Hartree Hamiltonian (Eq.~\ref{eq:HHq0_linear}) self-consistently. This means that for a given external potential $V$, we calculate the screened interlayer potential $U(V)$ that satisfies
\begin{equation}\label{eq:UV}
    U(V)=V+V_\text{int}[\delta P_{\text{s.p}}(U(V))],
\end{equation}
such that $P_{\text{s.p}}(U(V))$ is the $\nu=0$ self-consistent ground state of $\hat{H}_{\text{H},\mbf{q}=0}[P_{\text{s.p}}(U(V))]$ (see App.~\ref{secapp:screening} for details of how we perform the self-consistent calculation). Since $U(V)$ better approximates the screened interlayer potential in the many-body ground state of $\hat{H}$, constructing $\mathscr{H}_\text{act.}$ using the band basis of $\hat{H}_{\text{s.p}}(U(V))$ yields better convergence with respect to the projection band cutoffs, and allows us to perform projected calculations with smaller $n_v,n_c$.

We refer to to the construction of the active subspace $\mathscr{H}_\text{act.}$ using the band basis of $\hat{H}_0=\hat{H}_\text{s.p.}(V)$ as \emph{bare basis projection}, while the construction of $\mathscr{H}_\text{act.}$ using $\hat{H}_0=\hat{H}_\text{s.p.}(U(V))$ is called \emph{screened basis projection}. A comparison of the two choices of projection is provided in App.~\ref{secapp:basiscomparison}.

\subsection{Hartree-Fock calculations}

We perform self-consistent HF calculations on the projected Hamiltonian $\hat{H}_{\text{act}.}$ (see Eq.~\ref{eq:Hact}). Recall from Sec.~\ref{subsec:projection} that projection involves specifying the active subspace $\mathscr{H}_\text{act.}$ of single-particle states. Following Sec.~\ref{subsec:internal_screening}, we mostly use the screened basis projection, where $\mathscr{H}_\text{act.}$ is constructed from the lowest $n_v$ valence and $n_c$ conduction bands of the continuum model $\hat{H}_{\text{s.p.}}(U(V))$ evaluated at the screened interlayer potential $U(V)$ (Eq.~\ref{eq:UV}). 
We always consider systems of $N_1\times N_2$ moir\'e unit cells, where $N_1=N_2$ is a multiple of 6 to capture the near-degeneracy of multiple bands at the high-symmetry points $\tilde{K}_M,\tilde{K}'_M,\tilde{M}_M$, and $\tilde{M}'_M$ of the mBZ (see Fig.~\ref{fig:SP_bands}). We keep all spins and valleys in our calculation. We restrict to $S_z$-conserving states, but allow for intervalley coherence (IVC) that hybridizes mBZ momentum $\mbf{k}$ in valley $K$ with momentum $\mbf{k}$ in valley $K'$ ($\mbf{k}=0$ at $\tilde{\Gamma}_M$ for each valley). For the results presented in the main text, the HF state is constrained to preserve moir\'e translation symmetry so that the mBZ momentum $\mbf{k}$ remains a good quantum number. We retain the same mBZ and allow different RLV's $\mbf{G}$ to couple even if the hBN coupling is switched off and $\hat{H}$ has continuous translation symmetry. The HF state is uniquely parameterized by the one-body density matrix (projector)
\begin{equation}
    P_{m\eta,n\eta'}(\mbf{k},s)=\langle c^\dagger_{\mbf{k},m,\eta s} c_{\mbf{k},n,\eta' s} \rangle,
\end{equation}
where $c^\dagger_{\mbf{k},m,\eta s}$ are screened basis creation operators belonging to $\mathscr{H}_\text{act.}$, and the total occupation is fixed by the filling factor $\nu$. In our projected calculations, this translates to $\sum_{\mbf{k}m\eta s}P_{m\eta,m\eta}(\mbf{k},s)=4n_vN_1N_2+\nu\eqqcolon N_{e,\text{act}.}$, where $N_{e,\text{act}.}$ is the number of electrons in the projected calculation. 

The HF numerics consist of an iterative loop, starting with an initial seed density matrix $P^{(0)}$, where the projector $P^{(t)}$ from iteration $t$ is used to construct the HF Hamiltonian of $\hat{H}_{\text{act.}}$ for the next iteration
\begin{align}
\begin{split}\label{eq:HFactt+1}
    \hat{H}^{(t+1)}_{\text{HF,act.}}=&\left.\hat{H}_{\text{s.p.}}(V)\right|_{\text{act.}}\\
    &+\left.\hat{H}_{\text{HF,int}}[P^{(t)}-P^\text{ref}]\right|_{\text{act.}}+\hat{H}^\text{eff}_{\text{rem. val.}}.
\end{split}
\end{align}
We define ${P}^{(t+1)}$ by diagonalizing $\hat{H}^{(t+1)}_{\text{HF,act.}}$ and occupying the lowest $N_{e,\text{act}.}$ eigenstates. This procedure is repeated until the HF energy functional
\begin{align}
\begin{split}
    E_\text{HF}[P]=&\langle\left.\hat{H}_{\text{s.p.}}(V)\right|_{\text{act.}}\rangle_P+\langle\hat{H}^\text{eff}_{\text{rem. val.}}\rangle_P\\
    &+\frac{1}{2}\langle\left.\hat{H}_{\text{HF,int}}[P-P^\text{ref}]\right|_{\text{act.}}\rangle_{P-P^\text{ref}},
\end{split}
\end{align}
where $\langle\rangle_P$ denotes expectation value in the density matrix $P$, does not differ between consecutive iterations by more than $0.2\,\mu$eV per moir\'e unit cell.  For each parameter in the phase diagrams, we perform multiple self-consistent HF calculations involving at least 16 initial seeds for $P^{(0)}$, and show the results of the calculation that minimizes the HF energy (see App.~\ref{secapp:HFdetails} for more details). HF band structures are computed by diagonalizing the HF Hamiltonian at the final iteration to obtain the single-particle HF energies $\mathscr{E}_\alpha$, where $\alpha$ is a composite label for all quantum numbers.

At this stage, we point out a subtlety with the renormalization from remote bands in the average interaction scheme. Recall that in our projected calculations, the interaction-induced one-body potential generated from the filled remote valence bands and the reference density is $
    \left.\hat{H}_{\text{HF,int}}[\delta P_{\text{rem.~val.}}]\right|_{\text{act.}}$ (see Eq.~\ref{eq:Hact}), 
where we have defined $\delta P_{\text{rem.~val.}}=P_\text{rem.~val.}-P^\text{ref,avg.}$. Since $\delta P_{\text{rem.~val.}}$ takes non-vanishing values $1/2$ ($-1/2$) for all remote valence (conduction) projection bands, this means that the one-body potential felt by a state in the active subspace $\mathscr{H}_\text{act.}$, which for valley $K$ is predominantly constructed from plane waves near the graphene Dirac momentum $\mbf{K}_G$, receives corrections from remote states whose Bloch functions only have support at the edge of the plane wave cutoff used to specify the `total' single-particle Hilbert space $\mathscr{H}$. In particular, the Fock renormalization from these remote states has a suppression factor that is upper-bounded by $V_{ll}(\mbf{q})$, which only falls off as $\sim 1/q$ at large momentum transfer $q$. This means that the one-body term in the active Hamiltonian $\hat{H}_\text{act.}$ may weakly change as the plane wave cutoff in $\mathscr{H}$ is increased. However, the single-particle continuum Hamiltonian is not a reliable model for states with energies approaching the eV scale, so Fock renormalization from states around or beyond this scale is unphysical. To avoid this unphysical renormalization and reduce the computational time, we set radial momentum cutoffs on the Hilbert space $\mathscr{H}$ and the interaction potential at $4|\mbf{q}_1|$ and $3|\mbf{q}_1|$ respectively (see App.~\ref{secapp:cutoff} for a comparison of HF results with larger cutoffs).

\subsection{Time-dependent Hartree-Fock}\label{subsec:TDHF}

To understand the neutral excitation spectrum and collective modes in the HF ground state, we use the time-dependent Hartree-Fock (TDHF) method, which is equivalent to the random phase approximation (RPA) with exchange~\cite{ring2004nuclear,mclachlan1964TDHF,Casida2012TDDFT,khalaf2020soft,wu2020collective,kwan2021exciton,TBG5,wang2023magnets}. We simply summarize the formalism here, and defer to App.~\ref{secapp:TDHF} for a detailed derivation and discussion. 

By diagonalizing the HF Hamiltonian (Eq.~\ref{eq:HFactt+1}) at the end of a converged HF calculation, we obtain HF orbitals indexed by $\alpha$, which is a composite label for all the quantum numbers such as $\mathbf{k}$, and their corresponding HF eigenenergies $\mathscr{E}_\alpha$. The HF orbitals in our calculations comprise $\mathscr{H}_{\text{act}.}$ since we perform projected HF calculations, but the TDHF framework can just as well be applied to unprojected calculations. Depending on their occupation in the converged HF state, the HF orbitals are either assigned to the occupied subspace $\mathscr{H}_\text{occ.}$ (of dimension $N_\text{occ.}$) or the unoccupied subspace $\mathscr{H}_\text{unocc.}$ (of dimension $N_\text{unocc.}$). The objective of TDHF is to compute the collective mode creation operators $Q^\dagger_a$ and their corresponding excitation energies $\Omega^a$, where $a$ is an index for the collective mode. 

Consider the entire set of particle-hole (ph) labels $\phi=(\phi_\text{p},\phi_\text{h})$, where $\phi_\text{p}$ is an unoccupied HF orbital, and $\phi_\text{h}$ is an occupied HF orbital. There are $N_\text{occ.}N_{\text{unocc.}}$ such labels. Define the following matrices that act on the space of ph labels
\begin{align}
\begin{split}
    A_{\phi,\phi'}=&(\mathscr{E}_{\phi_\text{p}}-\mathscr{E}_{\phi_\text{h}})\delta_{{\phi_\text{p}}{\phi'_\text{p}}}\delta_{{\phi_\text{h}}{\phi'_\text{h}}}\\
    &+V_{{\phi_\text{p}}{\phi'_\text{h}},{\phi_\text{h}}{\phi'_\text{p}}}-V_{{\phi_\text{p}}{\phi'_\text{h}},{\phi'_\text{p}}{\phi_\text{h}}}
\end{split}
\end{align}
\begin{align}
    B_{\phi,\phi'}=V_{{\phi_\text{p}}{\phi'_\text{p}},{\phi_\text{h}}{\phi'_\text{h}}}-V_{{\phi_\text{p}}{\phi'_\text{p}},{\phi'_\text{h}}{\phi_\text{h}}}.
\end{align}
Above, $V_{\alpha\beta,\gamma\delta}$ are interaction matrix elements that appear when the normal-ordered four-fermion interaction term (which does not depend on the interaction scheme) is expressed in the basis of HF orbitals
\begin{equation}
    \left.\normOrd{\hat{H}_\text{int}}\right|_{\text{act.}}=\frac{1}{2}\sum_{\alpha\beta\gamma\delta}V_{\alpha\beta,\gamma\delta}d^\dagger_\alpha d^\dagger_\beta d_\delta d_\gamma
\end{equation}
where $d^\dagger_\alpha$ is the creation operator for HF orbital $\alpha$. 

In TDHF, $Q^\dagger_a$ and $\Omega^a$ are obtained by solving the eigenvalue problem
\begin{gather}
    \begin{pmatrix}
    A & B \\ -B^* & -A^*
    \end{pmatrix}\begin{pmatrix}X^a\\Y^a\end{pmatrix}=\Omega^a \begin{pmatrix}X^a\\Y^a\end{pmatrix}\\
    Q^\dagger_a=\sum_{\phi}\left(X^a_{\phi}d^\dagger_{\phi_\text{p}}d_{\phi_\text{h}}-Y^a_{\phi}d^\dagger_{\phi_\text{h}}d_{\phi_\text{p}}\right).\label{eq:Qop}
\end{gather}
We only choose the eigenvectors that satisfy $(X^a)^\dagger X^{a}-(Y^a)^\dagger Y^{a}=1$.

For R$L$G/hBN, the collective mode spectrum will form bands as a function of momentum transfer $\mbf{q}$ due to moir\'e translation invariance. Furthermore, the collective modes will split into sectors characterized by other conserved charges. We perform our TDHF calculations for gapped HF states, which we will find in Sec.~\ref{sec:HF_results} are spin-valley polarized. For such $\ket{\text{HF}}$, $Q^\dagger_a$ will carry definite flavor (i.e.~spin and valley) quantum numbers. We refer to collective modes that change the spin (such as the spin wave mode in a ferromagnet) but preserve the valley as interspin modes. Similarly, excitations that change the valley but preserve the spin are dubbed intervalley modes. We do not explicitly consider inter-spin-valley modes that change both the spin and valley, because they are degenerate with the intervalley modes owing to the $SU(2)_K\times SU(2)_{K'}$ spin-rotation symmetry. Finally, intraflavor modes are those that preserve both spin and valley. Goldstone modes due to a broken continuous symmetry will manifest as gapless $\Omega(\mbf{q})$.

\begin{table}
\centering
\newcommand{\colskip}{\hskip 0.15in}
\renewcommand{\arraystretch}{1.15}
\begin{tabular}{
l @{\hskip 0.2in} 
l @{\colskip} 
l  }\toprule[1.3pt]\addlinespace[0.3em]
{\textbf{sector}} & 
\textbf{ph/hp content} & 
\textbf{flavor content}
\\ \midrule[1.3pt]
intraflavor & $X_\phi\neq0,Y_\phi\neq0$ & $d^\dagger_{K\uparrow}d_{K\uparrow}$\\  \midrule
interspin & $X_\phi\neq0,Y_\phi=0$ & $d^\dagger_{K\downarrow}d_{K\uparrow}$ \\  \midrule
intervalley & $X_\phi\neq0,Y_\phi=0$ & $d^\dagger_{K'\uparrow}d_{K\uparrow}$ \\  \midrule
inter-spin-valley & $X_\phi\neq0,Y_\phi=0$ &  $d^\dagger_{K'\downarrow}d_{K\uparrow}$\\
\bottomrule[1.3pt]
\end{tabular}
\caption{Constraints on the neutral excitation operators for different collective mode sectors within TDHF theory. We assume that the HF state is fully spin-valley polarized in flavor $K\uparrow$, and obtained from a HF calculation projected onto only conduction bands. ph/hp content indicates the constraints on the particle-hole ($X_\phi$) and hole-particle ($Y_\phi$) coefficients in the excitation operator $Q_a^\dagger$ (Eq.~\ref{eq:Qop}). Flavor content shows the required flavor indices on the creation and annihilation operators in $Q_a^\dagger$.\label{tab:flavormodes}}
\end{table}

To lower the computational cost, we will reduce the size of the active subspace $\mathscr{H}_\text{act.}$ to include only conduction bands in the HF calculation when performing TDHF. For a fully spin-valley polarized state at $\nu=+1$, we can then, without loss of generality, choose the occupied HF orbitals to be all in the $K\uparrow$ flavor. Tab.~\ref{tab:flavormodes} indicates in this case the constraints on the excitation operator $Q_a^\dagger$ for the different sectors of excitations.

\section{Hartree-Fock results}\label{sec:HF_results}

\begin{figure*}[t]
\includegraphics[width=1.0\linewidth]{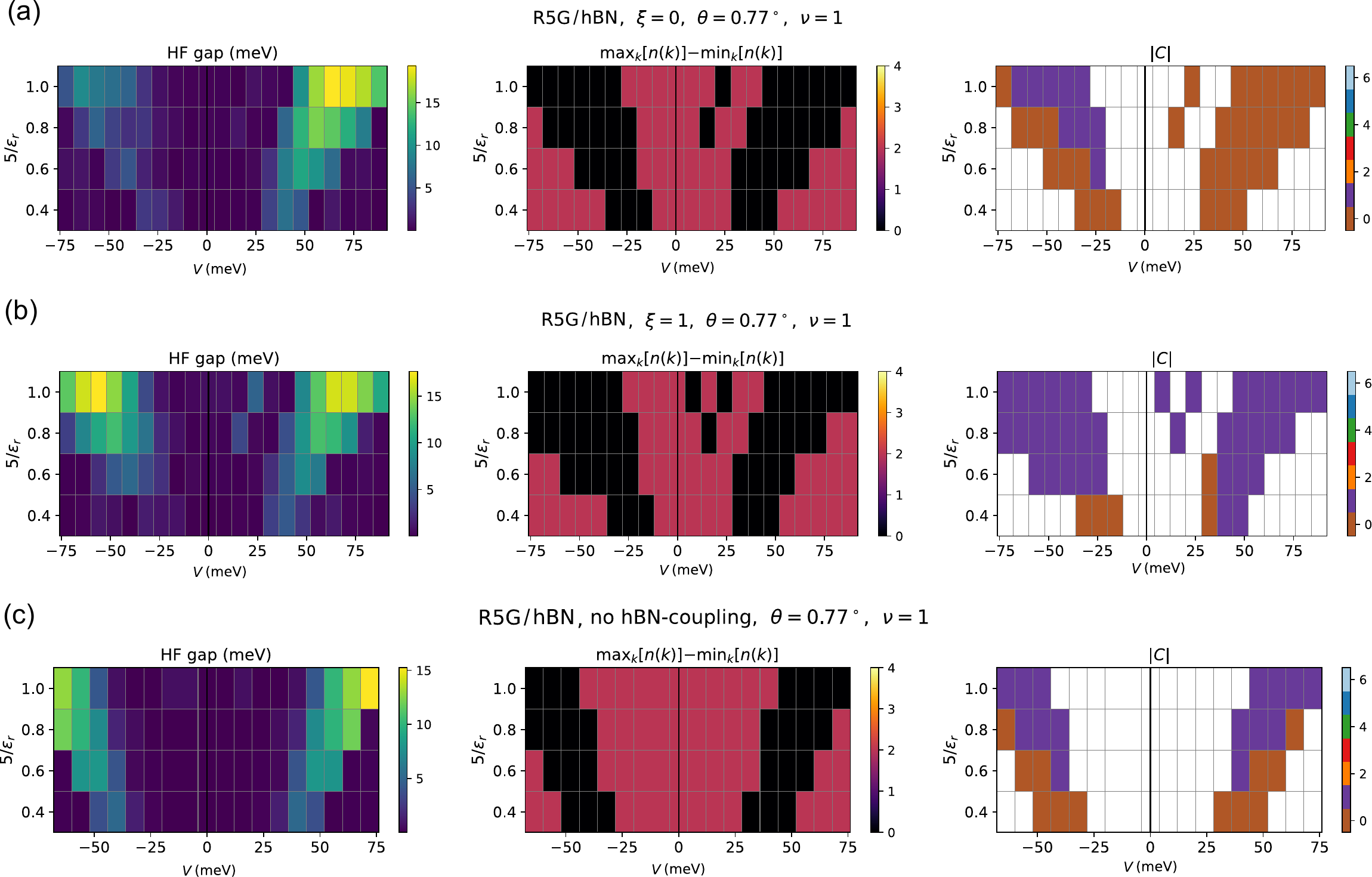}
\caption{HF phase diagram at $\nu=1$ for R5G/hBN in the average interaction scheme. (a) Results for $\xi=0$ stacking. The HF gap indicates the energy difference of the HF eigenvalues between the lowest unoccupied and highest occupied orbital, and is equivalent to the indirect gap for an insulating state. A non-zero $\max_\mbf{k}[n(\mbf{k})]-\min_\mbf{k}[n(\mbf{k})]$, where $n(\mbf{k})$ counts the occupation number at momentum $\mbf{k}$, rules out an insulating state. $C$ is the Chern number of the HF state. (b) Results for $\xi=1$ stacking. (c) Results for zero hBN coupling $\kappa_\text{hBN}=0$, which are independent of $\xi$. The HF calculations are performed with $(4+4)$ bands per spin/valley using the screened basis projection, average interaction scheme, $\theta=0.77^\circ$, and system size $N_1\times N_2=12\times 12$.}
\label{fig_main:main_phase_R5G}
\end{figure*}

\subsection{R5G/hBN at $\theta=0.77^\circ$}
In Fig.~\ref{fig_main:main_phase_R5G}, we show projected HF results for the $\nu=1$ phase diagram of $\theta=0.77^\circ$ R5G/hBN for the two stacking configurations $\xi=0,1$. This calculation is performed using the \emph{average interaction scheme} and 3D layer-dependent interactions $V_{ll'}(\mbf{q})$. We use the screened basis projection with projection band cutoffs $n_c=n_v=4$. We study a range of external interlayer potentials $V$, as well as interaction strengths (stronger interactions for larger $5/\epsilon_r$). The HF gap is defined as the difference between the lowest unoccupied single-particle HF energy, and the highest occupied single-particle HF energy. $n(\mbf{k})$ is the occupation of momentum $\mbf{k}$ in the HF state, such that non-zero $\max_\mbf{k}[n(\mbf{k})]-\min_\mbf{k}[n(\mbf{k})]$ rules out an insulating state. $C$ is the Chern number of the HF state, which is only shown for insulating states where the Berry curvature flux does not exceed $0.2\pi$ for any plaquette in the mBZ (a large plaquette flux indicates a high concentration of Berry curvature, which requires a bigger system size to reliably compute $C$). We find that all the HF phases in Fig.~\ref{fig_main:main_phase_R5G} have zero intervalley coherence, and are fully spin and valley polarized. This means that three flavors are at their charge neutrality point, while the remaining flavor is electron-doped to $\nu=1$.

\begin{figure*}[t]
\includegraphics[width=1.0\linewidth]{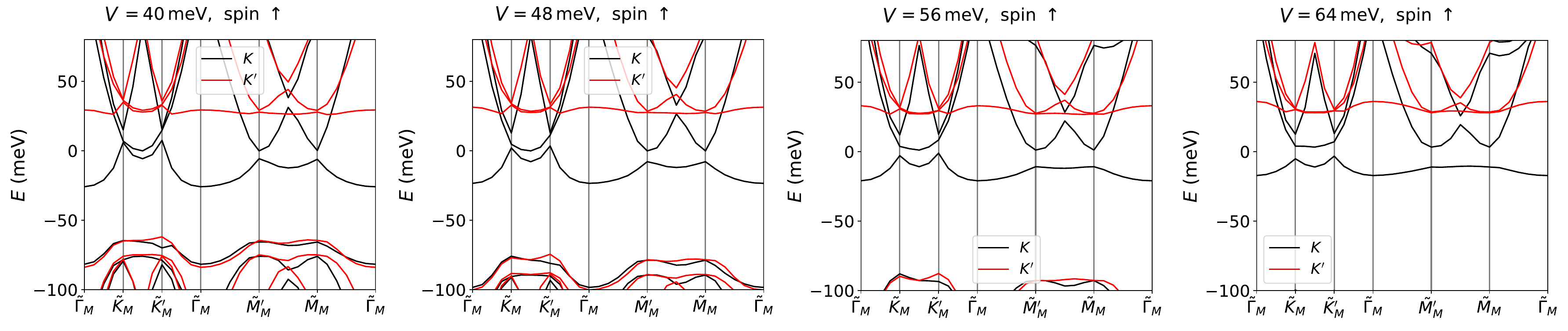}
    \caption{Band structures of the self-consistent $\nu=1$ HF ground states for $\xi=1$ R5G/hBN as a function of $V$. We only show the fully polarized spin-$\uparrow$ sector. Black (red) shows the dispersion in microscropic valley $K$ ($K'$). The HF calculations are performed with $(4+4)$ bands per spin/valley using the screened basis projection, average interaction scheme, $\theta=0.77^\circ$, $\epsilon_r=6.25$, and system size $N_1\times N_2=12\times 12$.
    }
\label{fig_main:main_HFbandstruct_U}
\end{figure*}

\begin{figure*}[t]
\includegraphics[width=0.9\linewidth]{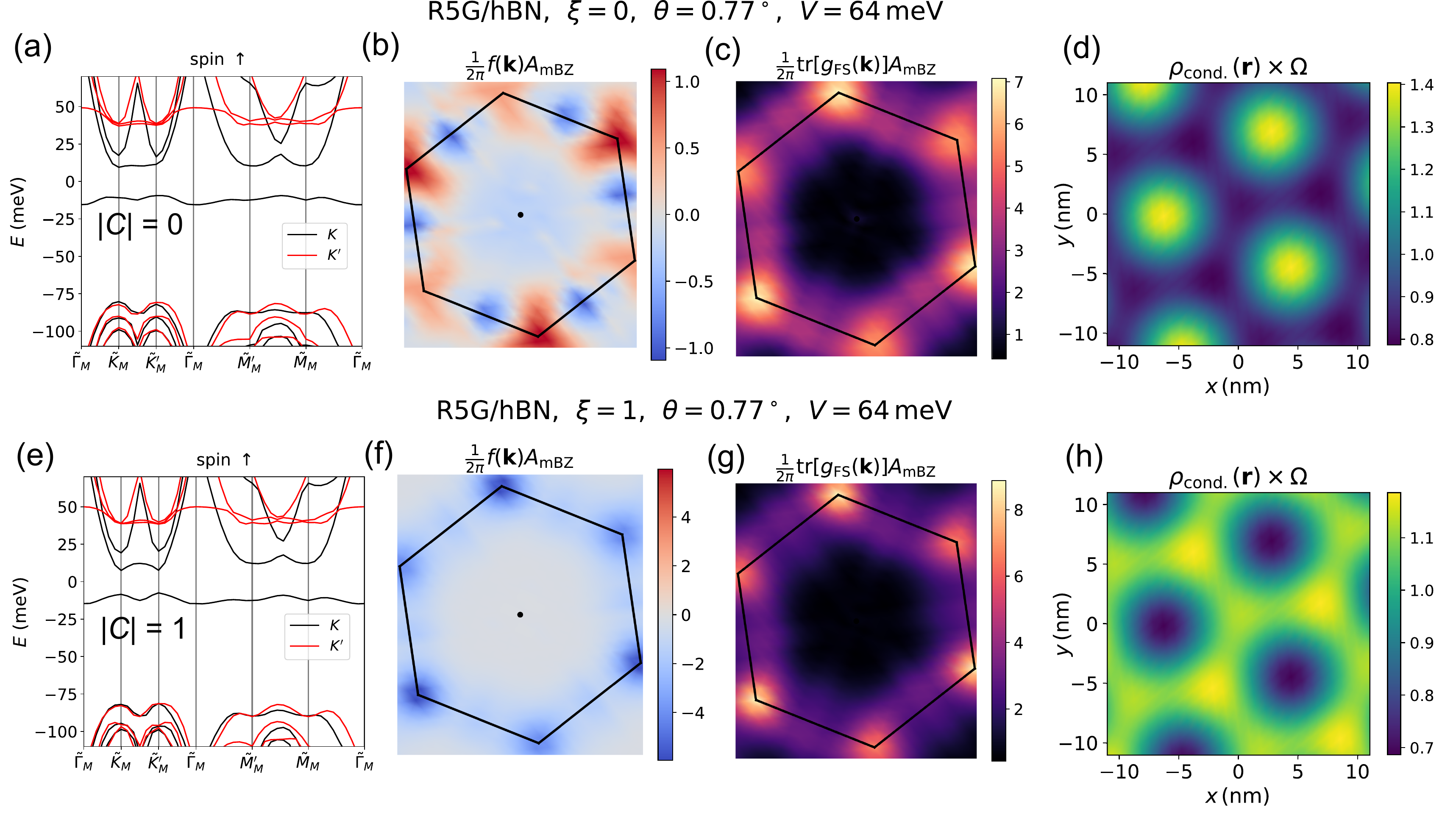}
    \caption{Details of the self-consistent $\nu=1$ HF ground states for R5G/hBN. (a-d) Results for $\xi=0$ stacking. (a) HF band structure for the fully-polarized spin-$\uparrow$ sector. Black (red) shows the dispersion in microscropic valley $K$ ($K'$).  The HF state has Chern number $|C|=0$. (b) Berry curvature $f(\mbf{k})$ of the occupied conduction band, normalized so that a uniform value of $1$ corresponds to a $C=1$ insulator ($A_{\text{mBZ}}$ is the momentum-space area of the mBZ). (c) Trace of the Fubini-Study metric of the occupied conduction band, which satisfies $\mathrm{tr}[g_\mathrm{FS}(\mbf{k}]>|f(\mbf{k})|$. (d) Dimensionless real-space density of the occupied conduction band, normalized so that the mean value in the moir\'e unit cell is 1 ($\Omega$ is the real-space area of the moir\'e unit cell).
    (e-h) Same as (a-d) but for $\xi=1$ stacking, where the HF state has $|C|=1$. The HF calculations are performed with $(3+3)$ bands per spin/valley using the screened basis projection, average interaction scheme, $\theta=0.77^\circ$, $\epsilon_r=5$, $V=64\,\text{meV}$ and system size $N_1\times N_2=18\times 18$.
    }
\label{fig_main:HF_bandstruct_QG}
\end{figure*}

\subsubsection{Gapless vs.~gapped states}

We first comment on general features of the competition between metallic and insulating states in Fig.~\ref{fig_main:main_phase_R5G}. Near $V=0$, the non-interacting band structure is either gapless or has a small  indirect gap at neutrality (Fig.~\ref{fig:SP_bands}). Since we include both conduction and valence bands in our active subspace $\mathscr{H}_\text{act.}$, our HF setup can capture the physics in this regime where mixing between conduction and valence subspaces may be significant. Consistent with the experiments in Ref.~\cite{LongJu2023FCIPentalayerGraphenehBN}, we find a large window of gapless states around $V=0$. As shown in Fig.~\ref{fig_main:main_phase_R5G}, depending on the interaction strength, this extends to $V\simeq 25-50\,\text{meV}$ on the $V>0$ side (where the conduction band electrons are polarized away from the hBN), and slightly smaller absolute values for the $V<0$ side. The size of the gapless region is similar in the two stackings.  While stronger interactions may be expected to more easily induce correlated gaps, we find that the gapless window is actually larger (Fig.~\ref{fig_main:main_phase_R5G}). However, the internal electrostatic screening of the external potential $V$ is stronger for smaller $\epsilon_r$, such that the effective interlayer potential $U$ experienced in R5G/hBN is suppressed. For example in our self-consistent $\mbf{q}=0$ screening calculation at $\nu=0$ (Sec.~\ref{subsec:internal_screening}), for $V=48.0\,\text{meV}$ and $\xi=0$, we find $U=28.3\,\text{meV}$ ($38.8\,\text{meV}$) for $\epsilon_r=5.00$ (12.50). That the screening of the external potential is non-negligible can be seen in the HF band structures of Fig.~\ref{fig_main:HF_bandstruct_QG}(a,d). In particular at $V=64\,\text{meV}$, focusing on the flavors at charge neutrality (e.g.~$K\downarrow$), we observe that the lowest conduction band in the HF band structure still has a large gap to the next conduction band at $\tilde{\Gamma}_M$. In contrast, the lowest conduction band in the non-interacting dispersion (Fig.~\ref{fig:SP_bands}) is about to collide with the higher conduction bands at $\tilde{\Gamma}_M$ already at $V=48\,\text{meV}$ (Fig.~\ref{fig:SP_bands}).

\begin{figure*}[t]
\includegraphics[width=1.0\linewidth]{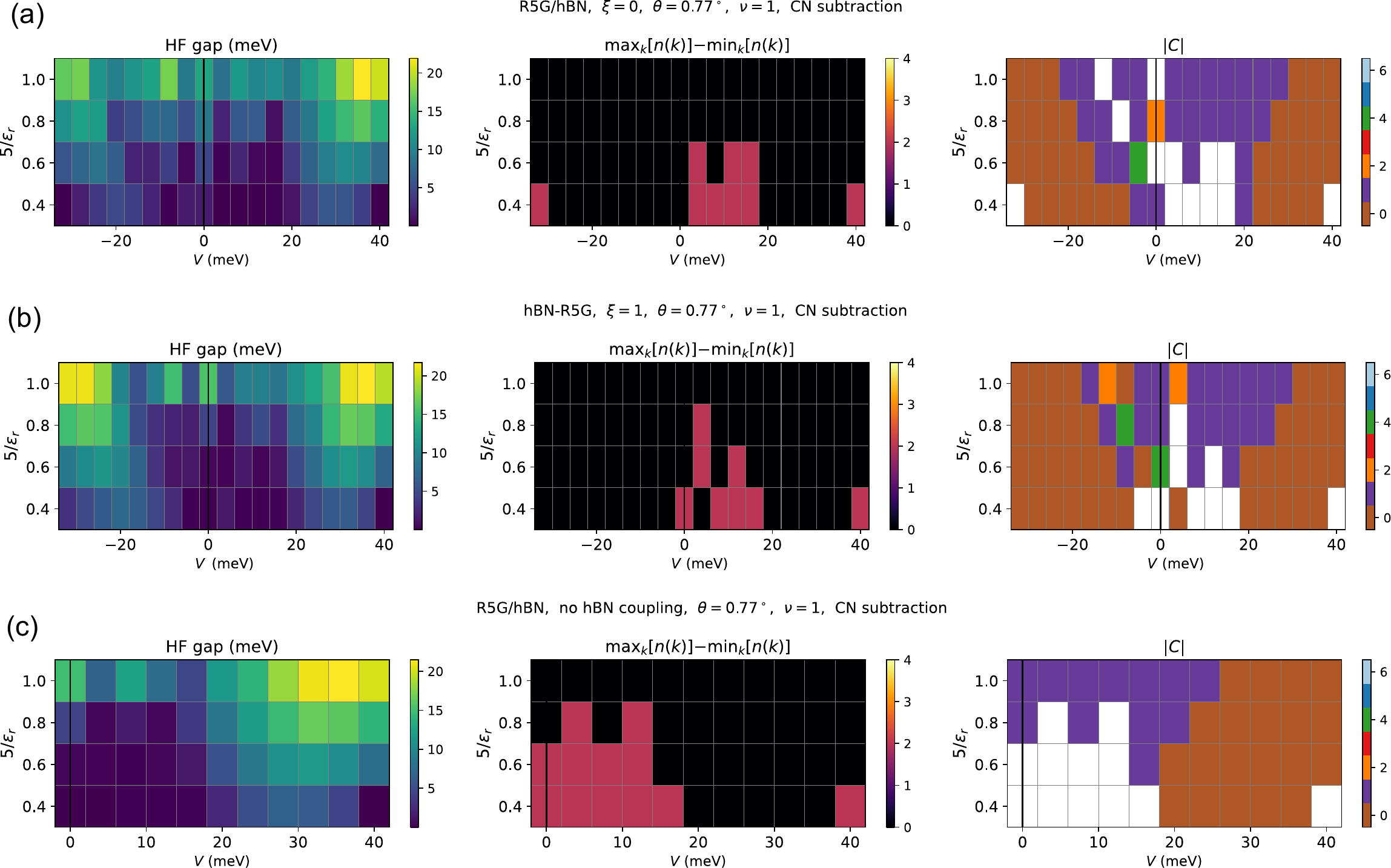}
\caption{HF phase diagram at $\nu=1$ for R5G/hBN in the CN interaction scheme. (a) Results for $\xi=0$ stacking. The HF gap indicates the energy difference of the HF eigenvalues between the lowest unoccupied and highest occupied orbital, and is equivalent to the indirect gap for an insulating state. A non-zero $\max_\mbf{k}[n(\mbf{k})]-\min_\mbf{k}[n(\mbf{k})]$, where $n(\mbf{k})$ counts the occupation number at momentum $\mbf{k}$, rules out an insulating state. $C$ is the Chern number of the HF state. (b) Results for $\xi=1$ stacking. (c) Results for zero hBN coupling $\kappa_\text{hBN}=0$, which are independent of $\xi$. The HF calculations are performed with $(3+3)$ bands per spin/valley using the screened basis projection, CN interaction scheme, $\theta=0.77^\circ$, and system size $N_1\times N_2=12\times 12$.}
\label{fig_main:main_phase_R5G_CN}
\end{figure*}

Recall that in the CN interaction scheme, the influence of the interactions from filled valence bands (including remote bands), including their contributions to internal screening, is greatly suppressed due to the choice of reference density $P^\text{ref}=P^\text{ref,CN}$ (see Sec.~\ref{subsec:schemes} and \ref{subsec:internal_screening}). We find that in the CN interaction scheme (Fig.~\ref{fig_main:main_phase_R5G_CN}), the gapless region is either significantly smaller, or disappears entirely for $\epsilon_r=5$.  Hence, the choice of interaction scheme has a qualitative impact on the positions of the gapless and gapped phases, and it is clear that the CN interaction scheme is not appropriate for small $|V|$.

For larger $|V|$ in the average interaction scheme, we find the emergence of HF states with finite indirect gaps (Fig.~\ref{fig_main:main_phase_R5G}). This occurs because as $|V|$ is increased, the lowest conduction band peels off from the next conduction band around the mBZ boundary, especially near the $\tilde{K}_M$ and $\tilde{K}'_M$ points. At the same time, the single occupied conduction band flattens. This is shown for $V>0$ in Fig.~\ref{fig_main:main_HFbandstruct_U}. For $\epsilon_r=5$, we find that the gapped state onsets at around $V=50\,\text{meV}$, which translates to a displacement field of $D/\epsilon_0=\frac{\epsilon_r V}{ed}=0.76\,\text{V/nm}$. This value is consistent with the measurements of Ref.~\cite{LongJu2023FCIPentalayerGraphenehBN}, though we caution that a more careful comparison of $D$ requires a detailed treatment of the anisotropy of the dielectric environment in R5G/hBN.

\subsubsection{Nature of the gapped states}

While the positions of the gapped states in the phase diagram are similar for both stackings, our calculations show that the nature of the insulating state is different in the two cases. For $\xi=0$, the HF gap for $V<0$ is significantly smaller than at $V>0$. Furthermore, the insulator for $V>0$ is always topologically trivial with $C=0$. On the other hand for $\xi=1$, the sizes of the HF gaps are comparable for both signs of $V$, and the $V>0$ gapped state is predominantly a Chern insulator with $|C|=1$.  The relative preference for $C=0$ states for $\xi=0$ is also present for $V<0$. In Fig.~\ref{fig_main:HF_bandstruct_QG}(a,d), we show the band structure of the HF states for the two stackings at $V=64\,\text{meV}$. In both cases, the occupied conduction band is narrow with bandwidth $\lesssim 10\,\text{meV}$, but the Chern number is $C=0$ ($|C|=1$) for $\xi=0$ ($\xi=1$).  The ground state phases obtained for $\xi=1$ and $5/\epsilon_r=0.4$ appear to be consistent with Ref.~\cite{LongJu2023FCIPentalayerGraphenehBN}: there is a gapless region around $V=0$, a $|C|=1$ region for $V>0$ that is preceded by a small window of $|C|=0$, and an absence of Chern insulators for $V<0$. The transition between the $|C|=0$ and $|C|=1$ states for $V>0$ and $5/\epsilon_r=0.4,0.6$ is first-order in the HF numerics. 

In Fig.~\ref{fig_main:HF_bandstruct_QG}(b,f), we show the Berry curvature $f(\mbf{k})$ of the occupied HF conduction band for the $|C|=0$ and $|C|=1$ states at stackings $\xi=0$ and $\xi=1$ respectively, for the same interaction strength and interlayer potential. In both cases, $|f(\mbf{k})|$ is largest around the mBZ boundary. For the Chern insulator, the Berry curvature is peaked at the $\tilde{K}_M$ point. In Fig.~\ref{fig_main:HF_bandstruct_QG}(c,g), we also show the trace of the Fubini-Study metric $g_\text{FS}(\mbf{k})$~\cite{Parameswaran13} of the occupied conduction band, which satisfies $\text{tr}[g_\text{FS}(\mbf{k})]>|f(\mbf{k})|$. We find that the modulation of $\text{tr}[g_\text{FS}(\mbf{k})]$ closely tracks that of the Berry curvature in the $|C|=1$ state, and the average violation of the trace condition is $\frac{A_\text{mBZ}}{2\pi}\int d\mbf{k}\,\int \text{tr}[g_\text{FS}(\mbf{k})]-|f(\mbf{k})|=1.2$.

\subsubsection{Role of the valence bands}

For the $V>0$ gapped region, it is surprising that the two stackings yield different Chern numbers if one only considers the conduction band dispersion and wavefunctions of the continuum model. This is because for large $V$, the lowest conduction bands are polarized away from the hBN layer, and do not directly feel the hBN-induced moir\'e coupling. Indeed, it was shown in Ref.~\cite{MFCI-II} that at $V=16.65\,\text{meV}$, the projector onto the lowest conduction band has 99.9\% overlap with the corresponding projector when the moir\'e coupling is switched off. In addition, the dispersion of the lowest conduction bands closely resembles that of the folded bands of isolated R5G. The non-interacting band structure and wavefunctions of the lowest conduction bands are hence only very weakly dependent on $\xi$. 

Therefore, the discrepancy in the Chern numbers of the two stackings in Fig.~\ref{fig_main:HF_bandstruct_QG} must arise from the influence of the valence bands. In the average interaction scheme, the filled valence bands (including remote bands not in $\mathscr{H}_\text{act.}$) renormalize the one-body potential felt by the lowest conduction bands by imparting effective Hartree and Fock potentials (see Sec.~\ref{subsec:projection}). Fig.~\ref{fig_main:chargedensity} illustrates the real-space density fluctuation $\Delta \rho(\mbf{r})$ of the filled valence subspace of $H_\text{s.p.}(V)$, which directly factors into the valence-subspace-induced Hartree background experienced by the conduction electrons. $\Delta \rho(\mbf{r})$ differs in the two stackings: the density for $\xi=0$ has pronounced minima which form a triangular moir\'e lattice, while the regions of low $\Delta \rho(\mbf{r})$ for $\xi=1$ are more spread out over the moir\'e unit cell. This suggests that the background Hartree potential landscape experienced by the conduction electrons for $\xi=0$ is stronger, which may tip the balance between competing insulating states. In fact, the charge density $\rho_\text{cond.}(\mbf{r})$ of the occupied conduction band in the gapped HF state closely mirrors the density fluctuation $\Delta\rho (\mbf{r})$ of the filled valence bands. In Fig.~\ref{fig_main:HF_bandstruct_QG}(d), $\rho_\text{cond.}(\mbf{r})$ of the $|C|=0$ insulator for $\xi=0$ possesses strong localized peaks on a triangular moir\'e lattice which precisely coincides with the troughs in  $\Delta\rho (\mbf{r})$ [Fig.~\ref{fig_main:chargedensity}(a,b)]. For the $\xi=1$ Chern insulator [Fig.~\ref{fig_main:HF_bandstruct_QG}(h)], the corresponding HF conduction band density is relatively more homogeneous (as expected for a topologically non-trivial state), forming a honeycomb mesh with pockets of lower density. These real-space features again line up with the corresponding charge background of the filled valence subspace in Fig.~\ref{fig_main:chargedensity}(c,d).

The hypothesis that the valence bands are important for selecting the ground state is further supported by HF calculations using the CN interaction scheme [Fig.~\ref{fig_main:main_phase_R5G_CN}(a,b)], where the renormalization of the potential due to the filled valence bands is largely cancelled out.
In this case for $V>0$, the insulator has consistent Chern number for both stackings, and the phase boundaries are quantitatively similar. 
Furthermore, as shown in Fig.~\ref{fig_main:main_phase_R5G_CN}(c), the phase diagram in the CN scheme for large $V$ in the absence of hBN-coupling ($\kappa_{\text{hBN}}=0$) remains nearly unchanged, highlighting the inability for the hBN potential to impact the conduction band physics in this interaction scheme. 

Returning to the average interaction scheme, when we artificially remove the hBN-coupling $\kappa_{\text{hBN}}=0$ (and thus any dependence on $\xi$) in Fig.~\ref{fig_main:main_phase_R5G}(c), we find that the insulating region of the phase diagram has competing $|C|=0$ and $|C|=1$ phases, with stronger interactions favoring $|C|=1$. In our HF calculations for the $V>0$ region with sizable HF gap, we find that the HF states with $|C|=0$ and $|C|=1$ both evolve continuously from $\kappa_{\text{hBN}}=1$ to $\kappa_{\text{hBN}}=0$, in the sense that HF gap does not close. The above observations suggest that in the average interaction scheme, both states are competing with each other in the moir\'eless limit ($\kappa_{\text{hBN}}=0$), and the hBN-coupling (which depends on $\xi$) plays a critical role in favoring one of the insulating states via the renormalized potential generated by the filled valence bands. Our results emphasize the inequivalence between $\xi=0,1$ stackings, and its consequences for the correlated phase diagram. 

\subsection{Other twist angles $\theta$ and number of layers $L$}

We have also computed $\nu=1$ phase diagrams in the average interaction scheme for R5G/hBN at different twist angles $\theta=0.6^\circ$ and $1.1^\circ$ (see Figs.~\ref{app_L5_4p4xi0_t0.60_nu1} and \ref{app_L5_4p4xi0_t1.10_nu1} in App.~\ref{secapp:HFphase}). For the larger twist angle $\theta=1.1^\circ$, the value of $|V|$ required to obtain gapped phases increases. For $V>0$, this can be understood by considering how the twist angle affects the size of the mBZ, and consequently the folding of the bands of isolated R5G (which closely approximates the moir\'e band structure for large $V$). Recall that isolated R5G at $V=0$ has a dispersion that scales as $\sim k^5$. At larger $\theta$, the mBZ is larger in size, which means that the lowest conduction band in the mBZ has a higher bandwidth. Therefore, a greater displacement field is required to sufficiently flatten the band in order to allow interactions to open an indirect gap.   There is a greater propensity towards $|C|=1$ states, as now even $\xi=0$ is exclusively $|C|=1$ for the gapped $V>0$ phase. We also find that the gapped regions shrink in area in the phase diagram, and parts of the gapless phase lose full spin and valley polarization. On the other hand, the gapped regions move to lower $V$ for the smaller twist angle $\theta=0.6^\circ$. Furthermore, the gapless region near $V=0$ shrinks, or even disappears completely. There is a tendency towards $|C|=0$, such that both stackings are in the $|C|=0$ state in the $V>0$ gapped region.

The results for different twist angles suggest that there is a `magic angle' window for realizing topological phases in R5G/hBN. If $\theta$ is too small, then the $|C|=0$ phase is lower in energy compared to the Chern insulator. If $\theta$ is too large, then stronger interactions are required to open a gap, and the larger displacement fields necessary to realize the Chern insulator at $\nu=1$ may be challenging to obtain experimentally where the maximum accessible fields are $D/\epsilon_0\simeq 1\,\text{V\,nm}^{-1}$. 

In App.~\ref{secapp:HFphase}, we present $\nu=1$ phase diagrams for R$L$G/hBN for $L=3,4,6,7$ layers and both stackings. At fixed $\theta=0.77^\circ$, the gapped regions appear at larger $V$ for smaller $L$, in agreement with the fact that the low-energy dispersion of isolated R$L$G at $V=0$ is $\sim k^L$, and hence a larger displacement field is needed to sufficiently flatten the lowest conduction band.  For larger numbers of layers $L=6,7$, the gapless region near $V=0$ shrinks, and disappears for strong enough interactions. For all $L$, we find a relative preference towards $|C|=1$ states for the $\xi=1$ stacking, similar to the case of R5G/hBN.

\section{Collective modes}

\begin{figure*}[t]
\includegraphics[width=1.0\linewidth]{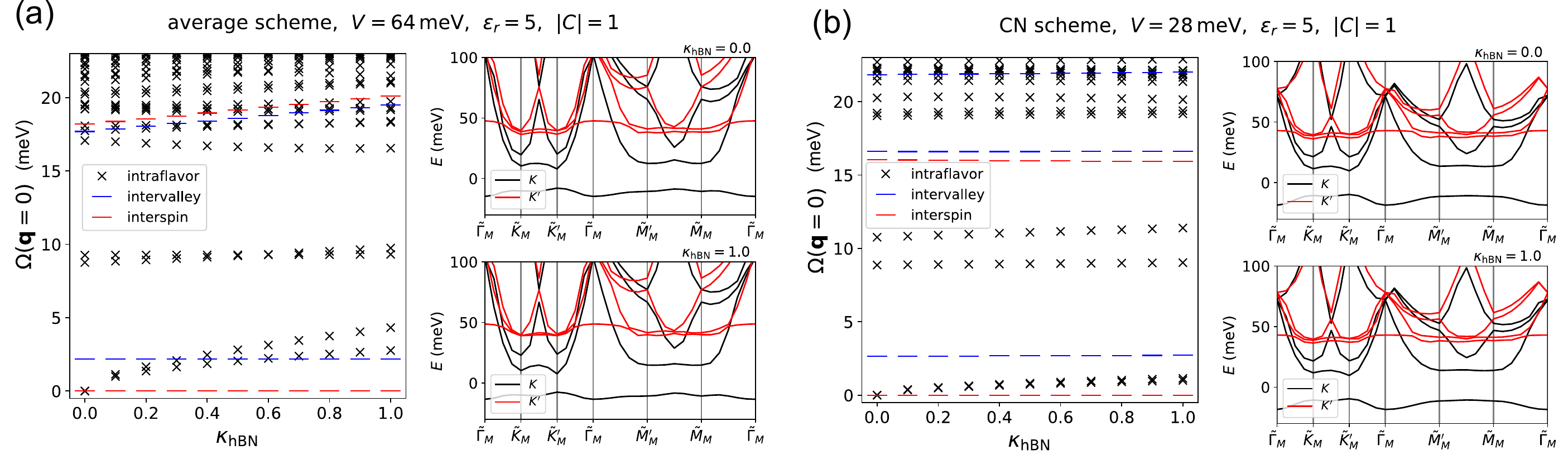}
\caption{$\mbf{q}=0$ collective modes of the $|C|=1$ Chern insulator at $\nu=1$ in R5G/hBN with $\xi=1$ stacking, computed by applying TDHF to the HF state. (a) Results using the average interaction scheme. Left shows the $\mbf{q}=0$ collective modes, measured in meV, and filtered by whether they carry valley or spin charge. Intraflavor means that the mode preserves both spin and valley (for example the pseudophonons). The collective modes are computed function of the strength of hBN coupling, where $\kappa_\text{hBN}=1$ corresponds to the physical model and the $\kappa_\text{hBN}=0$ limit has continuous translation symmetry and zero moir\'e potential. Right shows the HF band structure for $\kappa_\text{hBN}=0,1$ in the spin sector with finite filling (the other spin sector is at charge neutrality). (b) Results using the CN interaction scheme. Parameters have been chosen in (a,b) so that the interacting gap is similar in the two interaction schemes. The HF and TDHF calculations are performed with $(0+4)$ bands per spin/valley using the screened basis projection,  $\theta=0.77^\circ$, and system size $N_1\times N_2=12\times 12$.}
\label{fig_main:collq0}
\end{figure*}

\begin{figure}[t]
\includegraphics[width=1.0\linewidth]{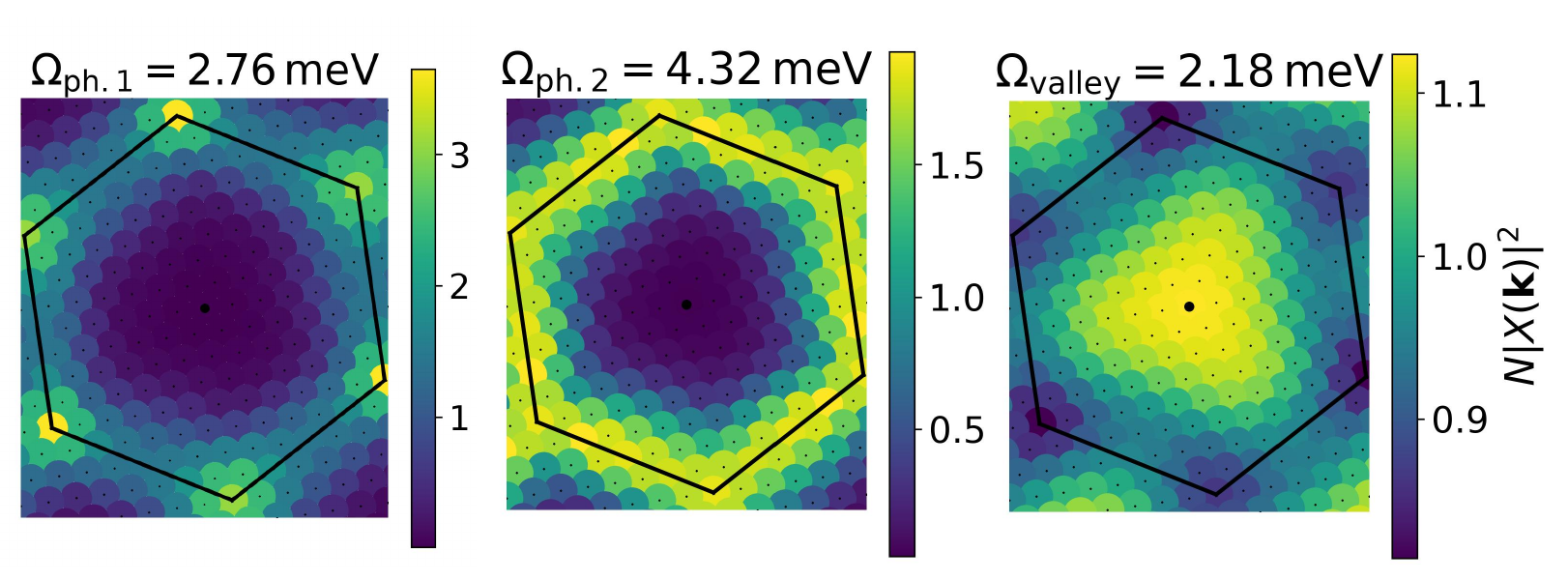}
\caption{Momentum-resolved particle-hole (ph) content of the $\mbf{q}=0$ collective modes of the $|C|=1$ insulator in $\xi=1$ R5G/hBN. We show results for the two pseudophonons $\Omega_\text{ph. 1},\Omega_\text{ph. 2}$ and the lowest intervalley mode $\Omega_\text{valley}$. The data here use the same system parameters as Fig.~\ref{fig_main:collq0}(a). $N|X(\mbf{k})|^2$ indicates the total probability amplitude of $\mbf{q}=0$ ph operators at momentum $\mbf{k}$ in the collective mode operator (see Eq.~\ref{appeq:QRPA} in App.~\ref{secapp:TDHF}), and has been normalized so that uniform probability (in the absence of any hole-particle components proportional to $Y(\mbf{k})$) across the mBZ corresponds to a value of 1.}
\label{fig_main:collwfn}
\end{figure}

\subsection{Motivation}

The HF calculations in Sec.~\ref{sec:HF_results} demonstrate that the phase diagram of R5G/hBN depends sensitively on how long-range electron interactions are accounted for in the many-body Hamiltonian $\hat{H}$. As reviewed in Sec.~\ref{subsec:schemes}, the different interaction schemes can be characterized by the choice of reference density $P^\text{ref}$ from which interactions are measured from. Focusing on the large $V>0$ regime of relevance to the observed FCIs and $|C|=1$ Chern insulator in Ref.~\cite{LongJu2023FCIPentalayerGraphenehBN}, we found that the interaction-induced moir\'e potential, or lack thereof, from the valence bands was a deciding factor in the competition between the topological and trivial gapped states. Already for $V=16.65\,\text{meV}$, the non-interacting low-energy conduction bands of $\hat{H}_\text{s.p.}(V)$ barely feel the moir\'e potential, and are well described by the folded continuum bands of isolated R5G with continuous translation symmetry, as evidenced by the 99.9\% overlap of the projectors onto the lowest non-interacting conduction band with and without the hBN coupling~\cite{MFCI-II}. In the CN interaction scheme where the density matrix of filled valence bands is cancelled by the reference density $P^\text{ref}=P^\text{ref,CN}$, and hence does not influence the conduction electrons, the physics at $\nu=1$ is largely unaffected by the hBN coupling and the stacking orientation. By contrast, in the average interaction scheme, the occupied valence subspace, which includes bands which strongly feel the hBN coupling, can propagate the effects of the moir\'e potential onto the conduction electrons, leading to a $|C|=0$ ($|C|=1$) state for $\xi=0$ ($\xi=1$).

For $V>0$, even when the hBN coupling is completely switched off ($\kappa_\text{hBN}=0$), both competing states appear in the phase diagram, with stronger interactions favoring the Chern insulator. As $\kappa_\text{hBN}$ is increased to the physical limit $\kappa_\text{hBN}=1$, the phase boundaries remain mostly unchanged in the CN interaction scheme, but shift significantly in the average interaction scheme depending on the stacking.
Interestingly, for both interaction schemes, our HF calculation at fixed $V$ and $\epsilon_r$ is often able to converge to both the $|C|=0$ and $|C|=1$ gapped states (one of them is therefore a local minimum) for all values of $0\leq\kappa_\text{hBN}\leq 1$, without the HF gap closing within this range of hBN coupling strengths.
For the average interaction scheme,  this raises the question of whether the extrinsic valence-bands-induced moir\'e coupling merely tips the balance between competing states that exist in the moir\'eless limit, or if it can fundamentally change the nature of the phases. For the CN interaction scheme, the quantitative similarity of the phase diagram to the $\kappa_\text{hBN}=0$ limit with continuous translation symmetry suggests that the gapped HF phases at $\kappa_\text{hBN}=1$ should be viewed as (topological) Wigner crystal-like states. Then the questions turn to (i) quantifying the proximity to the crystalline limit where there is actual spontaneous symmetry-breaking of an exact continuous translation symmetry, and (ii) identifying the source of the moir\'e pinning potential necessary to explain the sharpness of the fractional and integer topological states along the filling axis in Ref.~\cite{LongJu2023FCIPentalayerGraphenehBN}.

These subtle questions regarding the interplay of extrinsic moir\'e potentials and intrinsic translation symmetry-breaking tendencies are challenging to answer directly in HF calculations. This is because the self-consistent HF state strongly breaks the (approximate) symmetries due to interaction effects, such that the action of the symmetry generators connects HF orbitals are separated in energy by the large HF gap. Therefore in this section, we use TDHF calculations to understand how the broken symmetries enter the low-energy physics of the phases.

\subsection{Evolution of collective modes with $\kappa_\text{hBN}$}
In Fig.~\ref{fig_main:collq0}, we first compute the self-consistent HF state for different values of $\kappa_\text{hBN}$, and then apply TDHF theory to extract the collective modes at zero momentum transfer $\mbf{q}=0$. Consider Fig.~\ref{fig_main:collq0}(a), which was computed for the $|C|=1$ insulator in R5G/hBN with $\xi=1$ stacking using the average interaction scheme. Recall from Sec.~\ref{subsec:TDHF} and Tab.~\ref{tab:flavormodes} that since the converged HF state preserves spin and valley $U(1)$ symmetries, the collective modes are classified into four types. In the interspin channel (red lines), we find a single exact zero-energy mode for all $\kappa_\text{hBN}$, which disperses quadratically for small $q$. This is simply the Goldstone mode (spin wave) arising from the broken $SU(2)_S$ symmetry.
In the intervalley channel (blue lines), we find a low-lying mode at energy around $\Omega_\text{valley}=2.2\,\text{meV}$ that barely evolves with $\kappa_\text{hBN}$. The fact that there is a finite valley gap reflects the lack of $SU(2)$ symmetry in valley space. The system has easy-axis anistropy and valley polarizes, which preserves the continuous $U(1)_V$ symmetry. The wavefunction of the $\mbf{q}=0$ intervalley mode is delocalized around the mBZ (Fig.~\ref{fig_main:collwfn}), leading to its interpretation as a (gapped) valley magnon. Note that the intervalley modes are degenerate with the inter-spin-valley modes, so we do not plot the latter.

In the intraflavor channel (black crosses), there are four distinct modes below the particle-hole continuum that starts around 17\,meV. The two modes around 10\,meV are interband excitons which are localized at the mBZ corners. Below these, there are two modes which start off gapless at $\kappa_{\text{hBN}}=0$, and develop a gap of $\Omega_{\text{ph.},1}=2.8\,\text{meV}$ and $\Omega_{\text{ph.},2}=4.3\,\text{meV}$ at the physical $\kappa_\text{hBN}=1$ limit. In the $\kappa_\text{hBN}=0$ limit, these are the gapless pseudophonons of the interaction-induced electronic lattice which spontaneously breaks continuous translation symmetry (of which there are two generators). They become gapped pseudophonons as soon as a finite extrinsic moir\'e potential is applied so that the Hamiltonian only retains a discrete moir\'e translation symmetry. At the same time, the HF band structures [Fig.~\ref{fig_main:collq0}(a)] in the two limits look nearly identical, underscoring the need to utilize post-HF methods such as done here. The pseudophonons are localized near the mBZ boundary (Fig.~\ref{fig_main:collwfn}), where the presence of several nearly degenerate low-lying bands in the moir\'eless non-interacting band structure allows the moir\'e potential to hybridize them.

The low-lying neutral excitation spectrum provides more realistic estimates of the stability of the correlated states against perturbations~\cite{Lin2022exciton}, like thermal fluctuations, than the HF charge gap, whose typically large values $\sim 10\,\text{meV}$ correspond to temperatures $\sim 100\,\text{K}$ that are unrealistically large for correlated topological phenomena in experiments on moir\'e systems. Furthermore, the nature of the low-energy collective modes gives insight into the dominant instabilities that may act to degrade the symmetry-breaking order and any concomitant phenomena. The Mermin-Wagner theorem in 2d precludes any spin-ordering at finite temperature $T$ due to the $SU(2)_S$ symmetry. This is not an issue for the topological response of the Chern insulator, which is unaffected by the spin fluctuations, so we will not consider them further. 

This leaves the next lowest modes, which are the pseudophonons and the valley magnon. The pseudophonon gap reflects the extrinsic moir\'e potential scale relevant to the Chern insulator. Below this scale, the state is pinned to the moir\'e lattice whose lengthscale is determined by $\theta$. The valley gap determines the stability against intervalley excitations. In our HF calculations, we find that if we obtain a Chern insulator with Chern number $C$ with valley polarization in valley $\eta$, we do not find a competing state in the other valley $-\eta$ with the same $C$. This suggests that proliferation of such valley magnons will adversely affect the topological properties of the Chern insulator. Therefore, $\Omega_\text{valley}$ sets an important scale for the robustness of the Chern insulator. In the average interaction scheme calculation of Fig.~\ref{fig_main:collq0}a, we find $\Omega_\text{valley}<\Omega_{\text{ph.},1},\Omega_{\text{ph.},2}$. Thus, we argue that at this parameter, the Chern insulator should not be considered a Wigner crystal-like phase, since its stability as a topological phase is more tied to protection against valley magnons.

In Fig.~\ref{fig_main:collq0}(b), we perform a similar calculation in the CN interaction scheme, choosing parameters such that the HF charge gap is similar to that in Fig.~\ref{fig_main:collq0}(a). While the valley gap remains similar $\Omega_\text{valley}=2.7\,\text{meV}$, the pseudophonons are noticeably lower in energy with $\Omega_{\text{ph.},1}=1.0\,\text{meV}$ and $\Omega_{\text{ph.},2}=1.1\,\text{meV}$ at the physical limit $\kappa_\text{hBN}=1$. The suppression of the pseudophonon gaps is not surprising since filled valence bands in the CN interaction scheme cannot communicate the moir\'e potential to the conduction bands via interaction-induced renormalization. Since $\Omega_\text{valley}>\Omega_{\text{ph.},1},\Omega_{\text{ph.},2}$ for this particular calculation, there is justification, at least from the perspective of TDHF theory, for labelling the Chern insulator a topological Wigner crystal-like state. 

\begin{figure*}[t]
\includegraphics[width=1.0\linewidth]{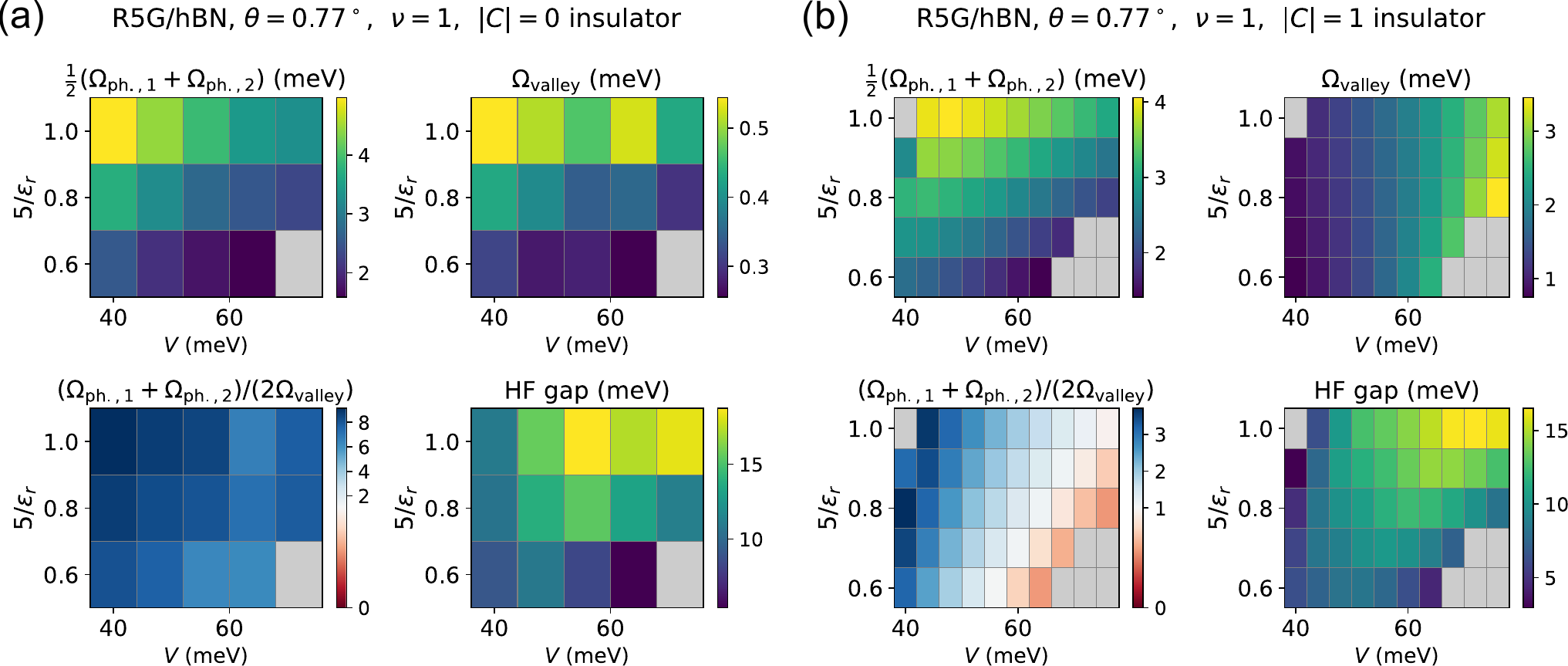}
\caption{Low-energy $\mbf{q}=0$ intervalley and  pseudophonon modes of the $|C|=0,1$ gapped states in R5G/hBN with $\xi=1$ stacking, as a function of $V$ and $\epsilon_r$. Note that for almost all parameters, our HF calculations are able to converge to both $|C|=0,1$ states. (a) Results for the $|C|=0$ HF state. $\Omega_\text{ph.,1},\Omega_\text{ph.,2}$ are the energies of the two lowest excitations (pseudophonons) in the intraflavor channel, which evolve into the two gapless Goldstone modes corresponding to continuous translation symmmetry in the limit of zero moir\'e potential $\kappa_\text{hBN}=0$. $\Omega_\text{valley}$ is the energy of the lowest intervalley excitation. Grey regions denote parameters where there are no gapped HF states. (b) Results for the $|C|=1$ HF state. The HF and TDHF calculations are performed with $(0+4)$ bands per spin/valley using the screened basis projection, average interaction scheme, $\theta=0.77^\circ$, and system size $N_1\times N_2=12\times 12$.}
\label{fig_main:collphase}
\end{figure*}

We have also compared the $\mbf{q}=0$ collective mode spectrum between the $|C|=0$ and $|C|=1$ states in R5G/hBN with $\xi=1$ stacking in the average interaction scheme (see Fig.~\ref{collq0_C0_vs_C1} in App.~\ref{secapp:collq0}). We find that while the pseudophonon gaps are similar, $\Omega_\text{valley}$ is appreciably smaller in the $|C|=0$ state (see also Fig.~\ref{fig_main:collphase}). This suggests that the valley magnetism is significantly more fragile in the topologically trivial phase. A relative suppression of the valley magnon gap in the $|C|=0$ state compared  to the $|C|=1$ Chern insulator has previously been pointed out in twisted TMD homobilayers~\cite{wang2023magnets}.

\subsection{Collective mode phase diagram}

In Fig.~\ref{fig_main:collphase}, we perform HF calculations targeting the competing gapped $|C|=0,1$ states for R5G/hBN with $\xi=1$ stacking in the average interaction scheme, and compute the $\mbf{q}=0$ valley magnon and pseudophonon gaps for a range of external potentials and interaction strengths. We use $\gamma=\frac{\Omega_{\text{ph.},1}+\Omega_{\text{ph.},2}}{2\Omega_\text{valley}} $ as a heuristic indicator for the relative energy scales for valley vs.~translational fluctuations. For the $|C|=0$ state (Fig.~\ref{fig_main:collphase}a), $\gamma>1$ for all parameters. On the other hand, the ratio $\gamma$ in the Chern insulator state (Fig.~\ref{fig_main:collphase}b) can be either greater than or less than one, depending on the parameters. As expected, the pseudophonon gaps decrease for large $V$, as the charge background of the filled valence bands becomes less inhomogeneous in the moir\'e unit cell (see Fig.~\ref{fig_main:chargedensity}). The pseudophonons are also lower in energy for weaker interactions, since $\epsilon_r$ controls the strength of the interaction between the filled valence bands and the low-energy conduction electrons. Combined with the fact that the valley gap increases with $V$, we find that $\gamma$ is larger (i.e.~valley stability is the main concern) for small external potentials and strong interactions, but still remains greater than 1 for most parameters.

We have repeated the calculation for $L=4,6$ layers  in the average interaction scheme (see Fig.~\ref{collphase_L4_t0.77_C1} and \ref{collphase_L6_t0.77_C1} in App.~\ref{secapp:coll_phase}).  Similar to the $L=5$ case, increasing the interaction strength and reducing $V$ increases $\gamma$. We find that that the pseudophonon gaps are larger for smaller numbers of layers, which is expected because the low-energy conduction bands that are polarized away from the aligned hBN layer are now vertically closer to the occupied valence bands that are polarized towards the hBN. This reduces the finite-momentum suppression of the layer-dependent interaction potential (Eq.~\ref{eq:Vint_layer}) between layers $l=0$ and $l'=L-1$ at momentum transfer $\mbf{q}=\mbf{g}_i$, and hence enhances the strength of the moir\'e pinning potential experienced by the conduction bands due to the charge background of the filled valence bands.

\section{Conclusions}

\subsection{Summary of the results}

Through HF and TDHF calculations on the single-particle continuum model derived in Ref.~\cite{MFCI-II}, our work provides a detailed characterization of the $\nu=+1$ phase diagram of rhombohedral $L$-layer graphene ($L=3,\ldots,7$) singly-aligned to hBN (R$L$G/hBN) as a function of the external interlayer potential $V$ and interaction strength $1/\epsilon_r$, as well as the low-lying collective modes of the competing trivial ($C=0$) and topological ($|C|=1$) gapped states. Crucially, our analysis incorporates the 3D nature of the Coulomb interaction, the inequivalence of the two stacking alignments $\xi=0,1$, the influence of the valence bands, and the possibility of different `interaction schemes' (in particular, the average scheme and the charge neutrality (CN) scheme). We demonstrate that all these ingredients play a key role in shaping the phase diagram and the qualitative properties of the insulating phases. 

This is best illustrated at $\theta=0.77^\circ$ in the regime of large positive $V>0$ relevant to the recent observation in R5G/hBN of a Chern insulator at $\nu=+1$ and multiple FCIs at $\nu=\frac{2}{3},\frac{3}{5},\frac{4}{7},\frac{4}{9},\frac{3}{7}$ and $\frac{2}{5}$~\cite{LongJu2023FCIPentalayerGraphenehBN}, where the filling factor is defined with respect to the hBN-induced moir\'e unit cell. For this direction of displacement field, the low-lying non-interacting conduction bands are strongly polarized away from the aligned hBN layer, leading to the natural inference that the hBN-alignment and associated moir\'e potential only weakly influences the physics at $\nu>0$. This is seemingly at odds with the robust quantization of the aforementioned correlated topological phases according to the moir\'e unit cell. 

We show that the choice of interaction scheme leads to starkly contrasting conclusions regarding the role of the extrinsic moir\'e potential. In the CN scheme, where the interactions are normal-ordered with respect to the non-interacting direct band gap at charge neutrality, the occupied valence bands do not affect the physics of the conduction electrons. Consequently, the HF phase diagram at $\nu=+1$ is nearly identical between the two stackings $\xi$, and hardly changes when the hBN potential is switched off entirely ($\kappa_\text{hBN}=0$). 

On the other hand, in the average scheme, the interacting Hamiltonian is not fine-tuned in such a way as to decouple the valence bands from the conduction bands. Here, the charge density of the occupied valence bands, which inherits moir\'e-periodic modulations $\Delta \rho(\mbf{r})$ due to the proximity to the aligned hBN layer, can generate spatially inhomogeneous Hartree and Fock potentials that are experienced by the conduction bands. This interaction-induced injection of `moir\'eness' into the low-energy conduction bands, mediated by the charge background of the filled valence subspace, has significant ramifications for the $\nu=+1$ phase diagram. Our HF calculations reveal that for $V>0$, the $\xi=1$ ($\xi=0$) stacking has a preference towards the $|C|=1$ ($C=0$) gapped phase, which we trace to the properties of the background charge fluctuations $\Delta \rho(\mbf{r})$. This finding has potential consequences for the experimental reproducibility of the correlated topological phases, and provides falsifiable predictions to help narrow down the interaction scheme most appropriate for theoretical modelling of R$L$G/hBN. Our analysis underscores the importance of a careful specification of the Hamiltonian at both the non-interacting and interacting level. To help establish the correct theoretical model, it would be useful to also consider alternative interaction schemes to the two choices that were extensively studied in this work. For example, the `graphene' scheme has been utilized in twisted bilayer graphene~\cite{XieSub,Bultinck_2020}, and involves measuring interactions relative to the filled valence bands of moir\'eless R$L$G at zero displacement field. 

Because the interaction Hamiltonian in this work accounts for the layer-dependence of the electron interactions, our calculations in the average scheme capture the internal interlayer Hartree potentials that act to screen the externally-applied displacement field. As a result, we are able to quantitatively recover the experimental displacement fields required in Ref.~\cite{LongJu2023FCIPentalayerGraphenehBN} to access the Chern insulator, as well as the large window of gapless states for smaller $V$. 

Our investigations for different numbers of layers $L$ and twist angles $\theta$ illustrate general trends for the position and electronic topology of the gapped phase at $V>0$. We find that the gapped region shifts to higher $V$ and has a tendency towards $|C|=1$ for smaller $L$ and larger $\theta$. This suggests that similar phenomenology as in Ref.~\cite{LongJu2023FCIPentalayerGraphenehBN} may be experimentally obtainable only for a restricted subset of the total space of possible R$L$G/hBN systems. For large $L$/small $\theta$, the Chern insulator is outcompeted by the topologically trivial insulator, while for small $L$/large $\theta$, the Chern insulator appears for high displacement fields $D$ that are out of reach for experiments. To expand the list of related candidate materials for correlated topological phases, it would be interesting to also compute the phase diagrams for doubly-aligned hBN/R$L$G/hBN, for which Ref.~\cite{MFCI-II} derived the single-particle continuum models and found higher Chern number bands.

Beyond mean-field theory, this paper also studies the low-lying collective modes of the topological and trivial insulating HF phases using TDHF calculations. Knowledge of the soft mode energies improves our understanding of the stability of the gapped states, whose mean-field charge gap is often a significant overestimate of the robustness (e.g.~against temperature) of correlation-induced phenomena, such as quantized anomalous transport, in experiments. Furthermore, the flavor structure of the soft modes provides insight into the spontaneous breaking of (approximate) continuous symmetries and the energy scales governing symmetry restoration. 

For both $|C|=0,1$ states, we identify two types of low-lying neutral excitations that play a central role in the gapped states. The first is an intervalley magnon with gap $\Omega_\text{valley}\lesssim 0.5\,\text{meV}$ ($1<\Omega_\text{valley}<4\,\text{meV}$) for the $C=0$ ($|C|=1$) phase, which is significantly reduced compared to the HF charge gap ($10-20\,\text{meV}$). For the Chern insulator, we argue that $\Omega_\text{valley}$ sets an important scale, since the proliferation of valley magnons and degradation of valley polarization also leads to the deterioration of the topological response. The second consists of two pseudophonons, with gaps $\Omega_{\text{ph.}1}$ and $\Omega_{\text{ph.}2}$, which are connected to the gapless Goldstone modes in the limit of vanishing moir\'e potential $\kappa_\text{hBN}=0$ where the Hamiltonian has continuous translation symmetry. Hence, the pseudophonons reflect the pinning strength of the moir\'e potential, which is required to make the state incompressible. Here, we find again a significant dependence on the interaction scheme. For the CN scheme where the conduction electrons barely couple in any way to the moir\'e potential, $\Omega_{\text{ph.}1},\Omega_{\text{ph.}2}\lesssim1\,\text{meV}$ remain relatively small. Since $\Omega_{\text{ph.}1},\Omega_{\text{ph.}2}<\Omega_\text{valley}$ for the Chern insulator, it could then be regarded as a topological Wigner crystal-like state. On the other hand, the pseudophonon gaps are substantially enhanced in the average scheme owing to the interaction-induced moir\'e landscape created by the occupied valence bands, and we find large regions in parameter space where $\Omega_{\text{ph.}1},\Omega_{\text{ph.}2}>\Omega_\text{valley}$. In this case, the Chern insulator should not be considered a Wigner crystal-like phase, since there are important energy scales (i.e.~the valley magnon gap) that undercut the moir\'e pinning scale.

\subsection{Perspectives}

An interesting future direction would be to examine the impact of the low-lying collective modes directly in the finite-temperature regime, for example using the exponential tensor renormalization group (XTRG)~\cite{Lin2022exciton,lu2023thermodynamic}. We also comment on the possible connections between our TDHF results at $\nu=1$, and the gapped phases at non-integer filling $\nu<1$. The $\mbf{q}=0$ pseudophonon wavefunction in the Chern insulator is mostly localized in momentum space on the boundary of the mBZ. This is because there are multiple low-lying conduction bands near $\tilde{K}_M,\tilde{K}'_M,\tilde{M}_M,\tilde{M}'_M$  which can be hybridized by the moir\'e potential. If the $\nu=1$ insulator is the parent state for the FCIs at lower densities, then these fractional states share the same mBZ geometry. Since FCIs have a momentum occupation $n(\mbf{k})$ that is relatively homogeneous throughout the mBZ, we expect that they should also sense the moir\'e pinning potential, which would explain their quantization along the density axis. 

Curiously, Ref.~\cite{LongJu2023FCIPentalayerGraphenehBN} also observed an extended topologically-trivial insulating region for $\nu\lesssim \frac{1}{2}$ that spans a \emph{continuous} range of density. Furthermore, the density range overlaps with the quantized $\nu=\frac{2}{5}$ FCI which occurs at a higher displacement field, where the moir\'e pinning is actually expected to be weaker. One candidate state put forward in Ref.~\cite{LongJu2023FCIPentalayerGraphenehBN} to explain the extended insulating region is a Wigner crystal that is decoupled from the moir\'e lattice. This state could be related to the $|C|=0$ insulator in our HF calculations that is favored by a smaller $\theta$, which translates to a smaller mBZ size and electron density. Also, the charge density of the occupied conduction band in the $|C|=0$ insulator is sharply peaked and forms a triangular lattice. We note that the Wigner crystal operates with a smaller reconstructed Brillouin zone (rBZ) than the mBZ set by the hBN-alignment angle $\theta$, meaning that in the moir\'eless limit, the lowest non-interacting conduction band states in the first rBZ only sample a subset of those in the first mBZ. For intermediate $V$, the lowest energy conduction states have momenta around $\tilde{\Gamma}_M$, which do not efficiently couple to the (interaction-induced) moir\'e potential. Hence, it is expected that the Wigner crystal only weakly experiences the moir\'e pinning potential for intermediate $V$ and small densities where the relevant low-energy single-particle conduction states are localized around $\tilde{\Gamma}_M$. 

Beyond the experiments of Ref.~\cite{LongJu2023FCIPentalayerGraphenehBN} in R5G/hBN, multiple studies have observed a variety of correlated phenomena, such as superconductivity, in other rhomhedral graphene systems~\cite{ref29, ref43, ref44, ref45, ref46, ref47,ref48,ref49,ref50,ref51,ref52,ref53,ref54,ref55,ref56}, including those that are not aligned to hBN. It is clear that this materials family presents a rich arena where interactions, multi-band physics, and electronic topology and geometry~\cite{PhysRevLett.124.167002,2021Natur.600..641L,yankowitz2022moire,2022arXiv220702312Z,mak2022semiconductor,mai2023interaction,2023arXiv230914340S,PhysRevB.103.205415,2022PhRvL.129d7601S,Bultinck2019GroundSA,2023PhRvB.107x5145S,2022arXiv221200030H,Phillips2023DQMCHaldaneModel,2023arXiv230105588W,Phillips2023DQMC, ding2023particle,mai2023interaction,fu2021flat,crepel2023topological,PhysRevLett.123.237002,PhysRevLett.124.167002,PhysRevB.101.060505,rossi2021quantum} are all integral to a fundamental understanding of the strongly-correlated physics.

{\it Note added:} During the preparation of this manuscript, several related theoretical works appeared on the arXiv~\cite{dong2023theory,zhou2023fractional,dong2023anomalous,guo2023theory}, which performed HF calculations on R$L$G/hBN at $\nu=+1$, as well as one-band exact diagonalization (ED) calculations at fractional fillings of the single occupied conduction HF band (obtained at $\nu=+1$). We discuss the main differences of these other works with our paper, focusing on the gapped $V>0$ regime of R5G/hBN at $\nu=+1$. Refs.~\cite{dong2023theory,zhou2023fractional,dong2023anomalous} performed HF numerics on a conduction-bands-projected model in the CN interaction scheme, where the valence bands do not affect the physics. Refs.~\cite{dong2023theory,dong2023anomalous} used a 2D interaction potential. In agreement with our calculations in the CN scheme, Refs.~\cite{zhou2023fractional,dong2023anomalous} found that the $|C|=1$ state in HF survived as the hBN coupling $\kappa_{\text{hBN}}$ was tuned to zero, and suggested that the Chern insulator should be viewed fundamentally as a topological Wigner crystal-like phase that spontaneously breaks the continuous translation symmetry. In this work, we quantitatively assess the role of the (approximate) continuous translation symmetry by computing the pseudophonon gap $\Omega_\text{ph.}$ using TDHF theory in the physical limit $\kappa_\text{hBN}=1$, and find the extrinsic moir\'e pinning scale $\Omega_\text{ph.}\lesssim 1\,\text{meV}$ in the CN scheme, which is lower than that of the other non-trivial collective modes. However, we show that in the average interaction scheme, $\Omega_\text{ph.}$ is significantly enhanced and can be larger than the valley magnon gap, implying that the moir\'e potential is not a weak perturbation in this case. In the average scheme, our HF phase diagrams show a qualitative difference between the two inequivalent stackings $\xi=0,1$, which was not considered in Refs.~\cite{dong2023theory,zhou2023fractional,dong2023anomalous}. The authors of Ref.~\cite{guo2023theory} first utilized an RG procedure to obtain an effective continuum model, and then performed HF calculations projected onto the lowest $(3+3)$ valence and conduction bands per spin/valley. Within the projected effective model, which includes the effects of interlayer screening, their calculations used an interaction scheme where the interactions are measured relative to the vacuum of the active subspace. They considered both stackings at the non-interacting level, but only used one stacking for the interacting calculations.

\section{Acknowledgments}
B. A. B.’s work was primarily supported by the the Simons Investigator Grant No. 404513,  by the Gordon and Betty Moore Foundation through Grant No. GBMF8685 towards the Princeton theory program, Office of Naval Research (ONR Grant No. N00014-20-1-2303), BSF Israel US foundation No. 2018226 and NSF-MERSEC DMR-2011750, Princeton Global Scholar and the European Union’s Horizon 2020 research and innovation program under Grant Agreement No 101017733 and from the European Research Council (ERC).
N.R. also acknowledges support from the QuantERA II Programme that has received funding from the European Union’s Horizon 2020 research and innovation programme under Grant Agreement No 101017733 and from the European Research Council (ERC) under the European Union’s Horizon 2020 Research and Innovation Programme (Grant Agreement No. 101020833). 
J. Y. is supported by the Gordon and Betty Moore Foundation through Grant No. GBMF8685 towards the Princeton theory program and through the Gordon and Betty Moore Foundation’s EPiQS Initiative (Grant No. GBMF11070) and by 
DOE Grant No. DE-SC0016239.
Y.H.K is supported by a postdoctoral research fellowship at
the Princeton Center for Theoretical Science.
J. H.-A. is supported by a Hertz Fellowship, with additional support from DOE Grant No. DE-SC0016239 by the Gordon and Betty Moore Foundation through Grant No. GBMF8685 towards the Princeton theory program, Office of Naval Research (ONR Grant No. N00014-20-1-2303), BSF Israel US foundation No. 2018226 and NSF-MERSEC DMR-2011750, Princeton Global Scholar and the European Union’s Horizon 2020 research and innovation programme under Grant Agreement No 101017733 and from the European Research Council (ERC).

%

\clearpage

\appendix

\onecolumngrid

\tableofcontents

\setcounter{figure}{0}
\let\oldthefigure\thefigure
\renewcommand{\thefigure}{S\oldthefigure}

\section{Single-particle model}
\label{app:SP_model}

\subsection{Review of the continuum model}

In this work, we use the single-particle model  discussed in \refcite{MFCI-II} which includes the effect of relaxation, trigonal distortion, moir\'e potential, and internal and external electric fields. This is a generalized version of the models introduced in Refs.~\cite{MacDonald2014R5GhBN,Kushino2014R5GhBN,Park2023RMGhBNChernFlatBands}. The parameters determined in the continuum model of \refcite{MFCI-II} are directly fitted to {\it ab initio} calculations which also include relaxation and internal electric fields. To keep this paper self-contained, we briefly review this model here; the reader can find the full details in \refcite{MFCI-II}.

The full single-particle model consists of four parts: the pristine rhombohedral $L$-layer graphene (R$L$G), the hBN layer, the moir\'e coupling between hBN and the R$L$G, and the displacement field.
As discussed in the \refcite{MFCI-II}, we will only consider one hBN in order to simulate the single moir\'e potential in the experiment~\cite{Ju2023PentalayerGraphenehBN}.
The continuum model for the pristine R$L$G has as basis $ c^\dagger_{\bsl{r}, l\sigma \eta s} $, where $\bsl{r}$ is the in-plane continuum position, $l=0,1,2,...,L-1$ is the layer index for $L$ total layers, $\sigma=A,B$ represents the sublattice, $\eta=\pm \K$ labels the valley, and $s=\uparrow,\downarrow$ is the spin index. 
In this section, we will neglect the spin index unless specified otherwise, since the model has spin SU$(2)$ symmetry.

In the $\K$ valley, the matrix Hamiltonian for R$L$G reads
\eqa{ 
\label{eq:H_K}
H_{\K}(\bsl{p}) &= \bpm
v_F\mbf{p} \cdot \pmb{\sigma}  & t^\dag(\mbf{p}) & t'^\dagger &   &\\
t(\mbf{p}) & \ddots & \ddots & t'^\dagger \\
t' & \ddots & v_F\mbf{p} \cdot \pmb{\sigma} & t^\dagger(\mbf{p})\\
& t' & t(\mbf{p})  & v_F\mbf{p} \cdot \pmb{\sigma}
\epm + H_{ISP}, 
}
where $\bsl{p}=-\ii \nabla$, $\bsl{\sigma}=(\sigma_x,\sigma_y)$ are Pauli matrices in the sublattice subspace, $t(\mathbf{p})$ and $t'$ are also $2\times 2$ matrices that carry the sublattice index:
\eq{
t(\mbf{k}) = -\bpm v_4 p_+ & -t_1 \\ v_3 p_- &  v_4 p_+ \epm, \qquad  \qquad t' = \bpm 0 & 0 \\ t_2 & 0 \epm\ ,
}
where $p_\pm = p_x \pm \ii p_y$,  $v_F$ is the Fermi velocity, $t_1,v_3,v_4$ are parameters describing hopping between consecutive layers, $t_2$ describes hopping between next-nearest layers, and the $2\times 2$ blocks in $H_{\K}(\bsl{p})$ are arranged according to the layer index.
Throughout the work, we will choose $v_3=v_4$.
$H_{ISP}$ is the inversion symmetric polarization, and characterizes the local chemical potential environment of each graphene layer:
\eq{
[H_{ISP}]_{l l'} = V_{ISP} \delta_{ll'} \left| l - \frac{L-1}{2} \right| \sigma_0\ ,
}
where $V_{ISP} = 16.65$meV is determined by fitting to the DFT calculated bands in \refcite{MFCI-II}, and $\sigma_0$ is the identity $2\times 2$ matrix for the sublattice index.

\begin{table}[t]
\centering
\begin{tabular}{ c|c c c c| c c c } 
 & $v_F$& $v_3$  & $t_1$  & $t_2$  & $V_0$ & $V_1$ & $\psi_{\xi}$ \\ 
 \hline
 \text{$L=3$, $\xi $=1} & 542.1 & 34. & 355.16 & -7 & 0 & 5.54 & \text{16.55${}^{\circ}$} \\
 \text{$L=4$, $\xi $=1} & 542.1 & 34. & 355.16 & -7 & 1.44 & 6.91 & \text{16.55${}^{\circ}$} \\
 \text{$L=5$, $\xi $=1} & 542.1 & 34. & 355.16 & -7 & 1.50 & 7.37 & \text{16.55${}^{\circ}$} \\
 \text{$L=6$, $\xi $=1} & 542.1 & 34. & 355.16 & -7 & 1.56 & 7.80 & \text{16.55${}^{\circ}$} \\
 \text{$L=7$, $\xi $=1} & 542.1 & 34. & 355.16 & -7 & 1.47 & 7.93 & \text{16.55${}^{\circ}$} \\
 \hline
 \text{$L=3$, $\xi $=0} & 542.1 & 34. & 355.16 & -7 & 6.13 & 5.95 & \text{-136.55${}^{\circ}$} \\
 \text{$L=4$, $\xi $=0} & 542.1 & 34. & 355.16 & -7 & 7.16 & 6.65 & \text{-136.55${}^{\circ}$} \\
 \text{$L=5$, $\xi $=0} & 542.1 & 34. & 355.16 & -7 & 7.19 & 7.49 & \text{-136.55${}^{\circ}$} \\
 \text{$L=6$, $\xi $=0} & 542.1 & 34. & 355.16 & -7 & 7.12 & 7.16 & \text{-136.55${}^{\circ}$} \\
 \text{$L=7$, $\xi $=0} & 542.1 & 34. & 355.16 & -7 & 7.00 & 7.37 & \text{-136.55${}^{\circ}$} \\
 \hline
\end{tabular}
 \caption{Parameter values of the full model for $L=3,4,5,6,7$ layers determined in \refcite{MFCI-II}. Here $v_F, v_3=v_4$ are reported in meV$\cdot$nm, while $t_1, t_2, V_0$ and $V_1$ are in meV.  
 }
    \label{tab:parameters_full}
 \end{table}

The hBN part of the Hamiltonian can be approximated as
\eqa{
\label{eq:H_hBN}
H_{hBN, \xi} &= \sigma_1^\xi \bpm V_B  & \\ & V_N \epm \sigma_1^\xi, 
}
where $V_B = 3352 \text{meV}$ and $V_N =-1388 \text{meV}$, and $\xi = 0$ and $\xi = 1$ stand for the two orientations of the hBN related by 180$^\circ$ rotation (see Fig.~\ref{fig:moireconventions}).
Owing to the large onsite potential difference between the boron ($V_B$) and nitrogen ($V_N$), we neglect the momentum dependence of the hBN part of the Hamiltonian as it will have little effect on the low-energy physics of R$L$G which takes place around zero energy.

Combined with the moir\'e coupling between the R$L$G and hBN, the full Hamiltonian without displacement field has the form
\eqa{
\mat{ H_{hBN, \xi}  & \tilde{T}_b(\mbf{r}) \\
\tilde{T}_b^\dag(\mbf{r})  & H_{\K}(- i \pmb{\nabla})  }\ ,
}
where $\xi = 0$ and $\xi = 1$ correspond to the stacking configurations where carbon-A,B is nearly vertically aligned with B,N and N,B in the AA region, respectively.
$\tilde{T}_b(\mbf{r})$ is the moir\'e coupling which acts only on the bottom graphene layer and reads
\eq{
\label{eq:Ttilde_b}
\tilde{T}_b(\mbf{r}) = \mat{ T_b(\mbf{r}) & 0_{2 \times 2} & \cdots }\ .
}
where $T_b(\mbf{r})$ is the $2\times 2$ moir\'e coupling between one layer of hBN and one layer of graphene.

We note that we will eventually ``integrate out" the hBN to get an effective moir\'e potential that only couples to the R$L$G modes. 
To derive the form of the effective potential, we only need to explicitly write out the first harmonic terms in $T_b(\mbf{r})$
\eq{
\label{eq:T_b}
T_{b}(\mbf{r})  = \sum_{j=1}^3 e^{i \mbf{q}_j \cdot \mbf{r}} T_j + ...\ ,
}
where ``$...$" contains higher harmonics,
\eq{
\label{eq:qvecmain}
\mbf{q}_1 = \mbf{K}_G - \mbf{K}_{hBN} = \frac{4\pi}{3 a_G}\left(1 - \frac{R(-\th)}{1+0.01673} \right)\hat{x}
}
with its $C_3$ partners $\mbf{q}_{j+1} = R(\frac{2\pi}{3}) \mbf{q}_j$, $\th$ is the twist angle, $R(\th)$ is the counterclockwise rotation matrix, $\mbf{K}_G$ and $\mbf{K}_{hBN}$ are the valley $\eta=K$ Dirac momenta of graphene and hBN, $a_G = 2.46\AA$ is the graphene lattice constant, and $(1+0.01673)a_G$ is the hBN lattice constant.
As the low-energy states of the R$L$G are localized at the $A$ sublattice on the bottom layer, only one of the four matrix elements of $T_{b}(\mbf{r})$ is meaningful, allowing us to choose the following simplified form for $T_j$
\eqa{
\label{eq:T_T'}
& T_j = w \mat{ 1 & e^{-\ii \frac{2\pi (j-1)}{3}} \\ e^{\ii \frac{2\pi (j-1)}{3}} & 1 } .
}
The displacement field term reads
\eq{
\label{eq:H_D_hBN}
\null [H_{D, hBN}]_{l \sigma,l' \sigma'} =V \left(l - \frac{n-1}{2} \right) \delta_{ll'} \delta_{\sigma \sigma'}\ ,
}
where we choose $l=-1$ for the hBN layer, and $l=0,\ldots,L-1$ for the graphene layers.
The relation between $V$ and the displacement field $D$ reads $V= e D  d/\epsilon$, where $e$ is the charge of the electron, and $d\approx 3.33\AA$ is the interlayer distance in rhombohedral graphene.

\begin{figure}[t]
    \centering
    \includegraphics[width=\columnwidth]{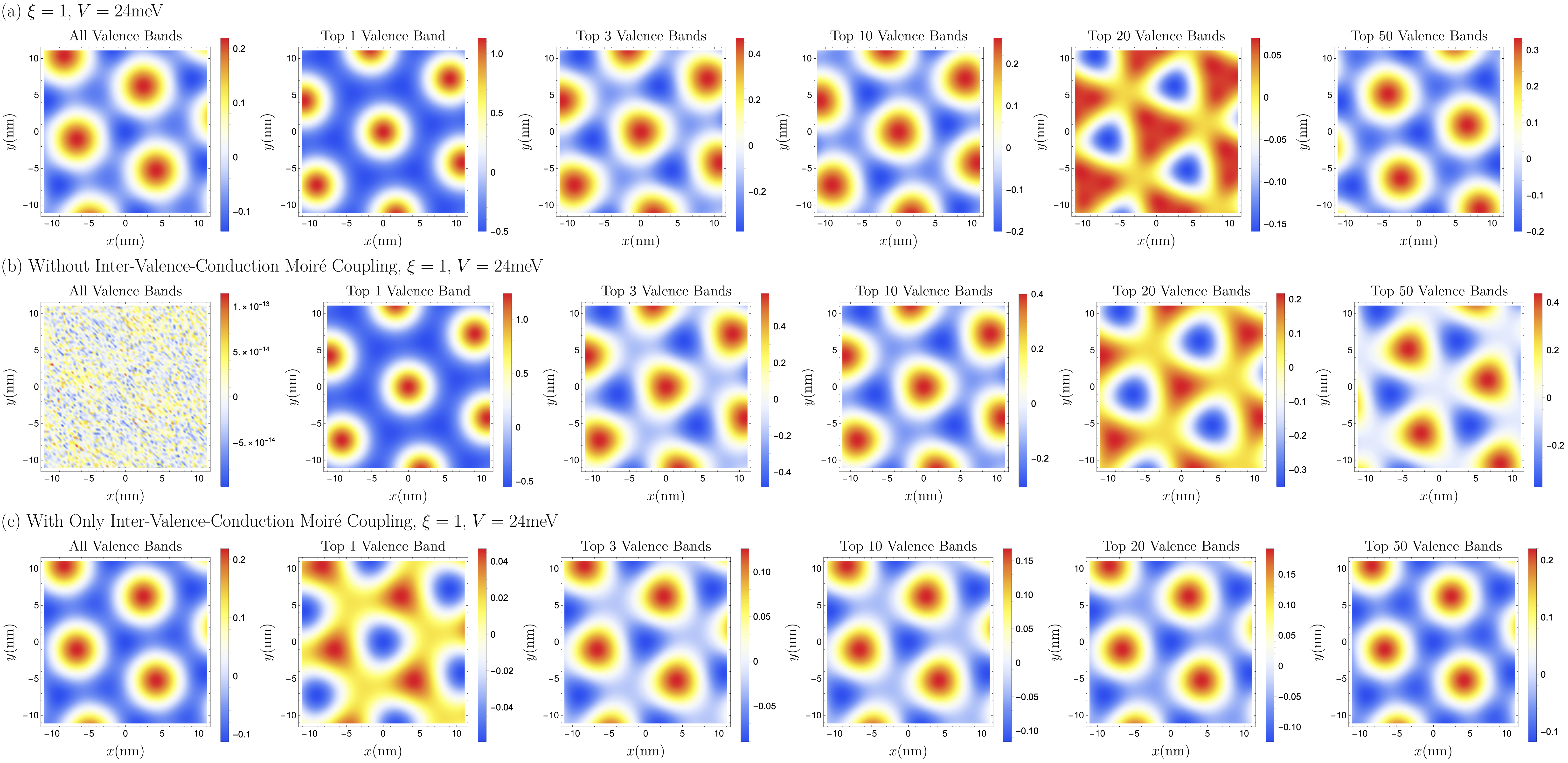}
    \caption{
    The dimensionless density fluctuation $\rho(\bsl{r})-\langle \rho(\bsl{r})$ (\eqnref{eq:density_profile}, color bar) of the various number of valence bands in the $\K$ valley for (a) the full $2 L\times 2 L $ continuum model (\eqnref{eq:H_K_nohBN}), for (b) the $2 L\times 2 L $ continuum model without the inter-valence-conduction moir\'e coupling, and for (c) the $2 L\times 2 L $ continuum model with only the inter-valence-conduction moir\'e coupling (See the definition of the intra-valence-conduction moir\'e coupling in \eqnref{eq:H_K_nohBN_moire_split}).
    We choose 19 reciprocal lattice vectors (i.e.~4 shells) for the plot, which gives $95$ valence bands per valley per spin.
    We choose $\xi=1$, $V=24$meV, $L=5$, and $\theta=0.77^\circ$ for this plot. Note that the colorbars are not the same for each plot. The rapidly-fluctuating features in the leftmost plot of (b) are due to machine-level numerical precision errors (see the corresponding colorbar scale).
    }
    \label{fig:charge_density_xi1}
\end{figure}

\begin{figure}[t]
    \centering
    \includegraphics[width=\columnwidth]{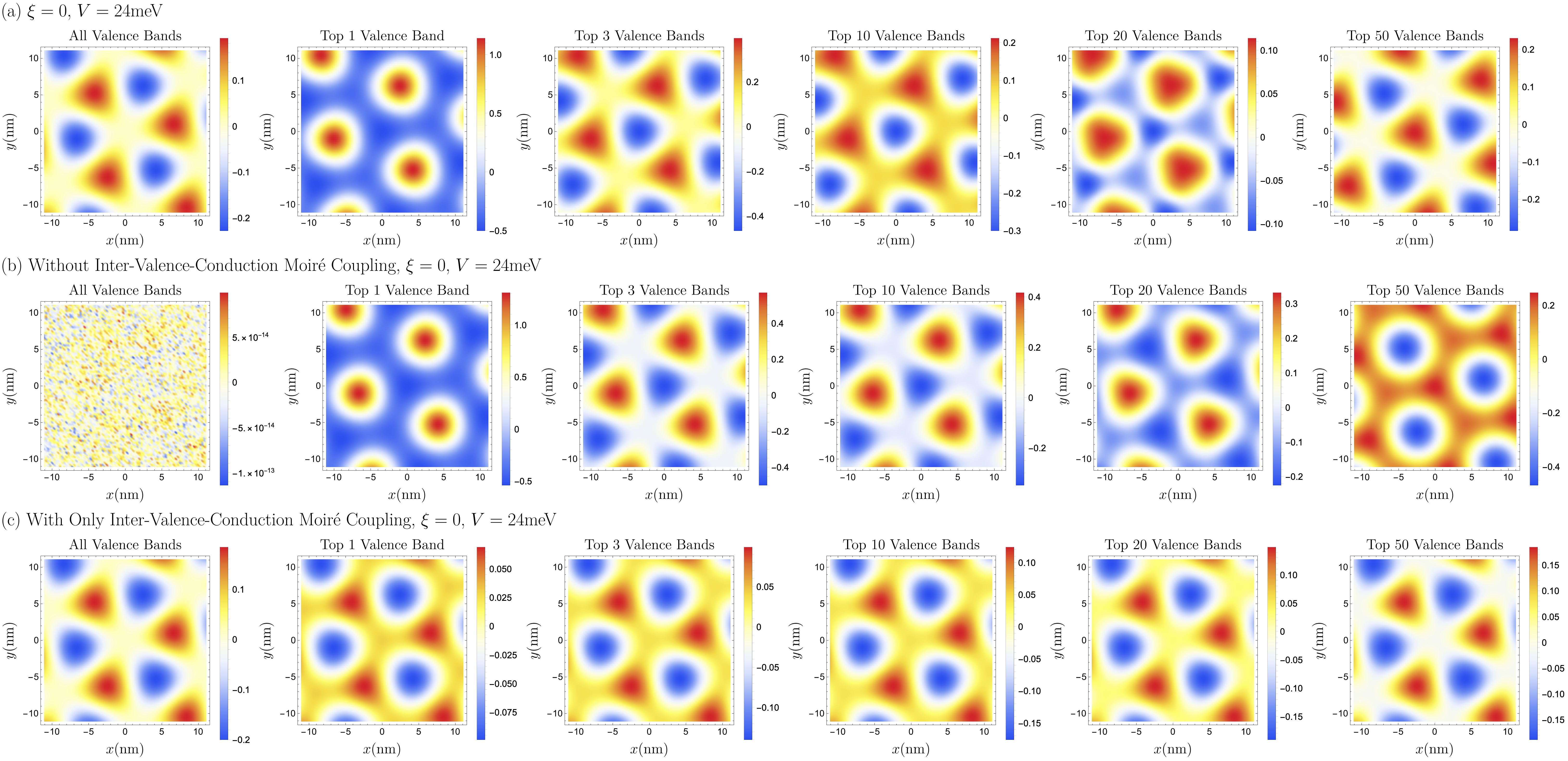}
    \caption{
     The dimensionless density profile (\eqnref{eq:density_profile}, color bar) of the various number of valence bands in the $\K$ valley for (a) the full $2 L\times 2 L $ continuum model (\eqnref{eq:H_K_nohBN}), for (b) the $2 L\times 2 L $ continuum model without the inter-valence-conduction moir\'e coupling, and for (c) the $2 L\times 2 L $ continuum model with only the inter-valence-conduction moir\'e coupling (See the definition of the intra-valence-conduction moir\'e coupling in \eqnref{eq:H_K_nohBN_moire_split}).
    We choose 19 reciprocal lattice vectors (i.e.~4 shells) for the plot, which gives $95$ valence bands per valley per spin.
    We choose $\xi=0$, $V=24$meV, $L=5$, and $\theta=0.77^\circ$ for this plot. Note that the colorbars are not the same for each plot. The rapidly-fluctuating features in the leftmost plot of (b) are due to machine-level numerical precision errors (see the corresponding colorbar scale).
    }
    \label{fig:charge_density_xi0}
\end{figure}

In summary, the single-particle continuum model in the $\K$ valley reads
\eq{
H_{\K, hBN, \xi}(\mbf{r}) = \mat{ H_{hBN, \xi}  & \tilde{T}_b(\mbf{r}) \\
\tilde{T}_b^\dag(\mbf{r})  & H_{\K}(- i \pmb{\nabla})  } + H_{D, hBN}\ , 
}
where $H_{hBN, \xi}$ is in Eq.\,\eqref{eq:H_hBN}, $H_{\K}(\bsl{p})$ is in \eqnref{eq:H_K}, $\tilde{T}_b(\mbf{r})$ (up to the first harmonics) is in \eqnref{eq:Ttilde_b} with \eqnref{eq:T_b} and \eqnref{eq:T_T'}, and $H_{D, hBN}$ is in \eqnref{eq:H_D_hBN}.

As hBN states have much higher energies, we can integrate them out to obtain an effective model with only R$L$G states.
Specifically, we integrate out the hBN substrate by  perturbation theory at zero displacement field, and obtain the following effective moir\'e potential:
\eqa{
\label{eq:Vxifinal}
V_\xi(\mbf{r}) &= V_0 + \left[V_1 e^{i\psi_\xi}\sum_{j=1}^3 e^{i \mbf{g}_j\cdot\mbf{r}}\bpm 1& \omega^{-j} \\ \omega^{j+1} &\omega \epm + h.c.\right]\ ,
}
which only acts on the bottom layer of graphene, and $\mbf{g}_j = R(\frac{2\pi}{3}(j-1)) (\mbf{q}_2-\mbf{q}_3)$ for $j = 1,2,3$. 
Here we only keep terms up to the first harmonics in the effective moir\'e potential $V_\xi(\mbf{r})$, which is justified because the momentum states of R$L$G that differ by larger $\mbf{G}$ have a higher difference in kinetic energies.
The relation between $V_\xi(\bsl{r})$, $T_b(\bsl{r})$ and $H_{hBN,\xi}$ is given by 
\eqa{
& V_0 = -3 |w|^2\lp \frac{1}{V_B}+\frac{1}{V_N} \rp + ... \\
& V_{1} e^{i \psi_{\xi=0}} = - |w|^2\lp\frac{1}{V_B}+ e^{-\ii 2\pi/3} \frac{1}{V_N}\rp +  ... \\
& V_{1} e^{i \psi_{\xi=1}} = - |w|^2\lp\frac{1}{V_N}+ e^{-\ii 2\pi/3} \frac{1}{V_B}\rp + ... \ ,
}
where ``$...$" indicates contributions to $V_\xi(\bsl{r})$ from the higher harmonics of $T_b(\bsl{r})$.
Note that the contribution from the higher harmonics of $T_b(\bsl{r})$ will essentially make $V_0$, $V_1$ and $\psi_{\xi}$ independent of each other, and \refcite{MFCI-II} effectively takes into account this contribution by directly fitting $V_0$, $V_1$ and $\psi_{\xi}$ to the {\it ab initio} band structure.
The resulting parameter values are listed in \tabref{tab:parameters_full}.
As a result, the Hamiltonian after integrating out hBN reads
\eq{
\label{eq:H_K_nohBN}
H_{\K, \xi}(\mbf{r}) = H_{\K}(- i \pmb{\nabla}) + H_{\text{moir\'e},\xi}(\bsl{r})+ H_{D} \ ,
}
where
\eq{
\label{eq:H_V}
\null [H_{\text{moir\'e},\xi}(\bsl{r})]_{l \sigma,l' \sigma'} = \left[ V_\xi(\mbf{r}) \right]_{\sigma\sigma} \delta_{l0}\delta_{ll'}\ ,
}
\eq{
\label{eq:H_D}
\null [H_{D}]_{l \sigma,l' \sigma'}= V_l\delta_{ll'}\delta_{\sigma\sigma'} = V \left(l - \frac{n-1}{2} \right) \delta_{ll'} \delta_{\sigma \sigma'}\ ,
}
$V_\xi(\mbf{r})$ is in \eqnref{eq:Vxifinal}, and $H_{\K}(\bsl{p})$ is in \eqnref{eq:H_K}.
$H_{\K, \xi}(\mbf{r})$ in \eqnref{eq:H_K_nohBN} has spinless $C_3$ symmetry, which is represented as
\eq{
C_3 c^\dagger_{\bsl{r},l\sigma \K s } C_3^{-1} = \sum_{\sigma'}c^\dagger_{C_3\bsl{r},l\sigma \K s } e^{i\frac{2\pi}{3}(l-1-\lfloor\frac{n}{2}\rfloor)} \left[ e^{-i\frac{\pi}{3}\sigma_3} \right]_{\sigma'\sigma}\ .
}
The Hamiltonian at $\K'$ valley can be obtained by the spinless time-reversal (TR) symmetry, which is represented as
\eq{
\TR c^\dagger_{\bsl{r},l\sigma \K s } \TR^{-1} = c^\dagger_{\bsl{r},l\sigma \K' s }\ .
}

In the practical calculation, we transform the Hamiltonian into the momentum space via
\eq{
c^\dagger_{\bsl{k},\bsl{G},l\sigma\eta s} = \frac{1}{\sqrt{\V}} \int d^2 r e^{\ii (\bsl{k}+\bsl{G})\cdot\bsl{r}} 
c^\dagger_{\bsl{r},l\sigma\eta s} \ ,
}
where $\V$ is the area of the whole sample, $\bsl{k}$ is in the first moir\'e Brillouin zone (mBZ), and $\bsl{G}$ labels the reciprocal moir\'e lattice vectors.
In the momentum space, the matrix Hamiltonian reads
\eq{
\label{eq:H_K_nohBN_k}
\left[ h^{\K}(\bsl{k}) \right]_{\bsl{G}\bsl{G}'} = H_{\K}(\bsl{k}+\bsl{G})\delta_{\bsl{G}\bsl{G}'} + \left[ H_{\text{moir\'e},\xi} \right]_{\bsl{G}\bsl{G}'}+ H_{D} \delta_{\bsl{G}\bsl{G}'}\ ,
}
where
\eq{
\left[H_{\text{moir\'e},\xi} \right]_{\bsl{G}\bsl{G}',ll',\alpha\beta} = \left[  V_0 \delta_{\bsl{G}\bsl{G}'} + V_1 e^{\ii\psi_\xi}\sum_{j=1}^3 \delta_{\bsl{G},\bsl{G}'+\bsl{g}_j} \bpm 1& \omega^{-j} \\ \omega^{j+1} &\omega \epm + V_1 e^{-\ii\psi_\xi}\sum_{j=1}^3 \delta_{\bsl{G},\bsl{G}'-\bsl{g}_j} \bpm 1& \omega^{-j-1} \\ \omega^{j} &\omega^* \epm  \right]_{\sigma\sigma'} \delta_{l0}\delta_{ll'}\ .
}
The Hamiltonian in the $-K$ valley can be obtained via time-reversal symmetry.
The eigenequation reads
\eq{
h^{\eta }(\bsl{k}) U^\eta_{n}(\mbf{k}) = E_{n}^\eta(\bsl{k}) U^\eta_{n}(\mbf{k}).
}

In this work, we will use \eqnref{eq:H_K_nohBN} for R$L$G/hBN with the parameter values in \tabref{tab:parameters_full}.
In \figref{fig:SP_bands} of the main text, we show the band structures of the R5G/hBN single-particle model (\eqnref{eq:H_K_nohBN}) in valley $K$ for the parameters in \tabref{tab:parameters_full} and for $\theta=0.77^\circ$.
In our continuum model, we always set the microscopic graphene Dirac momentum $\bsl{\K}_G$ to fold onto $\tilde{\Gamma}_M$ regardless of the twist angle $\theta$, while in the DFT calculation $\bsl{K}_D$ can be folded to $\Gamma_M,\K_M$ or $\K_M'$ of the DFT Brillouin zone depending on the twist angle as discussed in \refcite{MFCI-II}.

\subsection{Charge density and the issue of band cutoff}
\label{app:charge_density_band_cutoff}

Consider positive interlayer potentials $V>0$ where the lowest conduction bands are polarized away from the hBN layer and only weakly feel the hBN coupling at a non-interacting level. Indeed, the dispersion and wavefunctions of the lowest conduction bands are very similar to those of isolated R$L$G folded into the mBZ~\cite{MFCI-II}. It would be tempting to therefore conclude that the interacting physics at electron-doping $\nu=+1$ is only weakly dependent on the hBN coupling, and hence the stacking $\xi$. However, many of the valence bands, such as those closest to the conduction-valence band gap opened by $V$, are polarized near the hBN layer. As a result, the charge density of the subspace of filled valence bands, which we refer to as the \emph{charge background}, is spatially inhomogeneous with periodicity of the moir\'e unit cell (see Figs.~\ref{fig:charge_density_xi1} and \ref{fig:charge_density_xi0}). As will be discussed in App.~\ref{secapp:int_HF}, depending on how interactions are incorporated into the Hamiltonian, this charge background can generate an interaction-induced moir\'e potential acting on the conduction bands, for instance through the electrostatic Hartree contribution. Hence, it is important to accurately capture the moir\'e-periodic charge background of the valence bands in order to understand the phase diagram at $\nu=+1$, as well as the degree to which the low-energy conduction electrons are sensitive to the physics of the hBN moir\'e coupling.

When performing calculations involving a single-particle continuum Hamiltonian such as that described by Eq.~\ref{eq:H_K_nohBN_k}, we must first specify a single-particle Hilbert space $\mathscr{H}$ consisting of a set of $N_\mbf{G}$ continuum plane waves per mBZ momentum $\mbf{k}$ and flavor (i.e.~spin and valley). $N_\mbf{G}$ depends on the number of RLVs that fit within the plane wave cutoff centered on each microscopic valley $\eta$. Diagonalization of the continuum Hamiltonian leads to $2LN_\mbf{G}$ bands per flavor (the factor of 2 accounts for sublattice), of which half are valence bands. A natural question is whether it is possible to  accurately capture the total charge background of the valence bands by considering just a small number $n_\text{cut}\ll LN_\mbf{G}$ of low-energy valence bands, and ignoring the contributions from other remote bands. We will show below that this is not possible, and we must choose $n_\text{cut}=LN_\mbf{G}$, i.e.~include all valence bands within the plane wave cutoff. The reason is the ``intra-valence" moir\'e coupling as elaborated in the following.

According to \eqnref{eq:H_K_nohBN_k} in the momentum space, the matrix Hamiltonian can be split into a moir\'eless part $H_{\K,0}$ and the moir\'e part $H_{\text{moir\'e},\xi}$:
\eq{
h^{\K}(\bsl{k}) = H_{\K,0}(\bsl{k}) + H_{\text{moir\'e},\xi}\ .
}
At large interlayer potentials $V$, $H_{\K,0}(\bsl{k})$ is gapped at charge neutrality, and we can diagonalize it to obtain the projector onto the space of all moir\'eless valence bands $P_{\K,0,val}(\bsl{k})$ and the projector onto the space of all moir\'eless  conduction bands $P_{\K,0,cond}(\bsl{k})$.
As $P_{\K,0,val}(\bsl{k})+P_{\K,0,cond}(\bsl{k}) = \mathds{1}$, the moir\'e term can be expressed as 
\eqa{
\label{eq:H_K_nohBN_moire_split}
H_{\text{moir\'e},\xi} & = P_{\K,0,val}(\bsl{k}) H_{\text{moir\'e},\xi}  P_{\K,0,val}(\bsl{k}) + P_{\K,0,val}(\bsl{k}) H_{\text{moir\'e},\xi}  P_{\K,0,cond}(\bsl{k}) \\
& \quad + P_{\K,0,cond}(\bsl{k}) H_{\text{moir\'e},\xi}  P_{\K,0,val}(\bsl{k}) + P_{\K,0,cond}(\bsl{k}) H_{\text{moir\'e},\xi}  P_{\K,0,cond}(\bsl{k}),
}
where $P_{\K,0,val}(\bsl{k}) H_{\text{moir\'e},\xi}  P_{\K,0,val}(\bsl{k})$ is the intra-valence moir\'e coupling, $P_{\K,0,cond}(\bsl{k}) H_{\text{moir\'e},\xi}  P_{\K,0,cond}(\bsl{k})$ is the intra-conduction moir\'e coupling, and $P_{\K,0,val}(\bsl{k}) H_{\text{moir\'e},\xi}  P_{\K,0,cond}(\bsl{k}) + P_{\K,0,cond}(\bsl{k}) H_{\text{moir\'e},\xi}  P_{\K,0,val}(\bsl{k})$ is the inter-valence-conduction moir\'e coupling.

We define the charge fluctuation $\Delta \rho_{n_\text{cut}}(\mbf{r})$ as the real-space density profile of the charge background of valence bands, measured relative to its real-space average, i.e.~
\eq{
\label{eq:density_profile}
\Delta\rho_{n_\text{cut}}(\bsl{r}) = \rho_{n_\text{cut}}(\bsl{r}) - \langle \rho_{n_\text{cut}}(\bsl{r}) \rangle \ ,
}
where
\eq{\label{appeq:rhoncut}
\rho_{n_\text{cut}}(\bsl{r}) = \frac{1}{N}\sum_{\bsl{k},l,\sigma}\sum_n{}^{'}\left| \sum_{\bsl{G}} e^{\ii (\bsl{k}+\bsl{G}) \cdot\bsl{r}} U^{\K}_{n,\bsl{G}l\sigma}(\bsl{k})\right|^2
}
is the product of the real-space particle number density (per spin at $\K$ valley) and the area of the moir\'e unit cell $\Omega$, and
\eq{
\langle \rho_{n_\text{cut}}(\bsl{r}) \rangle = \frac{1}{\Omega} \int d^2 r \rho(\bsl{r}),
}
where the real space integral is taken over the moir\'e unit cell. The primed summation $\sum^{'}_n$ in Eq.~\ref{appeq:rhoncut} runs over the top $n_\text{cut}$ valence bands.

The real-space modulation of the charge background of all valence bands ($n_\text{cut}=LN_\mbf{G}$) relies on the inter-valence-conduction moir\'e coupling---without it, the valence band subspace is the same as that of the moir\'eless one, which has zero charge fluctuation $\Delta \rho_{n_\text{cut}}(\mbf{r})$ if we include all valence bands (see \figref{fig:charge_density_xi1}(b) and \figref{fig:charge_density_xi0}(b)).
However, if we only include a partial set of valence bands $(n_\text{cut}<LN_\mbf{G})$, we can still see a non-zero charge fluctuation $\Delta \rho_{n_\text{cut}}(\mbf{r})$ even if we neglect the inter-valence-conduction moir\'e coupling. We can intuitively understand this by first considering the folded valence bands of the moir\'eless part $H_{K,0}$, which contains many band crossings. The intra-valence moir\'e coupling gaps these band crossings, such that the Bloch functions near these gapped band crossings develop real-space inhomogeneity within the moir\'e unit cell. If we only consider the total charge density of a subset $n_\text{cut}$ of these reconstructed valence bands, then $\Delta \rho_{n_\text{cut}}(\mbf{r})$ is generally non-vanishing. To see this, consider the case where the band cutoff $n_\text{cut}$ of the charge fluctuation $\Delta \rho_{n_\text{cut}}(\mbf{r})$ intersects the middle of a gapped band crossing at some mBZ momentum $\mbf{k}_0$. Without any moir\'e coupling, at the cutoff $n_\text{cut}$, there are two degenerate states at $\mbf{k}_0$ which are plane waves carrying different RLV momenta and no charge fluctuations. When a small intra-valence moir\'e coupling is introduced, these two states hybridize to form moir\'e Bloch functions which individually have charge fluctuations, while the sum of their charge fluctuations vanish. This hybridization not only happens at the band touching point, but also happens for the momenta away from it as long as the gap between the two states is not much larger than the moir\'e coupling. However, because the band cutoff $n_\text{cut}$ on $\Delta \rho_{n_\text{cut}}(\mbf{r})$ only keeps one of the two states at each momentum with considerable hybridization, the resulting $\Delta \rho_{n_\text{cut}}(\mbf{r})$ is spatially inhomogeneous.

\begin{figure}[t]
    \centering
    \includegraphics[width=\columnwidth]{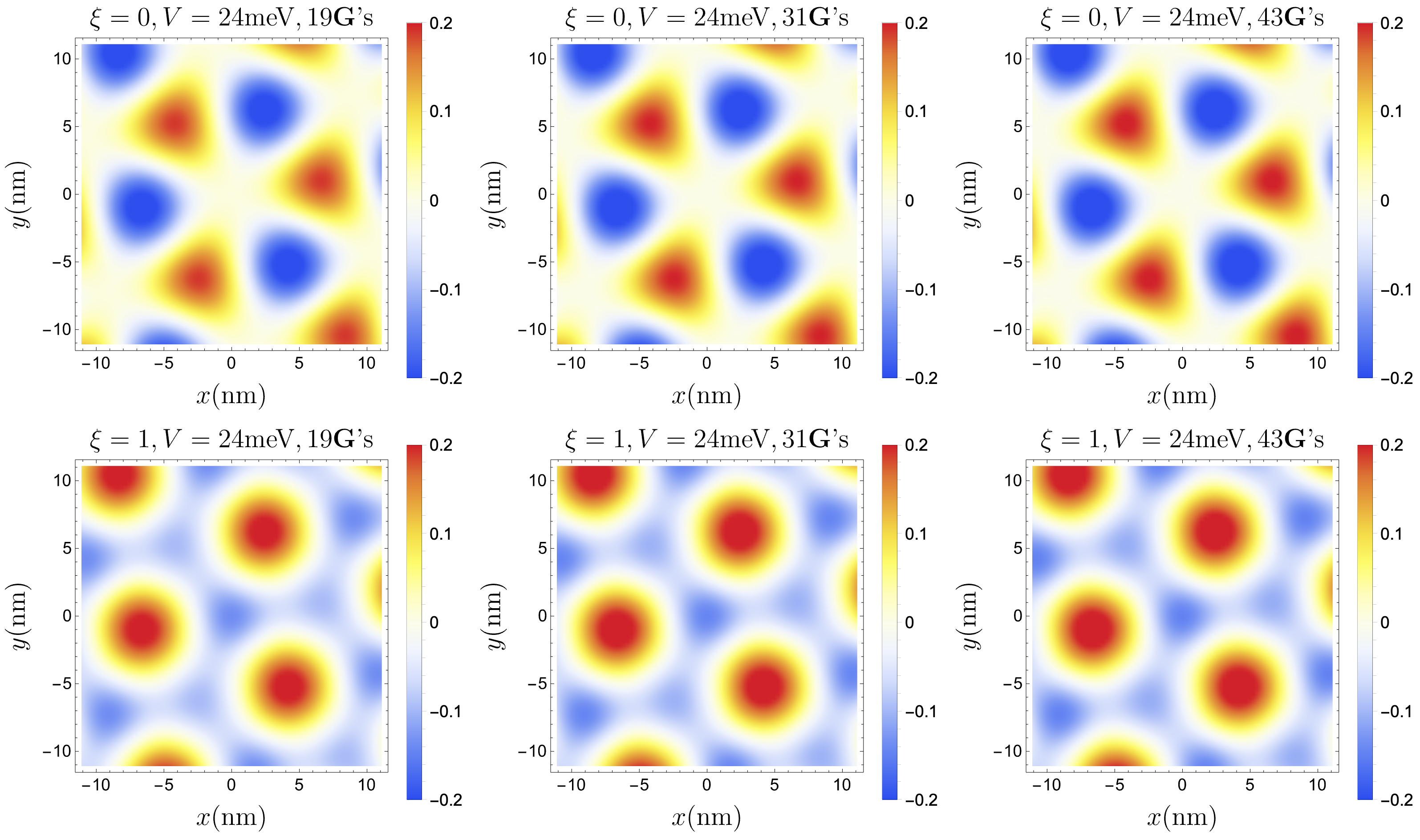}
    \caption{
     The dimensionless density profile (\eqnref{eq:density_profile}, color bar) of the various number of reciprocal lattice vectors $\bsl{G}$'s in the $\K$ valley.
   19, 31 and 43 $\bsl{G}$'s are for 4, 5 and 7 shells, respectively.
    We choose  $L=5$, and $\theta=0.77^\circ$ for this plot.
    }
    \label{fig:charge_density_Gnumber}
\end{figure}

Therefore, if we neglect the intra-valence and intra-conduction moir\'e couplings and only keep the inter-valence-conduction moir\'e coupling, we should expect the following results:
\begin{enumerate}
    \item The charge background of all valence bands is extremely similar to that of the complete model for considerably large $V$ because the intra-valence (intra-conduction) moir\'e coupling only approximately hybridizes states within the moir\'eless valence (conduction) subspace.
    \item The charge background of the top $n_\text{cut}$ valence bands converges to that of all valence bands quickly as $n_\text{cut}$ increases. This is because for large $n_\text{cut}$, the valence band states not included in the $n_\text{cut}$ band cutoff are far in energy from the conduction bands, and hence the inter-valence-conduction moir\'e coupling is weak in hybridizing them.
\end{enumerate}
Such expectations are verified numerically in \figref{fig:charge_density_xi1}(c) and \figref{fig:charge_density_xi0}(c).
In summary, due to the intra-valence moir\'e coupling, including all valence bands ($n_\text{cut}=LN_\mbf{G}$) is the most reliable way to compute the charge background. We have checked that the charge background of all valence bands converges quickly with respect to the plane wave cutoff which determines the number of plane waves $N_\mbf{G}$ in the single-particle Hilbert space $\mathscr{H}$, as shown in \figref{fig:charge_density_Gnumber}. This is expected because the deep remote valence bands are too far detuned in energy from the conduction bands to hybridize with them.

\section{Interaction Hamiltonian and projection}\label{secapp:int_HF}

In this section, we discuss how to define the interacting Hamiltonian, and how to perform, in our calculations, projection into a restricted set of bands. We first explain the `average' interaction scheme with purely two-dimensional interactions in App.~\ref{secapp:average_scheme}, and discuss how to perform projection in App.~\ref{secapp:projection}. We introduce more general interaction schemes in App.~\ref{secapp:general_scheme}. In App.~\ref{secapp:layer_int}, we discuss the generalization to layer-dependent interactions. Finally, in App.~\ref{secapp:screening} we explain the self-consistent screening of the interlayer potential.

\subsection{Average scheme}\label{secapp:average_scheme}
Consider the particle-hole symmetric interaction term
\bea\label{appeq:Hint_start}
\hat{H}_\text{int} &= \frac{1}{2 N} \sum_{\mbf{G} \mbf{q}} \frac{V(\mbf{q}+\mbf{G})}{\Omega} \delta \hat{\rho}_{\mbf{q}+\mbf{G}} \delta \hat{\rho}_{-\mbf{q}-\mbf{G}}
\eea
where $N$ is the number of moir\'e unit cells with area $\Omega$, $V(\mbf{q})$ is the Fourier transform of the screened Coulomb interaction and $\delta \hat{\rho}_{\mbf{q}+\mbf{G}}$ is the Fourier transformed particle-hole symmetrized density
\bea
\label{appeq:deltarhounproj}
\delta \hat{\rho}_{\mbf{q}+\mbf{G}} &= \sum_{\eta s,\mbf{G}'\al} \sum_{\mbf{k} } (c^\dag_{\mbf{k}+\mbf{q},\mbf{G}+\mbf{G}', \al \eta s} c_{\mbf{k},\mbf{G}', \al \eta s} - \frac{1}{2} \delta_{\mbf{q},0} \delta_{\mbf{G},\mbf{0}} ),
\eea
where $\alpha$ is a composite index for the microsopic layer $l$ and sublattice $\sigma$. $c^\dag_{\mbf{k},\mbf{G}, \al \eta s}$ is a creation operator for a plane wave with layer-sublattice $\alpha$, valley $\eta$, spin $s$, and momentum $\mbf{k}+\mbf{G}$, measured with respect to the R$L$G Dirac momentum $K_G$ in valley $\eta$.

We will rewrite the interaction in a basis of bands determined by some effective single-particle model $\hat{H}_0$ (not necessarily the physical non-interacting model $\hat{H}_\text{s.p.}$), which is
\bea
\hat{H}_0 = \sum_\mbf{k} \sum_{\mbf{G} \al,\mbf{G}' \be,\eta,s} c^\dag_{\mbf{k},\mbf{G},\al \eta s} h^\eta_{\mbf{G} \al , \mbf{G}' \be}(\mbf{k}) c_{\mbf{k},\mbf{G}',\be \eta s}
\eea
employing periodic boundary conditions. We diagonalize $\hat{H}_0$ to obtain the band basis
\bea
 \sum_{\mbf{G}\alpha, \mbf{G}'\beta} U^{\eta *}_{\mbf{G}\alpha, m} (\mbf{k}) h^\eta_{\mbf{G}\alpha, \mbf{G}'\beta}(\mbf{k}) U^\eta_{\mbf{G}'\beta,n}(\mbf{k}) &= \delta_{mn} E_n^\eta(\mbf{k})
\eea
thereby defining the band operators
\bea\label{appeq:U}
c^\dag_{\mbf{k},n,\eta,s} &= \sum_{\mbf{G}\alpha} c^\dag_{\mbf{k},\mbf{G},\alpha\eta s} U^\eta_{\mbf{G}\alpha,n}(\mbf{k})
\eea
which, via the embedding relation  $U_{\mbf{G}-\mbf{b}_i,\alpha,n}(\mbf{k}+\mbf{g}_i)=U_{\mbf{G},\alpha,n}(\mbf{k})$, are periodic and obey $c^\dag_{\mbf{k}+\mbf{b}_i,n,\eta,s} = c^\dag_{\mbf{k},n,\eta,s}$. $\hat{H}_0$  can then be written as
\bea
\hat{H}_0 &= \sum_{\mbf{k} n, \eta s} E_n^\eta(\mbf{k})  c^\dag_{\mbf{k},n,\eta,s} c_{\mbf{k},n,\eta,s}  \ .
\eea
To write the interaction Hamiltonian in the band basis, we use
\bea
\label{appeq:deltarhoproj}
\delta \hat{\rho}_{\mbf{q}+\mbf{G}} &=  \sum_{\mbf{k} mn, \eta s} M^\eta_{mn}(\mbf{k},\mbf{q}+\mbf{G})(c^\dag_{\mbf{k}+\mbf{q},m, \eta s} c_{\mbf{k},n,\eta s} - \frac{1}{2} \delta_{\mbf{q},0} \delta_{mn}) 
\eea
where we defined the form factor
\bea
M_{mn}^\eta(\mbf{k},\mbf{q}+\mbf{G}) = \sum_{\mbf{G}'\alpha}U^{\eta *}_{\mbf{G}+\mbf{G}',\alpha,m}(\mbf{k}+\mbf{q}) U^\eta_{\mbf{G}',\alpha,n}(\mbf{k}) = \sum_{\mbf{G}'\alpha}U^{\eta *}_{\mbf{G}',\alpha,m}(\mbf{k}+\mbf{q}+\mbf{G}) U^\eta_{\mbf{G}',\alpha,n}(\mbf{k}) = [U^{\eta \dag}(\mbf{k}+\mbf{q}+\mbf{G})U^\eta(\mbf{k})]_{mn} 
\eea
which is an overlap matrix of eigenvectors at different $\mbf{k}$, and satisfies $M_{mn}^\eta(\mbf{k},\mbf{q}+\mbf{G})=M_{nm}^{\eta*}(\mbf{k}+\mbf{q},-\mbf{q}-\mbf{G})$. Note that $\sum_{mn} M^\eta_{mn}(\mbf{k},\mbf{G}) \delta_{mn} = \sum_{\mbf{G}'\al} \delta_{\mbf{G},0}$, which is important in going from \Eq{appeq:deltarhounproj} to \Eq{appeq:deltarhoproj}. Note that all transformations so far are exact rewritings since they merely change the basis. 

Our next step is to rewrite the Hamiltonian such that the two-body interaction term is normal-ordered with respect to the full vacuum:
\bea\label{appeq:Hint_nord}
\hat{H}_{\text{int}} &=  \frac{1}{2 N} \sum_{\mbf{q},\mbf{G}} \frac{V(\mbf{q}+\mbf{G})}{\Omega} \delta \hat{\rho}_{\mbf{q}+\mbf{G}} \delta \hat{\rho}_{-\mbf{q}-\mbf{G}} \\
&= \frac{1}{2N}\sum_{\mbf{q},\mbf{G}} \sum_{\mbf{k} mn \eta s,\mbf{k}' m'n' \eta' s'} \frac{V(\mbf{q}+\mbf{G})}{\Omega} M^\eta_{mn}(\mbf{k},\mbf{q}+\mbf{G}) M^{\eta'}_{m'n'}(\mbf{k}',-\mbf{q}-\mbf{G})
c^\dag_{\mbf{k}+\mbf{q},m, \eta s} c_{\mbf{k},n,\eta s} c^\dag_{\mbf{k}'-\mbf{q},m', \eta' s'} c_{\mbf{k}',n',\eta' s'} \\
&\qquad - \frac{N_f}{2}\sum_{\mbf{k}, m n \eta s} \sum_{\mbf{G}} \frac{V(\mbf{G})}{2\Omega} M^\eta_{mn}(\mbf{k},\mbf{G}) \rho_{-\mbf{G}} c^\dag_{\mbf{k},m,\eta s} c_{\mbf{k},n,\eta s}, \qquad \rho_\mbf{G} = \frac{1}{N} \sum_\mbf{k} \Tr M^\eta(\mbf{k},\mbf{G})   \\
&= :H_{\text{int}}: + \tilde{H}_\text{H} + \tilde{H}_\text{F}
\eea
where $N_f=\sum_{s,\eta} =4$ is the number of flavors, and 
\bea\label{appeq:infinity_average}
:H_{\text{int}}: &= \frac{1}{2N}\sum_{\mbf{q},\mbf{G}} \sum_{\mbf{k} mn \eta s,\mbf{k}' m'n' \eta' s'} \frac{V(\mbf{q}+\mbf{G})}{\Omega} M^\eta_{mn}(\mbf{k},\mbf{q}+\mbf{G}) M^{\eta'*}_{n'm'}(\mbf{k}'-\mbf{q},\mbf{q}+\mbf{G})
c^\dag_{\mbf{k}+\mbf{q},m, \eta s}  c^\dag_{\mbf{k}'-\mbf{q},m', \eta' s'} c_{\mbf{k}',n',\eta' s'} c_{\mbf{k},n,\eta s}\\
\tilde{H}_\text{H} &=  - \frac{N_f}{2N}\sum_{\mbf{k}, m n \eta s} \lp \sum_{\mbf{G}} \frac{V(\mbf{G})}{\Omega} M^\eta_{mn}(\mbf{k},\mbf{G}) \sum_{\mbf{k}'m'}M^{+ *}_{m'm'}(\mbf{k}',\mbf{G})\rp c^\dag_{\mbf{k},m,\eta s} c_{\mbf{k},n,\eta s} \\
\tilde{H}_\text{F} 
&= \frac{1}{2N} \sum_{\mbf{k} mn \eta s} c^\dag_{\mbf{k},m, \eta s}  \lp \sum_{\mbf{q},\mbf{G}} \frac{V(\mbf{q}+\mbf{G})}{\Omega} \sum_{m'}M^{\eta *}_{m'm}(\mbf{k},\mbf{q}+\mbf{G}) M^{\eta}_{m'n}(\mbf{k},\mbf{q}+\mbf{G}) \rp 
c_{\mbf{k},n,\eta s}. \\
\eea
We can use $M^+$ in the innermost summation of $\tilde{H}_\text{H}$ because the charge density of the two valleys are equivalent from TRS. The band indices run over all bands of $\hat{H}_0$, and so the Hamiltonian above is still an exact rewriting of Eq.~\ref{appeq:Hint_start}. 

Given a spin-collinear moir\'e-translation-invariant one-body density matrix $P_{m\eta,n\eta'}(\mbf{k},s)=\langle c^\dagger_{\mbf{k},m,\eta s} c_{\mbf{k},n,\eta' s}\rangle$, we can perform a HF decoupling of $:H_{\text{int}}:$ to obtain the mean-field interaction term $H_{\text{HF,int}}[P]=H_\text{H,int}[P]+H_\text{F,int}[P]$ with
\bea\label{appeq:HFham}
H_\text{H,int}[P]&=\frac{1}{N}\sum_{\mbf{k}m n \eta s} \lp \sum_{\mbf{G}} \frac{V(\mbf{G})}{\Omega} M^\eta_{mn}(\mbf{k},\mbf{G}) \sum_{\mbf{k}'m'n'\eta's'}M^{\eta' *}_{n'm'}(\mbf{k}',\mbf{G})P_{m'\eta',n'\eta'}(\mbf{k}',s)  \rp c^\dag_{\mbf{k},m,\eta s} c_{\mbf{k},n,\eta s}\\
H_\text{F,int}[P]&=-\frac{1}{N} \sum_{\mbf{k} mn \eta\eta' s} c^\dag_{\mbf{k},m, \eta' s}  \lp \sum_{\mbf{q},\mbf{G},m',n'} \frac{V(\mbf{q}+\mbf{G})}{\Omega} M^{\eta' *}_{n'm}(\mbf{k},\mbf{q}+\mbf{G}) P_{m'\eta,n'\eta'}(\mbf{k}+\mbf{q},s)M^{\eta}_{m'n}(\mbf{k},\mbf{q}+\mbf{G}) \rp 
c_{\mbf{k},n,\eta s}.
\eea
$\tilde{H}_H+\tilde{H}_F$ is equivalent to $-1$ times the HF decoupling of $:H_{\text{int}}:$ using the `reference density matrix' $P^{\text{ref,avg.}}_{m\eta,n\eta'}(\mbf{k},s)=\frac{1}{2}\delta_{mn}\delta_{\eta\eta'}$. Hence we can write
\begin{equation}
        \hat{H}_{\text{int}}=\normOrd{\hat{H}_\text{int}}-\hat{H}_{\text{HF,int}}[P^\text{ref,avg.}]
\end{equation}
In the language of App.~\ref{secapp:general_scheme}, this corresponds to the $(\infty+\infty)$-average scheme (or just `average scheme' in the main text) with reference density matrix 
\begin{equation}\label{eqapp:Pref_average}
P^\text{ref,avg.}_{m\eta,n\eta'}(\mbf{k},s)=\frac{1}{2}\delta_{mn}\delta_{\eta\eta'}.
\end{equation} 
Note that $\tilde{H}_\text{H}$ and $\tilde{H}_\text{F}$ are actually just chemical potential terms. However, this will no longer be true when we generalize to layer-dependent interactions in App.~\ref{secapp:layer_int}.

\subsection{General projection}\label{secapp:projection}

In the following, we present an extended version of Sec.~\ref{subsec:schemes} in the main text, which discusses how to perform projection of the Hamiltonian.

Consider the many-body Hamiltonian
\begin{equation}\label{appeq:H_unproj}
    \hat{H}=\hat{H}_\text{s.p.}(V)+\hat{H}_\text{int}=\hat{H}_\text{s.p.}(V)+\normOrd{\hat{H}_\text{int}}-\hat{H}_{\text{HF,int}}[P^\text{ref}],
\end{equation}
where $\hat{H}_\text{s.p.}(V)$ is the non-interacting continuum model with interlayer potential $V$. In anticipation for other interaction schemes in Sec.~\ref{secapp:general_scheme}, we have generalized to a general reference density matrix $P^\text{ref}$. Since the correlation effects in the systems studied here primarily arise from degrees of freedom with energies near charge neutrality, we will project the Hamitonian $\hat{H}$ to a restricted number of bands around neutrality. Let $\mathscr{H}$ be the single-particle Hilbert space, from which the many-body Hilbert space that $\hat{H}$ acts on is constructed. In the first step of projection, we choose an \emph{active} set of bands $\mathscr{H}_{\text{act.}}$, corresponding to some number of valence bands $n_v$ and conduction bands $n_c$ (per flavor) closest to neutrality. The assumption is that the kinetic energies of the \emph{remote} (i.e.~not active) bands are detuned far enough from the Fermi level that to a good approximation, they can be frozen to be either fully occupied or unoccupied. Once the interaction has been projected, we can then perform HF by considering a variational calculation on the remaining active bands, i.e.~$(n_v+n_c)$-HF calculations. The projected interaction Hamiltonian can just as well be studied with other techniques besides HF, such as exact diagonalization (ED).

We now explain projection in more detail. We first need to choose some `active' subspace described by a set of single-particle states $\mathscr{H}_\text{act.}$. The active subspace is chosen by diagonalizing some effective single-particle Hamiltonian $\hat{H}_0$, yielding the Bloch coefficients $U^\eta_{\mbf{G}\alpha,n}(\mbf{k})$ in Eq.~\ref{appeq:U}.  Note that  $\hat{H}_0$ is not necessarily $\hat{H}_{\text{s.p.}}(V)$---as will be explained in App.~\ref{secapp:layer_int}, we will consider $\hat{H}_0=\hat{H}_{\text{s.p.}}(U)$, where $U$ accounts for internal screening of the interlayer potential. The band basis of $\hat{H}_0$ will referred to as the \emph{projection band basis}. We then pick the $n_v$ valence and $n_c$ conduction bands per flavor closest to neutrality. Crucially, we do not simply ignore the rest of the bands. All other bands are {\it frozen}, in the sense that remote valence bands below the active valence bands (which we denote by $\mathscr{H}_\text{rem.~val.}$) are fixed to be fully filled, while the remote conduction bands above the active conduction bands (which we denote by $\mathscr{H}_\text{rem.~con.}$) are forced to be empty. In a projected calculation, the many-body Hilbert space is effectively restricted to states of the form
\begin{equation}\label{appeq:psifillrem}
    \ket{\Psi}=\hat{\mathcal{O}}\prod_{\alpha\in\mathscr{H}_{\text{rem.~val.}}}c^\dagger_{\alpha}\ket{\text{vac}}
\end{equation}
where $\ket{\text{vac}}$ is the fermion vacuum, and $\hat{\mathcal{O}}$ is an operator consisting of an arbitrary combination of creation operators belonging to the active subspace $\mathscr{H}_\text{act.}$. 

Despite the fact that $\ket{\Psi}$ is still defined in the original full Hilbert space, the computational cost of numerical calculations can be reduced because the many-body Hilbert space has been restricted. The second step of projection exploits this by constructing a related many-body Hamiltonian $\hat{H}_\text{act.}$ which acts only on the many-body Hilbert space constructed from $\mathscr{H}_\text{act.}$, but gives equivalent answers to analyzing $\hat{H}$ within the variational manifold of Eq.~\ref{appeq:psifillrem}. To do this, we need to first account for the possible effects of the fully occupied remote valence bands on the active subspace. Consider the normal ordered four-fermion interaction term $\normOrd{\hat{H}_\text{int}}$ (Eq.~\ref{appeq:Hint_nord}) expressed in the basis of $\hat{H}_0$ used to specify $\mathscr{H}_\text{act.}$. In terms where one creation and one annihilation operator belong to $\mathscr{H}_{\text{rem. val.}}$, they can be replaced (after anticommuting to bring them together) by a delta function of their quantum numbers.
If the other two operators belong to $\mathscr{H}_\text{act.}$, then this generates a one-body contribution acting on active states in $\mathscr{H}_{\text{act.}}$. Collecting these contributions yields the one-body term 
\eq{
\hat{H}^\text{eff}_{\text{rem. val.}} = \left.\hat{H}_\text{HF,int}[P_\text{rem. val.}]\right|_{\text{act.}}\ ,
}
where
\eq{
\left[P_\text{rem. val.}(\bsl{k},s)\right]_{m\eta,n \eta'} = \begin{cases}
        \delta_{mn}\delta_{\eta\eta'}, & \text{if $m$ in } \mathscr{H}_\text{rem.~val.}\\
        0, &\text{otherwise,}
    \end{cases}
}
and for any second-quantized operator $\hat{O}$ acting on the many-body Hilbert space constructed from $\mathscr{H}$, $\left.\hat{O}\right|_{\text{act.}}$ denotes the truncation to only terms that solely involve creation/annihilation operators belonging to $\mathscr{H}_\text{act.}$. $\hat{H}^\text{eff}_{\text{rem. val.}}$ captures the renormalization from the filled remote valence states. We then obtain
\begin{align}\begin{split}\label{appeq:Hact}
    \hat{H}_{\text{act.}}=&\left.\hat{H}_{\text{s.p.}}(V)\right|_{\text{act.}}+\left.\normOrd{\hat{H}_\text{int}}\right|_{\text{act.}}\\
    &-\left.\hat{H}_{\text{HF,int}}[P^\text{ref}]\right|_{\text{act.}}+\hat{H}^\text{eff}_{\text{rem. val.}}.
\end{split}\end{align}
We can then perform computations using the projected interaction Hamiltonian $\hat{H}_{\text{act.}}$. Note that the energy expectation value of $\ket{\Psi_\text{act.}}=\hat{\mathcal{O}}\ket{\text{vac}}$ in $\hat{H}_\text{act.}$ is equal to that of $\ket{\Psi}=\hat{\mathcal{O}}\prod_{\alpha\in\mathscr{H}_{\text{rem.~val.}}}c^\dagger_{\alpha}\ket{\text{vac}}$ in $\hat{H}$, up to constants that do not depend on $\mathcal{O}$.

\subsection{More general schemes}\label{secapp:general_scheme}

In this subsection, we describe different interaction schemes (sometimes referred to as subtraction schemes) for the interaction Hamiltonian within a unified framework. Following the discussion in App.~\ref{secapp:average_scheme}, we write the full many-body Hamiltonian (before doing any projection of the previous App.~\ref{secapp:projection}) as
\begin{equation}\label{appeq:Hgen}
    \hat{H}=H_{\text{s.p.}}(V)+:H_{\text{int}}:-H_{\text{HF,int}}[P^{\text{ref}}],
\end{equation}
where $H_{\text{s.p.}}(V)$ is the physical non-interacting continuum Hamiltonian evaluated with interlayer potential $V$, $:H_{\text{int}}:$ is normal-ordered, and $H_{\text{HF,int}}[P^{\text{ref}}]$ is the mean-field decoupling of the normal-ordered interaction using the reference density matrix $P^{\text{ref}}$ (see Eq.~\ref{appeq:HFham}). The Hamiltonian in App.~\ref{secapp:average_scheme} corresponds to $P^\text{ref}=P^\text{ref,avg.}$ (Eq.~\ref{eqapp:Pref_average}). The physical interpretation of Eq.~\ref{appeq:Hgen} is that the interactions are measured relative to the mean-field state $P^\text{ref}$. In other words, the HF decoupling of $H$ in the state $P^\text{ref}$ is exactly $H_{\text{s.p.}}(V)$. We list some possible choices of $P^{\text{ref}}$---in all cases, $P^\text{ref}$ is the identity in spin space, valley diagonal, and preserves all other symmetries of $H_\text{s.p.}(V)$. In the below discussion, recall that the projection band basis refers to the band basis of $\hat{H}_0$ which was used to specify the active subspace $\mathscr{H}_\text{act.}$ in a projected calculation (App.~\ref{secapp:projection}).
\begin{itemize}
    \item The $(n'_v+n'_c)$-average scheme corresponds to $P^\text{ref,avg.}_{(n'_v+n'_c)}=\frac{1}{2}I$ within the highest $n'_v$ valence projection bands and lowest $n'_c$ conduction projection bands, $P^\text{ref,avg.}_{(n'_v+n'_c)}=I$ for all other valence bands, and $P^\text{ref,avg.}_{(n'_v+n'_c)}=0$ for all other conduction bands ($I$ is the identity matrix in band space). We will always take $n'_v=n'_c$ in the average scheme to respect the approximate PHS that exists in the absence of hBN and moire potential. The scheme is sometimes referred to as the infinite-temperature scheme, and is commonly utilized in twisted bilayer graphene and alternating-stacked twisted multilayer graphene~\cite{TBG3,2021PhRvB.103t5414L}. Note that $n'_v,n'_c$ are not necessarily equal to the number of active valence/conduction bands $n_v,n_c$ in $\mathscr{H}_\text{act.}$. In the main text, we take $n'_v,n'_c$ to be the total number of bands in the single-particle Hilbert space $\mathscr{H}$ used to construct the many-body Hilbert space that the many-body Hamiltonian $\hat{H}$ acts on. In the context of moir\'e continuum models where $\mathscr{H}$ is determined by a plane wave expansion cutoff, $n'_v,n'_c$ is taken to include all bands within this cutoff. In this case $P^\text{ref,avg.}$ is basis-independent (Eq.~\ref{eqapp:Pref_average}).

    A choice of small $n'_v,n'_c$ that does not encompass the entire plane wave cutoff is prone to band cutoff artifacts in the renormalized potential felt by the active subspace. This is because the remote bands have charge density variations within the moir\'e unit cell, and the precise real-space pattern of the summed density of multiple bands can be band-cutoff dependent. This is demonstrated in detail in App.~\ref{app:charge_density_band_cutoff}. Hence, unless otherwise stated, we use the term `average scheme' to encompass all bands within the plane wave cutoff. 

    One subtlety in the average scheme is that the low-energy single-particle states near charge neutrality experience a Fock renormalization from remote single-particle states near the edge of the Hilbert space plane wave cutoff (the size of the plane wave cutoff sets the dimension of the single-particle Hilbert space $\mathscr{H}$). This is because the far remote valence (conduction) states are fully occupied (empty) in the physical density matrix $P$, while the reference density $P^\text{ref,avg.}$ is $1/2$ for both remote valence and conduction states. Therefore, $P-P^\text{ref,avg.}$ does not vanish for remote states whose Bloch functions have support only near the Hilbert space plane wave cutoff. Consider the Fock renormalization of a single-particle state $\alpha$ due to another single-particle state $\beta$. Let $\alpha$ be a low-energy eigenstate of the non-interacting continuum model, so that its Bloch function is comprised of plane waves near the graphene Dirac momentum $K_G$, and $\beta$ be a high-energy remote band eigenstate, so that its Bloch function is mostly localized on a single plane wave near the edge of the Hilbert space plane wave cutoff. The magnitude of the Fock renormalization is affected by the form factor between $\alpha$ and $\beta$, and the suppression of the interaction potential at large momentum transfer $q$ (as argued above, the states $\alpha$ and $\beta$ have momenta near $K_G$ and the edge of the plane wave cutoff respectively). Since the interaction potential decays only as $1/q$ at large $q$, the convergence as we increase the plane wave cutoff is relatively slow compared to other schemes where $P-P^\text{ref,avg.}$ vanishes quickly for far remote bands. At the same time, the single-particle continuum Hamiltonian is not a reliable model for states with energies approaching the eV scale, so the Fock renormalization from states around or beyond this scale is unphysical. To avoid this unphysical renormalization and reduce the computational time, we set radial momentum cutoffs on the Hilbert space $\mathscr{H}$ and the interaction potential at $4|\mbf{q}_1|$ and $3|\mbf{q}_1|$ respectively. In App.~\ref{secapp:cutoff}, we compare numerical HF results using different cutoff radii.
    
    \item The charge neutrality (CN) scheme corresponds to $P^\text{ref,CN}=I$ for all non-interacting valence bands of $\hat{H}_\text{s.p.}(V)$ and $P^\text{ref,CN}=0$ for all non-interacting conduction bands. Written explicitly in the non-interacting band basis of $\hat{H}_\text{s.p.}(V)$, the reference density is
\begin{equation}
    P^\text{ref,CN}_{m\eta,n\eta'}(\mbf{k},s)=\begin{cases}
        \delta_{mn}\delta_{\eta\eta'}, &\text{if $m$ is a valence band}\\
        0, &\text{otherwise.}
    \end{cases}
\end{equation} 
Note that the non-interacting band structure and wavefunctions of $\hat{H}_\text{s.p.}(V)$ change with $V$. For example, the valence band subspace of $\hat{H}_\text{s.p.}(V)$ becomes increasingly polarized towards the hBN (which is adjacent to the lowest graphene layer $l=0$) for larger $V>0$. Therefore, the physical properties of $P^\text{ref,CN}$, such as its layer polarization, vary with $V$. 

Note that if there is finite energetic overlap between conduction and valence bands (in the sense that the indirect gap is less than 0), then $P^{\text{ref,CN}}$ is not equivalent to the zero-temperature density matrix of $H_\text{s.p.}(V)$ at charge neutrality (which would be a compensated semimetal). This is the scheme used in Refs.~\cite{dong2023anomalous,dong2023theory,zhou2023fractional}.

    \item The graphene scheme corresponds to choosing $P^\text{ref,graphene}$ to be the filled non-interacting valence bands of isolated rhombohedral graphene without any moir\'e potentials.
    
    \item The `zero-field' scheme $P^\text{ref,zero-field}$ corresponds to filling the valence bands of $\hat{H}_{\text{s.p.}}$ evaluated at $V=0$.

\end{itemize}

In this work, we perform calculations using the average interaction scheme and the CN interaction scheme. While the CN interaction scheme has been used in very recent  works~\cite{dong2023anomalous,dong2023theory,zhou2023fractional}, we argue that the average interaction scheme has several desirable features. First, the average scheme reference density $P^\text{ref,avg.}$, and hence the part of the Hamiltonian that depends on the interaction potential $V(\mbf{q})$, is independent of $V$. It is natural to expect that an interaction scheme should not depend on parameters that can be tuned {\it in situ} experimentally. In fact, $P^\text{ref,avg.}$ is independent of all other system parameters, such as twist angle. Second, the average scheme is not fine-tuned in such a way as to insulate the valence and conduction subspaces from each other for large $V$. This is unlike the CN scheme, where for $\nu>0$, the entire set of valence bands have their influence on the conduction bands artifically suppressed since $-H_{\text{HF,int}}[P^\text{ref,CN}]$ largely cancels it out. Third, the average scheme reference density is smooth, unlike the graphene scheme where the density matrix of filled graphene bands is discontinuous due to Dirac points.

\subsection{Layer-dependent interactions}\label{secapp:layer_int}
The discussion in App.~\ref{secapp:int_HF} so far has assumed a layer-independent interaction potential $V(\mbf{q})$, e.g. for dual-gate screening
\begin{equation}\label{eq:Vq_tanh}
    V_{2d}(\mbf{q})=\frac{e^2}{2\epsilon q}\tanh{(qd_\text{sc})}
\end{equation}
where the two metallic gates are at a distance $d_\text{sc}$ above and below the sample, which is treated as a two-dimensional plane.
As a result, the density operator (Eq.~\ref{appeq:deltarhounproj}) which appears in the interaction Hamiltonian involves an internal contraction over the layer index. However, multilayer systems have a finite width, and for large enough layers, the layer dependence of the interaction can have a qualitative effect on the physics. For example, finite wavevector interactions between bands polarized on opposite layers are suppressed to some extent by the vertical distance (we will give a numerical example of this suppression below Eq.~\ref{eq:Vq_layer}). Indeed, the in-plane 2D Fourier transform of the Coulomb interaction between two charges $-e$ at heights $z$ and $z'$ is
\begin{equation}\label{appeq:Vq_layer}
    U(\mbf{q},z,z')=\frac{e^2}{2\epsilon q}e^{-q|z-z'|}
\end{equation}
where we have assumed for simplicity an isotropic dielectric environment with dielectric constant $\epsilon=\epsilon_0\epsilon_r$. (We have used a different symbol $U$ to distinguish it from $V(\mbf{q},z,z')$, which is defined below and accounts for gate screening effects). Note that $\mbf{q}$ always refers to the 2d in-plane momentum. Another important effect of the layer degree of freedom is the internal screening of external displacement fields. Applying an external interlayer potential polarizes the electrons towards one side of the system, but such a vertical density imbalance also electrostatically sets up internal potential differences that act to counter the external field.

One work on R$L$G/hBN in the literature (e.g.~Ref.~\cite{zhou2023fractional}) accounts for the layer dependence in dual-gate screened devices by multiplying the purely 2D interaction with the exponential factor, leading to
\begin{equation}\label{appeq:Vq_exp}
    \tilde{V}(\mbf{q},z,z')=\frac{e^2}{2\epsilon q}\tanh{(qd)}e^{-q|z-z'|}.
\end{equation}
While this captures the finite-momentum suppression of interlayer interactions, it does not include internal electrostatic screening of interlayer potentials, since the $\mbf{q}=0$ limit is independent of $z,z'$. Thus, we will not use $\tilde{V}(\mbf{q},z,z')$ in our calculations. Below, we review a derivation of the full 3D interaction, following App.~\cite{kolar2023electrostatic}, which will be used in the calculations of this paper.

In the following, we use the term `sample' to refer to the multilayer material (i.e.~not including the gates). The goal is to derive an effective gate-screened interaction $V_{l,l'}(\mbf{q})$ between sample electrons. Consider sample and gate electrons confined to 2d layers at different heights $z_F$, where $F$ is a composite index that runs over both sample layers $l$, and gate layers $g$ (i.e.~the electrons that will screen the interactions between sample electrons are confined to 2d metallic plates at heights $z_g$). For dual-gate screening $g=t,b$, we have top $t$ and bottom $b$ gates positioned at $z=\pm d_\text{sc}$. All sample layers lie between the gates. The normal-ordered unprojected interaction Hamiltonian is
\begin{equation}
    :\tilde{H}_{\text{int}}:=\frac{1}{2N}\sum_{\mbf{q}FF'}\frac{U_{F,F'}(\mbf{q})}{\Omega}:\rho_{\mbf{q},F}\rho_{-\mbf{q},F'}:
\end{equation}
where $U_{F,F'}(\mbf{q})=U(\mbf{q},z_F,z_{F'})$. The effective one-body potential arising from the choice of interaction scheme can be obtained as in Sec.~\ref{secapp:general_scheme}, so we do not consider this point further. Note that at this point $:\tilde{H}_{\text{int}}:$ also involves the gate electrons. 

Since the Coulomb interaction Eq.~\ref{appeq:Vq_layer} diverges for $q\rightarrow 0$, we first separate out the term above corresponding to $\mbf{q}=0$. The leading divergent piece $\sim \frac{e^2}{2\epsilon q}$ vanishes if the total system (sample$+$gates) is charge neutral. The remaining finite piece reads
\begin{equation}
    :\tilde{H}_{\text{int},\mbf{q}=0}:=-\frac{1}{2N}\sum_{F,F'}\frac{e^2|z_F-z_{F'}|}{2\epsilon\Omega}:\rho_{\mbf{q}=0,F}\rho_{\mbf{q}=0,F'}:.
\end{equation}
Note that $\rho_{\mbf{q}=0,F}$ counts the total number of electrons on layer $F$. The above expression simply corresponds to the electrostatic energy of uniformly charged parallel planes.  If the total sample electron number is fixed as $N_e=\sum_{l}\langle\rho_{\mbf{q}=0,l}\rangle$, the condition of total charge neutrality leads to the following charges on the top and bottom gates
\begin{equation}
    \rho_{\mbf{q}=0,t}=-\frac{N_e}{2}+\delta, \quad\rho_{\mbf{q}=0,b}=-\frac{N_e}{2}-\delta 
\end{equation}
where we have allowed for a gate population imbalance $\delta$ to induce external interlayer potentials. Note that there is no hybridization between the gates and the sample, so the gate density operators are simply constants (conserved). Substituting them into the zero-momentum Hamiltonian leads to 
\begin{equation}
    :\tilde{H}_{\text{int},\mbf{q}=0}:=-\frac{1}{2N}\sum_{l,l'}\frac{e^2|z_l-z_{l'}|}{2\epsilon\Omega}:\rho_{\mbf{q}=0,l}\rho_{\mbf{q}=0,l'}:+\frac{1}{N}\sum_{l}\frac{e^2z_l}{\epsilon\Omega}\rho_{\mbf{q}=0,l}+\frac{1}{N}\frac{e^2}{2\epsilon\Omega}\delta^2
\end{equation}
The second term is a potential that depends linearly on $z$, and can be absorbed into the single-particle Hamiltonian as an external interlayer potential. The final constant piece does not depend on the physics of the sample electrons and will be discarded. The first term is identified as $\normOrd{\hat{H}_{\text{int},\mbf{q}=0}}$, and corresponds to an effective $\mbf{q}=0$ limit of the gate-screened interaction potential between sample electrons
\begin{gather}\label{eq:Vq0}
    V_{ll'}(\mbf{q}=0)=-\frac{e^2|z_l-z_{l'}|}{2\epsilon}\\
    \normOrd{\hat{H}_{\text{int},\mbf{q}=0}}=\frac{1}{2N}\sum_{l,l'}\frac{V_{ll'}(\mbf{q}=0)}{\Omega}:\rho_{\mbf{q}=0,l}\rho_{\mbf{q}=0,l'}:
\end{gather}

We now return to the remaining finite-momentum components of the interaction Hamiltonian. The gate electrons can be eliminated using the method of image charges. (The internal energy of the gate electrons does not depend on their finite-momentum density components since the metallic gates effectively have $\epsilon_{\parallel}\rightarrow\infty$.) Consider a test charge $-e$ positioned at in-plane coordinate $\mbf{r}$ and vertical coordinate $z_0$. To ensure that the top gate plane $z=d_\text{sc}$ is kept at a uniform electric potential $V=0$, this requires an image charge of $+e$ at $\mbf{r}=0$ and $z=2d_\text{sc}-z_0$. A similar argument for the bottom gate plane at $z=-d_\text{sc}$ leads to another $+e$ image charge at $\mbf{r}=0$ and $z=-2d_\text{sc}-z_0$. But the image charge at $z=2d_\text{sc}-z_0$ needs to be compensated by an image charge to maintain $V=0$ on the bottom gate, and similarly the image charge at $z=-2d_\text{sc}-z_0$ needs to be compensated by an image charge to maintain $V=0$ on the top gate. Repeating this argument generates a recursion relation, leading to an infinite series of image charges at positions 
\begin{gather}
z^t_n=2nd_\text{sc}+(-1)^nz_0\\
z^b_n=-2nd_\text{sc}+(-1)^nz_0
\end{gather}
with charge $(-1)^n(-e)$ for $n\geq1$. Including the original test charge, the interaction felt by a charge $-e$ at some radial distance $r$ and height $z$ is
\begin{equation}
    V(z,z_0,r)=\frac{e^2}{4\pi\epsilon}\lp
    \frac{1}{\sqrt{r^2+(z-z_0)^2}}+\sum_{n\geq1}\frac{(-1)^n}{\sqrt{r^2+(z-2nd_\text{sc}-(-1)^nz_0)^2}}+\sum_{n\geq 1}\frac{(-1)^n}{\sqrt{r^2+(z+2nd_\text{sc}-(-1)^nz_0)^2}}
    \rp.
\end{equation}
Using the integral
\begin{equation}
    \int d\mbf{r}\,\frac{e^{i\mbf{k}\cdot\mbf{r}}}{\sqrt{r^2+r_0^2}}=\frac{2\pi}{k}e^{-kr_0}
\end{equation}
we obtain the Fourier transform of the interaction potential
\begin{align}
    V(\mbf{q},z,z_0)=&\frac{e^2}{2\epsilon q}\lp
    e^{-q|z-z_0|}+\sum_{n\geq 1}(-1)^ne^{-q|z-2nd_\text{sc}-(-1)^nz_0|}
    +\sum_{n\geq 1}(-1)^ne^{-q|z+2nd_\text{sc}-(-1)^nz_0|}
    \rp\\
    =&\frac{e^2}{2\epsilon q}\lp
    e^{-q|z-z_0|}+\sum_{n\geq 0}\left(
    -e^{q(z+z_0-4(n+\frac{1}{2})d_\text{sc})}
    +e^{q(z-z_0-4(n+1)d_\text{sc})}
    -e^{-q(z+z_0+4(n+\frac{1}{2})d_\text{sc}}
    +e^{-q(z-z_0+4(n+1)d_\text{sc})}
    \right)\rp\\
    =&\frac{1}{2\epsilon q}\left[ e^{-q|z-z_0|}+\frac{e^{-q(z+z_0)}\left(-e^{2q(d_\text{sc}+z+z_0)}-e^{2qd_\text{sc}}+e^{2qz}+e^{2qz_0}\right)}{e^{4qd_\text{sc}}-1}\right].
\end{align}

This results in
\begin{equation}\label{eq:Vq_layer}
    V_{ll'}(\mbf{q})=\frac{e^2}{2\epsilon q}\left[\frac{e^{-q(z_l+z_{l'})}\left(-e^{2q(d_\text{sc}+z_l+z_{l'})}-e^{2qd_\text{sc}}+e^{2qz_l}+e^{2qz_{l'}}\right)}{e^{4qd_\text{sc}}-1}+e^{-q|z_l-z_{l'}|}\right].
\end{equation}
We provide an example of the interlayer suppression of the interaction for finite wavevector. The interlayer suppression is $V_{0,4}(\mbf{g}_1)/V_{2,2}(\mbf{g}_1)=43\%$ at finite wavevector $\mbf{q}=\mbf{g}_1$ for R5G/hBN at $\theta=0.77^\circ$.

The interaction can be written in the band basis using layer-dependent form factors $M^{l\eta}_{mn}(\mbf{k},\mbf{q}+\mbf{G})$
\begin{gather}\label{appeq:form_ll}
M_{mn}^{l\eta}(\mbf{k},\mbf{q}+\mbf{G}) = \sum_{\mbf{G}'\sigma}U^{\eta *}_{\mbf{G}+\mbf{G}',l\sigma,m}(\mbf{k}+\mbf{q}) U^\eta_{\mbf{G}',l\sigma,n}(\mbf{k})\\
    :H_{\text{int}}: = \frac{1}{2N}\sum_{\mbf{q},\mbf{G}} \sum_{\mbf{k} mn l\eta s,\mbf{k}' m'n' l'\eta' s'} \frac{V_{ll'}(\mbf{q}+\mbf{G})}{\Omega} M^{l\eta}_{mn}(\mbf{k},\mbf{q}+\mbf{G}) M^{l'\eta'*}_{n'm'}(\mbf{k}'-\mbf{q},\mbf{q}+\mbf{G})
c^\dag_{\mbf{k}+\mbf{q},m, \eta s}  c^\dag_{\mbf{k}'-\mbf{q},m', \eta' s'} c_{\mbf{k}',n',\eta' s'} c_{\mbf{k},n,\eta s}.
\end{gather}
For convenience, we also provide the expressions for the Hartree and Fock decoupling of the interaction in the density matrix $P$ with layer-dependent interactions, as well as the layer-resolved density operator expressed in the band basis
\begin{align}\begin{split}\label{eqapp:HHint_ll}
\hat{H}_\text{H,int}[P]=&\frac{1}{N}\sum_{\mbf{k}l l' m n \eta s}  \sum_{\mbf{G}} \frac{V_{ll'}(\mbf{G})}{\Omega}M^{l\eta}_{mn}(\mbf{k},\mbf{G}) \\
&\times  \lp\sum_{\mbf{k}'m'n'\eta's'}M^{l'\eta' *}_{n'm'}(\mbf{k}',\mbf{G})P_{m'\eta',n'\eta'}(\mbf{k}',s')\rp  \\
&\times c^\dag_{\mbf{k},m,\eta s} c_{\mbf{k},n,\eta s}
\end{split}\end{align}
\begin{align}\begin{split}
\hat{H}_\text{F,int}[P]=&-\frac{1}{N} \sum_{\mbf{k} mn \eta\eta' s} \sum_{\mbf{q},\mbf{G},m',n'} \frac{V_{ll'}(\mbf{q}+\mbf{G})}{\Omega} \\
&\times M^{l'\eta' *}_{n'm}(\mbf{k},\mbf{q}+\mbf{G}) M^{l\eta}_{m'n}(\mbf{k},\mbf{q}+\mbf{G})\\
&\times P_{m'\eta,n'\eta'}(\mbf{k}+\mbf{q},s)c^\dag_{\mbf{k},m, \eta' s}c_{\mbf{k},n,\eta s}
\end{split}\end{align}
\begin{equation}
\delta\rho_{\mbf{q+G},l}=\sum_{\mbf{k}mn, \eta s} M^{l\eta}_{mn}(\mbf{k},\mbf{q}+\mbf{G})(c^\dag_{\mbf{k}+\mbf{q},m, \eta s} c_{\mbf{k},n,\eta s} - \frac{1}{2} \delta_{\mbf{q},0} \delta_{mn}).
\end{equation}

\subsection{Self-consistent layer screening}\label{secapp:screening}

In this section, we consider the internal screening of the external interlayer potential due to electrostatic effects arising from the use of the layer-dependent potential $V_{ll'}(\mbf{q})$ defined in the previous subsection. Consider the physical one-body density matrix $P$ defined in the full Hilbert space, i.e.~$\mathscr{H}$. Our objective is to isolate the $\mbf{q}=0$ Hartree part of all interacting terms in the Hamiltonian, since these reflect the interaction-induced layer potentials set up by the charges on the layers. In particular, we require the $\mbf{q}=0$ Hartree decoupling of $\normOrd{\hat{H}_\text{int.}}$ (which is the $\mbf{G}=0$ part of Eq.~\ref{eqapp:HHint_ll}) and also the $\mbf{q}=0$ Hartree part of $-\hat{H}_{\text{HF,int}}[P^\text{ref}]$ in Eq.~\ref{appeq:Hgen} (which is the $\mbf{G}=0$ part of Eq.~\ref{eqapp:HHint_ll} with the replacement $P\rightarrow -P^\text{ref}$). These contributions can be combined into a single term $\hat{H}_{\text{HF,int}}[\delta P]$, where $\delta P=P-P^\text{ref}$, leading to
\begin{align}
\begin{split}
    \hat{H}_{\text{H,int},\mbf{q}=0}[\delta P]=&\frac{1}{N}\sum_{\mbf{k}l l' m n \eta s} \frac{V_{ll'}(\mbf{q}=0)}{\Omega}M^{l\eta}_{mn}(\mbf{k},\mbf{0}) \\
&\times  \lp\sum_{\mbf{k}'m'n'\eta's'}M^{l'\eta' *}_{n'm'}(\mbf{k}',\mbf{0})\delta P_{m'\eta',n'\eta'}(\mbf{k}',s')\rp\\
&\times c^\dag_{\mbf{k},m,\eta s} c_{\mbf{k},n,\eta s}.
\end{split}
\end{align}
To make the above expression more physically transparent, we express it in the plane wave basis, which is naturally used to define $\mathscr{H}$. In other words, we use Eq.~\ref{appeq:U} to transform from the band basis $c^\dagger_{\mbf{k},n,\eta,s}$ to the plane wave basis $c^\dagger_{\mbf{k},\mbf{G},l\sigma\eta s}$. Using Eq.~\ref{appeq:form_ll} and the orthonormality of the basis transformation in Eq.~\ref{appeq:U}, we have
\begin{equation}
    \sum_{mn}M^{l\eta}_{mn}(\mbf{k},\mbf{0})c^\dag_{\mbf{k},m,\eta s} c_{\mbf{k},n,\eta s}=\sum_{\mbf{G}\sigma}c^\dagger_{\mbf{k},\mbf{G},l\sigma \eta s}c_{\mbf{k},\mbf{G},l\sigma \eta s}.
\end{equation}
Consider the expression for the expectation value $N_l[\delta P]$ of the total electron number operator on layer $l$ in the density matrix $\delta P$
\begin{equation}
    N_{l'}[\delta P]=\sum_{\mbf{k}',\mbf{G},\eta',s',\sigma}\langle c^\dagger_{\mbf{k}+\mbf{G},l\sigma \eta s}c_{\mbf{k}+\mbf{G},l\sigma \eta s}\rangle_{\delta P}
    =\sum_{\mbf{k}'m'n'\eta's'}M^{l'\eta' *}_{n'm'}(\mbf{k}',\mbf{0})\delta P_{m'\eta',n'\eta'}(\mbf{k}',s'),
\end{equation}
where we have used Eqs.~\ref{appeq:U} and \ref{appeq:form_ll}. We can then write $\hat{H}_{\text{H,int},\mbf{q}=0}[\delta P]$ as
\begin{equation}
    \hat{H}_{\text{H,int},\mbf{q}=0}[\delta P]=\frac{1}{N}\sum_{ll'}\frac{V_{ll'}(\mbf{q}=0)}{\Omega}N_{l'}[\delta P]\sum_{\mbf{k}\mbf{G}\sigma\eta s}c^\dagger_{\mbf{k},\mbf{G},l\sigma \eta s}c_{\mbf{k},\mbf{G},l\sigma \eta s}.
\end{equation}
Since $\sum_{\mbf{k}\mbf{G}\sigma\eta s}c^\dagger_{\mbf{k},\mbf{G},l\sigma \eta s}c_{\mbf{k},\mbf{G},l\sigma \eta s}$ is just the number operator for layer $l$, we see that $\hat{H}_{\text{H,int},\mbf{q}=0}[\delta P]$ acts as an interaction-induced interlayer potential with layer-dependent potentials
\begin{equation}\label{appeq:Vint_l}
V_{\text{int},l}[\delta P]=\frac{1}{N}\sum_{l'}\frac{V_{ll'}(\mbf{q}=0)}{\Omega}N_{l'}[\delta P].
\end{equation}

Combining $\hat{H}_{\text{H,int},\mbf{q}=0}[\delta P]$ with the non-interacting continuum model $\hat{H}_\text{s.p.}(V_l)$, where we have explicitly indicated the external layer potential $V_l$ on each layer, leads to the following Hamiltonian which captures the non-interacting physics as well as the $\mbf{q}=0$ interlayer Hartree screening
\begin{equation}\label{appeq:HHq0_l}
    \hat{H}_{\text{H},\mbf{q}=0}[P]\equiv\hat{H}_\text{s.p.}(V_l)+\hat{H}_{\text{H,int},\mbf{q}=0}[\delta P]=\hat{H}_\text{s.p.}(V_l+V_{\text{int},l}[\delta P])=\hat{H}_\text{s.p.}(U_l).
\end{equation}
Note that for R$L$G, $V_l$ is linear in layer $l$ since the interlayer distance between adjacent layers is constant. $\hat{H}_{\text{H},\mbf{q}=0}[P]$ is equivalent to the non-interacting Hamiltonian with internally screened interlayer potential
\begin{equation}\label{appeq:U_l}
    U_l=V_{l}+V_{\text{int},l}[\delta P].
\end{equation}
Application of an external layer-dependent potential $V_{l}$ will act to polarize electrons such that $N_l[P]$ is larger in layers with lower potential. However, this imbalance of layer population will generate an interaction-induced layer potential $V_{\text{int},l}[\delta P]$ that will generally counteract $V_{l}$. For example, consider the case where $V>0$ so that the system polarizes towards the bottom layer near the hBN. As a result $N_{l}[\delta P]$ is larger for smaller $l$, so that $V_{\text{int},l}[\delta P]$ (Eq.~\ref{appeq:Vint_l}) is smaller for higher $l$ away from the hBN.

Define $P_\text{s.p.}(U_l)$ to be the density matrix constructed by filling the valence bands of $\hat{H}_\text{s.p.}(U_l)$. $P_\text{s.p.}(U_l)$ is a self-consistent ground state of the non-interacting model with $\mbf{q}=0$ interlayer Hartree interactions, if $P_\text{s.p.}(U_l)$ is the same as the density matrix constructed by occupying the valence bands of $\hat{H}_{\text{H},\mbf{q}=0}[P_\text{s.p.}(U_l)]$. This is equivalent to the condition
\begin{equation}\label{appeq:U_sc}
    U_l=V_{l}+V_{\text{int},l}[\delta P_\text{s.p.}(U_l)],
\end{equation}
which implicitly defines the self-consistent internally-screened potentials $U_l(V_l)$ as a function of the external layer potentials $V_{l}$.

One way to solve this is via full-band self-consistent Hartree calculations, which we outline below. We first start with some initial guess for $U_l^{(0)}$, e.g.~$U_l^{(0)}=V_l$. We then generate the density matrix $P_\text{s.p.}^{(0)}$, defined by occupying the valence bands of $\hat{H}_{\text{s.p.}}(U_l^{(0)})$. This is the zeroth step of the iterative loop. For the next step, we compute $U_l^{(1)}=V_l+V_{\text{int},l}[\delta P_\text{s.p.}(U_l^{(0)})]$ (see Eq.~\ref{appeq:U_l}), which is then used to generate $P_\text{s.p.}^{(1)}$ by filling the valence bands of $\hat{H}_{\text{s.p.}}(U_l^{(1)})$. This procedure is iterated until convergence of $U^{(n)}_l$ is reached within some tolerance at step $n$. The final value $U^{(n)}_l$ is then the self-consistent internally screened interlayer potential.

\begin{figure*}
    \centering
    \includegraphics[width=0.6\linewidth]{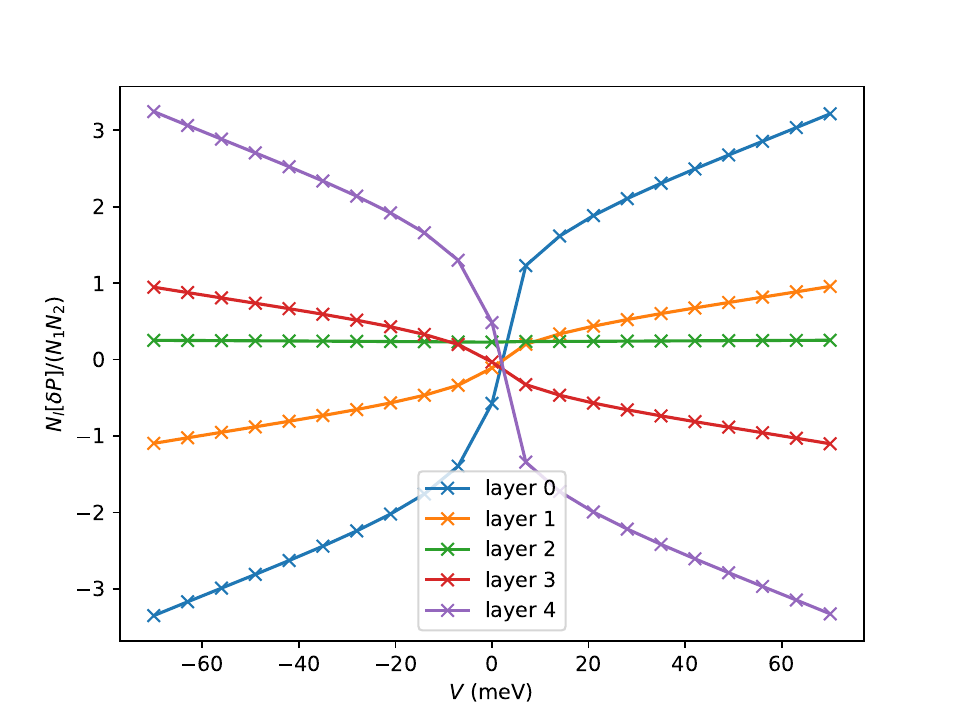}
    \caption{Layer electron population per moir\'e unit cell $N_l[\delta P_\text{s.p.}(V)]$, where $\delta P_\text{s.p.}(V)=P_\text{s.p.}(V)-P^\text{ref,avg.}$, and $P_\text{s.p.}(V)$ is the density matrix of filled valence bands of the non-interacting Hamiltonian $H_\text{s.p.}(V)$. R5G/hBN, $\xi=1$, $\theta=0.77^\circ$.}
    \label{Nl_L5}
\end{figure*}

Since $V_{l}$ varies linearly with layer $l$ in R$L$G/hBN, we have a single energy $V$ which gives the external potential energy difference between adjacent layers. In isolated R$L$G at $V=0$, the low-energy single-particle wavefunctions are superpositions of the top and bottom layers. When a finite $V$ is applied, these resolve into layer polarized wavefunctions, so that the charge imbalance in $\delta P_\text{s.p.}(V)$, obtained by occupying the filled valence bands of $\hat{H}_\text{s.p.}(V)$, is mostly localized on the outer layers with opposite signs (see Fig.~\ref{Nl_L5}). Hence, the interlayer Hartree potential $V_{\text{int},l}$ is also approximately linear, and we make an approximation by replacing this with a single variable $V_{\text{int}}=(V_{\text{int},L-1}-V_{\text{int},0})/(L-1)$. In this case, we can more easily obtain the self-consistent effective field $U(V)$ without performing the iterative loop explicitly. To do this, we compute $V_{\text{int}}[\delta P_\text{s.p.}(U)]$ (Eq.~\ref{appeq:Vint_l}) on a dense one-dimensional grid of values of $U$. From Eq.~\ref{appeq:U_sc}, we have the condition $U-V_{\text{int}}[\delta P_\text{s.p.}(U)]=V$ at the self-consistent field $U(V)$. We can therefore plot $U-V_{\text{int},l}[\delta P_\text{s.p.}(U)]-V$, and extract the value of $U(V)$ where this intersects the $x$-axis.

\begin{figure*}
    \centering
    \includegraphics[width=1.0\linewidth]{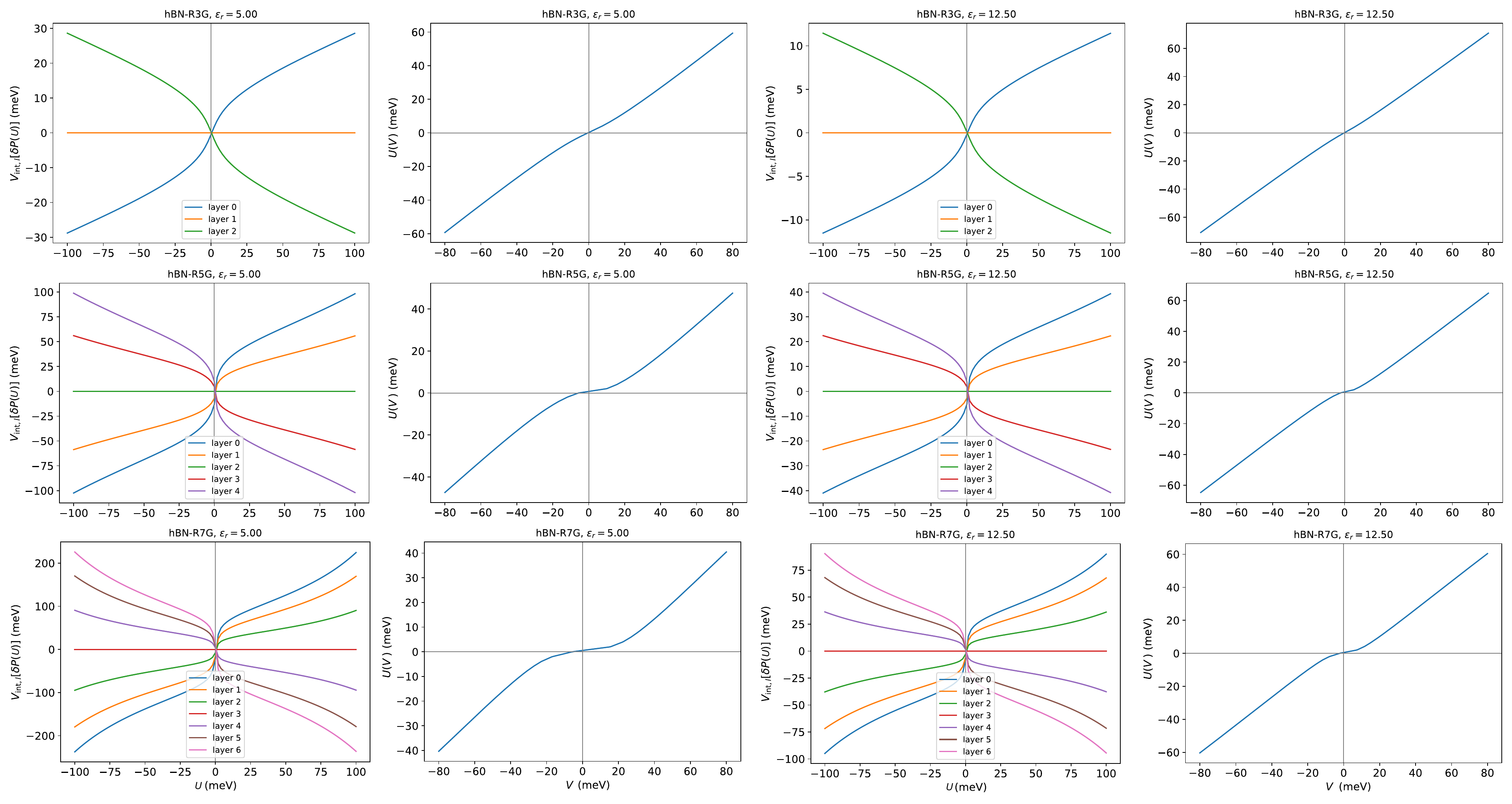}
    \caption{$\mbf{q}=0$ interlayer Hartree screening calculation for R$L$G/hBN with $L=3$ (top row), $L=5$ (middle row), $L=7$ (bottom row), and $\epsilon_r=5.0,12.5$. $V_{\text{int},l}[\delta P(U)]$ is the interaction-induced layer potential generated in the density matrix $\delta_\text{s.p.} P(U)$ constructed by filling all valence bands of $\hat{H}_{\text{s.p.}}(U)$ and subtracting off the reference density $P^\text{ref}$ in the average scheme. Note that we have zeroed the potential on the middle layer. $U(V)$ is the self-consistent effective interlayer potential corresponding to an externally applied interlayer potential $V$ within the approximation that all potentials vary linearly with layer. $\theta=0.77^\circ$ for all plots.}
    \label{self_screening}
\end{figure*}

We show examples of the behavior of $V_{\text{int},l}[\delta P_\text{s.p.}(U)]$ and the self-consistent interlayer potential $U(V)$ in Fig.~\ref{self_screening}. $V_{\text{int},l}[\delta P_\text{s.p.}(U)]$ (recall that this is the interlayer Hartree potential generated by occupying the valence bands of $\hat{H}_\text{s.p.}(U)$) is indeed approximately linear in layer index. $\frac{d}{dV}\left(V_{\text{int},l}[\delta P(V)]\right)$ is sharply peaked near $V=0$. This reflects the large layer polarizability of R$L$G near $V=0$, where the lowest valence and conduction bands, which form layer superposition states in isolated free R$L$G, become localized onto opposite extremal layers. This enhancement is more pronounced for larger numbers of layers, which is due to a combination of the greater interlayer distance between extremal layers, and the fact that low-energy states are flatter for larger $L$. Note that the center of the enhancement is slightly offset from $V=0$, since the single-alignment of hBN already introduces a slight layer potential on the bottom layer.  The shape of $V_{\text{int},l}[\delta P(U)]$ is also reflected in the self-consistent field $U(V)$ (computed according to the procedure in the paragraph above), which is relatively flat near $V=0$. For strong interactions (small $\epsilon_r$), application of a small displacement field $D$ only weakly changes $U(V)$.

Any projected calculation inevitably incurs errors when compared to calculations using the full unprojected many-body Hilbert space (constructed using the single-particle Hilbert space $\mathscr{H}$). The single-particle basis obtained by diagonalizing $\hat{H}_\text{s.p.}(U)$ at the self-consistent layer potential $U(V)$ (which we call the \emph{screened basis}) provides a more physically sensible projection basis for specifying the active subspace $\mathscr{H}_\text{act.}$ (see App.~\ref{secapp:projection} for details of how projection is defined), because the final many-body state in an interacting calculation for the ground state would have an effective layer potential that is closer to $U(V)$ than $V$. This means that a projected calculation on the projected Hamiltonian $\hat{H}_{\text{act.}} $ (Eq.~\ref{appeq:Hact}) would converge faster with projection band cutoffs $n_v,n_c$, compared to if we had instead specified $\mathscr{H}_\text{act}.$ using the eigenbasis of $\hat{H}_\text{s.p.}(V)$ (which we call the \emph{bare basis}). We call this choice of using the eigenbasis of  $\hat{H}_\text{s.p.}(U(V))$ as the projection basis as \emph{screened basis projection}, to contrast with the \emph{bare basis projection} which instead uses the bands of $\hat{H}_\text{s.p.}(V)$ to specify $\mathscr{H}_\text{act.}$. For the HF and TDHF computations in this paper, we will specify whether we used screened basis projection or bare basis projection. See App.~\ref{secapp:basiscomparison} for a comparison within a HF calculation between screened basis projection and bare basis projection, which shows significantly improved convergence using screened basis projection. 

More generally, for fixed physical parameters (such as the bare interlayer potential $V$) and projection basis band cutoffs $n_v,n_c$, we could consider repeating a projected calculation (such as a projected HF calculation) but with additional possibilities for the projection eigenbasis used to construct $\mathscr{H}_\text{act.}$. For instance, we could choose the eigenbasis of $\hat{H}_\text{s.p.}(\tilde{V})$ for a range of $\tilde{V}$ (where $\tilde{V}$ is not necessarily the bare potential $V$ or the screened interlayer potential $U(V)$ computed above). The quality of the projection basis could then be judged by the occupation of the highest projection conduction and valence bands (which should be close to 0\% and 100\% respectively --- see App.~\ref{secapp:basiscomparison}), and the many-body energy computed in the original unprojected Hamiltonian $\hat{H}$ (i.e.~the energy expectation value of Eq.~\ref{appeq:H_unproj} when the projected wavefunction is embedded into the unprojected wavefunction according to the parameterization of Eq.~\ref{appeq:psifillrem}). We do not use this generalization for the computations in this paper.

\clearpage

\section{Details of Hartree-Fock numerics}\label{secapp:HFdetails}

\subsection{Phase diagrams}\label{secapp:HFphase}

In this subsection, we present HF phase diagrams at $\nu=1$. We keep all spins and valleys in our calculation. We restrict to $S_z$-conserving states, but allow for intervalley coherence (IVC) that hybridizes mBZ momentum $\mbf{k}$ in valley $K$ with momentum $\mbf{k}$ in valley $K'$ ($\mbf{k}=0$ at $\tilde{\Gamma}_M$ for each valley). The HF state is constrained to preserve moir\'e translation symmetry so that the mBZ momentum $\mbf{k}$ remains a good quantum number. The HF state is uniquely parameterized by the one-body density matrix (projector)
\begin{equation}
    P_{m\eta,n\eta'}(\mbf{k},s)=\langle c^\dagger_{\mbf{k},m,\eta s} c_{\mbf{k},n,\eta' s} \rangle,
\end{equation}
where $c^\dagger_{\mbf{k},m,\eta s}$ are screened basis creation operators belonging to $\mathscr{H}_\text{act.}$ which consists of $n_v$ valence projection bands and $n_c$ conduction projection bands (see Sec.~\ref{secapp:projection}), and the total occupation is fixed by the filling factor $\nu$. In our projected calculations, this translates to $\sum_{\mbf{k}m\eta s}P_{m\eta,m\eta}(\mbf{k},s)=4n_vN_1N_2+\nu\eqqcolon N_{e,\text{act}.}$, where $N_{e,\text{act}.}$ is the number of electrons in the projected calculation.

Each point in the phase diagrams is the result of minimizing the energy over at least 16 HF calculations with different initial seeds $P^{(0)}_{m\eta,n\eta'}(\mbf{k},s)$ which we now describe. For the initial seeds, we use a selection of completely random states, spin-valley polarized states, and spin-valley unpolarized ground states of the $H_\text{s.p.}(U)$ evaluated at the screened interlayer potential $U(V)$ used for constructing $\mathscr{H}_\text{act.}$ (see Sec.~\ref{secapp:screening}). In more detail:
\begin{itemize}
    \item \underline{Random:} In our HF calculations, the good quantum numbers are the mBZ momentum $\mbf{k}$ and spin $s$. We first randomly choose the total (integer) occupation numbers $N_{(\mbf{k},s)}$ of each symmetry sector $(\mbf{k},s)$, consistent with the total number of electrons $N_{e,\text{act}.}=4n_vN_1N_2+\nu$ corresponding to filling $\nu=1$. For each symmetry sector $(\mbf{k},s)$, we then (using the \texttt{scipy.stats.unitary\_group} function in the Python module \texttt{scipy}) generate a Haar-random unitary matrix $\mathscr{U}_{(m\eta),(n\eta')}(\mbf{k},s)$ with linear dimension $2(n_v+n_c)$ (the factor of 2 accounts for the valleys). The initial projector for $(\mbf{k},s)$ is then initialized as
    \begin{equation}
        P^{(0)}_{m\eta,n\eta'}(\mbf{k},s)=\sum_{n'n''\eta''\eta'''}\mathscr{U}_{(m\eta),(n'\eta'')}(\mbf{k},s)\mathscr{P}_{(n'\eta''),(n''\eta''')}(\mbf{k},s)\mathscr{U}^*_{(n\eta'),(n''\eta''')}(\mbf{k},s)
    \end{equation}
    where $\mathscr{P}_{(n'\eta''),(n''\eta''')}(\mbf{k},s)$ is a diagonal matrix with $1$'s for the first $N_{(\mbf{k},s)}$ diagonal entries, and $0$'s for all other entries. From the above equation, it can be seen that the purpose of the random unitary is to randomly rotate the occupied subspace of $\mathscr{P}_{(n'\eta''),(n''\eta''')}(\mbf{k},s)$.
    \item \underline{Spin-valley polarized:} For spin-valley polarized seeds, the initial projector $P^{(0)}_{m\eta,n\eta'}(\mbf{k},s)$ is valley-diagonal. In flavors $(\eta,s)$ other than $(K,\uparrow)$, $P^{(0)}_{m\eta,n\eta}(\mbf{k},s)$ corresponds to occupying the active valence projection bands (recall that the projection band basis refers to the basis used to specify $\mathscr{H}_\text{act.}$), i.e.~$P^{(0)}_{m\eta,m\eta}(\mbf{k},s)=1$ for $m$ a valence band index, and 0 otherwise. These flavors are hence at charge neutrality. 
    
    In the remaining flavor  $(\eta,s)=(K,\uparrow)$, we first again occupy the active valence projection bands. We then need to occupy an additional orbital for each momentum $\mbf{k}$ in order to reach the required filling $\nu=1$. We consider two different options in our calculations. The first corresponds to fully occupying the lowest active conduction projection band in flavor $(K,\uparrow)$. The second corresponds to occupying a random conduction orbital at each $\mbf{k}$. This means that at each $\mbf{k}$, we construct a $n_c\times n_c$ Haar random unitary $\mathscr{U}_{m,n}(\mbf{k})$, and set the initial projector in the conduction subspace and flavor $(K,\uparrow)$ to be
    \begin{equation}
        P^{(0)}_{mK,nK}(\mbf{k},\uparrow)=\sum_{n'n''}\mathscr{U}_{mn'}(\mbf{k})\mathscr{P}_{n'n''}\mathscr{U}^*_{nn''}(\mbf{k}),\quad \text{for conduction indices }m,n,n',n'',
    \end{equation}
    where $\mathscr{P}_{n'n''}$ is an $n_c\times n_c$ matrix with $1$ in the top-left corner, and $0$'s everywhere else.
    \item \underline{Spin-valley unpolarized:} Recall that the projection basis used to define the active subspace $\mathscr{H}_\text{act.}$ is constructed by diagonalizing $\hat{H}_\text{s.p.}(U)$ at the internally screened interlayer potential $U(V)$ (see Sec.~\ref{secapp:projection} and Sec.~\ref{secapp:screening}), which also yields non-interacting kinetic energies $\tilde{E}^\eta_{n}(\mbf{k})$ for $\hat{H}_\text{s.p.}(U)$. We determine the initial density matrix $P^{(0)}_{m\eta,n\eta'}(\mbf{k},s)$ by occupying the $N_{e,\text{act.}}$ states with lowest $\tilde{E}^\eta_{n}(\mbf{k})$. Constructed this way, $P^{(0)}_{m\eta,n\eta'}(\mbf{k},s)$ is spin-valley diagonal and unpolarized. Note that $\tilde{E}^\eta_{n}(\mbf{k})$ is non-degenerate at a given momentum $\mbf{k}$ and valley $\eta$, though there may be near degeneracies.
\end{itemize}
For the spin-valley polarized and spin-valley unpolarized initial states, we also add a small amount of random noise (which breaks valley-$U(1)$ symmetry). For each symmetry sector $(\mbf{k},s)$, we add a random Hermitian matrix $\mathscr{M}(\mbf{k},s)$ of linear dimension $2(n_v+n_c)$ to $P^{(0)}_{m\eta,n\eta'}(\mbf{k},s)$. We set the average absolute value of entries in $\mathscr{M}(\mbf{k},s)$ to $0.1$.

Unless otherwise stated, we use a gate screening distance $d_\text{sc}=10\,\text{nm}$ in the interaction potential.

The HF gap indicates the energy difference of the HF eigenvalues between the lowest unoccupied and highest occupied HF orbitals. $n(\mbf{k})$ is the occupation of momentum $\mbf{k}$ in the HF state, such that non-zero $\max_\mbf{k}[n(\mbf{k})]-\min_\mbf{k}[n(\mbf{k})]$ rules out an insulating state. If $\max_\mbf{k}[n(\mbf{k})]-\min_\mbf{k}[n(\mbf{k})]=0$, then the HF gap is equivalent to the indirect HF band gap. $|C|$ is the absolute value of the Chern number, evaluated by computing gauge-invariant Berry fluxes through each plaquette of the momentum mesh. In our figures, $|C|$ is not shown (whited out) if the state is gapless, or the Berry flux through any plaquette exceeds $0.2\pi$ (a large Berry flux indicates a large concentration of Berry curvature, which requires a larger system size to ensure that $|C|$ is correctly computed). Valley polarization counts the absolute value of the net excess of electrons in valley $K$ compared to valley $K'$, divided by the number of moir\'e cells: $|(N_{e,K}-N_{e,K'})/N|$. Similarly, spin polarization is given by $|(N_{e,\uparrow}-N_{e,\downarrow})/N|$. Intervalley coherence (IVC) is given by the Frobenius norm of the inter-valley matrix $\sqrt{\sum_{m,n,s}|\langle c^\dagger_{\mbf{k},m,K,s}c_{\mbf{k},n,K',s}\rangle|^2}$, averaged over the mBZ.

\subsubsection{Phase diagrams for R$L$G/hBN ($L=3,4,5,6,7$) with $\xi=0,1$ and 3D interactions in the average interaction scheme}

In this subsection, we show HF phase diagrams for R$L$G/hBN ($L=3,4,5,6,7$) with $\xi=0,1$ and 3D interactions (App.~\ref{secapp:layer_int}) in the average interaction scheme (App.~\ref{secapp:general_scheme}), as a function of interaction strength $5/\epsilon_r$ and external interlayer potential $V$.

\begin{figure*}
    \centering
    \includegraphics[width=1.0\linewidth]{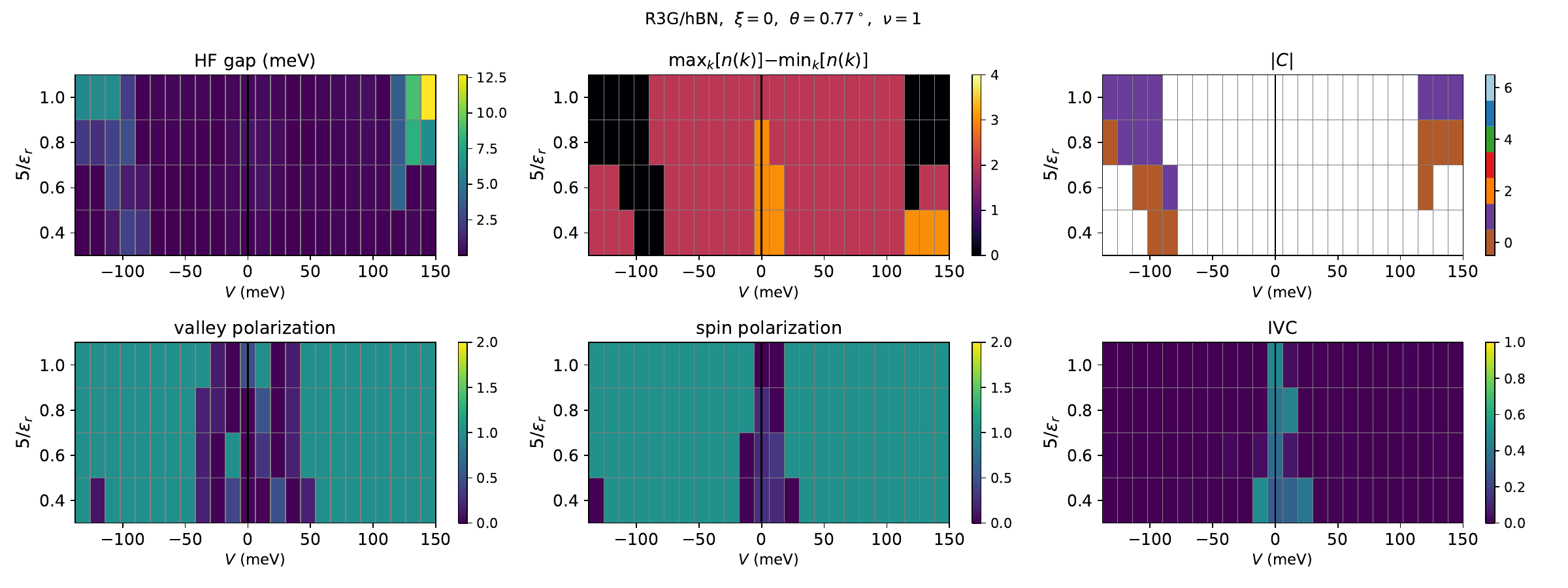}
    \caption{HF phase diagram for R3G/hBN, $\xi=0$, $\theta=0.77^\circ$, $\nu=1$. System size is $12\times 12$ and $(4+4)$ bands kept per spin/valley. }
    \label{app_L3_4p4xi0_t0.77_nu1}
\end{figure*}

\begin{figure*}
    \centering
    \includegraphics[width=1.0\linewidth]{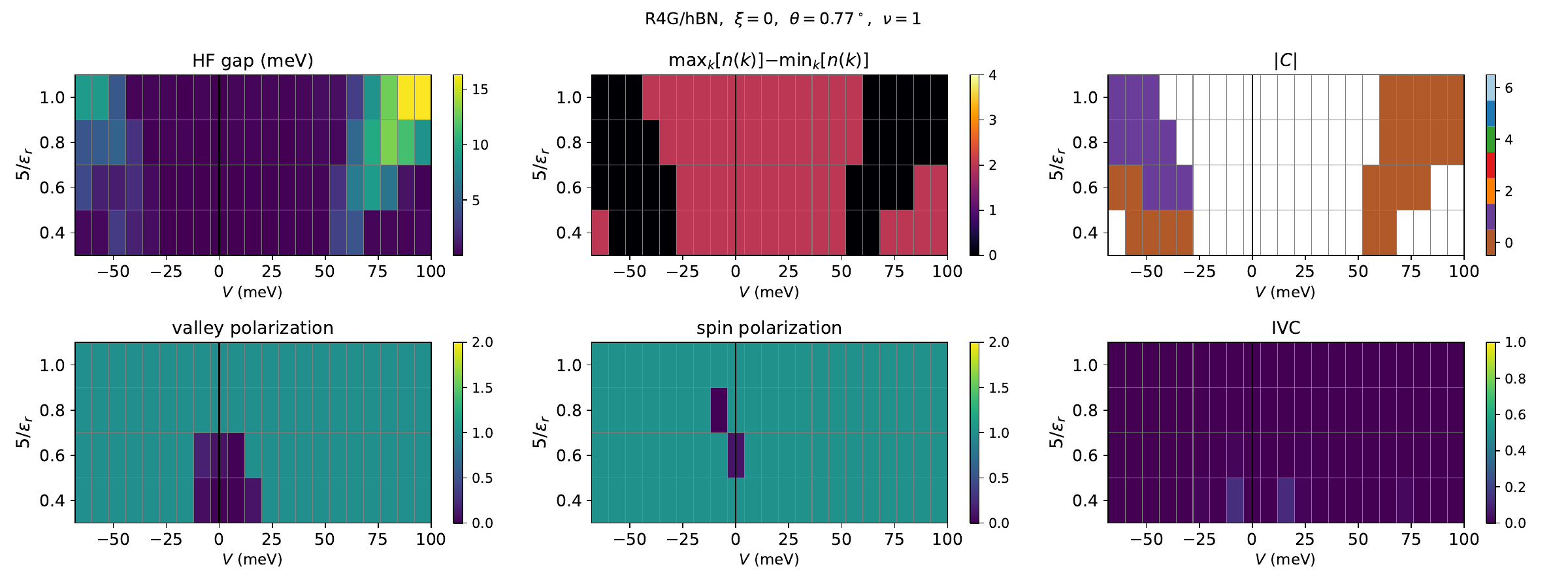}
    \caption{HF phase diagram for R4G/hBN, $\xi=0$, $\theta=0.77^\circ$, $\nu=1$. System size is $12\times 12$ and $(4+4)$ bands kept per spin/valley. }
    \label{app_L4_4p4xi0_t0.77_nu1}
\end{figure*}

\begin{figure*}
    \centering
    \includegraphics[width=1.0\linewidth]{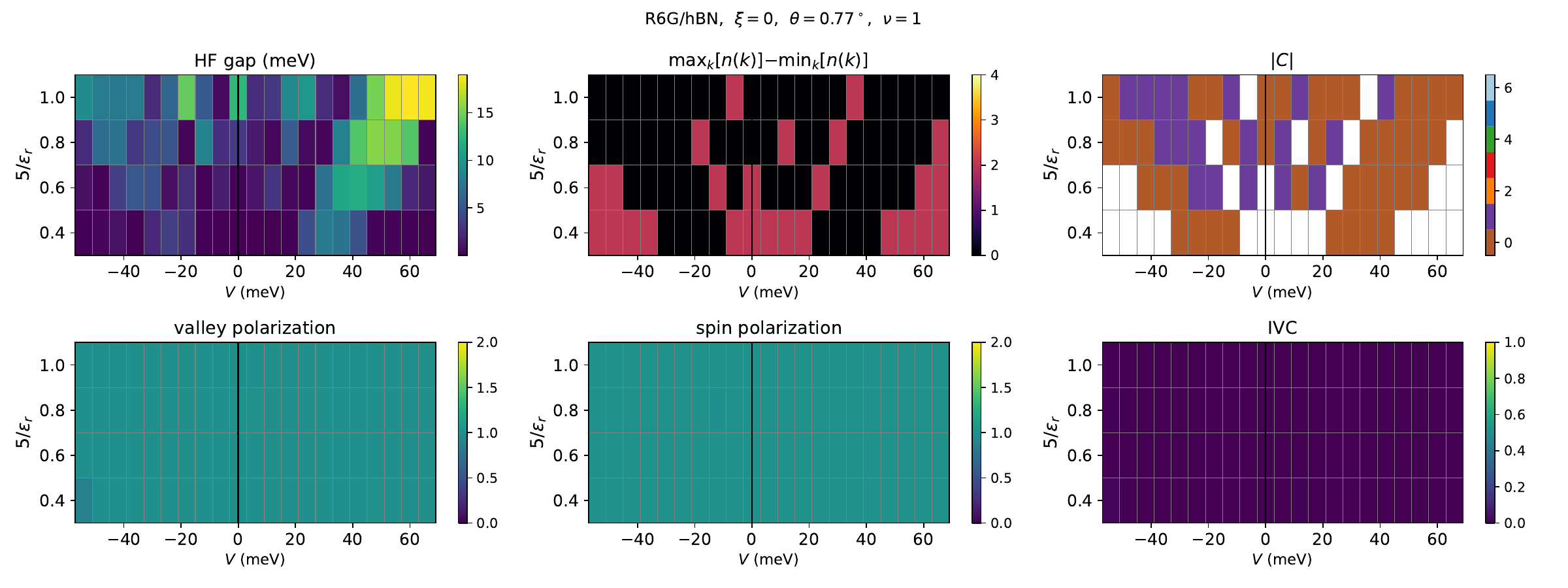}
    \caption{HF phase diagram for R6G/hBN, $\xi=0$, $\theta=0.77^\circ$, $\nu=1$. System size is $12\times 12$ and $(4+4)$ bands kept per spin/valley. }
    \label{app_L6_4p4xi0_t0.77_nu1}
\end{figure*}

\begin{figure*}
    \centering
    \includegraphics[width=1.0\linewidth]{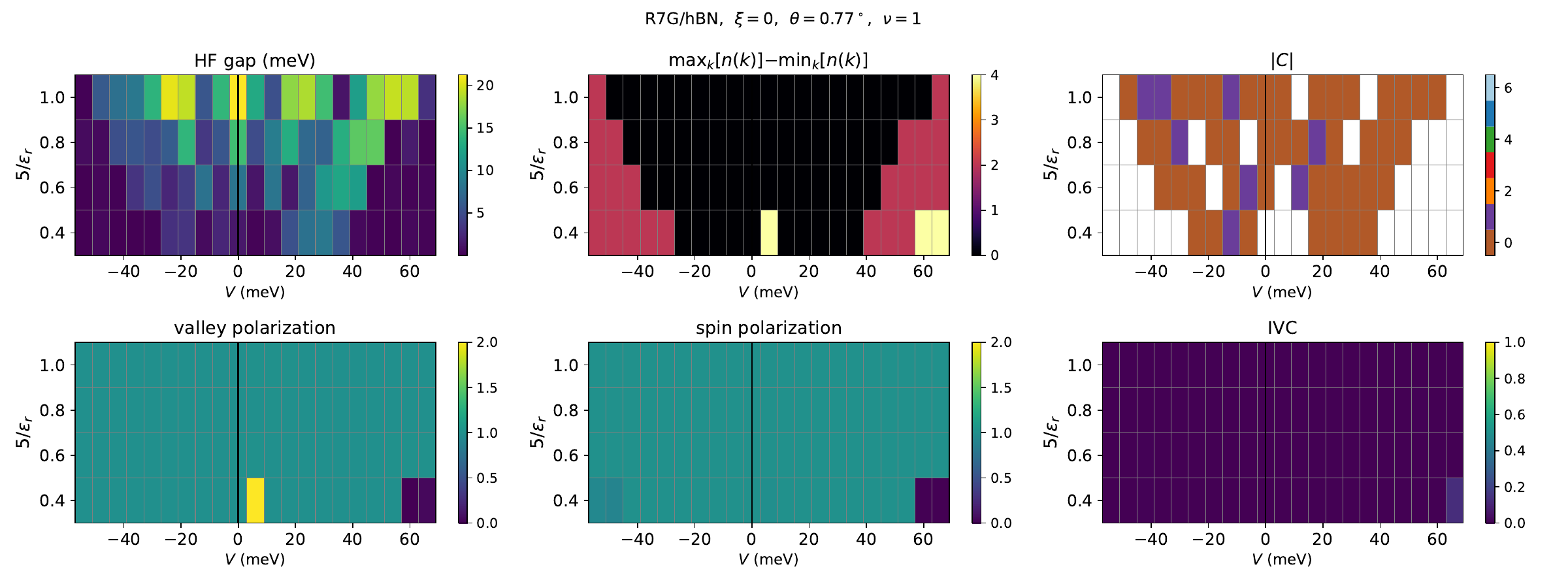}
    \caption{HF phase diagram for R7G/hBN, $\xi=0$, $\theta=0.77^\circ$, $\nu=1$. System size is $12\times 12$ and $(4+4)$ bands kept per spin/valley. }
    \label{app_L7_4p4xi0_t0.77_nu1}
\end{figure*}

Figs.~\ref{app_L3_4p4xi0_t0.77_nu1}, \ref{app_L4_4p4xi0_t0.77_nu1}, \ref{app_L5_4p4xi0_t0.77_nu1}, \ref{app_L6_4p4xi0_t0.77_nu1}, \ref{app_L7_4p4xi0_t0.77_nu1} show results for $\xi=0$ stacking and $\theta=0.77^\circ$, for $L=3,4,5,6,7$ layers respectively. We first comment on broad trends in the competition between gapped and gapless 
HF phases. For $L=3,4,5$, there is a sizable window of gapless states around $V=0$ that persists for all interaction strengths studied. This window is larger for a lower number of layers $L$. Near $V=0$ in the gapless region, the HF state can have finite IVC and imperfect spin-valley polarization for smaller $L$. For $L=6,7$, there appear to be gapped states around $V=0$ for stronger interactions. Our HF calculations find several competing states with full spin-valley polarization (i.e.~one flavor is at $\nu=+1$ while the others are at charge neutrality) but different Chern numbers for small $V$ and $L=6,7$, so it is likely that the large number of low-energy bands is making it challenging for HF to find the global energy minimum. The above observations regarding the widths of the gapless regions (or absence thereof for large $L$) is consistent with our understanding of the non-interacting band structure of R$L$G. For $V=0$, pristine R$L$G (i.e.~without hBN alignment) has a dispersion that is approximately $\sim k^L$ for small $k$. Since the band structure is more dispersive for smaller $L$, a large interlayer potential is required to sufficiently flatten the low-energy states in the conduction band in order to allow interactions to open an indirect gap at $\nu=1$. For larger $|V|$ and general $L$, we find two regions of gapped states, one each for $V<0$ and $V>0$, with the onset of the gapped region occuring at larger $|V|$ for stronger interactions. The gapped region at negative $V<0$ occurs at a lower value of $|V|$ and has a smaller HF gap than that at positive $V>0$. The gapped region at positive $V>0$ (where the conduction bands are polarized away from the hBN) has exclusively $C=0$, except for a small region of $|C|=1$ for $L=3$. On the other hand for the gapped region for $V<0$, there is a competition between $|C|=0,1$ states, with $|C|=1$ favored for smaller $|V|$.

\begin{figure*}
    \centering
    \includegraphics[width=1.0\linewidth]{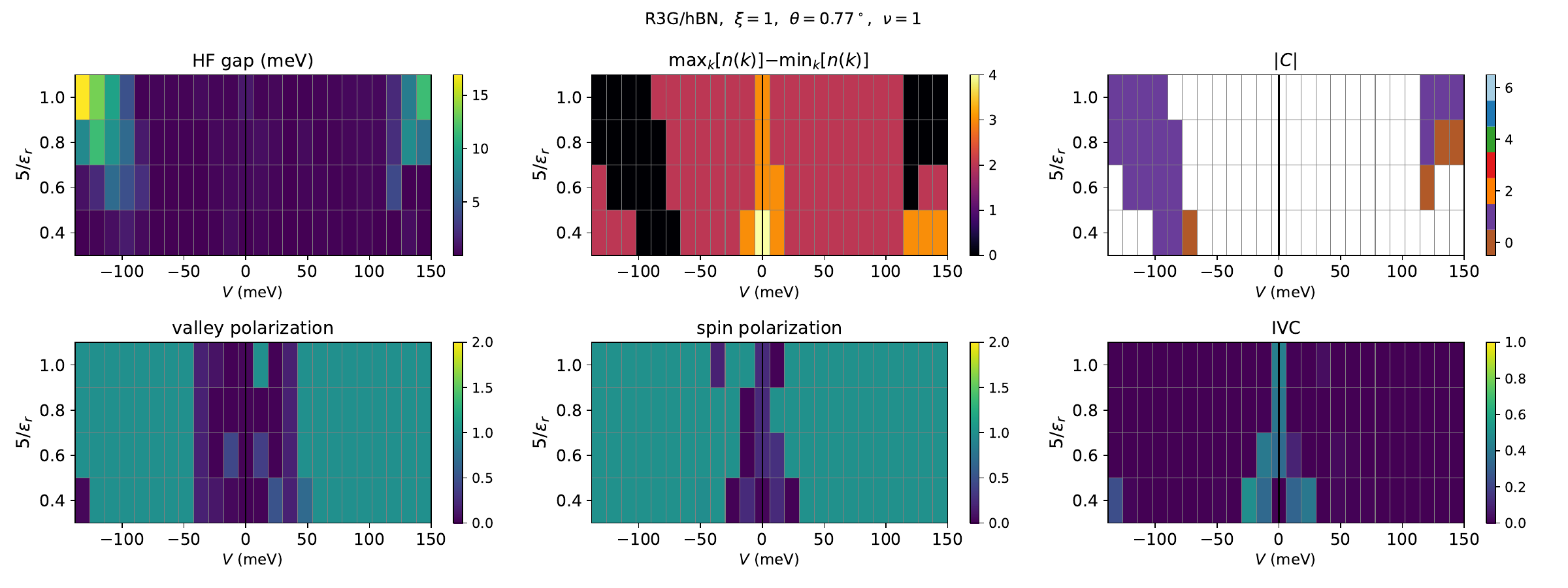}
    \caption{HF phase diagram for R3G/hBN, $\xi=1$, $\theta=0.77^\circ$, $\nu=1$. System size is $12\times 12$ and $(4+4)$ bands kept per spin/valley.}
    \label{app_L3_4p4xi1_t0.77_nu1}
\end{figure*}

\begin{figure*}
    \centering
    \includegraphics[width=1.0\linewidth]{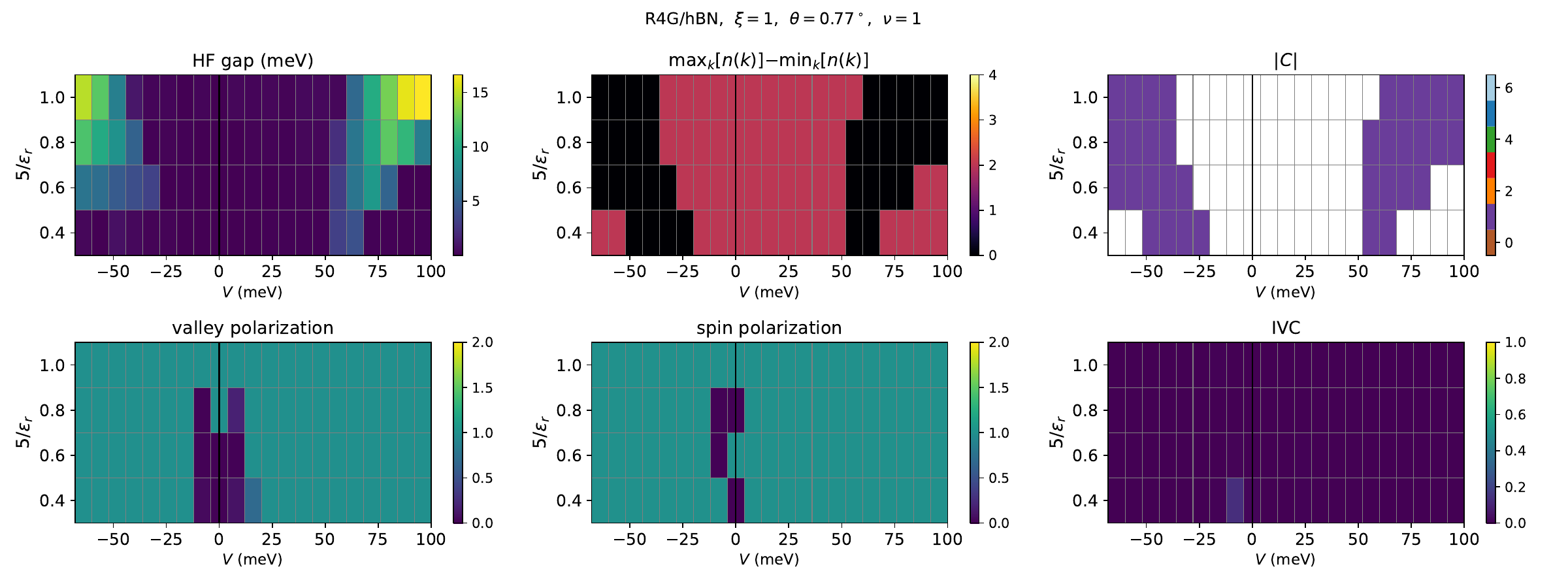}
    \caption{HF phase diagram for R4G/hBN, $\xi=1$, $\theta=0.77^\circ$, $\nu=1$. System size is $12\times 12$ and $(4+4)$ bands kept per spin/valley.}
    \label{app_L4_4p4xi1_t0.77_nu1}
\end{figure*}

\begin{figure*}
    \centering
    \includegraphics[width=1.0\linewidth]{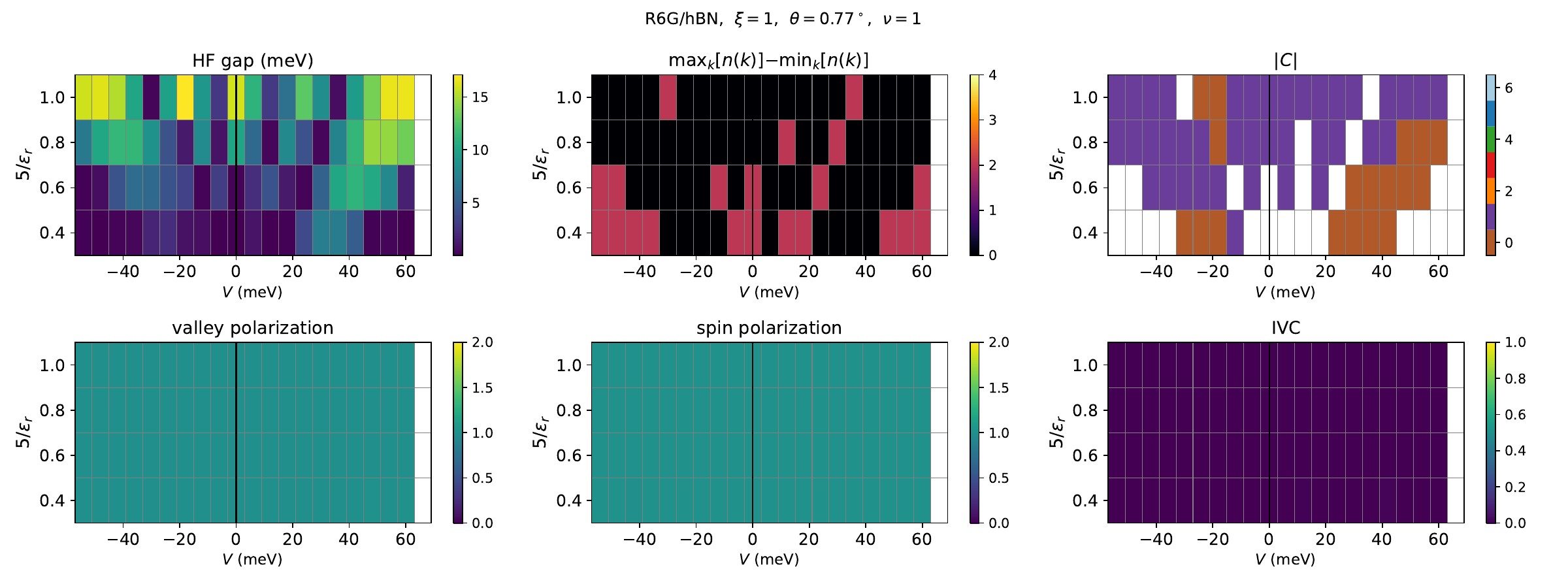}
    \caption{HF phase diagram for R6G/hBN, $\xi=1$, $\theta=0.77^\circ$, $\nu=1$. System size is $12\times 12$ and $(4+4)$ bands kept per spin/valley. }
    \label{app_L6_4p4xi1_t0.77_nu1}
\end{figure*}

\begin{figure*}
    \centering
    \includegraphics[width=1.0\linewidth]{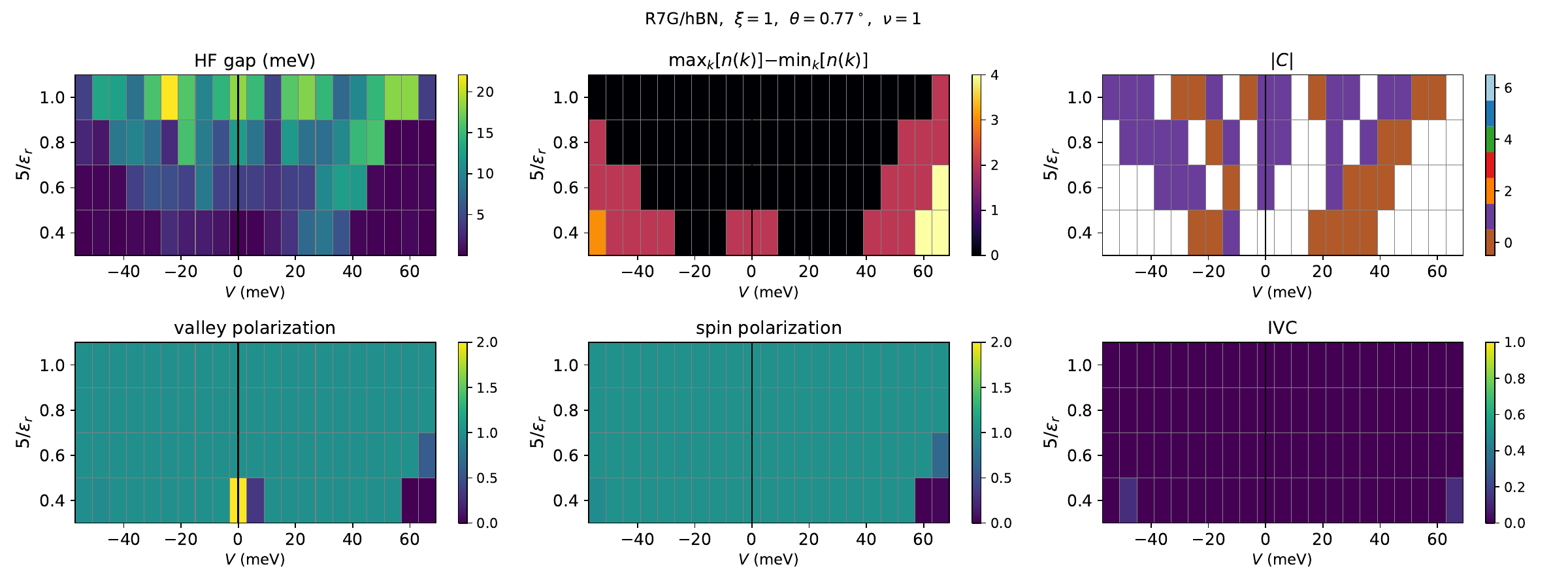}
    \caption{HF phase diagram for R7G/hBN, $\xi=1$, $\theta=0.77^\circ$, $\nu=1$. System size is $12\times 12$ and $(4+4)$ bands kept per spin/valley. }
    \label{app_L7_4p4xi1_t0.77_nu1}
\end{figure*}

Figs.~\ref{app_L3_4p4xi1_t0.77_nu1}, \ref{app_L4_4p4xi1_t0.77_nu1}, \ref{app_L5_4p4xi1_t0.77_nu1}, \ref{app_L6_4p4xi1_t0.77_nu1}, \ref{app_L7_4p4xi1_t0.77_nu1} show results for $\xi=1$ stacking and $\theta=0.77^\circ$, for $L=3,4,5,6,7$ layers respectively. The broad trends of the competition between gapped and gapless regions are the same as in the $\xi=0$ stacking. The most striking difference between the two stackings is that $\xi=1$ stacking has a greater tendency towards $|C|=1$ states compared to $\xi=0$ stacking. For example, the $V>0$ gapped region is mostly $|C|=1$ for $L=3,4,5$. For $L=6,7$ there is a small sliver of $|C|=1$, compared to the $\xi=0$ where the HF calculations instead find almost exclusively $|C|=0$ states. Hence, at least in the average interaction scheme, the electronic topology of the gapped phase at $\nu=+1$ can help distinguish between the inequivalent stackings $\xi=0,1$. This is true even for the $V>0$ regime, where the low-energy conduction bands are polarized away from the hBN, and hence only weakly affected by the hBN-induced moir\'e coupling at the non-interacting level. The qualitative difference between the two stackings must therefore come from the effect of the filled valence bands, which can directly couple to the hBN. The ground state phases obtained for $\xi=1$ and $5/\epsilon_r=0.4$ appear to be consistent with Ref.~\cite{LongJu2023FCIPentalayerGraphenehBN}: there is a gapless region around $V=0$, a $|C|=1$ region for $V>0$ that is preceded by a small window of $|C|=0$, and an absence of Chern insulators for $V<0$.

\begin{figure*}
    \centering
    \includegraphics[width=1.0\linewidth]{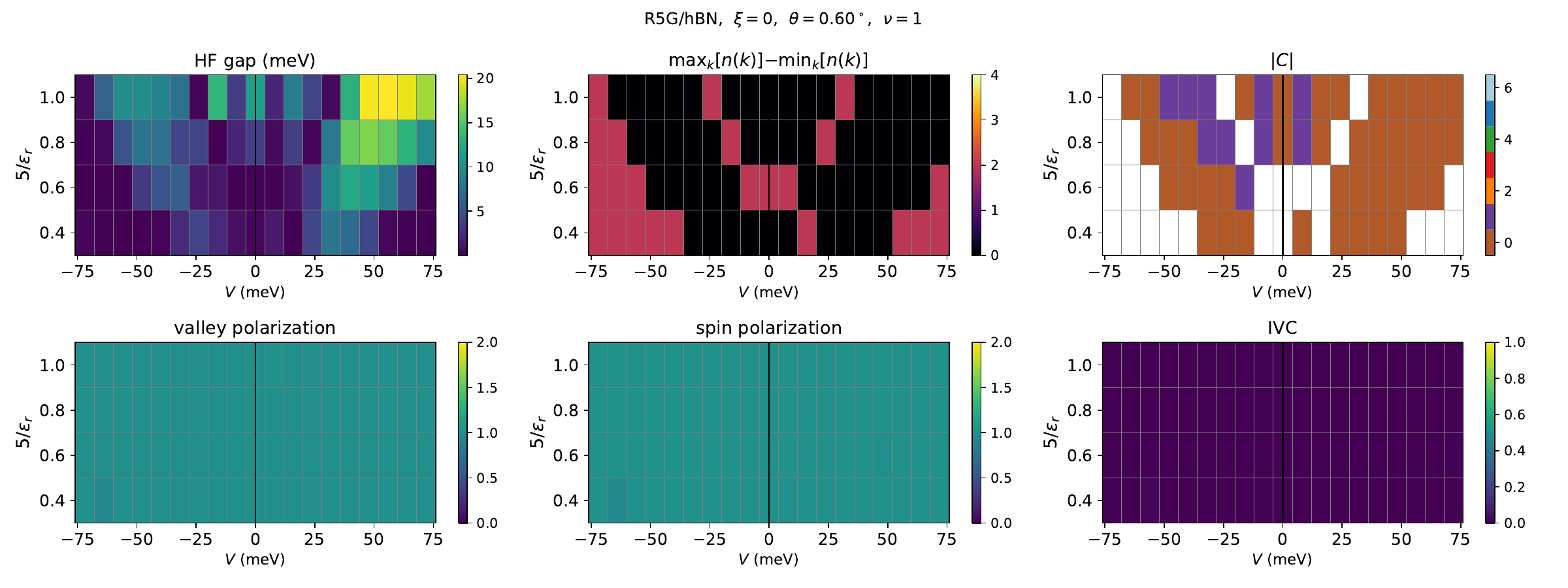}
    \caption{HF phase diagram for R5G/hBN, $\xi=0$, $\theta=0.60^\circ$, $\nu=1$. System size is $12\times 12$ and $(4+4)$ bands kept per spin/valley. }
    \label{app_L5_4p4xi0_t0.60_nu1}
\end{figure*}

\begin{figure*}
    \centering
    \includegraphics[width=1.0\linewidth]{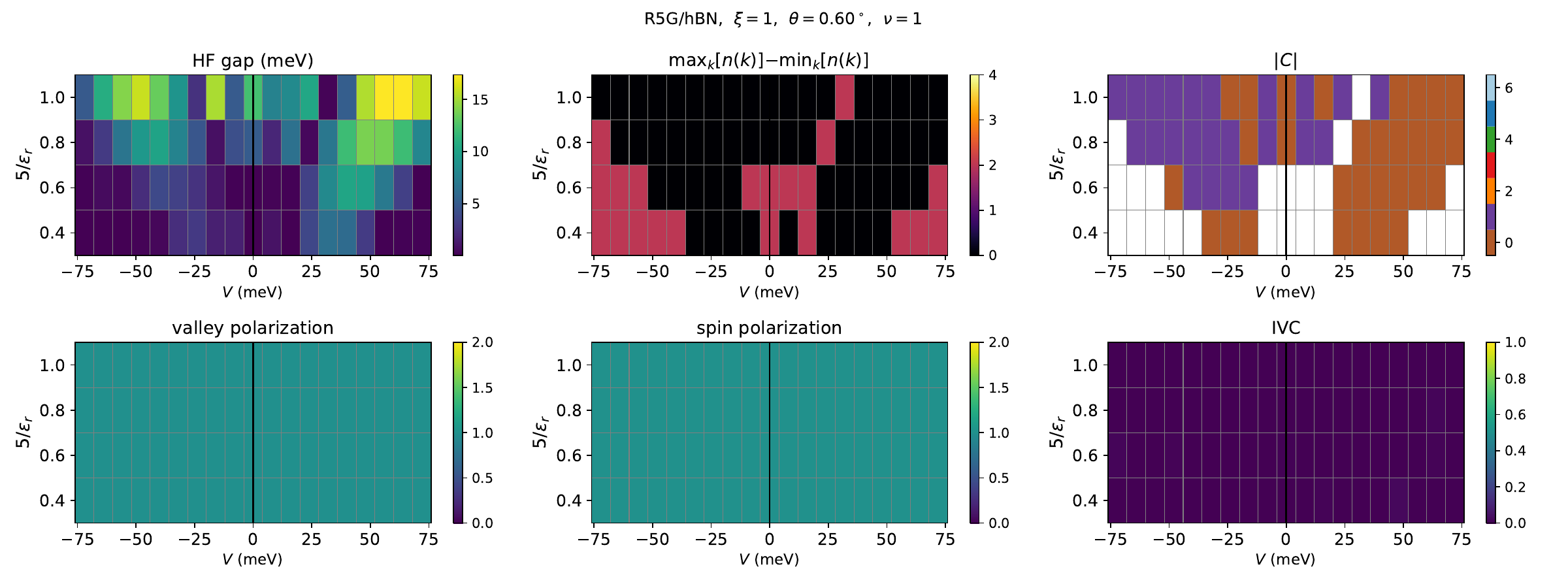}
    \caption{HF phase diagram for R5G/hBN, $\xi=1$, $\theta=0.60^\circ$, $\nu=1$. System size is $12\times 12$ and $(4+4)$ bands kept per spin/valley. }
    \label{app_L5_4p4xi1_t0.60_nu1}
\end{figure*}

\begin{figure*}
    \centering
    \includegraphics[width=1.0\linewidth]{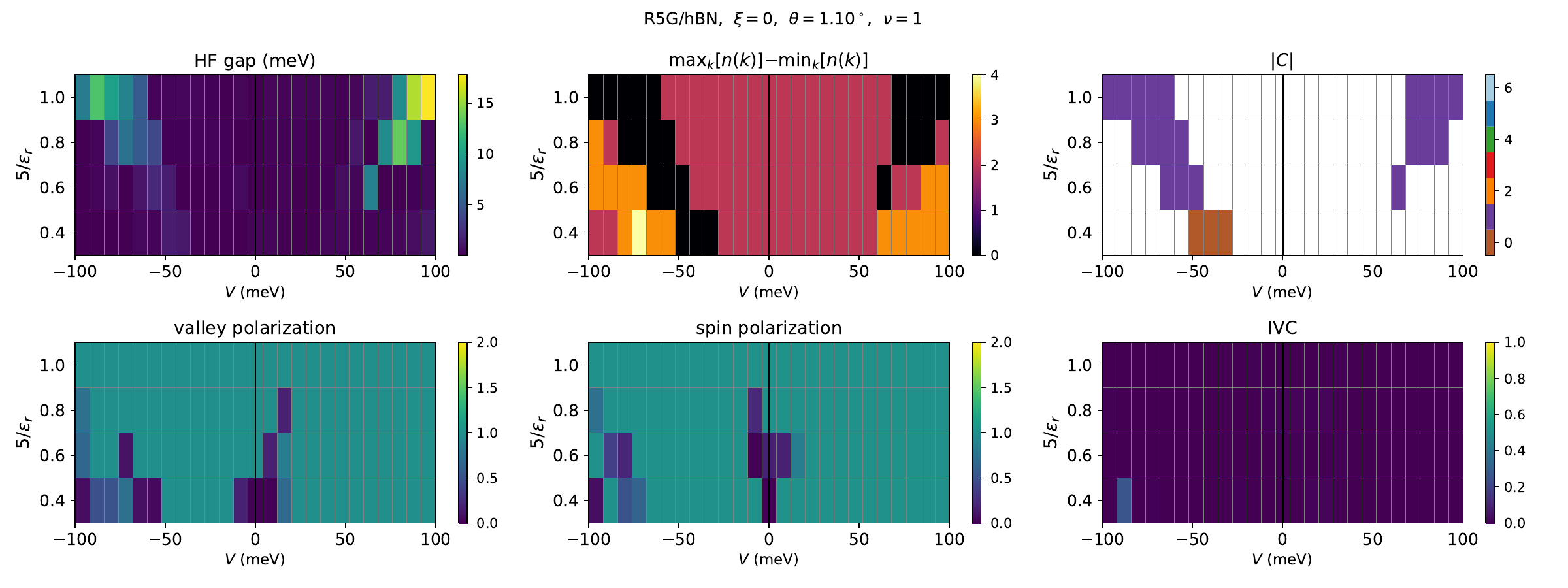}
    \caption{HF phase diagram for R5G/hBN, $\xi=0$, $\theta=1.10^\circ$, $\nu=1$. System size is $12\times 12$ and $(4+4)$ bands kept per spin/valley. }
    \label{app_L5_4p4xi0_t1.10_nu1}
\end{figure*}

\begin{figure*}
    \centering
    \includegraphics[width=1.0\linewidth]{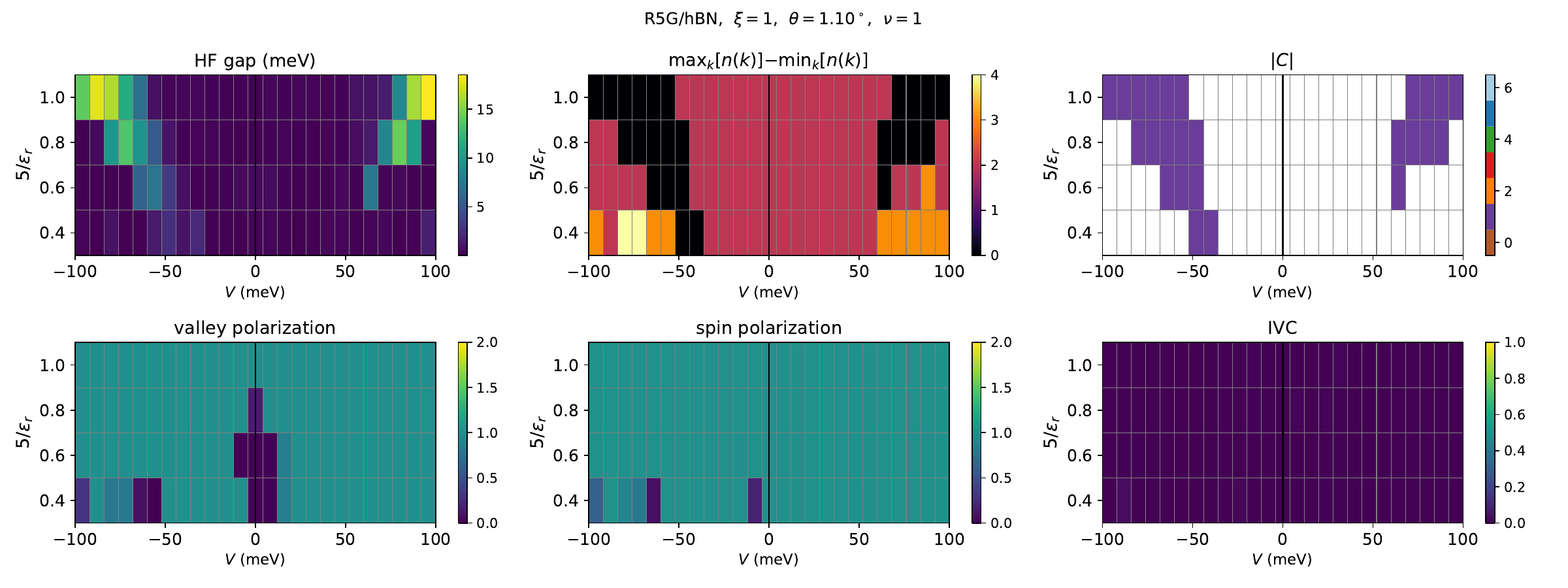}
    \caption{HF phase diagram for R5G/hBN, $\xi=1$, $\theta=1.10^\circ$, $\nu=1$. System size is $12\times 12$ and $(4+4)$ bands kept per spin/valley. }
    \label{app_L5_4p4xi1_t1.10_nu1}
\end{figure*}

Figs.~\ref{app_L5_4p4xi0_t0.60_nu1}, \ref{app_L5_4p4xi1_t0.60_nu1}, \ref{app_L5_4p4xi0_t0.77_nu1}, \ref{app_L5_4p4xi1_t0.77_nu1}, \ref{app_L5_4p4xi0_t1.10_nu1}, \ref{app_L5_4p4xi1_t1.10_nu1} show phase diagrams for R5G/hBN for stackings $\xi=0,1$ and twist angles $\theta=0.60^\circ,0.77^\circ,1.10^\circ$. For the largest twist angle $1.10^\circ$ (Figs.~\ref{app_L5_4p4xi0_t1.10_nu1} and \ref{app_L5_4p4xi1_t1.10_nu1}), we find for both stackings that the positions of gapped phases increase to higher $|V|$. This can be understood by considering how the twist angle affects the size of the mBZ, and consequently the folding of the bands of isolated R5G (which closely approximates the moir\'e band structure for large $V$). Recall that isolated R5G at $V=0$ has a dispersion that scales as $\sim k^5$. At larger $\theta$, the mBZ is larger in size, which means that the lowest conduction band in the mBZ has a higher bandwidth. Therefore, a greater displacement field is required to sufficiently flatten the band in order to allow interactions to open an indirect gap. For larger $\theta$, there is a greater tendency towards $|C|=1$ states, as now $\xi=0$ is exclusively $|C|=1$ for the gapped $V>0$ region. We also find that the gapped regions shrink in area in the phase diagram, and parts of the gapless phase around $V=0$ lose full spin and valley polarization. On the other hand, the gapped regions move to lower $V$ for the smaller twist angle $\theta=0.6^\circ$ (Figs.~\ref{app_L5_4p4xi0_t0.60_nu1} and \ref{app_L5_4p4xi1_t0.60_nu1}). In addition, the gapless region near $V=0$ shrinks, or even disappears completely. There is a tendency towards $|C|=0$, such that both stackings are in the $|C|=0$ state in the $V>0$ gapped region. We observe that the trends of the gapped vs.~gapless competition and the Chern numbers for increasing twist angle at fixed $L$ are similar to that for decreasing $L$ at fixed $\theta$. We also comment that our results for different twist angles for $L=5$ suggest that there is a `magic angle' window for realizing correlated topological phases in R5G/hBN. If $\theta$ is too small, then the $|C|=0$ phase is lower in energy compared to the Chern insulator at $\nu=+1$, which also rules out FCIs at lower fractional fillings if they descend from the parent Chern insulator at $\nu=+1$. If $\theta$ is too large, then stronger interactions are required to open a gap, and the larger displacement fields necessary to realize the Chern insulator at $\nu=1$ may be challenging to reach experimentally where the maximum accessible displacement field is around $D/\epsilon_0\simeq 1\,\text{V\,nm}^{-1}$. For example at $\theta=1.10^\circ$ and $\epsilon_r=5$ for $\xi=1$ stacking, the HF gap in the Chern insulator is maximal at around $V=100\,\text{meV}$. Using the relation $V= e D  d/\epsilon_0\epsilon_r$, this interlayer potential corresponds to $D/\epsilon_0\simeq 1.5\,\text{V\,nm}^{-1}$.

\subsubsection{Other phase diagrams for R5G/hBN}
In this subsection, we present additional phase diagrams for R5G/hBN, where we change some combination of the moir\'e potential, interaction scheme, and interaction potential.

\begin{figure*}
    \centering
    \includegraphics[width=1.0\linewidth]{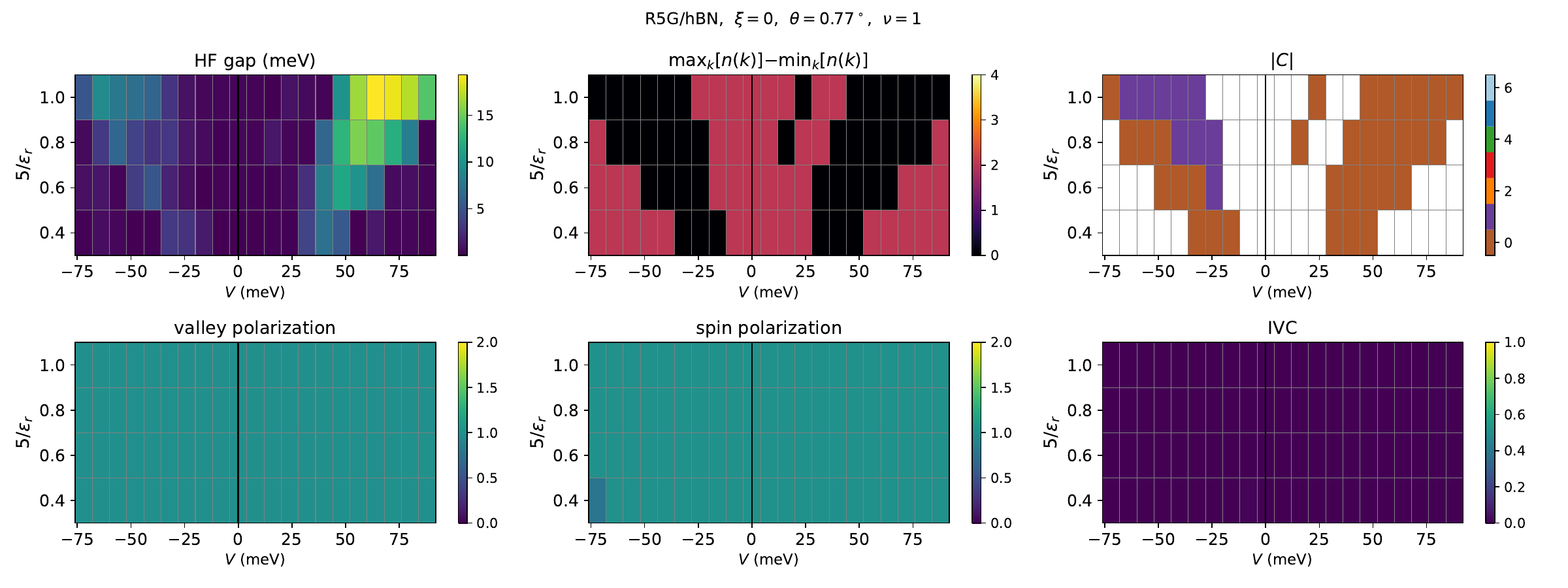}
    \caption{HF phase diagram for R5G/hBN, $\xi=0$, $\theta=0.77^\circ$, $\nu=1$. System size is $12\times 12$ and $(4+4)$ bands kept per spin/valley. }
    \label{app_L5_4p4xi0_t0.77_nu1}
\end{figure*}

\begin{figure*}
    \centering
    \includegraphics[width=1.0\linewidth]{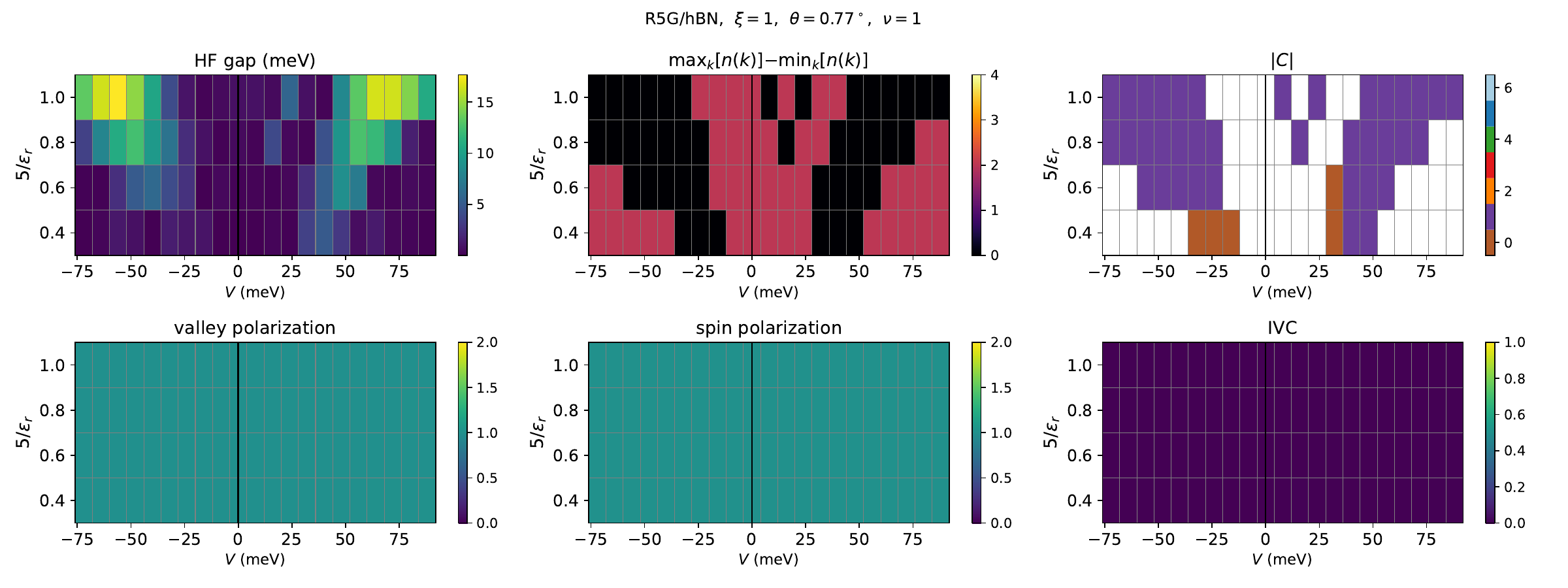}
    \caption{HF phase diagram for R5G/hBN, $\xi=1$, $\theta=0.77^\circ$, $\nu=1$. System size is $12\times 12$ and $(4+4)$ bands kept per spin/valley. }
    \label{app_L5_4p4xi1_t0.77_nu1}
\end{figure*}

\begin{figure*}
    \centering
    \includegraphics[width=1.0\linewidth]{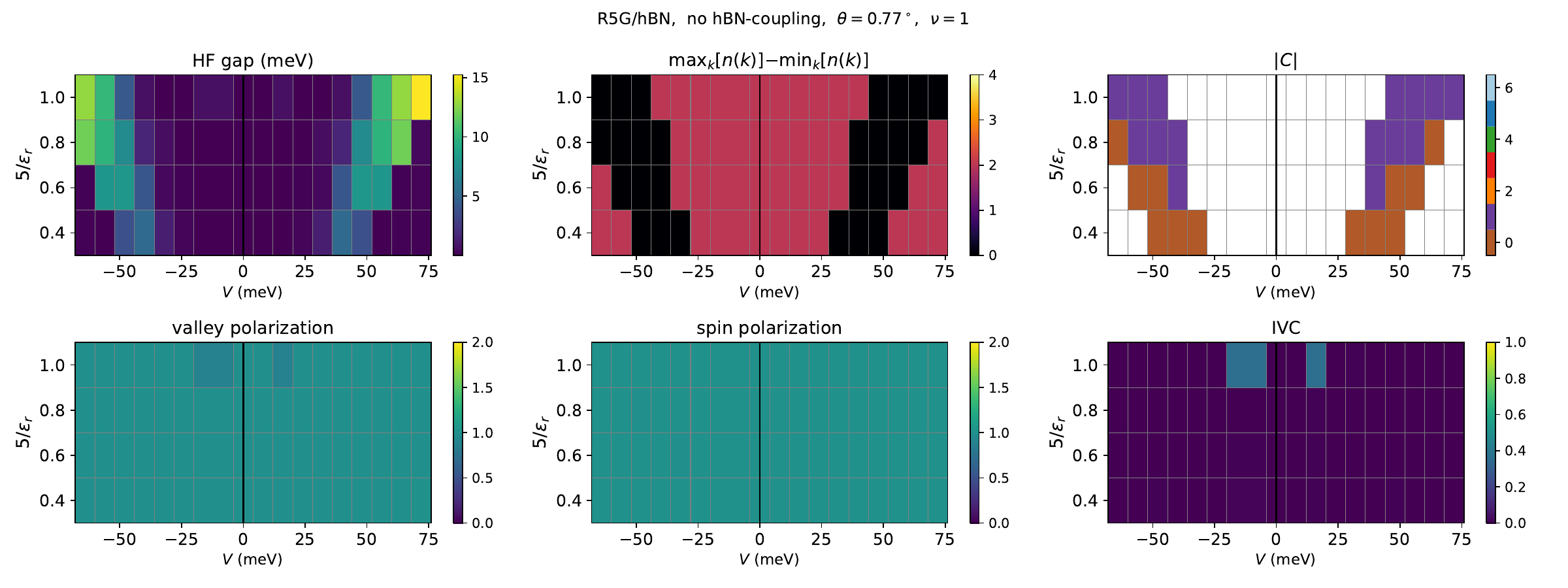}
    \caption{HF phase diagram for R5G/hBN, $\theta=0.77^\circ$, $\nu=1$. Note that the hBN-coupling has been switched off such that there is continuous translation symmetry. However, we allow spontaneous breaking of such symmetry consistent with the moir\'e cell at $\theta=0.77^\circ$. System size is $12\times 12$ and $(4+4)$ bands kept per spin/valley. }
    \label{app_L5_3p3_nomoire_t0.77_nu1}
\end{figure*}

Fig.~\ref{app_L5_3p3_nomoire_t0.77_nu1} shows the phase diagram for R5G/hBN where the coupling to the hBN has been switched off (i.e.~$\kappa_\text{hBN}=0$, so that there is no dependence on stacking $\xi$). In this case, the phase diagram is symmetric under $V\rightarrow -V$. The competition between gapless and gapped regions in the phase diagram is similar to the physical limit $\kappa_\text{hBN}=1$ with finite hBN coupling (compare to Figs.~\ref{app_L5_4p4xi0_t0.77_nu1} and \ref{app_L5_4p4xi1_t0.77_nu1}). In the gapped region, we find a competition between $|C|=0$ and $|C|=1$ states, with stronger interactions and smaller $V$ favoring the Chern insulator. These findings suggest that the $|C|=0$ and $|C|=1$ states are competing states in the absence of the moir\'e potential, and the hBN coupling plays a decisive role in tipping the energetic balance between these states.

\begin{figure*}
    \centering
    \includegraphics[width=1.0\linewidth]{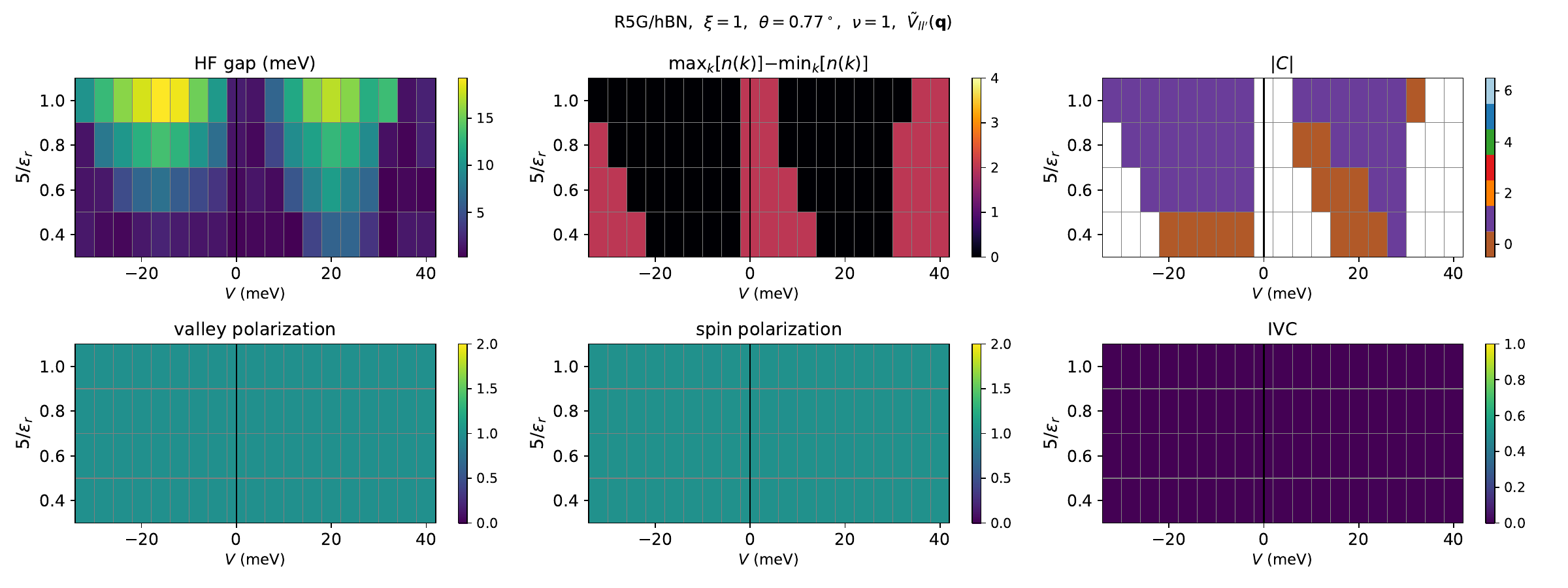}
    \caption{HF phase diagram for R5G/hBN, $\theta=0.77^\circ$, $\nu=1$. We have used an alternative layer-dependent interaction potential (Eq.~\ref{appeq:Vq_exp}). System size is $12\times 12$ and $(3+3)$ bands kept per spin/valley. }
    \label{app_L5_3p3avg_xi1_t0.77_nu1_exp}
\end{figure*}

Fig.~\ref{app_L5_3p3avg_xi1_t0.77_nu1_exp} shows the phase diagram for $\xi=1$ R5G/hBN using the alternative layer-dependent interaction potential in Eq.~\ref{appeq:Vq_exp}. Since $\tilde{V}(\mbf{q},z,z')$ is layer independent for $\mbf{q}=0$, there is no interlayer Hartree screening of the external interlayer potential. As such, the gapped phases occur at a significantly smaller $V$ (for positive interlayer potentials, compare $V\simeq 20\,\text{meV}$ in Fig.~\ref{app_L5_3p3avg_xi1_t0.77_nu1_exp} to $V\simeq 60\,\text{meV}$ in Fig.~\ref{app_L5_4p4xi1_t0.77_nu1}), demonstrating the importance of the $\mbf{q}=0$ layer dependence of the interaction for a quantitative treatment of the external displacement field.

\begin{figure*}
    \centering
    \includegraphics[width=1.0\linewidth]{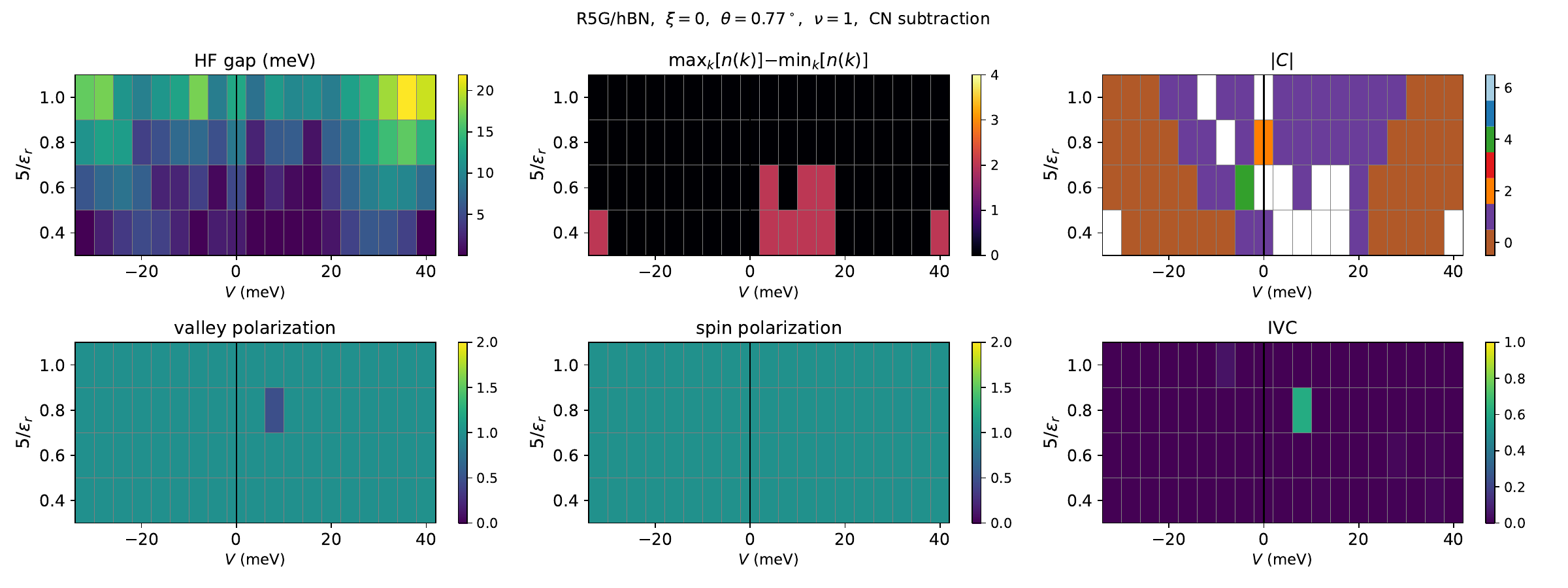}
    \caption{HF phase diagram for R5G/hBN, $\xi=0$, $\theta=0.77^\circ$, $\nu=1$. Note that we use the CN interaction scheme. (In the CN interaction scheme, our definitions for the screened basis and bare basis coincide). System size is $12\times 12$ and $(3+3)$ bands kept per spin/valley. }
    \label{app_L5_3p3CN_xi0_t0.77_nu1}
\end{figure*}

\begin{figure*}
    \centering
    \includegraphics[width=1.0\linewidth]{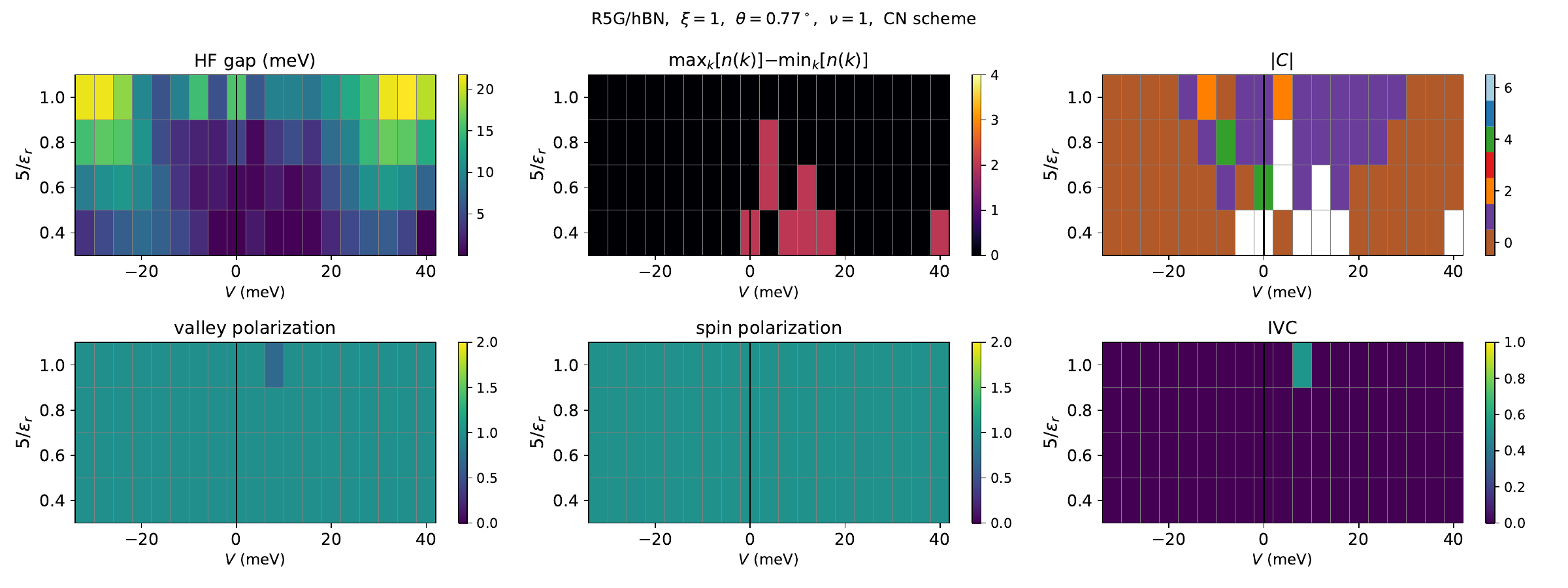}
    \caption{HF phase diagram for R5G/hBN, $\xi=1$, $\theta=0.77^\circ$, $\nu=1$. Note that we use the CN interaction scheme. (In the CN interaction scheme, our definitions for the screened basis and bare basis coincide). System size is $12\times 12$ and $(3+3)$ bands kept per spin/valley. }
    \label{app_L5_3p3CN_xi1_t0.77_nu1}
\end{figure*}

\begin{figure*}
    \centering
    \includegraphics[width=1.0\linewidth]{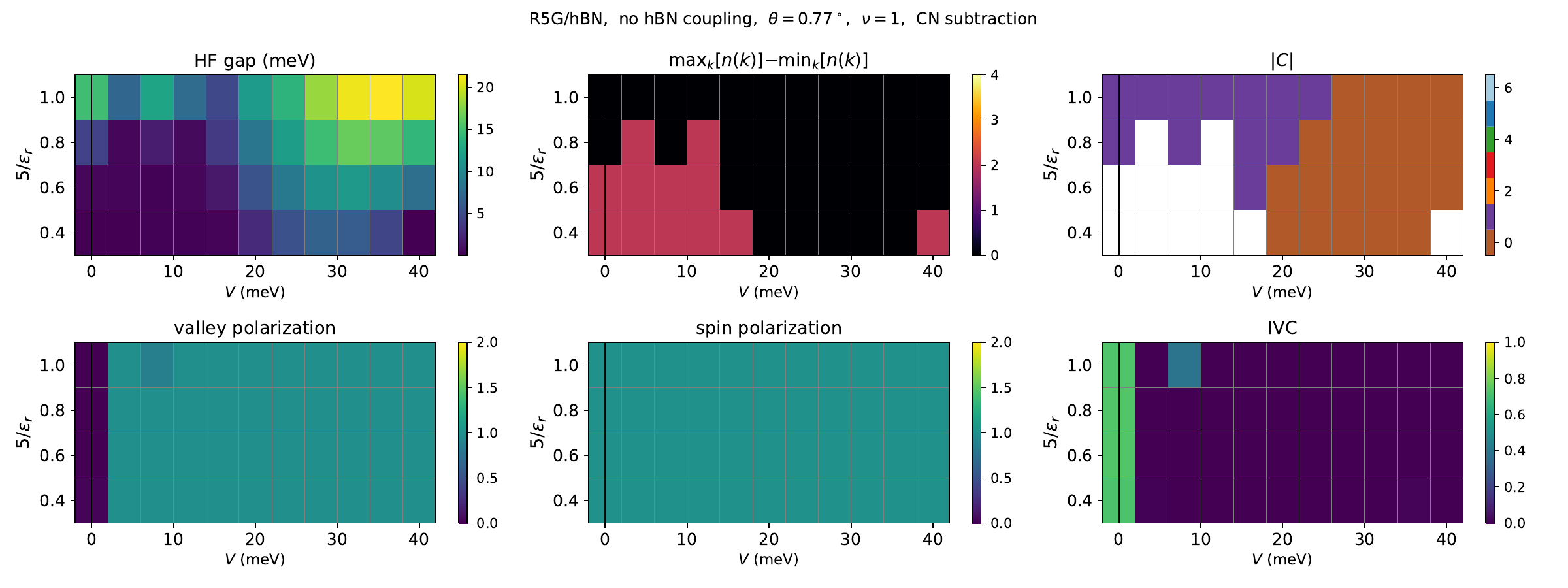}
    \caption{HF phase diagram for R5G/hBN, $\theta=0.77^\circ$, $\nu=1$. Note that we use the CN interaction scheme. (In the CN interaction scheme, our definitions for the screened basis and bare basis coincide). Also note that the hBN-coupling has been switched off suc that there is continuous translation symmetry. System size is $12\times 12$ and $(3+3)$ bands kept per spin/valley. }
    \label{app_L5_3p3CN_t0.77_nu1_nomoire}
\end{figure*}

Figs.~\ref{app_L5_3p3CN_xi0_t0.77_nu1} and \ref{app_L5_3p3CN_xi1_t0.77_nu1} show the phase diagram for $\xi=0,1$ R5G/hBN in the CN interaction scheme. Since the CN scheme does not including the physics of internal interlayer screening, the gapped phases appear at smaller $V$ compared to the average interaction scheme. We also find that the gapless window for small $V$ either shrinks considerably, or vanishes for larger interaction strengths, suggesting that the CN scheme is less appropriate in this regime for quantitatively capturing the large gapless window observed experimentally. For large $V$, we find little difference between the two stackings $\xi$ because the low-energy conduction bands are polarized away from the hBN, and the filled valence bands (whose charge density carries an imprint of the underlying hBN moir\'e coupling, see App.~\ref{app:charge_density_band_cutoff}) only weakly affects the conduction bands because its charge density is cancelled by the reference density. Fig.~\ref{app_L5_3p3CN_t0.77_nu1_nomoire} shows the phase diagram for R5G/hBN in the CN interaction scheme where the coupling to the hBN has been switched off (i.e.~$\kappa_\text{hBN}=0$, so that there is no dependence on stacking $\xi$), but the moir\'e unit cell is kept the same size. We only show positive $V$ since negative $V$ is related by symmetry. The results for large $V$ are similar to those with the physical hBN coupling strength $\kappa_\text{hBN}=1$ (Figs.~\ref{app_L5_3p3CN_xi0_t0.77_nu1} and \ref{app_L5_3p3CN_xi1_t0.77_nu1}) because the low-energy conduction electrons barely experience the moir\'e potential. The findings above illustrate the lack of dependence on the results for large $V>0$ on the stacking orientation $\xi$ in the CN scheme, in contrast to the results for the average scheme. This could provide a route for experiments to help narrow down the most appropriate interaction scheme.

\begin{figure*}
    \centering
    \includegraphics[width=1.0\linewidth]{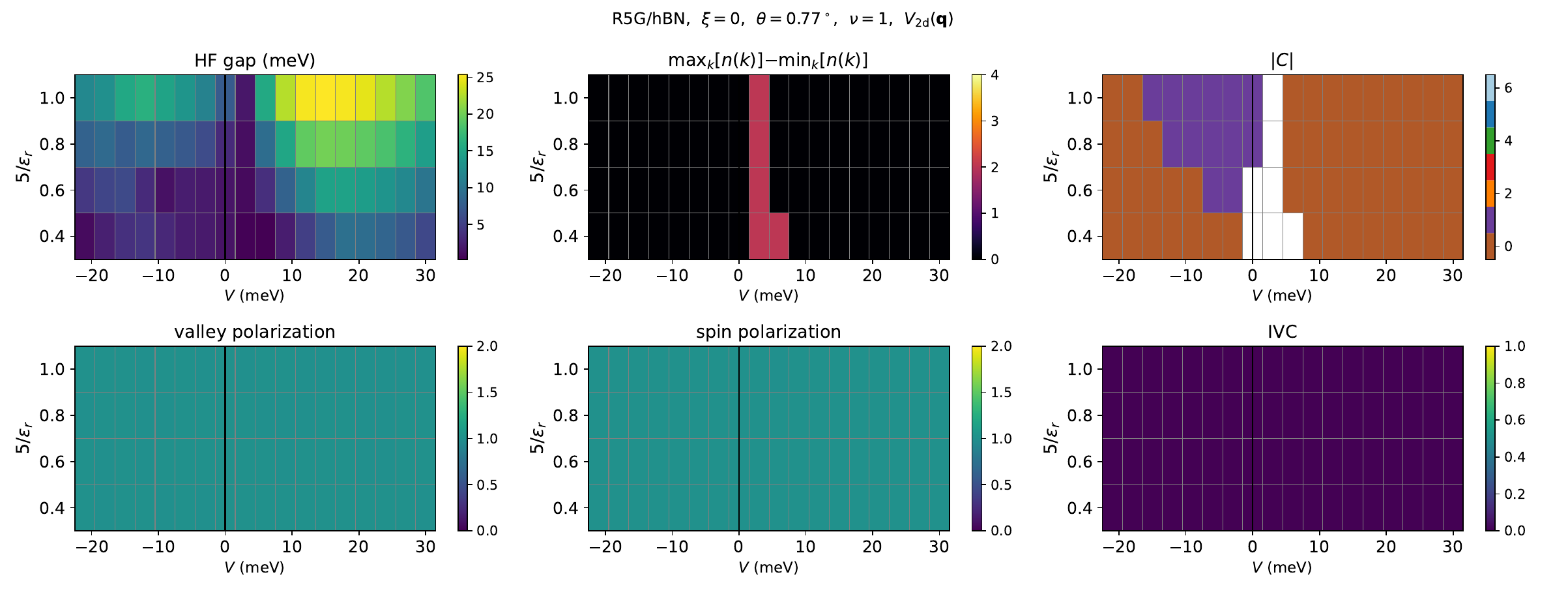}
    \caption{HF phase diagram for R5G/hBN, $\xi=0$, $\theta=0.77^\circ$, $\nu=1$. Furthermore, we use the purely 2d interaction $V_{2d}(\mbf{q})$, i.e.~we set the interlayer distance $d=0$. System size is $12\times 12$ and $(4+4)$ bands kept per spin/valley. }
    \label{app_L5_4p4xi0_t0.77_nu1_2d}
\end{figure*}

\begin{figure*}
    \centering
    \includegraphics[width=1.0\linewidth]{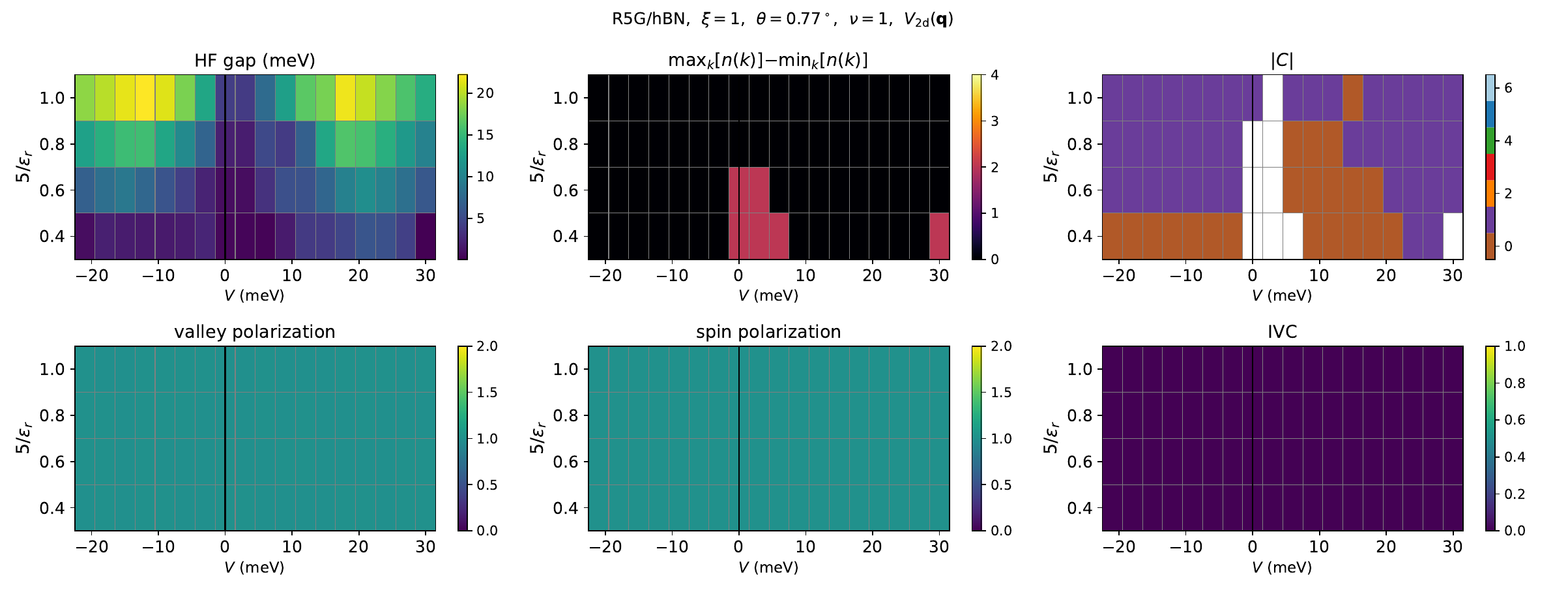}
    \caption{HF phase diagram for R5G/hBN, $\xi=1$, $\theta=0.77^\circ$, $\nu=1$. Furthermore, we use the purely 2d interaction $V_{2d}(\mbf{q})$, i.e.~we set the interlayer distance $d=0$. System size is $12\times 12$ and $(4+4)$ bands kept per spin/valley. }
    \label{app_L5_4p4xi1_t0.77_nu1_2d}
\end{figure*}

In Fig.~\ref{app_L5_4p4xi0_t0.77_nu1_2d} ($\xi=0$) and Fig.~\ref{app_L5_4p4xi1_t0.77_nu1_2d} ($\xi=1$), we use a purely 2d interaction potential by setting the rhombohedral graphene interlayer distance $d=0$, and use the average interaction scheme. Compared to the analogous calculations with the 3D interaction (Figs.~\ref{app_L5_4p4xi0_t0.77_nu1} and \ref{app_L5_4p4xi1_t0.77_nu1}), the Chern number dependence on the stacking (higher propensity for $|C|=1$ for $\xi=1$) is similar. This is expected, because the charge density of the valence bands affects the conduction electrons in the average scheme, regardless of the dimensionality of the interaction. 
However, due to the absence of interlayer screening, the gapped phases occur at much smaller $V$, and the the gapless region near $V=0$ is either small or non-existent.

\begin{figure*}
    \centering
    \includegraphics[width=1.0\linewidth]{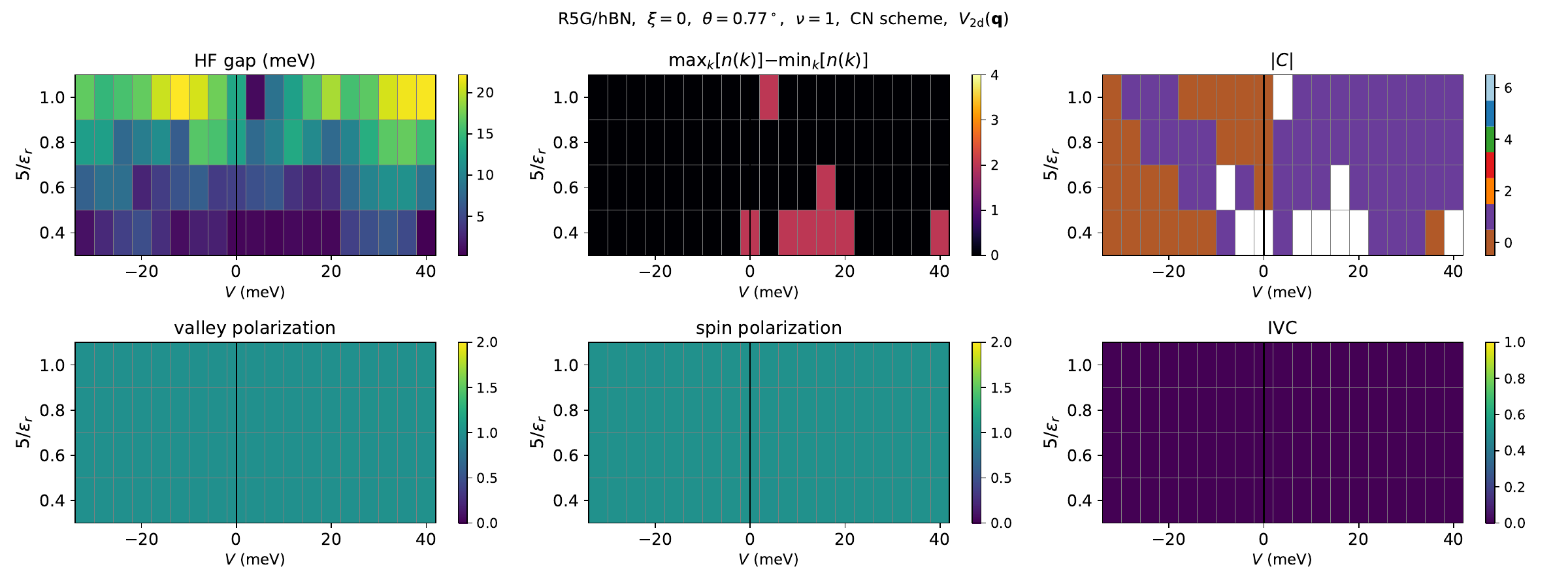}
    \caption{HF phase diagram for R5G/hBN, $\xi=0$, $\theta=0.77^\circ$, $\nu=1$. Note that we use the CN interaction scheme. Furthermore, we use the purely 2d interaction $V_{2d}(\mbf{q})$, i.e.~we set the interlayer distance $d=0$. System size is $12\times 12$ and $(3+3)$ bands kept per spin/valley. }
    \label{app_L5_3p3CN_xi0_t0.77_nu1_V2d}
\end{figure*}

\begin{figure*}
    \centering
    \includegraphics[width=1.0\linewidth]{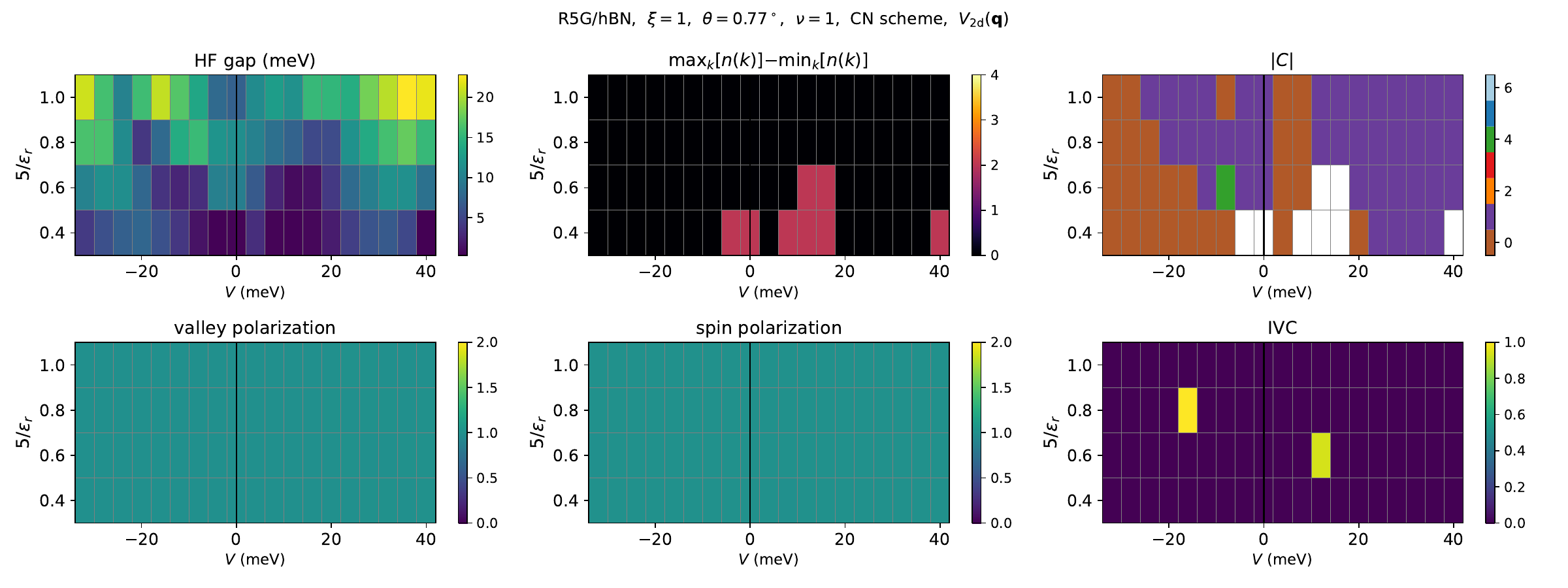}
    \caption{HF phase diagram for R5G/hBN, $\xi=1$, $\theta=0.77^\circ$, $\nu=1$. Note that we use the CN interaction scheme. Furthermore, we use the purely 2d interaction $V_{2d}(\mbf{q})$, i.e.~we set the interlayer distance $d=0$. System size is $12\times 12$ and $(3+3)$ bands kept per spin/valley. }
    \label{app_L5_3p3CN_xi1_t0.77_nu1_V2d}
\end{figure*}

In Fig.~\ref{app_L5_3p3CN_xi0_t0.77_nu1_V2d} ($\xi=0$) and Fig.~\ref{app_L5_3p3CN_xi1_t0.77_nu1_V2d} ($\xi=1$), we use a purely 2d interaction potential and the CN interaction scheme. The two stackings have similar phase diagrams at large $V$, where both are in the Chern insulator phase. The gapless region near $V=0$ is either small or non-existent.


\clearpage
\subsection{Comparison between screened basis and bare basis within HF}
\label{secapp:basiscomparison}

\begin{figure*}
    \centering
    \includegraphics[width=1.0\linewidth]{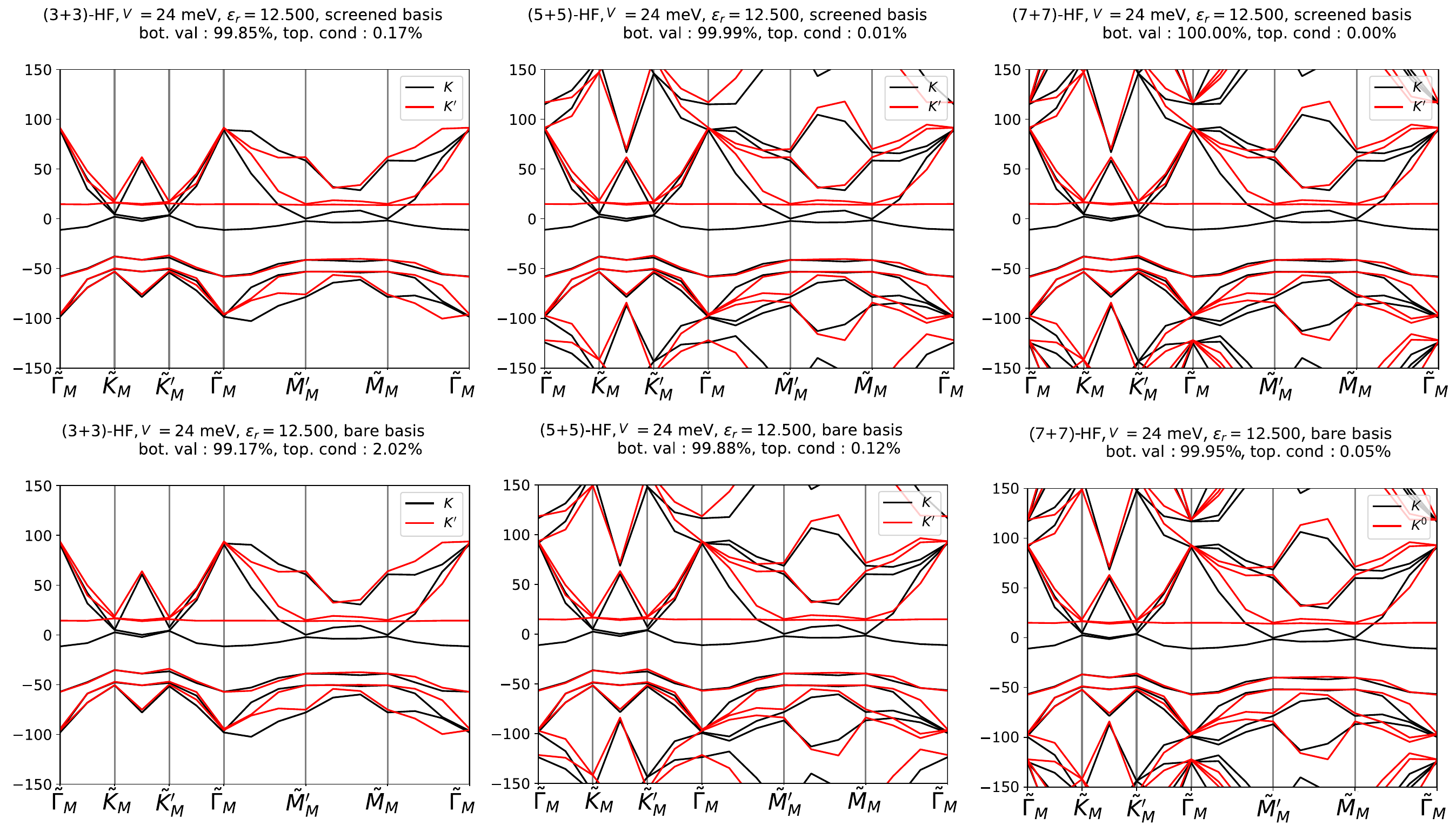}
    \caption{Comparison of screened basis (top) vs bare basis (bottom) R5G/hBN HF calculations at $\nu=1$ for different numbers of active bands $(n_c+n_v)$ per spin/valley. We have chosen to polarize in valley $K$ and spin $\uparrow$, and we show only the spin-$\uparrow$ HF band structure. We also list the occupation of the bottom valence and highest conduction bands, in the basis used to define the active subspace $\mathscr{H}_\text{act.}$.
    $\theta=0.77^\circ,V=24\,\text{meV},\epsilon_r=12.50$, and system size is $6\times 6$.}
    \label{bench_U0.024_epsr12.50}
\end{figure*}

\begin{figure*}
    \centering
    \includegraphics[width=1.0\linewidth]{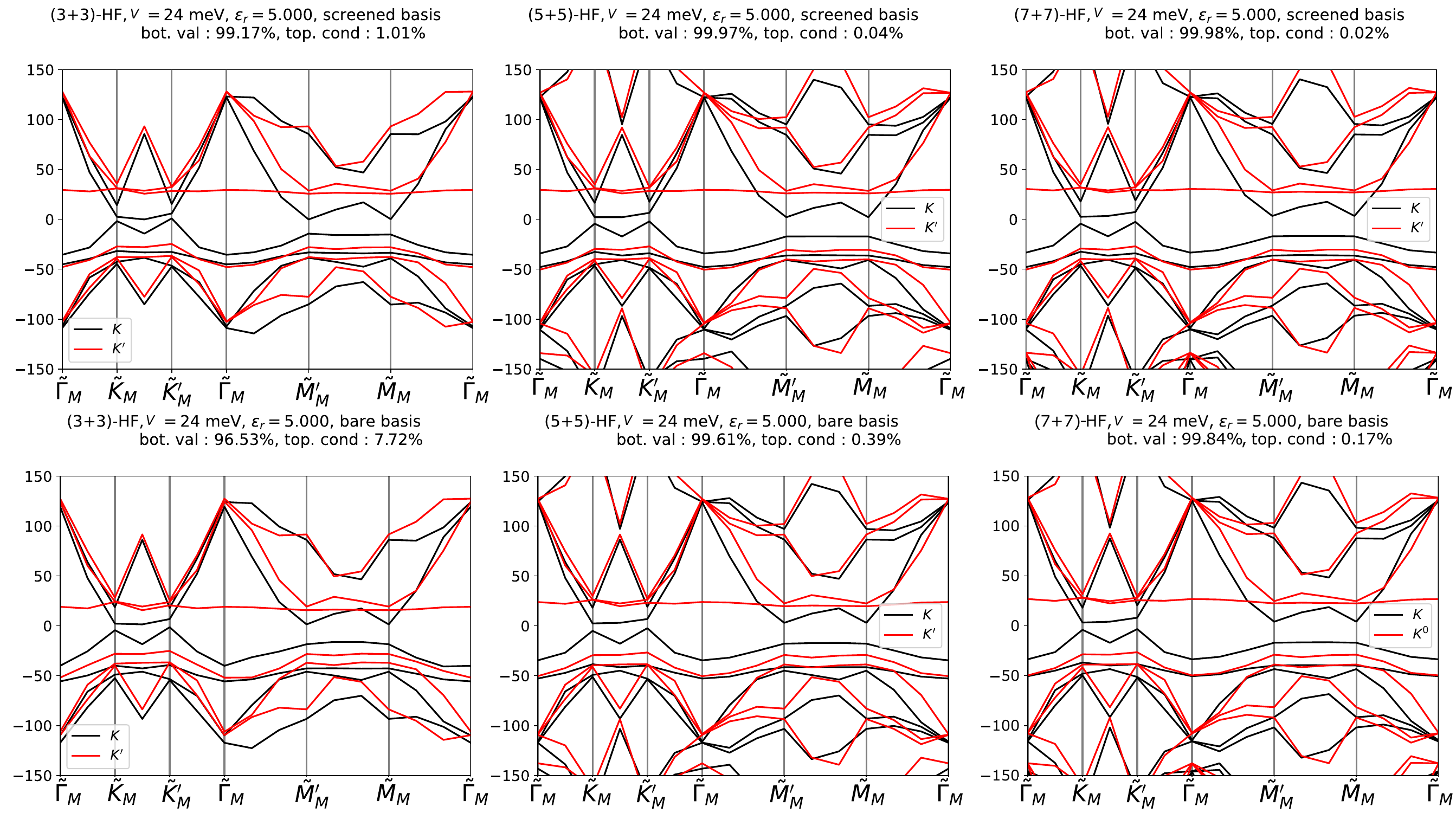}
    \caption{Comparison of screened basis (top) vs bare basis (bottom) R5G/hBN HF calculations at $\nu=1$ for different numbers of active bands $(n_c+n_v)$ per spin/valley. We have chosen to polarize in valley $K$ and spin $\uparrow$, and we show only the spin-$\uparrow$ HF band structure. We also list the occupation of the bottom valence and highest conduction projection bands, in the basis used to define the active subspace $\mathscr{H}_\text{act.}$. 
    $\theta=0.77^\circ$, $V=24\,\text{meV}$, $\xi=1$, $\epsilon_r=5.00$, and system size is $6\times 6$.}
    \label{bench_U0.024_epsr5.00}
\end{figure*}

\begin{figure*}
    \centering
    \includegraphics[width=1.0\linewidth]{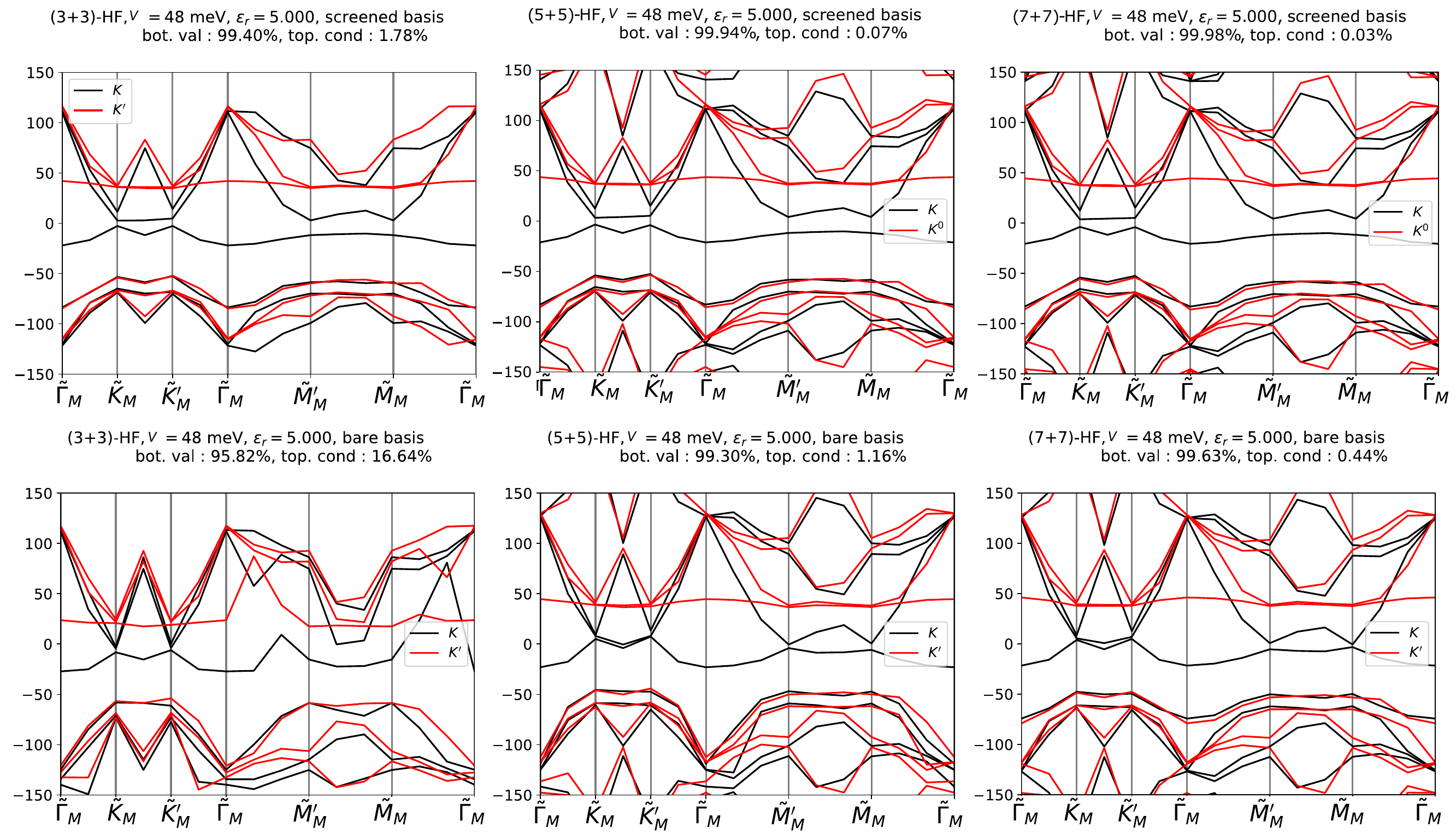}
    \caption{Comparison of screened basis (top) vs bare basis (bottom) R5G/hBN HF calculations at $\nu=1$ for different numbers of active bands $(n_c+n_v)$ per spin/valley. We have chosen to polarize in valley $K$ and spin $\uparrow$, and we show only the spin-$\uparrow$ HF band structure. We also list the occupation of the bottom valence and highest conduction bands, in the basis used to define the active subspace $\mathscr{H}_\text{act.}$.
    $\theta=0.77^\circ,V=48\,\text{meV},\epsilon_r=5.00$, and system size is $6\times 6$.}
    \label{bench_U0.048_epsr5.00}
\end{figure*}

\begin{figure*}
    \centering
    \includegraphics[width=1.0\linewidth]{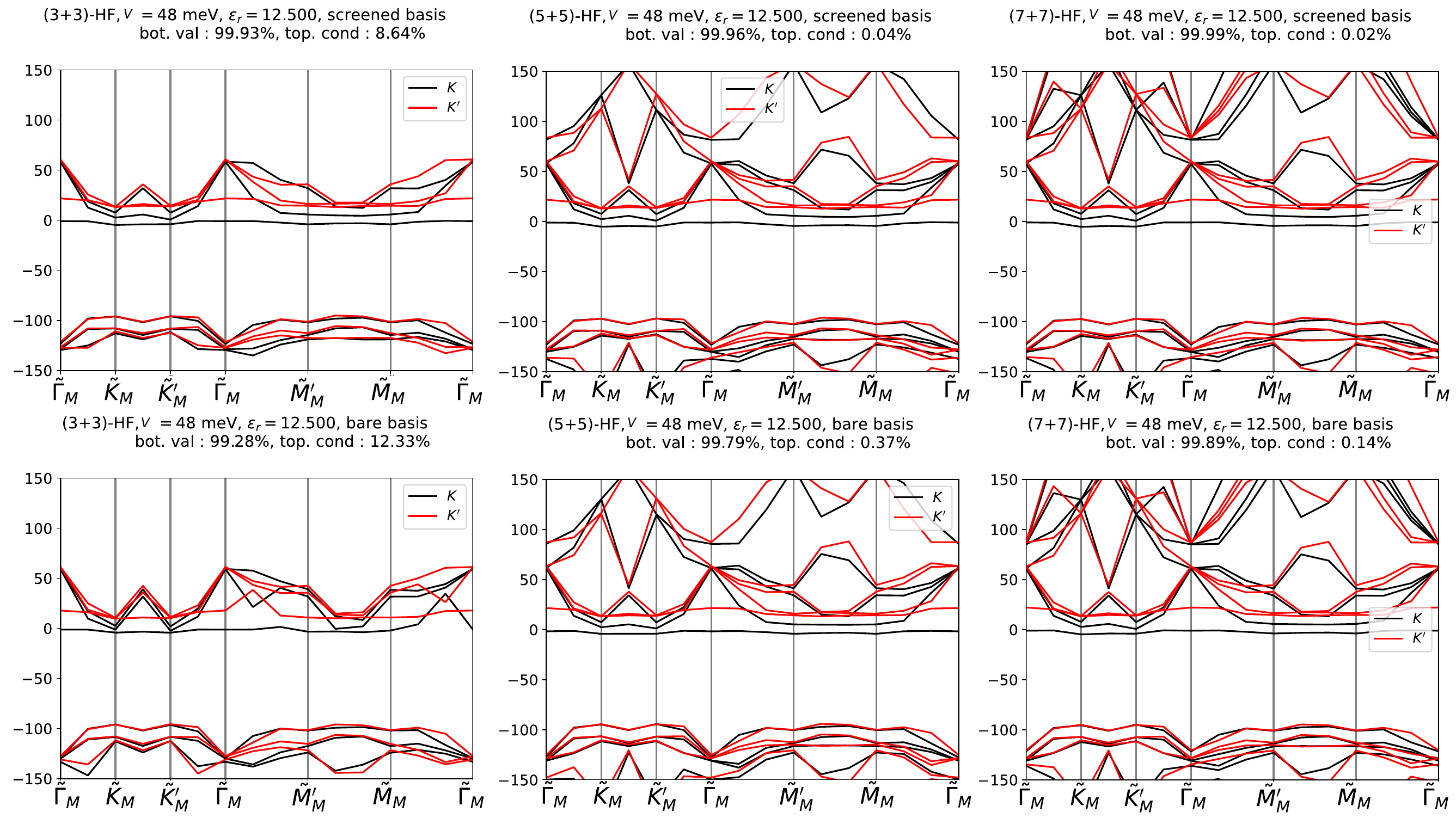}
    \caption{Comparison of screened basis (top) vs bare basis (bottom) R5G/hBN HF calculations at $\nu=1$ for different numbers of active bands $(n_c+n_v)$ per spin/valley. We have chosen to polarize in valley $K$ and spin $\uparrow$, and we show only the spin-$\uparrow$ HF band structure. We also list the occupation of the bottom valence and highest conduction bands, in the basis used to define the active subspace $\mathscr{H}_\text{act.}$.
    $\theta=0.77^\circ,V=48\,\text{meV},\epsilon_r=12.50$, and system size is $6\times 6$.}
    \label{bench_U0.048_epsr12.50}
\end{figure*}

\begin{figure*}
    \centering
    \includegraphics[width=1.0\linewidth]{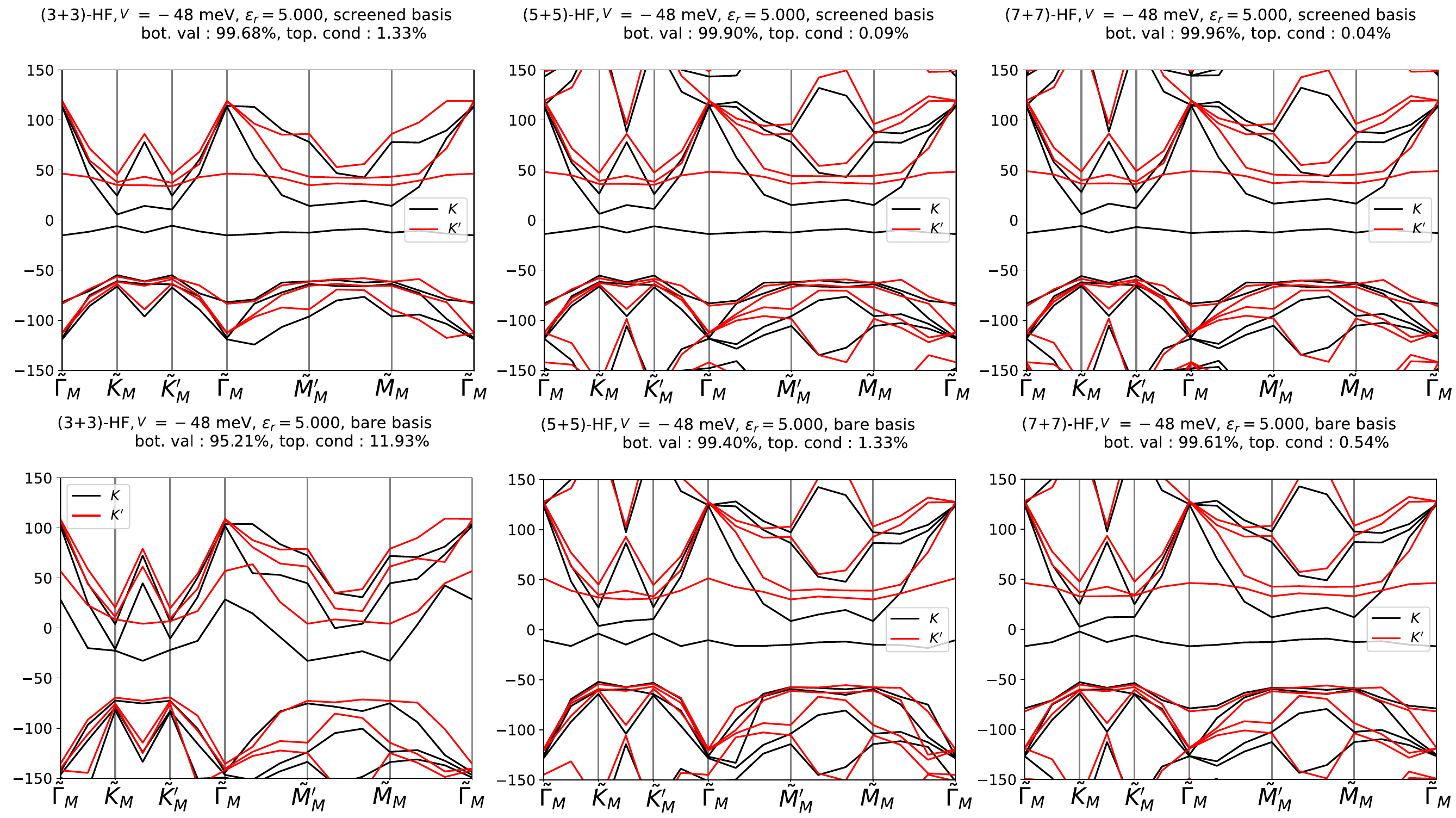}
    \caption{Comparison of screened basis (top) vs bare basis (bottom) R5G/hBN HF calculations at $\nu=1$ for different numbers of active bands $(n_c+n_v)$ per spin/valley. We have chosen to polarize in valley $K$ and spin $\uparrow$, and we show only the spin-$\uparrow$ HF band structure. We also list the occupation of the bottom valence and highest conduction bands, in the basis used to define the active subspace $\mathscr{H}_\text{act.}$.
    $\theta=0.77^\circ,V=-48\,\text{meV},\epsilon_r=5.00$, and system size is $6\times 6$.}
    \label{bench_U-0.048_epsr5.00}
\end{figure*}

In Figs.~\ref{bench_U0.024_epsr12.50}-\ref{bench_U-0.048_epsr5.00}, we compare the results of self-consistent HF calculations at $\nu=1$ using screened basis projection vs.~bare basis projection (see App.~\ref{secapp:projection} and App.~\ref{secapp:screening} for details of how these choices of projection are defined). Recall that the bare basis is obtained by diagonalizing $\hat{H}_\text{s.p.}(V)$ with the externally applied layer potential 
$V$. On the other hand, the screened basis is generated by first performing a self-consistent $\mbf{q}=0$ interlayer Hartree calculation to obtain the effective interlayer potential $U(V)$ (see App.~\ref{secapp:screening}). This is then used to diagonalize the continuum model $\hat{H}_\text{s.p.}(U)$ and identify a more suitable set of active bands $\mathscr{H}_\text{act.}$. Both methods should eventually converge to each other (and the model `ground truth' of the unprojected calculation) as we increase the active subspace band cutoffs $n_c,n_v$. However, it was argued in App.~\ref{secapp:layer_int} that the screened basis is a more sensible choice, as it captures the fact that the layers screen the externally applied displacement field. Hence, the non-interacting band basis of $\hat{H}_\text{s.p.}(U)$ should better reflect the interacting physics of the many-body Hamiltonian $\hat{H}$ with external field $V$.

Our objective in this subsection is to compare the two bases and assess convergence with band cutoff, rather than understand the HF phase diagram (which is done elsewhere in App.~\ref{secapp:HFphase}). As such, we will use a smaller system size $6\times6$ for computational convenience. We can judge the quality of projection by examining if the low-energy band structure changes as the active band cutoffs $n_c,n_v$ (which determines the size of the active subspace $\mathscr{H}_\text{act.}$) are increased, and checking the occupation of the lowest valence and highest conduction projection bands (i.e.~those bands that would be removed from $\mathscr{H}_\text{act.}$ if $n_v$ and $n_c$ are reduced by 1), which should ideally be close to 100\% and 0\% respectively. 

In Fig.~\ref{bench_U0.024_epsr12.50}, we show a comparison for $V=24\,\text{meV}$ and $\epsilon_r=12.5$. While the band structures for both the screened and bare basis are adequately converged for $n_c=n_v=3$ already, it is clear by looking at the occupations of the highest projection bands that the screened basis converges better. Upon increasing the interaction strength by going to $\epsilon_r=5.0$ (Fig.~\ref{bench_U0.024_epsr5.00}), the bare basis band structure for $n_c=n_v=3$ worsens (e.g.~around between $\tilde{M}_M$ and $\tilde{M}'_M$), while the screened basis remains good. This is expected as internal screening effects are larger for stronger interactions. Increasing the external field to $V=48\,\text{meV}$ and staying at $\epsilon_r=5.0$ (Fig.~\ref{bench_U0.048_epsr5.00}), we find that the screened basis remains converged, but the bare basis band structure for $n_c=n_v=3$ is problematic. There are discontinuities, for instance near $\tilde{\Gamma}_M$. This can be understood by tracking the energy of non-interacting band that is the lowest conduction band (for $\hat{H}_\text{s.p.}(V)$) for small $V$. Since for small $V$ this band becomes strongly layer polarized on the top layer, it rises rapidly in energy as $V$ is increased, compared to the higher conduction bands which are less layer polarized. At around $V\simeq 44\,\text{meV}$, this lowest conduction band collides with the higher conduction bands near $\tilde{\Gamma}_M$, and above this $V$ and at the momenta where this collision occurs, these formerly lowest-energy conduction states actually become the fourth-lowest-energy conduction states. Therefore, these states are projected out in a projected calculation using the bare basis with $n_c=n_v=3$. However the actual interacting physics reflects a smaller $U$ where this collision does not happen.
Hence in the bare basis, we have `lost' some of the relevant low energy degrees of freedom by projecting using the bare basis. For such large bare potentials $V$, this is not remedied by decreasing the interaction strength (Fig.~\ref{bench_U0.048_epsr12.50}). A similar conclusion holds for negative $V$ (Fig.~\ref{bench_U-0.048_epsr5.00}). Hence it is clear that the screened basis performs significantly better than the bare basis. We notice that the fourth valence/conduction bands approach the lowest six bands quite closely (e.g.~between $\tilde{K}_M$ and $\tilde{K}'_M$), while the fifth valence/conduction bands do not do so. This motivates our choice of predominantly using $(4+4)$ screened basis projection in our calculations.

\clearpage

\subsection{Dependence of average scheme on momentum cutoffs}\label{secapp:cutoff}

\begin{figure*}
    \centering
    \includegraphics[width=1.0\linewidth]{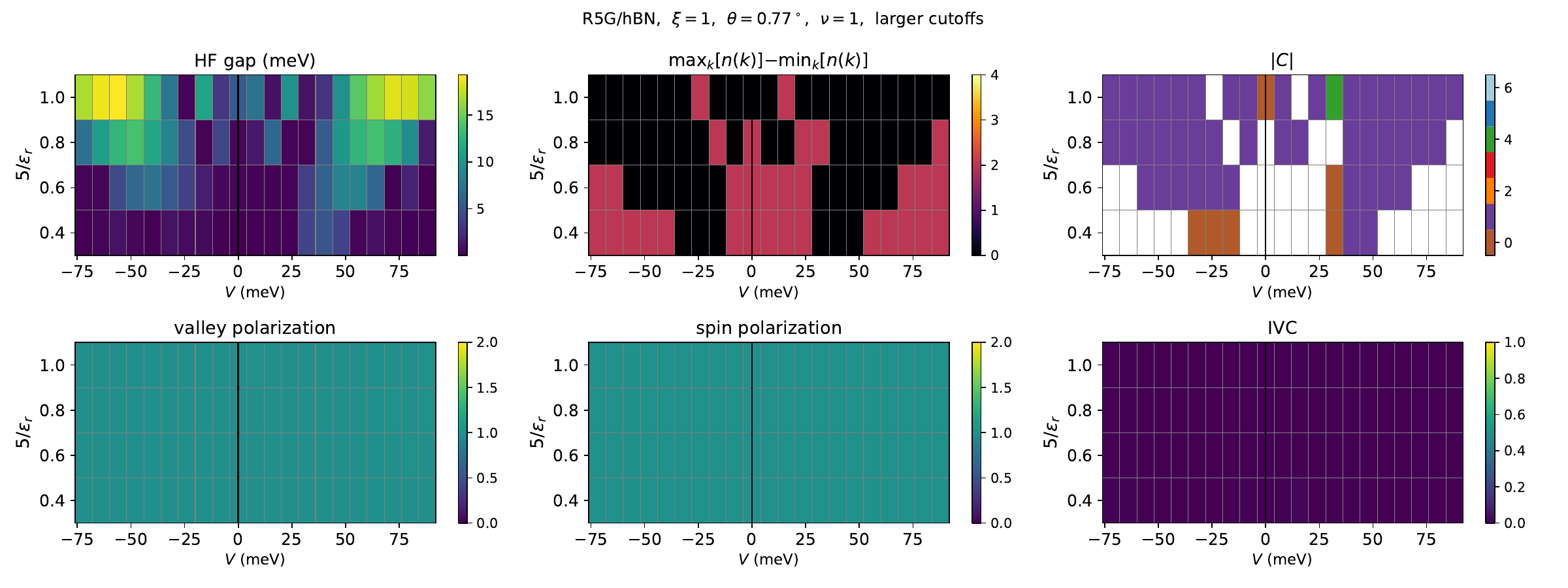}
    \caption{HF phase diagram for R5G/hBN, $\xi=1$, $\theta=0.77^\circ$, $\nu=1$. System size is $12\times 12$ and $(4+4)$ bands kept per spin/valley. Plane wave cutoff and interaction cutoff have radii $5.5|\mbf{q}_1|$ and $6|\mbf{q}_1|$ respectively.}
    \label{app_L5_4p4xi1_t0.77_nu1_Ng4NG4}
\end{figure*}

As discussed in App.~\ref{secapp:general_scheme}, the average scheme has a subtlety in that high-energy remote single-particle states near the Hilbert space plane wave cutoff (that controls the dimension of $\mathscr{H}$) contribute to the Fock renormalization of the states in the active subspace $\mathscr{H}_\text{act.}$, and such contributions are suppressed by the interaction potential which only decays as $1/q$ for large momentum transfer $q$. In Fig.~\ref{app_L5_4p4xi1_t0.77_nu1_Ng4NG4}, we show the phase diagram of R5G/hBN at $\xi=1$ stacking and $\theta=0.77^\circ$ in the average scheme. Compared to Fig.~\ref{app_L5_4p4xi1_t0.77_nu1} where the plane wave cutoff and interaction cutoff have radii $4|\mbf{q}_1|$ and $3|\mbf{q}_1|$ respectively, in Fig.~\ref{app_L5_4p4xi1_t0.77_nu1_Ng4NG4}, the plane wave cutoff and interaction cutoff have radii $5.5|\mbf{q}_1|$ and $6|\mbf{q}_1|$ respectively. It can be seen that while there are quantitative differences such as the size of the gapless region near $V=0$ and the precise positions of the phase boundaries of the gapped states at $|V|>0$, the qualitative structure of the phase diagram remains the same.

\clearpage

\section{Time-dependent Hartree-Fock theory}\label{secapp:TDHF}

\subsection{General formalism}

In this section, we review the formalism of time-dependent Hartree-Fock (TDHF) theory, which we use to compute the collective mode spectrum of a HF state~\cite{ring2004nuclear}, and which was sketched in Sec.~\ref{subsec:TDHF} in the main text. We first consider a general Hamiltonian with orbitals indexed by $a$, before specializing to R$L$G/hBN. Consider the Hamiltonian in the original ``orbital'' basis
\begin{equation}\label{appeq:TDHF_startH}
    \hat{H}=\sum_{ij}T_{ij}c^\dagger_ic_j+\frac{1}{2}\sum_{ijkl}V_{ij,kl}c^\dagger_jc^\dagger_ic_kc_l,
\end{equation}
which acts on the many-body Hilbert space constructed using the single-particle Hilbert space $\mathscr{H}$. [The formalism here could just as well be applied to a projected calculation (see Sec.~\ref{secapp:projection}) with the replacement $\hat{H}\rightarrow\hat{H}_\text{act.}$ and $\mathscr{H}\rightarrow\mathscr{H}_\text{act.}$.] This has exact ground state $\ket{\text{GS}}$ with energy $\tilde{E}_0$ and excited states $\ket{a}$ with energies $\tilde{E}_{a}$. Assume we have solved the HF equations (yielding the HF state $\ket{\text{HF}}$ with energy $E_{\text{HF}}$ --- note that $\ket{\text{HF}}$ does not necessarily correspond to the global minimum of the variational manifold of single Slater determinants) and found a self-consistent HF basis with creation operators 
\begin{equation}
d^\dagger_{\alpha}=\sum_{i}v_{\alpha,i}c^\dagger_i
\end{equation}
where $\alpha$ indexes the HF orbitals. The HF eigenenergies are $\mathscr{E}_\alpha$. The HF orbitals belong to either the occupied subspace $\mathscr{H}_\text{occ.}$ (with dimension $N_\text{occ.}$) or the unoccupied subspace $\mathscr{H}_\text{unocc.}$ (with dimension $N_\text{unocc.}$) depending on their occupation in $\ket{\text{HF}}$. We also consider the entire set of particle-hole (ph) labels $\phi=(\phi_\text{p},\phi_\text{h})$, where $\phi_\text{p}$ is an unoccupied HF orbital, and $\phi_\text{h}$ is an occupied HF orbital. There are $N_\text{occ.}N_{\text{unocc.}}$ such labels.

Our goal is to understand the neutral excitations of the system. We can define mode creation/annihilation operators $Q_a^\dagger,Q_a$ such that $\ket{a}=Q_a^\dagger\ket{\text{GS}}$, $Q_a\ket{\text{GS}}=0$, and $a$ indexes the neutral excitations. This is always possible, since we have the option $Q^\dagger_a=\ket{a}\bra{\text{GS}}$, though this choice might be non-local. From the time-independent Schr\"odinger equation $\hat{H}\ket{a}=\tilde{E}_a\ket{a}$, we obtain 
\begin{equation}
    [\hat{H},Q^\dagger_a]\ket{\text{GS}}=(\tilde{E}_a-\tilde{E}_0)Q^\dagger_a\ket{\text{GS}}.
\end{equation}
Contracting with a general state $\bra{\text{GS}}\delta Q$, where $\delta Q$ is an arbitrary operator, and using $\bra{\text{GS}}Q_a^\dagger=\bra{\text{GS}}\hat{H}Q^\dagger_a=0$, leads to
\begin{equation}\label{appeq:doublecom}
    \bra{\text{GS}}[\delta Q,[\hat{H},Q^\dagger_a]]\ket{\text{GS}}=(\tilde{E}_a-\tilde{E}_0)\bra{\text{GS}}[\delta Q,Q^\dagger_a]\ket{\text{GS}}.
\end{equation}
In the random phase approximation (RPA), we make an approximation by choosing the following form of the mode creation operator
\begin{equation}\label{appeq:Q_gen}
    Q^\dagger_a=\sum_{\phi}\left(X^a_{\phi}d^\dagger_{\phi_\text{p}}d_{\phi_\text{h}}-Y^a_{\phi}d^\dagger_{\phi_\text{h}}d_{\phi_\text{p}}\right).
\end{equation}
Note that while the terms proportional to $X^a_\phi$ create particle-hole (ph) pairs, the hole-particle (hp) terms proportional to $Y^a_\phi$ destroy them. Hence, the state $\ket{\text{RPA}}$ implicitly defined by the condition $Q_a\ket{\text{RPA}}=0$ is in general different from $\ket{\text{HF}}$. This can be seen from the fact that $\sum_\phi Y_\phi^{a*} d^\dagger_{\phi_\text{p}}d_{\phi_\text{h}}\ket{\text{HF}}\neq 0$.

To fix the coefficients $X^a_{\phi},Y^a_{\phi}$ in Eq.~\ref{appeq:Q_gen}, we consider $\delta Q$ (introduced in Eq.~\ref{appeq:doublecom}) constructed from  variations of the $X^a_{\phi},Y^a_{\phi}$ coefficients. Explicitly, we write
\begin{equation}
    \delta Q_a=\sum_{\phi}\left(\delta X^a_{\phi}d^\dagger_{\phi_\text{h}}d_{\phi_\text{p}}-\delta Y^a_{\phi}d^\dagger_{\phi_\text{p}}d_{\phi_\text{h}}\right)
\end{equation}
for a set of arbitrary coefficients $\delta X^a_{\phi},\delta Y^a_{\phi}$. Inserting into Eq.~\ref{appeq:doublecom} leads to
\begin{align}
    \bra{\text{RPA}}[\sum_{\phi}\left(\delta X^a_{\phi}d^\dagger_{\phi_\text{h}}d_{\phi_\text{p}}-\delta Y^a_{\phi}d^\dagger_{\phi_\text{p}}d_{\phi_\text{h}}\right),[\hat{H},Q^\dagger_a]]\ket{\text{RPA}}=(\tilde{E}_a-\tilde{E}_0)\bra{\text{RPA}}[\sum_{\phi}\left(\delta X^a_{\phi}d^\dagger_{\phi_\text{h}}d_{\phi_\text{p}}-\delta Y^a_{\phi}d^\dagger_{\phi_\text{p}}d_{\phi_\text{h}}\right),Q^\dagger_a]\ket{\text{RPA}}.
\end{align}
Since Eq.~\ref{appeq:doublecom} should hold for arbitrary $\delta Q$, then the above equation should hold for arbitrary $\delta X^a_{\phi},\delta Y^a_{\phi}$. As a result, we can isolate the terms corresponding to a given coefficient $\delta X^a_{\phi}$ or $\delta Y^a_{\phi}$. By doing this for all possible particle-hole labels $\phi$, we obtain the following conditions (one pair of equations for every possible $\phi$)
\begin{gather}
    \bra{\text{RPA}}\left[d^\dagger_{\phi_\text{h}}d_{\phi_\text{p}},\left[\hat{H},Q^\dagger_a\right]\right]\ket{\text{RPA}}=\Omega^a \bra{\text{RPA}}\left[d^\dagger_{\phi_\text{h}}d_{\phi_\text{p}},Q^\dagger_a\right]\ket{\text{RPA}}\\
        \bra{\text{RPA}}\left[d^\dagger_{\phi_\text{p}}d_{\phi_\text{h}},\left[\hat{H},Q^\dagger_a\right]\right]\ket{\text{RPA}}=\Omega^a \bra{\text{RPA}}\left[d^\dagger_{\phi_\text{p}}d_{\phi_\text{h}},Q^\dagger_a\right]\ket{\text{RPA}},
\end{gather}
where we have defined the excitation energies $\Omega_a=\tilde{E}_a-\tilde{E}_0$. At this stage,  we use the quasi-boson approximation to evaluate the above expectation values of four-fermion operators. This means that expectation values of commutators of fermion bilinears are evaluated as
\begin{equation}\label{appeq:QBA}
    \bra{\text{RPA}}[d^\dagger_{\phi_\text{h}} d_{\phi_\text{p}},d^\dagger_{\phi'_\text{p}} d_{\phi'_\text{h}}]\ket{\text{RPA}}\simeq \bra{\text{HF}}[d^\dagger_{\phi_\text{h}} d_{\phi_\text{p}},d^\dagger_{\phi'_\text{p}} d_{\phi'_\text{h}}]\ket{\text{HF}}=\delta_{{\phi_\text{h}}{\phi'_\text{h}}}\delta_{{\phi_\text{p}}{\phi'_\text{p}}}.
\end{equation}
This leads to the block-matrix equation
\begin{gather}\label{appeq:L}
\mathcal{L}\begin{pmatrix}X^a\\Y^a\end{pmatrix}=\begin{pmatrix}
    A & B \\ -B^* & -A^*
    \end{pmatrix}\begin{pmatrix}X^a\\Y^a\end{pmatrix}=\Omega^a\begin{pmatrix}X^a\\Y^a\end{pmatrix}
\end{gather}
where each block acts on the set particle-hole labels, and we have defined the $A$ and $B$ matrices
\begin{align}
\begin{split}\label{appeq:Amatrix}
    A_{\phi,\phi'}=&(\mathscr{E}_{\phi_\text{p}}-\mathscr{E}_{\phi_\text{h}})\delta_{{\phi_\text{p}}{\phi'_\text{p}}}\delta_{{\phi_\text{h}}{\phi'_\text{h}}}\\
    &+V_{{\phi_\text{p}}{\phi'_\text{h}},{\phi_\text{h}}{\phi'_\text{p}}}-V_{{\phi_\text{p}}{\phi'_\text{h}},{\phi'_\text{p}}{\phi_\text{h}}}\\
    B_{\phi,\phi'}=&V_{{\phi_\text{p}}{\phi'_\text{p}},{\phi_\text{h}}{\phi'_\text{h}}}-V_{{\phi_\text{p}}{\phi'_\text{p}},{\phi'_\text{h}}{\phi_\text{h}}},
\end{split}
\end{align}
which are Hermitian and symmetric respectively. Note that no direct reference is made to the one-body term $T_{ij}$ in Eq.~\ref{appeq:TDHF_startH}. All kinetic terms and interaction scheme subtleties are accounted for in the HF calculation itself, so that the only inputs needed for TDHF are the final HF Hamiltonian (from which one obtains the HF basis and the HF eigenenergies) and the four-fermion matrix elements $V_{ij,kl}$ (to be transformed to the HF basis).

$\mathcal{L}$ has eigenvectors $w^a=(X^a,Y^a)$ and eigenvalues $\Omega^a$.  Note that $\mathcal{L}$ satisfies 
\begin{equation}
    \mathcal{L}=-\sigma_x \mathcal{L}^* \sigma_x
\end{equation}
which means that an eigenvector $w^a=(X^a,Y^a)$ with eigenvalue $\Omega^a$ implies an eigenvector $w^a=-(Y^{a*},-X^{a*})$ with eigenvalue $-\Omega^{a*}$. Hence, finite real eigenvalues necessarily come in plus/minus pairs. Consider overlap between two excited states $\ket{a},\ket{a'}$
\begin{equation}
    \langle a | a'\rangle=\bra{\text{RPA}}Q_a Q_{a'}^\dagger\ket{\text{RPA}}=\bra{\text{RPA}}[Q_a ,Q_{a'}^\dagger]\ket{\text{RPA}}=(X^a)^\dagger X^b-(Y^a)^\dagger Y^b
\end{equation}
where we used the quasi-boson approximation (Eq.~\ref{appeq:QBA}) in the last equality. 
Since $\langle a | a'\rangle=\delta_{aa'}$ for the excited eigenstates, the physical excitation operators have coefficients $X,Y$ that satisfy $(X^a)^\dagger X^{a'}-(Y^a)^\dagger Y^{a'}=\delta_{aa'}$.

In the quasi-boson approximation, the particle-hole (ph) and hole-particle (hp) content of the excited states is
\begin{gather}
    \kappa_{\phi_\text{p}\phi_\text{h}}^a=\bra{\text{GS}}c_{\phi_\text{h}}^\dagger c_{\phi_\text{p}}\ket{a}\simeq X^a_{\phi_\text{p}\phi_\text{h}}\\
    \kappa_{\phi_\text{h}\phi_\text{p}}^a=\bra{\text{GS}}c_{\phi_\text{p}}^\dagger c_{\phi_\text{h}}\ket{a}\simeq Y^a_{\phi_\text{p}\phi_\text{h}}.
\end{gather}

We consider the physical interpretation of the various parts of the $A$ and $B$ matrices:
\begin{itemize}
\item The first $A$-term (A0) is the sum of Koopman addition/removal energies of a single particle-hole (ph) pair state $\ket{\phi_\text{p},\phi_\text{h}}\equiv d^\dagger_{\phi_\text{p}}d_{\phi_\text{h}}\ket{\text{HF}}$. 
\item The second $A$-term (A1) can be interpreted as ``direct scattering'' of ph pairs, since it encodes the matrix element corresponding to $\phi_\text{p}\rightarrow \phi_\text{h},\phi_\text{h}'\rightarrow \phi_\text{p}'$. In other words, the ph pair $(\phi_\text{p},\phi_\text{h})$ is annihilated and $(\phi_\text{p}',\phi_\text{h}')$ is created.
\item The third $A$-term (A2) can be interpreted as ``exchange scattering'' involving $\phi_\text{p}\rightarrow \phi_\text{p}',\phi_\text{h}'\rightarrow \phi_\text{h}$.
\item The first $B$-term (B1) involves the direct creation or destruction of two ph pairs since it involves $\phi_\text{p}\rightarrow \phi_\text{h},\phi_\text{p}'\rightarrow \phi_\text{h}'$.
\item The second $B$-term (B2) involves the ``exchange'' creation or destruction of two ph pairs since it involves $\phi_\text{p}\rightarrow \phi_\text{h}',\phi_\text{p}'\rightarrow \phi_\text{h}$.
\end{itemize}

\subsection{Application to R$L$G/hBN}\label{secapp:TDHF_apptoRLG}
Since the correlated insulator states of interest in our phase diagrams have no IVC, we consider HF states with conserved spin-$S_z$ and valley-$U(1)_v$ symmetry. We hence use $f=(s,\eta)$ as a composite index for the conserved flavor. In our projected HF calculations, the original degrees of freedom are labelled by $(\mbf{k},n,f)$, where $n$ is a projection band label (i.e.~$n$ indexes the projection bands used to construct $\mathscr{H}_\text{act.}$, see App.~\ref{secapp:projection}). The result of a HF calculation yields HF orbitals labelled by $(\mbf{k},\alpha,f)$, where $\alpha$ labels the HF bands. The two bases are related by a unitary transformation
\begin{equation}
    d^\dagger_{\mbf{k},\alpha,f}=\sum_{n}v_{\alpha,n}(\mbf{k},f)c^\dagger_{\mbf{k},n,f}
\end{equation}
where $c^\dagger$ ($d^\dagger$) is a creation operator in the original (HF) basis. From this, we can compute the interaction matrix elements (which continue to preserve flavor indices) in the HF basis
\begin{align}
\begin{split}
    V_{(\mbf{k},\alpha,f),(\mbf{k'+q},\gamma,f');(\mbf{k+q},\beta,f),(\mbf{k'},\delta,f')}=&\frac{1}{N}\sum_{\mbf{G}}\sum_{mnm'n',ll'}\frac{V_{ll'}(\mbf{q})}{\Omega}M^{l\eta}_{mn}(\mbf{k},\mbf{q+G})M^{l'\eta'*}_{n'm'}(\mbf{k'},\mbf{q+G})\\
    &\times v^*_{\alpha,m}(\mbf{k},f)v^*_{\gamma,m'}(\mbf{k'+q},f')v_{\delta,n'}(\mbf{k'},f')v_{\beta,n}(\mbf{k+q},f).
\end{split}
\end{align}

We use the following parameterization of the mode creation operator
\begin{equation}\label{appeq:QRPA}
    Q^\dagger_a(\mbf{q})=\sum_{\mbf{k}}\sum_{f_p\alpha_pf_h\alpha_h}\left(X^a_{\alpha_pf_p;\alpha_hf_h}(\mbf{k},\mbf{q})d^\dagger_{\mbf{k+q},\alpha_pf_p}d_{\mbf{k},\alpha_hf_h}-Y^a_{\alpha_pf_p;\alpha_hf_h}(\mbf{k},\mbf{q})d^\dagger_{\mbf{k},\alpha_hf_h}d_{\mbf{k-q},\alpha_pf_p}\right),
\end{equation}
where $f_p,\alpha_p$ run over unoccupied HF orbitals and $f_h,\alpha_h$ run over occupied HF orbitals.  

The collective mode spectrum $\Omega^a(\bm{q})$ forms bands in the mBZ. Furthermore, since our HF states conserve flavor $U(1)$, each collective mode band is associated with excitation operators that carry a definite flavor charge. By flavor charge, we mean the effect that the corresponding excitation operator $Q^\dagger_a(\mbf{q})$ has on the flavor of a state $\ket{\psi}$ with definite flavor $f$. For instance, we say that $Q^\dagger_a(\mbf{q})$ carries a valley charge of 1 ($-1$) if it maps a state in valley $K'$ ($K$) to a state in valley $K$ ($K'$). Similarly, $Q^\dagger_a(\mbf{q})$ carries a spin charge of 1 ($-1$) if it maps a state with spin $\downarrow$ ($\uparrow$) to a state with spin $\uparrow$ ($\downarrow$). An excitation can carry both non-zero spin and valley charge.

A significant simplification occurs when the projection is chosen so that only one flavor is partially filled. In particular,  consider $(0+n_c)$ projection for $n_c>1$, and consider a spin-valley polarized state at $\nu=1$ that is fully polarized in valley $\eta=K$ and spin $s=\uparrow$. This means that all flavors are completely unoccupied, except the flavor $f=(K,\uparrow)$ which is partially occupied. The modes can be classified as intraflavor, intervalley, interspin, or inter-spin-valley. The intervalley modes (which carry a non-zero valley charge) and inter-spin-valley modes (which carry a non-zero valley charge and spin charge) are degenerate due to the $SU(2)_K\times SU(2)_{K'}$ symmetry of our Hamiltonian. Hence, we will not need to explicitly refer to inter-spin-valley modes in the following.  In the intraflavor channel, the particle and hole states all reside in the partially filled flavor $f=(K,\uparrow)$. In the intervalley channel, the particle and hole states both have spin $\uparrow$, but valleys $K'$ and $K$ respectively. Furthermore, the $B$ matrix vanishes and only the first and last terms of $A$ (Eq.~\ref{appeq:Amatrix}) survive. Similarly, in the interspin channel, the particle and hole states are both of valley $K$ but have spins $\downarrow$ and $\uparrow$ respectively. Furthermore only the first and last terms of $A$ are non-vanishing. See Tab.~\ref{tab:flavormodes} for a summary of the constraints on $Q^\dagger_{a}(\mbf{q})$ for the different types of excitations.

\clearpage

\section{Details of TDHF numerics}

\subsection{Collective modes as a function of hBN coupling}\label{secapp:collq0}

In this subsection, we present results of $\mbf{q}=0$ TDHF calculations (see App.~\ref{secapp:TDHF} for a review of the formalism of TDHF theory) of the collective modes of the insulating spin-valley polarized HF states at $\nu=1$, for different values of the hBN coupling factor $\kappa_\text{hBN}$. Recall from App.~\ref{secapp:TDHF_apptoRLG} that for spin-valley diagonal states, the collective modes can be classified as intraflavor, intervalley, interspin, or inter-spin-valley. Since the intervalley and inter-spin-valley modes are degenerate with each other, we do not explicitly refer to the inter-spin-valley modes below.

\begin{figure*}
    \centering
    \includegraphics[width=0.8\linewidth]{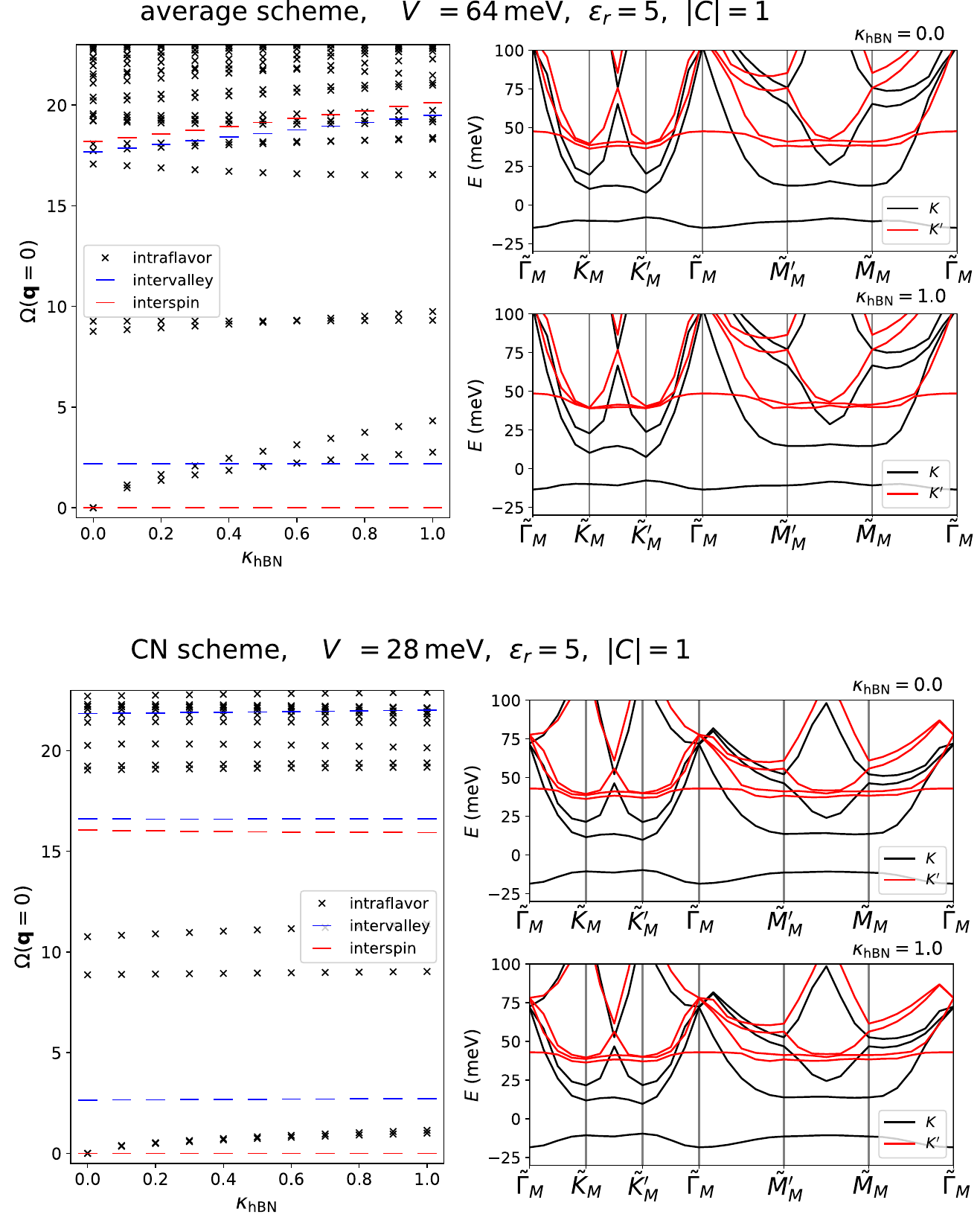}
    \caption{$\mbf{q}=0$ collective modes of the R5G/hBN $|C|=1$ Chern insulator. Top (bottom) row shows results for the average (CN) scheme. Left shows the $\mbf{q}=0$ collective modes, filtered by whether they carry valley or spin charge, as a function of the strength of hBN coupling, where $\kappa_\text{hBN}=1$ corresponds to the physical model and the $\kappa_\text{hBN}=0$ limit has continuous translation symmetry. Right shows the HF band structure for $\kappa_\text{hBN}=0,1$ in the spin sector with finite filling (the other spin sector is fully unoccupied). Parameters have to chosen so that the interacting gap is similar in the calculations. The HF and TDHF calculations are performed with $(0+4)$ screened basis projection, $\xi=1$, $\theta=0.77^\circ$, and system size $12\times 12$.}
    \label{collq0_avg_vs_CN}
\end{figure*}

Fig.~\ref{collq0_avg_vs_CN} shows the $\mbf{q}=0$ collective modes $\Omega(\mbf{q}=0)$ for the spin-valley-polarized $|C|=1$ Chern insulator in $\xi=1$ R5G/hBN at $\nu=+1$, as a function of the hBN coupling factor $\kappa_\text{hBN}$ which multiplies the potentials arising from the hBN alignment in the continuum model. The $\kappa_\text{hBN}=0$ limit possesses continuous translation symmetry, while $\kappa_\text{hBN}=1$ corresponds to the physical model. We first discuss some general properties. We always find a single quadratically-dispersing gapless spin magnon. For generic $\kappa_\text{hBN}$, spin $SU(2)_S$ symmetry is the only spontaneously broken continuous symmetry. Furthermore, we have two pseudophonon branches (corresponding to the two generators of continuous translation) which are gapless at $\kappa_{\text{hBN}}=0$, but develop a gap as soon as the continuous translation symmetry is broken by the moir\'e potential. We that find the lowest-energy modes apart from the spin magnon and pseudophonons are intervalley modes with moderately small excitation gap $\lesssim5\,\text{meV}$ in the Chern insulating regime, as well as two branches of intraflavor excitons that lie below the particle-hole continuum. In Fig.~\ref{collq0_avg_vs_CN}, we perform the TDHF calculation for both the average scheme and the CN scheme. Since the two schemes have drastically different layer screening properties, it is difficult to directly compare bare parameters. We choose different $V$ so that their HF gaps are similar. While the lowest intervalley modes have similar energy, it is clear that the average scheme induces significantly larger pseudophonon gaps. For the parameters in Fig.~\ref{collq0_avg_vs_CN}, both pseudophonons have a greater energy than the lowest intervalley mode at the physical limit $\kappa_\text{hBN}=1$. The reason for this difference between the schemes is that the average scheme allows the (remote) valence bands to affect the low-energy conduction bands through interaction-induced corrections to the effective one-body potential. So while the lowest conduction bands in the non-interacting limit have relative little moir\'eness for large $V>0$, the moir\'e-periodic charge density in the filled valence bands imparts an effective moir\'e potential landscape. On the other hand, the pseudophonon gaps in the CN interaction scheme are smaller, because in this scheme, any possible influence of the moir\'e charge density of the filled valence bands is cancelled out by the reference density (which also consists of filled valence bands). We find that the pseudophonon energies at $\mbf{q}=0$ do not exceed the lowest intervalley mode. We note that it is not possible to make such inferences directly from the HF band structures. Techniques beyond HF, such as TDHF employed here, are required for a quantitative assessment of the extrinsic `moir\'eness'. 

\begin{figure*}
    \centering
    \includegraphics[width=0.8\linewidth]{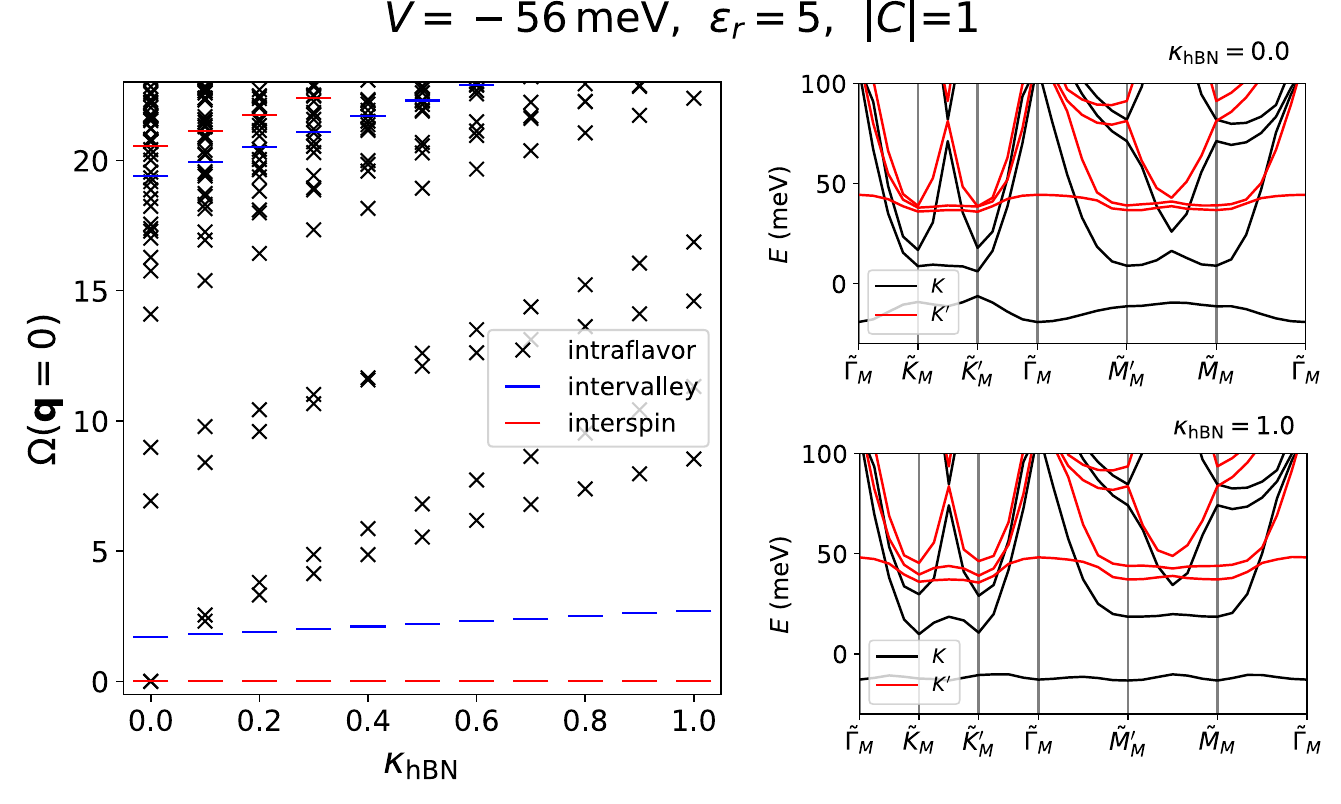}
    \caption{$\mbf{q}=0$ collective modes of the R5G/hBN $|C|=1$ Chern insulator. Left shows the $\mbf{q}=0$ collective modes, filtered by whether they carry valley or spin charge, as a function of the strength of hBN coupling, where $\kappa_\text{hBN}=1$ corresponds to the physical model and the $\kappa_\text{hBN}=0$ limit has continuous translation symmetry. Right shows the HF band structure for $\kappa_\text{hBN}=0,1$ in the spin sector with finite filling (the other spin sector is fully unoccupied). The HF and TDHF calculations are performed with $(0+4)$ screened basis projection, average interaction scheme, $\xi=1$, $\theta=0.77^\circ$, and system size $12\times 12$.}
    \label{collq0_negU}
\end{figure*}

In Fig.~\ref{collq0_negU}, we show the collective modes of the $|C|=1$ insulator for $\xi=1$ R5G/hBN for $V=-56\,\text{meV}$ in the average interaction scheme. Since $V$ is negative, the lowest conduction bands are polarized towards the hBN layer and hence strongly experience the moir\'e hBN coupling at the non-interacting level. As a result, the pseudophonon energies are significantly larger, and approach $\simeq 10\,\text{meV}$ for $\kappa_\text{hBN}=1$.

\begin{figure*}
    \centering
    \includegraphics[width=0.8\linewidth]{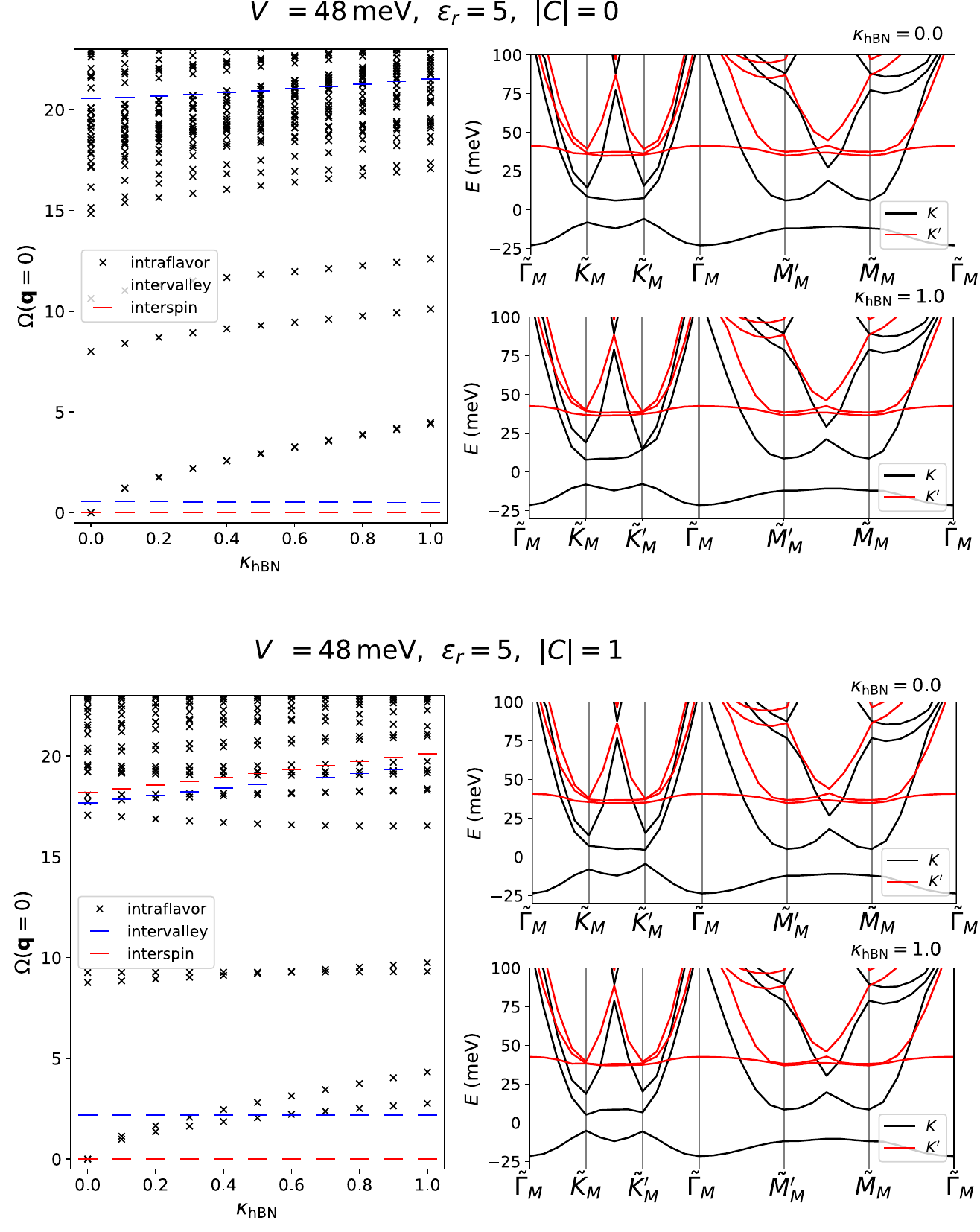}
    \caption{$\mbf{q}=0$ collective modes of the $|C|=0$ and $|C|=1$ states in R5G/hBN. Both HF states are obtained as converged solutions in our HF calculations. Top (bottom) row shows results for $|C|=0$ ($|C|=1$). Left shows the $\mbf{q}=0$ collective modes, filtered by whether they carry valley or spin charge, as a function of the strength of hBN coupling, where $\kappa_\text{hBN}=1$ corresponds to the physical model and the $\kappa_\text{hBN}=0$ limit has continuous translation symmetry. Right shows the HF band structure for $\kappa_\text{hBN}=0,1$ in the spin sector with finite filling (the other spin sector is fully unoccupied). The HF and TDHF calculations are performed with $(0+4)$ screened basis projection, average interaction scheme, $\xi=1$, $\theta=0.77^\circ$, and system size $12\times 12$.}
    \label{collq0_C0_vs_C1}
\end{figure*}

In Fig.~\ref{collq0_C0_vs_C1}, we compare the collective modes of the $|C|=0$ and $|C|=1$ insulators for $\xi=1$ R5G/hBN (our HF calculations can converge to either state for the same Hamiltonian).  Both insulators have pseudophonon modes that develop comparable gaps. However, the $|C|=0$ state has a significantly smaller valley gap. The large difference in the valley gap of $|C|=0$ and $|C|=1$ insulators has previously been pointed out in twisted TMD bilayers~\cite{wang2023magnets}, where the difference was attributed to the underlying electronic topology in the HF state.

\begin{figure*}
    \centering
    \includegraphics[width=0.8\linewidth]{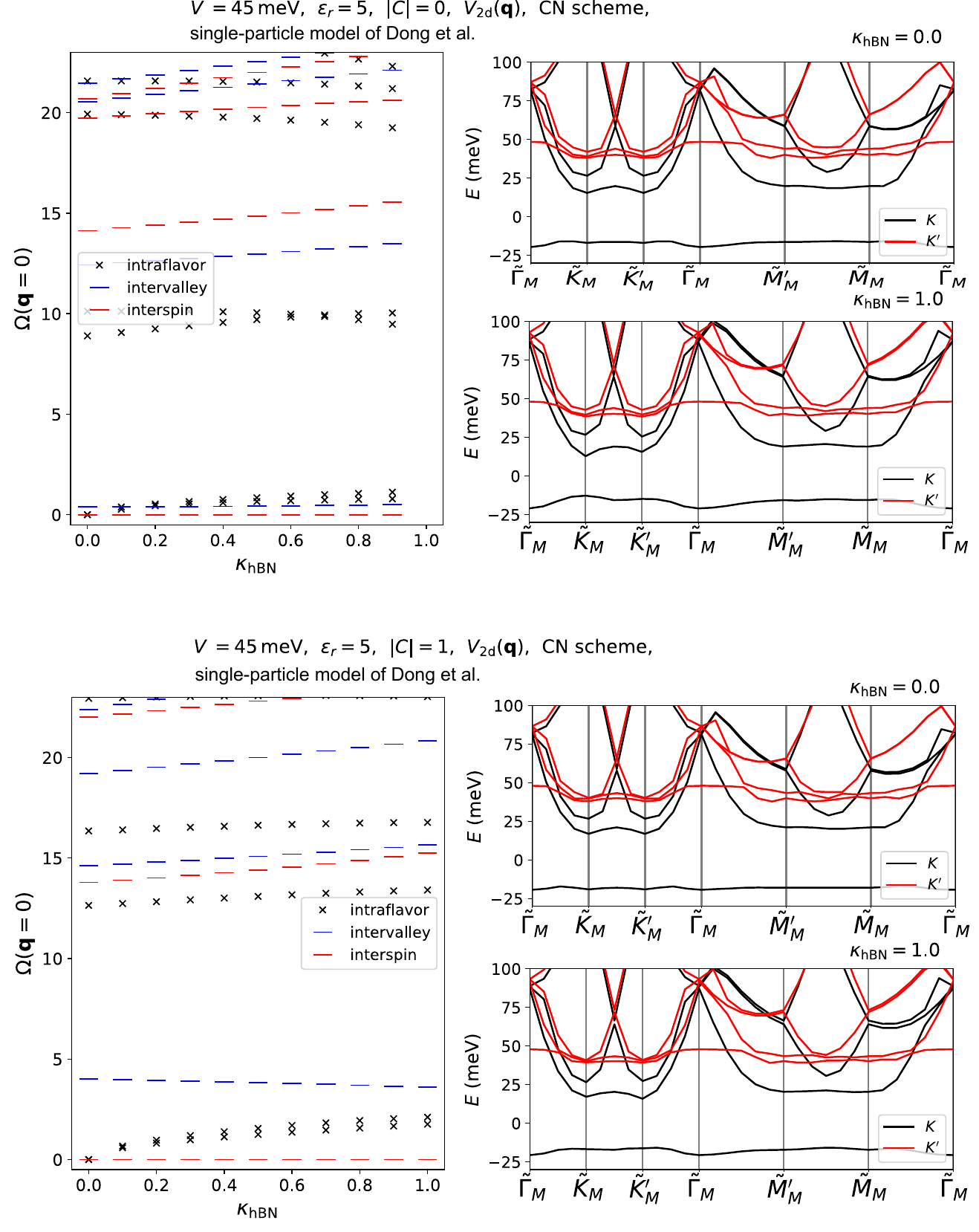}
    \caption{$\mbf{q}=0$ collective modes of the $|C|=0$ and $|C|=1$ states in R5G/hBN using the single-particle model of Ref.~\cite{dong2023anomalous} and the CN interaction scheme with bare basis projection. Both HF states are obtained as converged solutions in our HF calculations. Top (bottom) row shows results for $|C|=0$ ($|C|=1$). Left shows the $\mbf{q}=0$ collective modes, filtered by whether they carry valley or spin charge, as a function of the strength of hBN coupling, where $\kappa_\text{hBN}=1$ corresponds to the physical model and the $\kappa_\text{hBN}=0$ limit has continuous translation symmetry. Right shows the HF band structure for $\kappa_\text{hBN}=0,1$ in the spin sector with finite filling (the other spin sector is fully unoccupied). The HF and TDHF calculations are performed with $(0+4)$ bands, $\theta=0.60^\circ$, a 2d interaction $V_{\text{2d}}(\mbf{q})$ with $d_\text{sc}=25\,\text{nm}$, and system size $12\times 12$.}
    \label{collq0_Parker2d_CN}
\end{figure*}

\begin{figure*}
    \centering
    \includegraphics[width=0.8\linewidth]{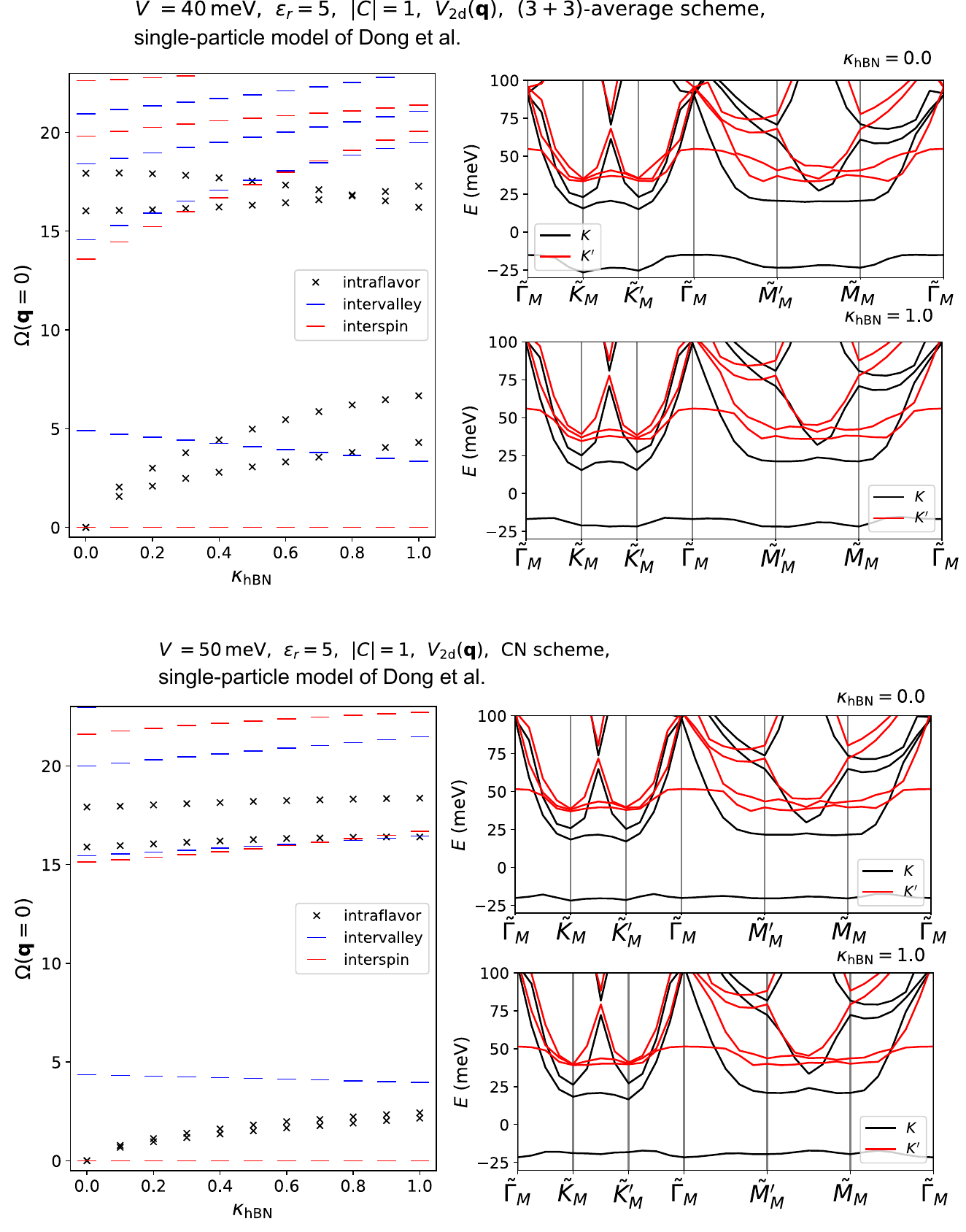}
    \caption{$\mbf{q}=0$ collective modes of the $|C|=1$ state in R5G/hBN using the single-particle model of Ref.~\cite{dong2023anomalous} with bare basis projection. Top row shows results for the $(3+3)$-average, while bottom row shows results for the CN scheme. Left shows the $\mbf{q}=0$ collective modes, filtered by whether they carry valley or spin charge, as a function of the strength of hBN coupling, where $\kappa_\text{hBN}=1$ corresponds to the physical model and the $\kappa_\text{hBN}=0$ limit has continuous translation symmetry. Right shows the HF band structure for $\kappa_\text{hBN}=0,1$ in the spin sector with finite filling (the other spin sector is fully unoccupied). The HF and TDHF calculations are performed with $(0+4)$ bands, $\theta=0.77^\circ$, a 2d interaction $V_{\text{2d}}(\mbf{q})$ with $d_\text{sc}=25\,\text{nm}$, and system size $12\times 12$.}
    \label{collq0_Parker2d_avg_vs_CN}
\end{figure*}

In Fig.~\ref{collq0_Parker2d_CN}, we show representative results using the single-particle model of Ref.~\cite{dong2023anomalous}. Note that following Ref.~\cite{dong2023anomalous}, we use a 2d interaction and the CN interaction scheme, and project our calculations using the bare band basis. The pseudophonon gaps are smaller due to the use of the CN scheme, and the contrast of the valley gap between $|C|=0$ and $|C|=1$ persists here as well. In Fig.~\ref{collq0_Parker2d_avg_vs_CN}, we further compare the results between the CN scheme and the $(3+3)$-average scheme (see Sec.~\ref{secapp:general_scheme}). As expected, the pseudophonons develop a larger gap in the latter calculation. This suggests that the enhancement of the pseudophonon gap in the average interaction scheme relative to the CN interaction scheme is not unique to the single-particle model constructed in Ref.~\cite{MFCI-II}.

\clearpage
\subsection{Collective mode phase diagrams}\label{secapp:coll_phase}

In this subsection, we repeat the TDHF calculation for a range of $V$ and interaction strengths $\epsilon_r$. For each parameter, we consider the energies $\Omega_\text{phonon,1}$ and $\Omega_\text{phonon,2}$ of the lowest two intraflavor modes (pseudophonons), and the energy $\Omega_\text{valley}$ of the lowest intervalley mode. We also show ratios of these energy scales. $\Omega_{\text{phonon},1}/\Omega_{\text{valley},1}>1$ suggests that the stability of the state is more tied to stability against valley fluctuations rather than translational fluctuations.

\begin{figure*}
    \centering
    \includegraphics[width=1.0\linewidth]{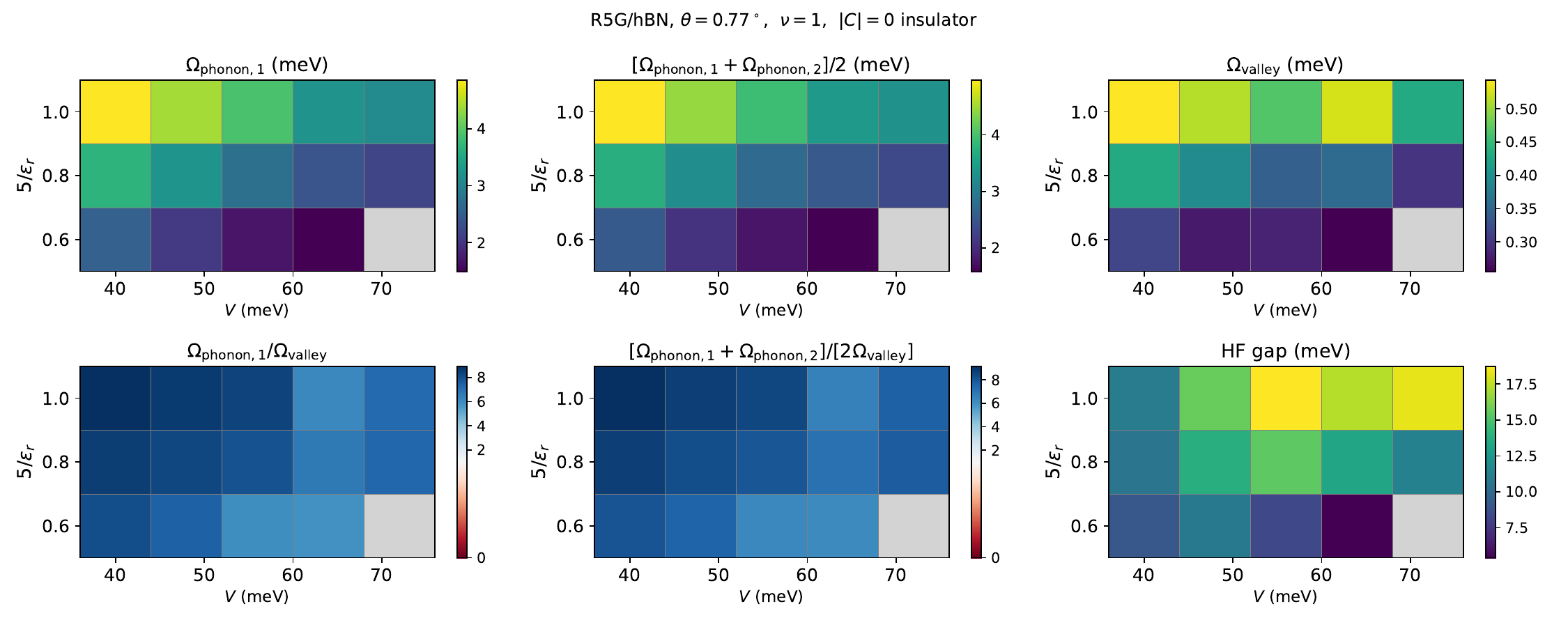}
    \caption{$\mbf{q}=0$ collective modes of the $|C|=0$ insulator in R5G/hBN as a function of $V$ and $\epsilon_r$. $\Omega_\text{phonon,1},\Omega_\text{phonon,2}$ are the energies of the lowest excitations in the intraflavor channel, and are adiabatically connected to the two gapless Goldstone modes in the limit of zero moir\'e potential. $\Omega_\text{valley}$ is the energy of the lowest intervalley excitation. Grey regions denote parameters where there are no gapped HF states. The HF and TDHF calculations are performed with $(0+4)$ screened basis projection, $\xi=1$, $\theta=0.77^\circ$, and system size $12\times 12$. }
    \label{collphase_t0.77_C0}
\end{figure*}

\begin{figure*}
    \centering
    \includegraphics[width=1.0\linewidth]{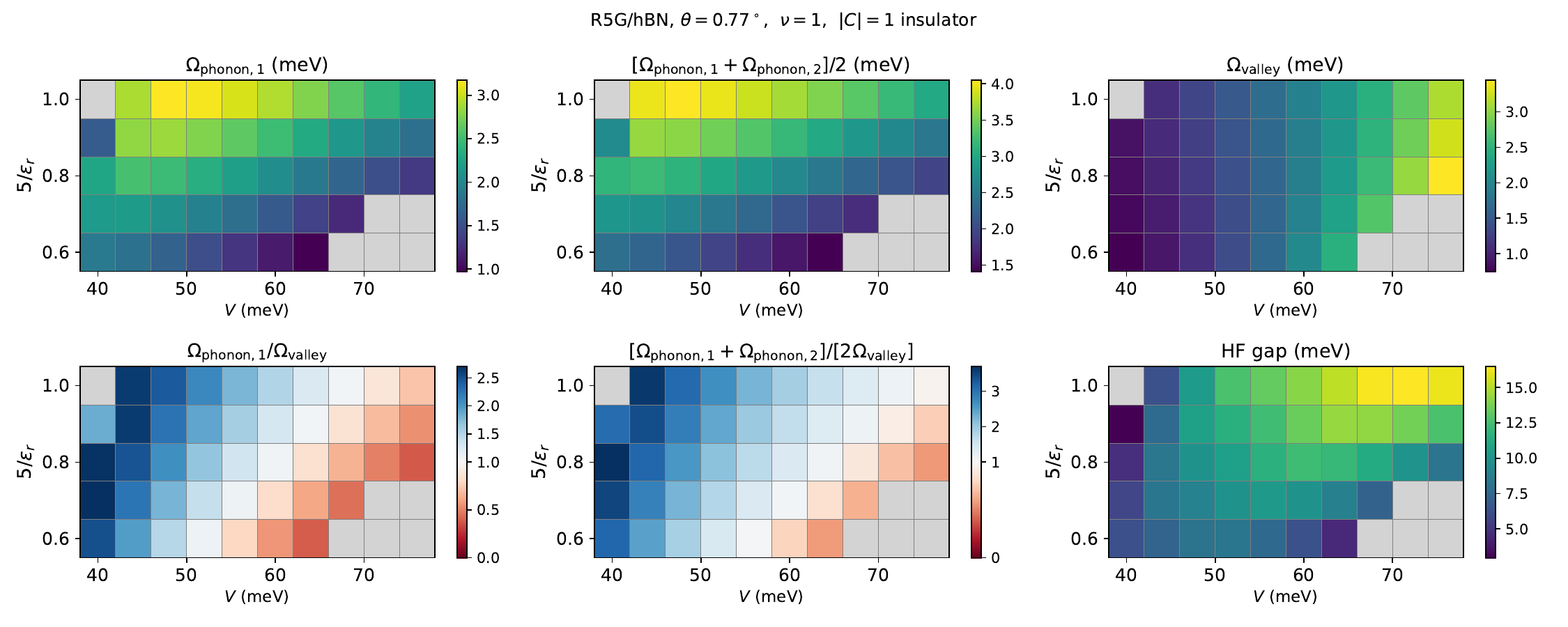}
    \caption{$\mbf{q}=0$ collective modes of the $|C|=1$ insulator in R5G/hBN as a function of $V$ and $\epsilon_r$. $\Omega_\text{phonon,1},\Omega_\text{phonon,2}$ are the energies of the lowest excitations in the intraflavor channel, and are adiabatically connected to the two gapless Goldstone modes in the limit of zero moir\'e potential. $\Omega_\text{valley}$ is the energy of the lowest intervalley excitation. Grey regions denote parameters where there are no gapped HF states. The HF and TDHF calculations are performed with $(0+4)$ screened basis projection, $\xi=1$, $\theta=0.77^\circ$, and system size $12\times 12$. }
    \label{collphase_t0.77_C1}
\end{figure*}

In Figs~\ref{collphase_t0.77_C0} and \ref{collphase_t0.77_C1}, we show results for $\xi=1$ R5G/hBN for the $|C|=0$ and $|C|=1$ insulator respectively. The valley gap of the $|C|=0$ state is significantly lower than that of the $|C|=1$ state throughout the entire phase diagram (a similar phenomenon was pointed out for twisted TMD homobilayers in Ref.~\cite{wang2023magnets}, which attributed the difference to the electronic topology of the HF state). The pseudophonon gap is smaller for larger $V$, which conforms with the expectation that a higher displacement field drives the conduction electrons further away from the hBN, and reduces the moir\'e charge density variation in the filled valence bands.

\begin{figure*}
    \centering
    \includegraphics[width=1.0\linewidth]{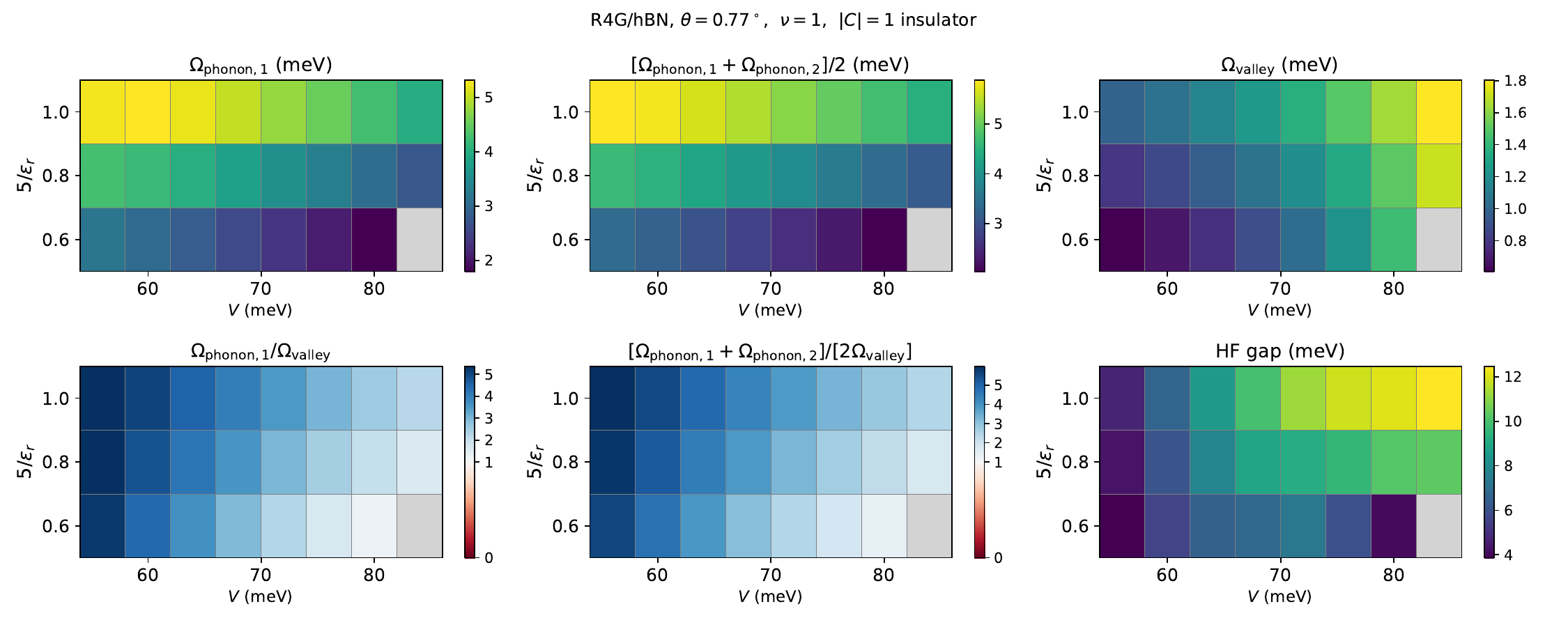}
    \caption{$\mbf{q}=0$ collective modes of the $|C|=1$ insulator in R4G/hBN as a function of $V$ and $\epsilon_r$. $\Omega_\text{phonon,1},\Omega_\text{phonon,2}$ are the energies of the lowest excitations in the intraflavor channel, and are adiabatically connected to the two gapless Goldstone modes in the limit of zero moir\'e potential. $\Omega_\text{valley}$ is the energy of the lowest intervalley excitation. Grey regions denote parameters where there are no gapped HF states. The HF and TDHF calculations are performed with $(0+4)$ screened basis projection, $\xi=1$, $\theta=0.77^\circ$, and system size $12\times 12$. }
    \label{collphase_L4_t0.77_C1}
\end{figure*}

\begin{figure*}
    \centering
    \includegraphics[width=1.0\linewidth]{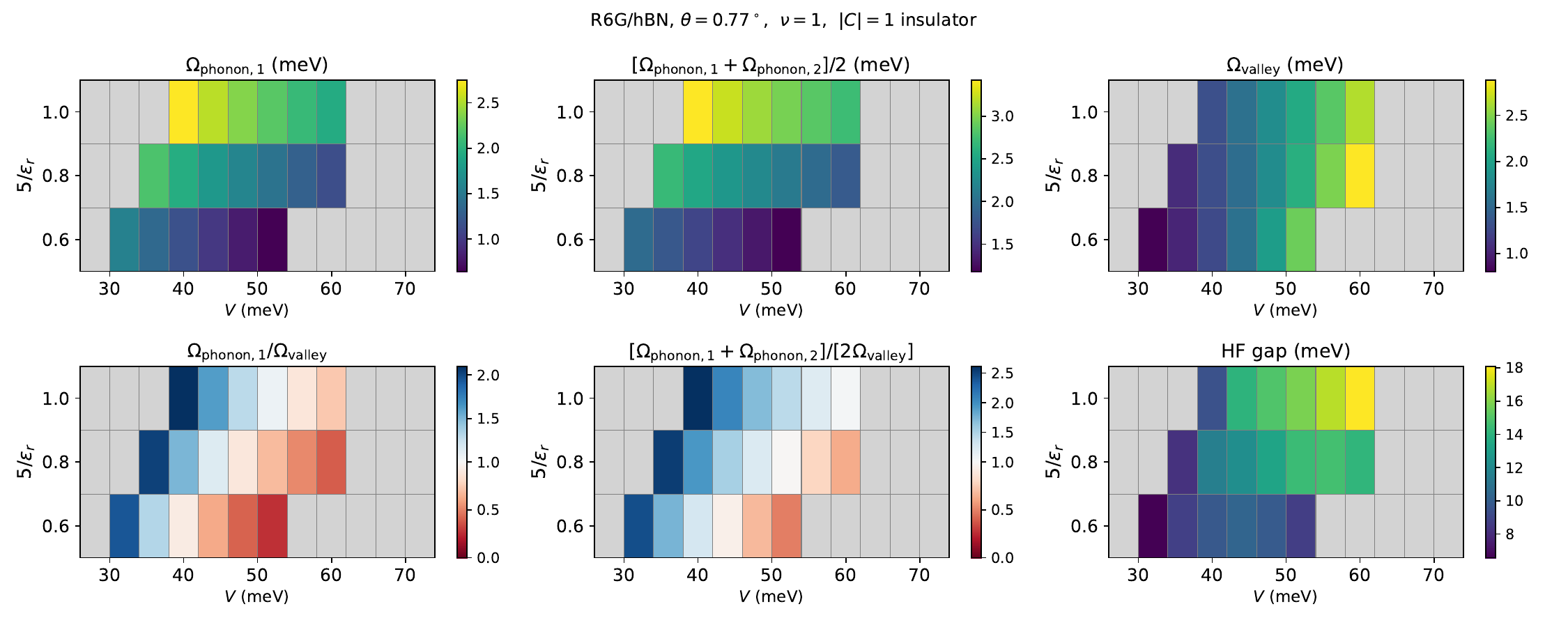}
    \caption{$\mbf{q}=0$ collective modes of the $|C|=1$ insulator in R6G/hBN as a function of $V$ and $\epsilon_r$. $\Omega_\text{phonon,1},\Omega_\text{phonon,2}$ are the energies of the lowest excitations in the intraflavor channel, and are adiabatically connected to the two gapless Goldstone modes in the limit of zero moir\'e potential. $\Omega_\text{valley}$ is the energy of the lowest intervalley excitation. Grey regions denote parameters where there are no gapped HF states. The HF and TDHF calculations are performed with $(0+4)$ screened basis projection, $\xi=1$, $\theta=0.77^\circ$, and system size $12\times 12$. }
    \label{collphase_L6_t0.77_C1}
\end{figure*}

In Figs~\ref{collphase_L4_t0.77_C1} and \ref{collphase_L6_t0.77_C1}, we show results for the $|C|=1$ state in R4G/hBN and R6G/hBN respectively (both for $\xi=1$). The reduced pseudophonon gaps for R6G/hBN is consistent with the intuition that the conduction bands feel the moir\'e potential of the valence bands less strongly due to the increased vertical spacing between them. 

\clearpage

\subsection{Collective mode wavefunctions and dispersion}

\begin{figure*}
    \centering
    \includegraphics[width=1.0\linewidth]{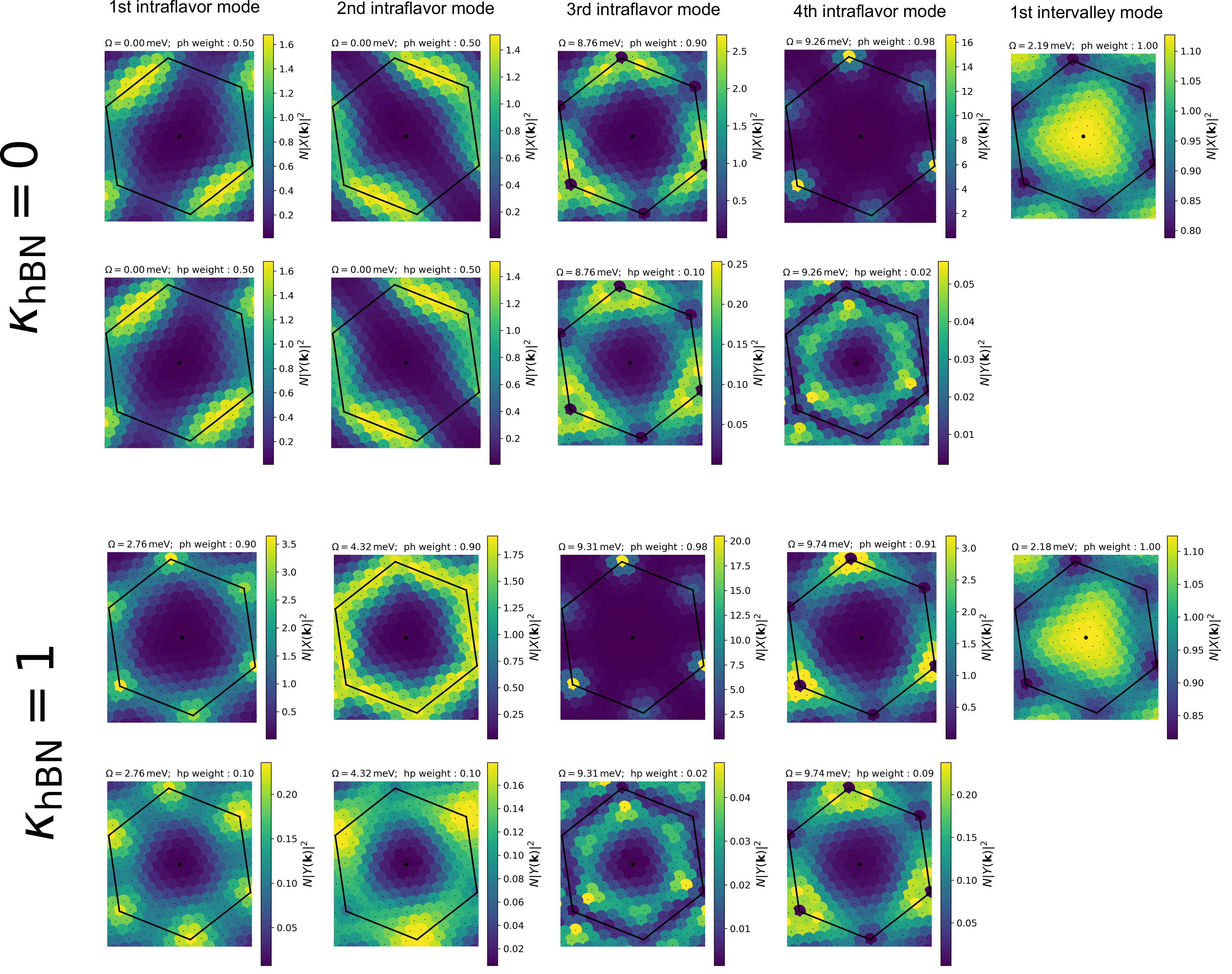}
    \caption{Momentum-resolved particle-hole (ph) and hole-particle (hp) content of the $\mbf{q}=0$ collective modes of the $|C|=1$ insulator in R5G/hBN for hBN-coupling strength $\kappa_\text{hBN}=0$ (top two rows) and $\kappa_\text{hBN}=1$ (bottom two rows). The data here use the same system parameters as the top row of Fig.~\ref{collq0_avg_vs_CN}, i.e.~the HF and TDHF calculations are performed with $(0+4)$ screened basis projection, $\xi=1$, $\theta=0.77^\circ$, $V=64\,\text{meV}$, $\epsilon_r=5$, and system size $12\times 12$. We show results for the lowest four intraflavor modes and the lowest interavalley mode. $N|X(\mbf{k})|^2$ indicates the total probability amplitude of $\mbf{q}=0$ ph operators at momentum $\mbf{k}$ in the collective mode operator Eq.~\ref{appeq:QRPA}, and has been normalized so that uniform probability across the mBZ corresponds to a value of 1. Similarly, $N|Y(\mbf{k})|^2$ indicates the total probability of $\mbf{q}=0$ hp operators at momentum $\mbf{k}$. Note that the intervalley modes do not have hp content. The mBZ is indicated by the black hexagon, whose center corresponds to $\tilde{\Gamma}_M$.}
    \label{coll_wfn}
\end{figure*}

In Fig.~\ref{coll_wfn}, we show $\mbf{q}=0$ collective mode wavefunctions for the $|C|=1$ insulator in $\xi=1$ R5G/hBN for hBN coupling factor $\kappa_\text{hBN}=0,1$. The parameters are identical to the top row of Fig.~\ref{collq0_avg_vs_CN}. In particular, we focus on the lowest four intraflavor modes [two pseudophonons and two excitons] and the lowest intervalley mode. We plot the probability density of $X$ (ph) and $Y$ (hp) components of the mode operator (Eq.~\ref{appeq:QRPA}) across the mBZ. The (pseudo)phonons are predominantly composed of ph and hp operators at the mBZ boundary. This is consistent with the non-interacting band structure (Fig.~\ref{fig:SP_bands}) where the lowest conduction bands have near-degeneracies near the $\tilde{K}_M,\tilde{K}'_M,\tilde{M}_M,\tilde{M}'_M$, while the lowest conduction band has a large gap to higher bands at $\tilde{\Gamma}_M$. (Note that the pseudophonon wavefunctions for $\kappa_{\text{hBN}}=0$ appear to break $C_3$-symmetry. This is an artifact of the fact that the pseudophonons are degenerate at $\kappa_{\text{hBN}}=0$, so the diagonalization routine yields arbitrary superpositions of the degenerate eigenfunctions.) The two lowest intraflavor excitons are either localized at $\tilde{K}_M$, or have a node there. Finally the lowest intervalley mode wavefunction, which has only ph components for our calculation (see App.~\ref{secapp:TDHF}), is nearly uniformly spread out over the mBZ. 

\begin{figure*}
    \centering
    \includegraphics[width=1.0\linewidth]{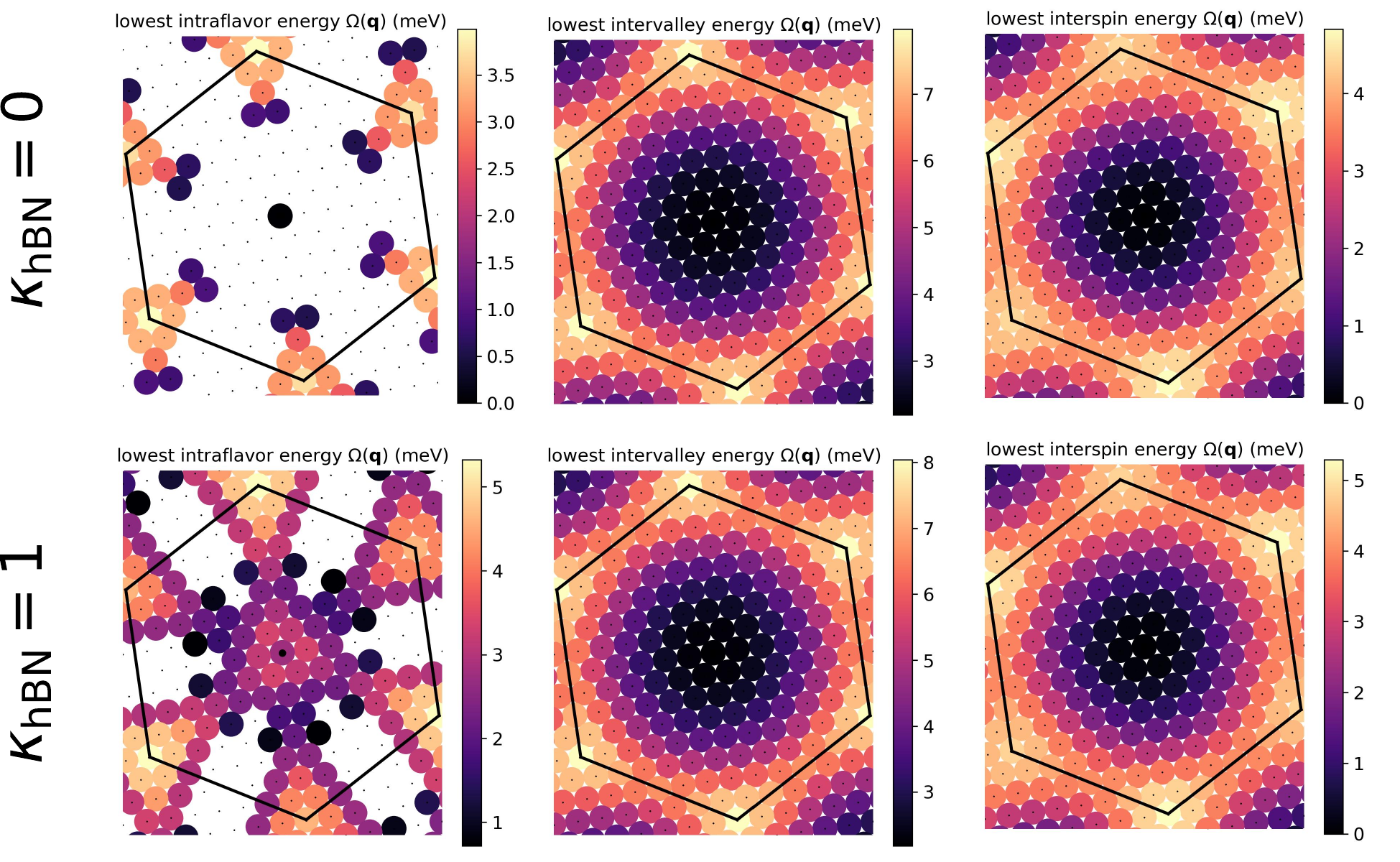}
    \caption{Collective mode dispersions $\Omega(\mbf{q})$ of the $|C|=1$ insulator in R5G/hBN for hBN-coupling strength $\kappa_\text{hBN}=0$ (top row) and $\kappa_\text{hBN}=1$ (bottom row). The data here use the same system parameters as the top row of Fig.~\ref{collq0_avg_vs_CN}, i.e.~the HF and TDHF calculations are performed with $(0+4)$ screened basis projection, $\xi=1$, $\theta=0.77^\circ$, $V=64\,\text{meV}$, $\epsilon_r=5$, and system size $12\times 12$. We show results for the lowest energy intraflavor, intervalley and interspin mode. White regions indicate where the collective mode energies are negative or complex, signalling an instability. The mBZ is indicated by the black hexagon, whose center corresponds to $\tilde{\Gamma}_M$.}
    \label{coll_dispersion}
\end{figure*}

In Fig.~\ref{coll_dispersion}, we plot the dispersion of the lowest intraflavor, intervalley, and interspin collective modes for the $|C|=1$ insulator in $\xi=1$ R5G/hBN for hBN coupling factor $\kappa_\text{hBN}=0,1$. The parameters are again identical to the top row of Fig.~\ref{collq0_avg_vs_CN}. The lowest interspin mode is a magnon branch with quadratic dispersion at $\mbf{q}=0$. The intervalley mode is gapped, consistent with lack of $SU(2)$ symmetry in valley space, and has a relativelly narrow bandwidth $\simeq 5\,\text{meV}$. Interestingly, there are regions in momentum space where the lowest intraflavor mode has negative or complex eigenvalues. This signals a local instability within the variational Slater determinant manifold. From the $\kappa_\text{hBN}=1$ result, it appears that the instability corresponds to a wavevector around the $\tilde{M}_M$ and $\tilde{M}'_M$. We have performed HF calculations allowing for doubling of the moir\'e unit cell along one axis, and indeed find spin-valley polarized HF states with lower energy than the best translation-invariant solution. This occurs for both the average and CN interaction schemes. We leave a detailed analysis of such moir\'e translation-symmetry broken states to future work.

\end{document}